\documentclass[12pt,twoside]{book}
\pdfoutput=1
\usepackage{latexsym}
\usepackage{amssymb}
\usepackage{amsmath}
\usepackage{amsthm}
\usepackage{psfrag}
\usepackage{subfigure}
\usepackage[pdftex]{graphicx}
\usepackage{longtable}
\usepackage{natbib}

\usepackage{sectsty}

\allsectionsfont{\sffamily\bfseries}

\usepackage{fancyhdr}
\usepackage[Conny]{fncychap}

\begin{document}

\pagestyle{empty}

\begin{center}
\large{Faculty of Physics, Astronomy and Applied Computer Science\\
Department of Medical Physics\\
Jagiellonian University}
\end{center}

\vspace*{\stretch{3}}

\begin{center}
\large{\textbf{Doctoral dissertation}}
\end{center}

\vspace*{\stretch{2}}

\begin{center}
\Huge{\textbf{Track structure modelling for ion radiotherapy}}
\end{center}

\vspace*{\stretch{4}}

\begin{center}
\Large{Marta Korcyl}
\end{center}

\vspace*{\stretch{20}}

\begin{flushright}
\large{Supervisor \\ Prof. dr hab. Michael P. R. Walig\'orski}
\end{flushright}

\vspace*{\stretch{16}}

        \begin{center}
        \includegraphics[width = 0.17\textwidth]{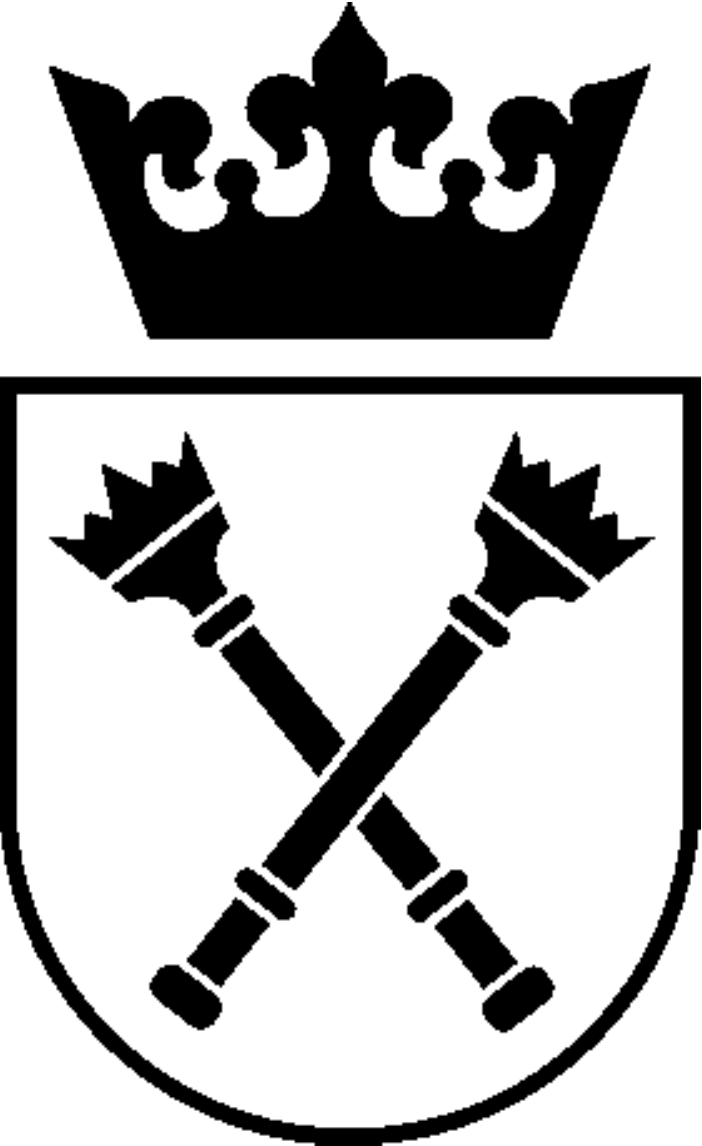}
        \end{center}

\vspace*{\stretch{1}}


\vspace*{\stretch{2}}

\begin{center}
\large{Krak\'ow, 2012\\}
\end{center}

\newpage
\vspace*{\stretch{20}}
\newpage

%
%

\newpage
\thispagestyle{empty}
\vspace*{\stretch{10}}
\begin{flushright}
\emph{Piotrowi}
\end{flushright}

\newpage
\vspace*{\stretch{20}}
\newpage

\newpage
\thispagestyle{empty}
\section*{Podzi\c{e}kowania}

\emph{} \\ \\
Chcia\l abym wyrazi\'c swoj\c{a} wdzi\c{e}czno\'s\'c wszystkim osobom, dzi\c{e}ki kt\'orym zako\'nczenie mojej pracy doktorskiej sta\l o si\c{e} mo\.{z}liwe.\\
Dzi\c{e}kuj\c{e} mojemu promotorowi prof. dr. hab. Michaelowi P. R. Walig\'orskiemu za wprowadzenie mnie we wszystkie tajniki Teorii Struktury \'Sladu. Za wskaz\'owki merytoryczne oraz liczne dyskusje. \\
Dzi\c{e}kuj\c{e} prof. Janowi Stankowi za podj\c{e}cie opieki nad moimi studiami doktoranckimi w Instytucie Fizyki Uniwersytetu Jagiello\'nskiego. \\
Dzi\c{e}kuj\c{e} \- prof. dr. hab. Paw\l owi Olko za stworzenie w Instytucie Fizyki J\c{a}drowej PAN warunk\'ow, umo\.{z}liwiaj\c{a}cych mi zako\'nczenie bada\'n w ko\'ncowej fazie doktoratu.\\ Dzi\c{e}kuj\c{e} mgr. Leszkowi Grzance za pomoc w usystematyzowaniu wiedzy dotycz\c{a}cej modelu oraz cenne wskaz\'owki dotycz\c{a}ce oblicze\'n numerycznych.\\

\emph{} \\
Szczeg\'olnie chcia\l abym podzi\c{e}kowa\'c mojemu m\c{e}\.{z}owi Piotrowi, za cierpliwo\'s\'c oraz ci\c{a}g\l \c{a} wiar\c{e} w moje mo\.{z}liwo\'sci, bez kt\'orej nie odwa\.{z}y\l abym si\c{e} doko\'nczy\'c mojej pracy doktorskiej.\\
Dzi\c{e}kuj\c{e} ca\l ej mojej Rodzinie i wszystkim moim Przyjacio\l om za obecno\'s\'c w chwilach rado\'sci, jak i wsparcie w chwilach zw\c{a}tpienia.

\vspace*{\stretch{2}}

\newpage
\vspace*{\stretch{20}}
\newpage


\setcounter{tocdepth}{5}
\thispagestyle{empty}
\setcounter{page}{1}
\tableofcontents

\newpage

%


\pagestyle{fancy}
\fancyhf{} \fancyhead[RE,LO]{\leftmark} \fancyfoot[CE,CO]{\thepage}

\chapter*{Motivation}
\addcontentsline{toc}{chapter}{\protect\numberline{}Motivation}
\label{ch. abstract}

Oncological radiotherapy is the process of killing cancer cells within the tumour volume by a prescribed dose of ionising radiation, while sparing to the extent possible the neighbouring healthy tissues. Several types of sources of ionising radiation have so far been used in radiotherapy, mainly of megavolt photons ($X$- or $\gamma$-rays) or electrons, produced either as external beams by medical accelerators, by sealed radioisotope sources, or by radioactively labelled compounds designed to be preferentially absorbed by malignant tissues. Typically, a modern external photon beam radiotherapy session consists of a sequence of 30 daily fractions, each delivering a dose of $2\;Gy$ to the tumour volume, to a total dose of about $60\;Gy$. Several modern techniques of delivering external photon beams from medical accelerators have been introduced, such as IMRT (intensity modulated RT), IGRT (image-guided RT), SRS (stereotactic radiosurgery), arc beam IMRT (Tomotherapy) or by robotic control of the movement of the accelerating assembly. Modern radiotherapy is image-based: three-dimensional reconstruction of the patient's geometry by computed tomography (CT) or better delineation of tumour volumes by functional imaging, such as Positron Emission Tomography (PET) or Magnetic Resonance Imaging (MRI) enable the design of the distribution of dose to the tumour volume to be exactly pre-planned, using dedicated therapy planning systems (TPS). The use of inverse planning techniques, whereby the TPS automatically optimises the dose delivery plan within appropriately chosen constraints is becoming a standard choice for the medical physicist involved in radiotherapy planning. 

Application of external beams of energetic ions (typically, protons or carbon, of energies around a few hundred $MeV/n$) has opened new avenues in oncological radiotherapy. The stimulus initially came from clinical experience gained by fast neutron radiotherapy in the 50's of last century. While now discontinued, mainly due to the difficulty of accurately shaping and focusing beams of fast neutrons, clinical application of these 'high-LET' beams demonstrated the potential advantages of such beams: enhanced biological effectiveness, improvement in treating poorly oxygenated tumour tissues, and the possibility of reducing the required number of fractions. 

Developments in accelerator technology made beams of ions of charges up to uranium available also to users other than nuclear physicists. Early trials of radiotherapy applications of beams of ions of inert gases (mainly neon and argon) at the Bevalac accelerator in Berkeley in the '70s and '80s of last century were unsuccessful, indicating the need to better understand the enhanced biological effectiveness of such beams and its dependence on physical and biological factors. Independently, development of the technique of culturing cells in laboratory conditions (\emph{in vitro}) made it possible to systematically study the biological effects of ion beams. Development of biophysical models was then necessary, to analyse experimental data from radiobiology and to better understand and possibly predict the biological effects of ionising radiation, including energetic ions, or 'high-LET' radiation. Among the models of radiation action developed at the time, Track Structure Theory, a biophysical model developed in the '80s of last century by Prof. Robert Katz, was extremely successful in analysing radiation effects in physical and biological systems, especially in cell cultures \emph{in vitro}. 

Despite the impressive advances of molecular biology, detailed knowledge of mechanisms of radiation damage at the molecular level of living organisms is still lacking, therefore phenomenological parametric biophysical models, such as the Track Structure model or the linear-quadratic approach applied to radiobiological studies of cells in culture, may still offer insight to radiotherapy planning, especially for ion beams, to analyse and predict their enhanced radiobiological effectiveness (RBE).

The main advantages of applying ion beams in radiotherapy stem from physical and radiobiological considerations. In their physical aspects, the well-specified energy-dependent range of ions and the Bragg peak effect, whereby most of the dose is delivered at the end of their range, enable better dose delivery to tumour volumes located at greater depths within the patient's body, better coverage of the tumour volume, and better sparing of the patient's skin than in the case of external photon beams. As for radiobiology, the enhancement of RBE, especially for ions heavier than protons or helium, and overcoming the enhanced radioresistance of poorly oxygenated cancerous cells (expressed by the so-called Oxygen Enhancement Ratio) raises hopes for achieving a better clinical effect by killing cancer cells more radically. The ability to correctly predict the complex behaviour of RBE and OER in ion radiotherapy is therefore of major importance in advancing this radiotherapy modality.

The major disadvantages of ion radiotherapy are presently the high cost of accelerating and delivering ion beams of the required energy by dedicated accelerators: cyclotrons or synchrotrons and the uncertainties in modelling and predicting the therapeutic effects of such beams. 

Of the presently operating ion radiotherapy facilities, most use proton beams of energy up to $250\;MeV$ to exploit the physical advantages of these beams. As based on the present clinical experience, the RBE of protons of such energy does not exceed $1.1$, so radiobiology aspects of proton beams are rather unimportant. A proton radiotherapy facility is currently operating in Poland at the Institute of Nuclear Physics PAN in Krak\'ow, based on the $60\;MeV$ AIC cyclotron, and a new facility which will use a $250\;MeV$ scanning proton beam is under construction and planned to begin clinical operation around 2015.

Presently, only three carbon ion radiotherapy facilities in the world operate clinically: one in Germany (at the HIT at Heidelberg) and two in Japan (the HIMAC at Chiba and the Hyogo facility), while several carbon beam therapy facilities are under construction.\\

In its broadest terms, this work is part of the supporting research background in the development of the ambitious proton radiotherapy project currently under way at the Institute of Nuclear Physics PAN in Krak\'ow. Another broad motivation was the desire to become directly involved in research on a topical and challenging subject of possibly developing a therapy planning system for carbon beam radiotherapy, based in its radiobiological part on the Track Structure model developed by Katz over 50 years ago.

Thus, the general aim of this work was, firstly, to recapitulate the Track Structure model and to propose an updated and complete formulation of this model by incorporating advances made by several authors who had contributed to its development in the past.

Secondly, the updated and amended (if necessary) formulation of the model should be presented in a form applicable for use in computer codes which would constitute the 'radiobiological engine' of the future therapy planning system for carbon radiotherapy, which the Krak\'ow ion radiotherapy research group wishes to develop.

Thirdly, currently available radiobiology data should be analysed in terms of Track Structure Theory to supply exemplary parameters for cell lines (preferably, exposed in normal and anoxic conditions) to be used as possible input for carbon ion radiotherapy planning studies.

Lastly, the general features of Track Structure Theory should be compared against biophysical models currently used in carbon radiotherapy planning (such as the LEM used by the German groups or the approach used by the Japanese groups), to indicate the possible advantages to be gained by applying the Track Structure approach in the future therapy planning system for carbon radiotherapy.\\ \\

There are many features of Katz's Track Structure Theory (TST) which make it a promising radiobiological model for purposes of carbon ion radiotherapy planning. The major one being its unquestionable success in quantitatively analysing RBE dependences in several physical and biological systems, and especially in many different mammalian cell lines cultured \emph{in vitro}, exposed to a variety of ion beams. Another distinct feature of this model is the requirement that energy-fluence spectra of all the primary and secondary charged particles along the beam range can be provided for model calculations, rather than depth-dose distributions. Using these energy-fluence spectra and four model parameters which characterise the radiobiological properties of a cell line, the Katz model is able to quantitatively predict the depth-survival dependences directly, without formally involving the product of local 'physical' dose and local value of RBE, thus obviating the need to evaluate the complex dependence of RBE on the properties of the ion beam and of the irradiated tissue. 

Scaling plays an important role in Track Structure Theory, therefore an analysis of the conditions under which such scaling may be achieved was yet another objective of this work. It is by exploiting the scaling properties of Track Structure Theory that robust and computer-efficient coding may be designed to make massive calculations possible, as required, e.g., in inverse planning techniques to be developed for carbon beam radiotherapy.

An interesting possibility offered by applying TST to treatment planning is to develop a 'kill' rather than the current 'dose' approach to optimising this planning. Namely, in the present 'classical' approach, dose distributions around the target volume are considered. However, due to RBE and OER considerations in ion radiotherapy, it may be more appropriate to optimise distributions of probability of target cell killing, which can be obtained directly from TST calculations. Comparison of 'iso-kill' distributions between 'classical' and ion radiotherapy treatment plans would permit direct transfer of the experience gained from 'classical' radiotherapy to ion radiotherapy, and perhaps lead to new optimisation tools, e.g., replacing dose-volume histograms by 'kill-volume' histograms. Development of a therapy planning systems based on the 'fluence approach' would allow cross-checking of some controversial issues, such as reporting ion therapy procedures, or the application of appropriate RBE values. Thus, introduction of track structure theory-based biophysical modelling may lead to the emergence of new concepts in ion therapy planning and its optimization.

\chapter[Introduction]{Introduction}
\label{ch. introduction}

\section{Interaction of photons with matter}
\label{ch. photoninteraction}

\begin{figure}[!ht]
\begin{center}
\includegraphics[width=0.75\textwidth]{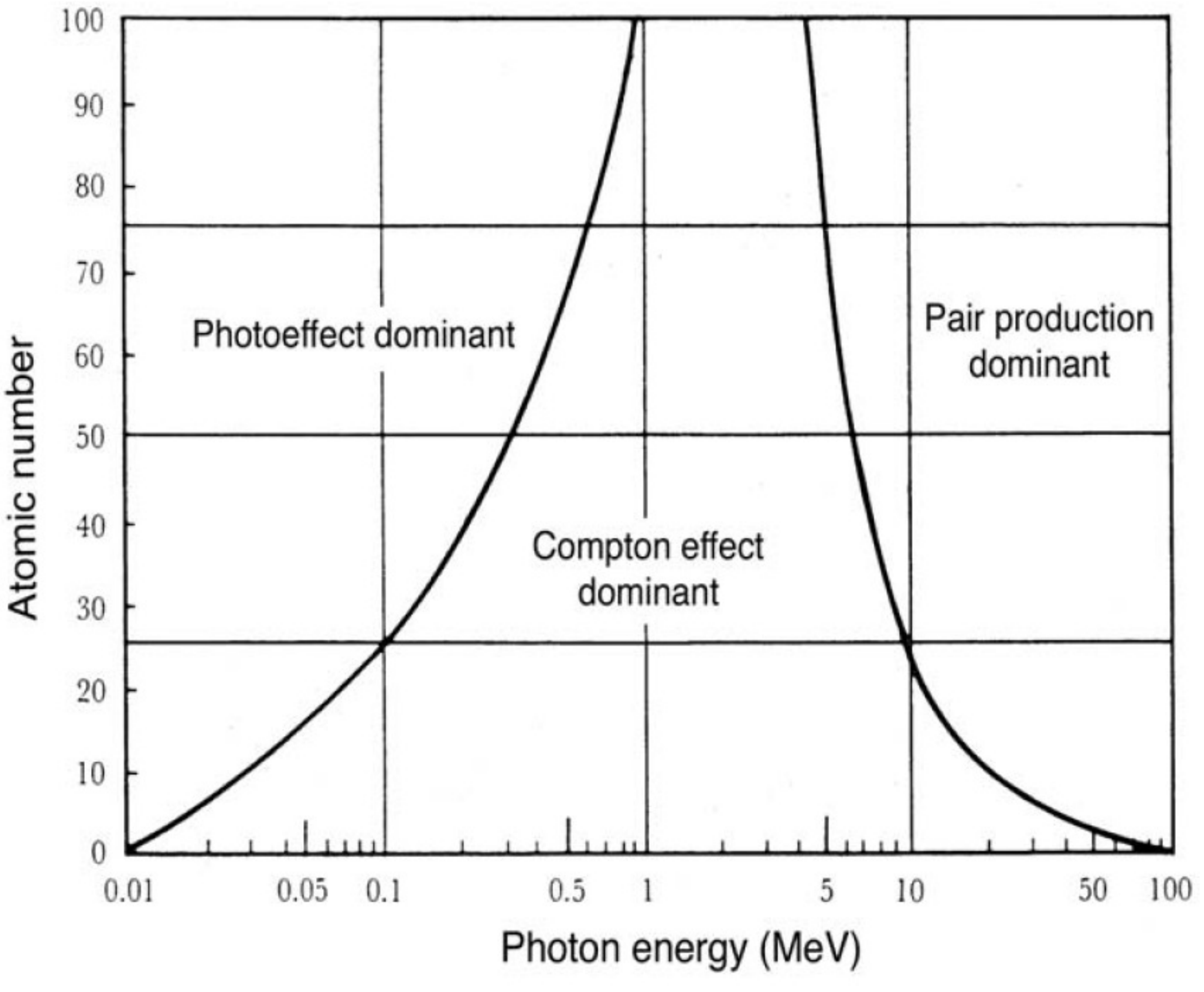}
\end{center}
\caption{Regions of relative predominance of the three main processes of photon interaction with matter. The left curve represents the region where the cross-sections for photoelectric effect and for Compton effect are equal, the right curve represents the region where Compton effect is equally probable to pair production. Figure reprinted from \cite{podgorsak2005}.}
\label{fig.photon}
\end{figure}

Photons can interact with matter through many different processes. The character of these processes depends on the energy of the photons concerned and on the chemical composition of the absorber. There are three major mechanisms involved in the energy loss by photons in the $MeV$ range, used in radiotherapy: photoelectric effect, Compton scattering and pair production. All these processes lead to partial or complete energy transfer of the photon to the electrons of the atoms of the absorber, removal of orbital electrons from the atoms (ionization) or to changing the internal state of the bound electrons from the ground state to a higher energy state (excitation). Electrons removed from atoms can obtain sufficient kinetic energy to cause secondary ionizations of other atoms of the absorber. Such electrons are called $\delta$-electrons or secondary electrons. As a beam of photons passes through the absorber, many different interactions may occur. Interaction mechanisms have different energy thresholds and regions of high cross-sections for different materials. Depending on the energy of photons and composition of the absorber different mechanisms dominate. An illustrative diagram presenting the regions of relative predominance of the three above-mentioned mechanisms as a function of atomic number of the absorber and photon energy, is shown in Fig. \ref{fig.photon}. Curves on the left and right side of this figure define photon energies and atomic numbers of the absorbers for which the probability of Compton scattering is equal to the probability of photoelectric effect and pair production. Interaction of photons used in radiotherapy with soft tissues, the atomic number of which is close to the effective atomic number of water $(Z = 7.4)$, is dominated by Compton scattering. Photoelectric effect dominates for photons of lower energy, whereas pair production dominates for photons of higher energy, both interacting with absorbers of higher atomic number.\\

\section{Interaction of ions with matter}
\label{ch. ioninteraction}

Depending on its velocity, the charged particle (projectile) may experience interaction with the target particles of the absorber by means of the following processes:
\begin{itemize}
\item excitation or ionization of target particles 
\item transfer of energy to target nuclei
\item changes in the internal state of the projectile
\item emission of radiation
\end{itemize}
As a result of these interactions the energy of the particles is reduced as they pass trough the absorber medium. The rate of energy lost by the particle depends on the energy, charge and atomic mass of the incident ion and on the composition of the absorber. In the literature the average loss of kinetic energy $E$ per path length $x$ is referred to as the Stopping Power, $-dE/dx$, of the material. The minus sign defines the stopping force as a positive quantity. 
The total stopping power is the sum of the electronic and nuclear stopping powers, and of the radiative stopping power, according to the expression:
\begin{displaymath}
-\frac{dE}{dx}\quad = \quad \left(\frac{dE}{dx}\right)_{el}\quad + \quad \left(\frac{dE}{dx}\right)_{nucl}+ \quad \left(\frac{dE}{dx}\right)_{rad},
\end{displaymath}
where electronic stopping power is the average rate of energy loss per unit ion path length, due to Coulomb collisions that result in the ionization and excitation of target atoms, and nuclear stopping power is the average rate of ion energy loss per unit ion path length due to the transfer of its energy to recoiling atoms in elastic collisions. This type of interaction occurs only for heavy charged particles. The radiative stopping power is the average rate of ion energy loss per ion's unit path length due to collisions with atoms and atomic electrons in which bremsstrahlung quanta are emitted. This type of interaction occurs only at extremely high ion velocities (which are outside of the range of ion radiotherapy) and light charged particles, such as electrons. Thus, ions interact with the matter mainly through the electromagnetic forces where the positive electric charge of the particle and the negative charge of the electrons of the atoms of absorber are mutually attractive. Along its path, the ion will then interact with several electrons of the absorber atoms, causing their excitation and, more frequently, their ionization.

\begin{figure}[!ht]
\begin{center}
\includegraphics[width=0.9\textwidth]{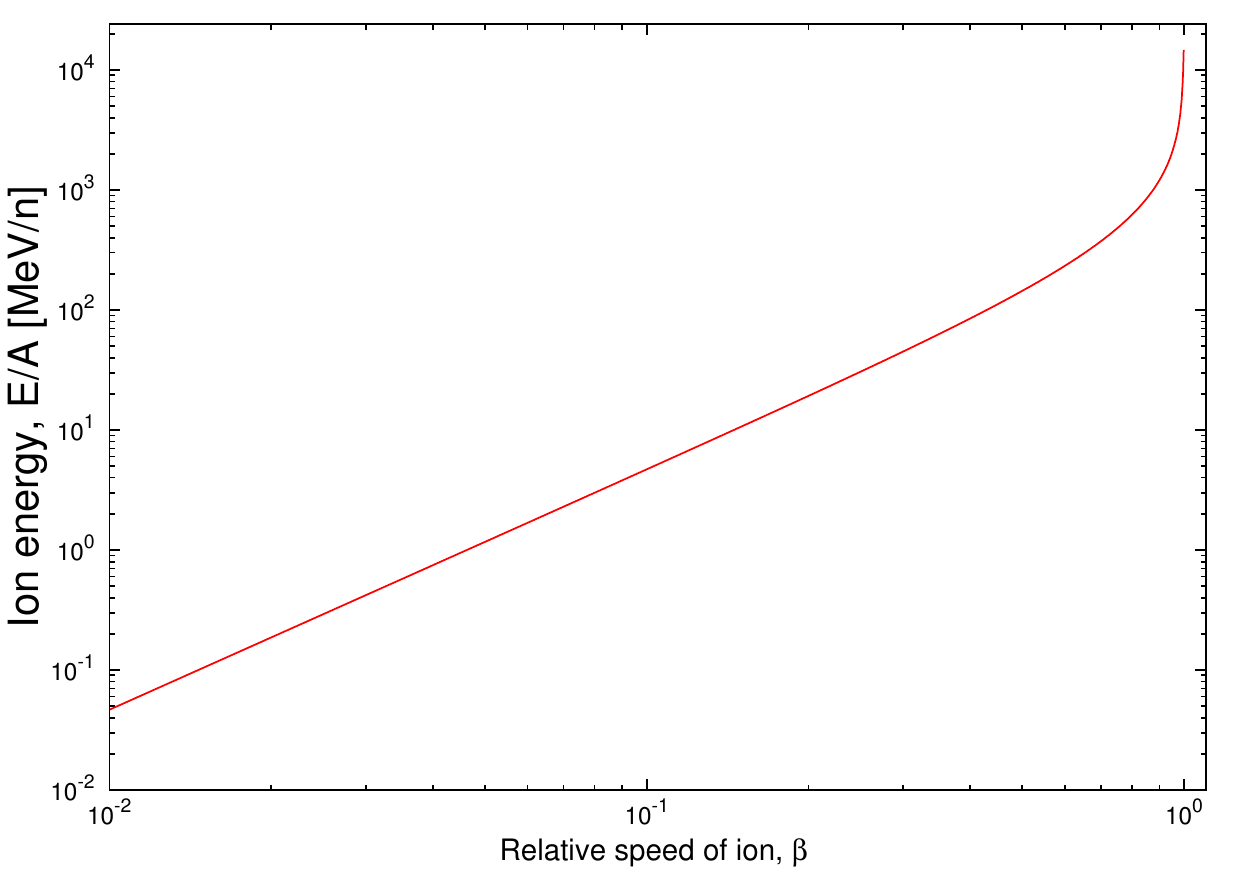}
\end{center}
\caption{The kinetic energy of an ion as a function of its relative speed, $\beta$, as given by eq.(\ref{eq.ionenergy}).}
\label{fig.ionenergy}
\end{figure}

\begin{figure}[!ht]
\begin{center}
\includegraphics[width=0.9\textwidth]{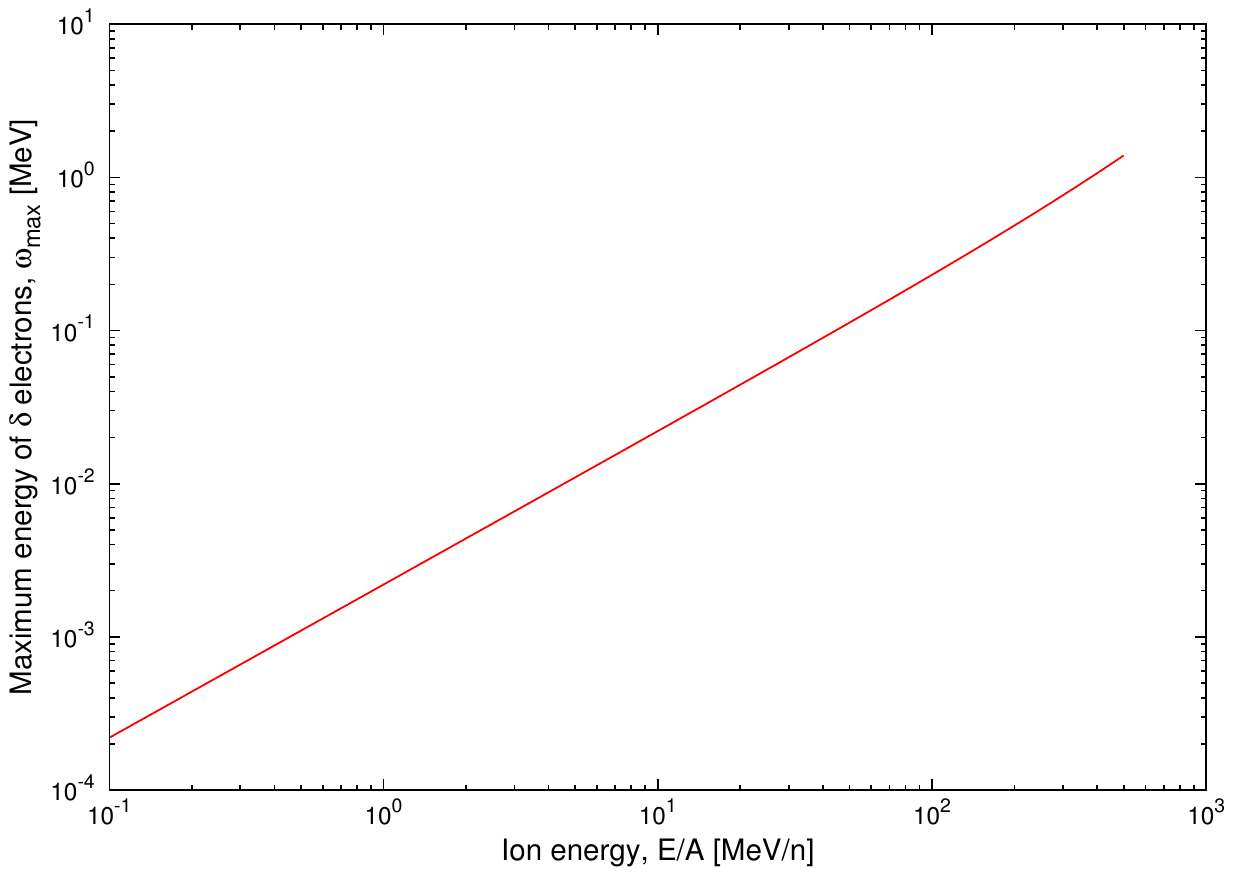}
\end{center}
\caption{Maximum energy of a $\delta$-electron transferred by an ion of energy $E/A$, as given by eq.(\ref{eq.maximum.energy}) and eq.(\ref{eq.ionenergy}).}
\label{fig.enel}
\end{figure}

In each Coulomb collision with an electron, the ion loses some part of its energy until it stops completely. Charged particles have a well-determined range in matter, which depends upon the type of particle, its initial energy and the atomic composition of the material it traverses. The maximum energy that an ion can transfer to a free electron (of rest energy $m_e c^2 = 0.511\;MeV$) is given by
\begin{equation}
\omega_{max} = 2m_e c^2 \frac{\beta^2}{1-\beta^2},
\label{eq.maximum.energy}
\end{equation}
where $\beta=v/c$ is the ion's relative velocity, related to the kinetic energy per nucleon or per atomic mass unit $E/A$ of the incident ion, through the relation (\cite{icru2005}):
\begin{equation}
\frac{E}{A} = uc^2(\gamma-1) = 931.6 \left( \frac{1}{\sqrt{1-\beta^2}}-1\right)\quad [MeV/n],
\label{eq.ionenergy}
\end{equation}
where $\gamma$ is the Lorentz factor, and $u$ is the atomic mass unit $(1.6605\times10^{-27}kg)$. Dependences given by eq.(\ref{eq.ionenergy}) and eq.(\ref{eq.maximum.energy}) are shown graphically in Fig. \ref{fig.ionenergy} and Fig. \ref{fig.enel}, respectively. Similarly to photon interactions, during some collisions, electrons interacting with the passing ion obtain kinetic energy sufficiently high to produce $\delta$-electrons which can transfer their received energy far away from the ion's path.
An analytical description of the ion's stopping power resulting from Coulomb collisions was first given by Hans Bethe (\cite{bethe1932}) and more recently, by \cite{sigmund2006}: 
\begin{equation}
-\frac{dE}{dx}=\frac{4\pi Z_{abs} N}{m_ec^2} \frac{Z^2}{\beta^2}\left(\frac{e^2}{4\pi \epsilon _0 }\right)^2 \left[\ln\frac{2m_ec^2\beta^{2}}{I(1-\beta^2)}-\beta^{2} \right],
\label{eq.let}
\end{equation}
where 
\begin{flushleft}
\begin{verse}
$Z_{abs}$ - atomic number of the absorber, \\
$N$ - electron density of the absorber, \\
$Z$ - atomic number of the incident particle, \\
$m_e$ - rest mass of the electron, \\
$c$ - speed of light in vacuum, \\
$e$ - charge of the electron, \\
$\epsilon _0$ - vacuum permittivity, \\
$I$ - mean excitation potential of the target, \\
$\beta$ - relative speed of the incident particle $\beta = v/c$.
\end{verse}
\end{flushleft}
Another concept related to the stopping power is the linear energy transfer (LET) of the ion, which is equivalent to the restricted collisional (electronic) stopping power. Additionally, the definition of LET may include only local collisions between the ion and electrons. Then the considered energy transfers should be less than a specified cut-off energy, $\Delta$ (usually expressed in $eV$): 
\begin{displaymath}
\textrm{LET}_{\Delta} = - \left( \frac{dE_{\Delta}}{dx} \right)_{el}.
\end{displaymath}
By including all possible energy transfers, one obtains the unrestricted LET$_{\infty}$ which is equivalent to the total electronic stopping power:
\begin{displaymath}
\textrm{LET}_{\infty} = - \left( \frac{dE}{dx} \right)_{el} .
\end{displaymath}
Linear energy transfer is an average quantity. In the case of a single particle of a given charge and velocity, LET denotes the amount of energy which a large number of such particles would on average transfer per unit path length (\cite{kempe2007}). Depending on the value of LET which characterizes a specific radiation, one may distinguish between high-LET and low-LET radiations. For example, $\alpha$-particles and charged particles heavier than He are called high-LET radiation, because they cause dense ionization along their tracks. In contrast, $X$- and $\gamma$-rays are recognized as low-LET radiations as they produce sparse and randomly distributed isolated ionization events.

The average energy imparted to the medium by any radiation per unit mass of this medium is called the dose, $D$. Given the LET and number of charged particles per $cm^2$ (fluence), $F$, one can calculate the dose delivered by a beam of ions (\cite{pathak2007}):
\begin{equation}
D = 1.602^{-10} \cdot F \cdot \textrm{LET} \cdot \frac{1}{\rho} \quad [Gy],
\label{eq.iondose}
\end{equation}
where LET is expressed in $[MeV/cm]$, $\rho$ is the density of the absorbing material (in the case of biological material, it is usually considered as being water equivalent, $\rho = 1 \; [g/cm^3]$), and the constant $1.602^{-10} \;[J]$ enables conversion of $MeV$ to $Gy$. 

In many publications concerning biological experiments also the 'dose-averaged' LET is reported. It is evaluated at the cell sample position if cells seeded into a Petri dish are irradiated with an ion beam. The averaged Linear Energy Transfer, $\overline{\textrm{LET}}_{dose}$ (\cite{belli2008}):
\begin{equation}
\overline{\textrm{LET}}_{dose} = \frac{\sum_{i} \textrm{LET}_i^2 \cdot F(\textrm{LET}_i)}{\sum_{i} \textrm{LET}_i \cdot F(\textrm{LET}_i)}.
\label{eq.doselet}
\end{equation}
assumes that in each unit of path length, there are particles (projectile as well as secondary particles) of specified LET$_i$ and fluence, $F(\textrm{LET}_i)$ which all contribute to given effect. The stopping power of each particle is thus weighted by its relative contribution to the total absorbed dose. \\

\begin{figure}[!ht]
\begin{center}
\includegraphics[width=1.0\textwidth]{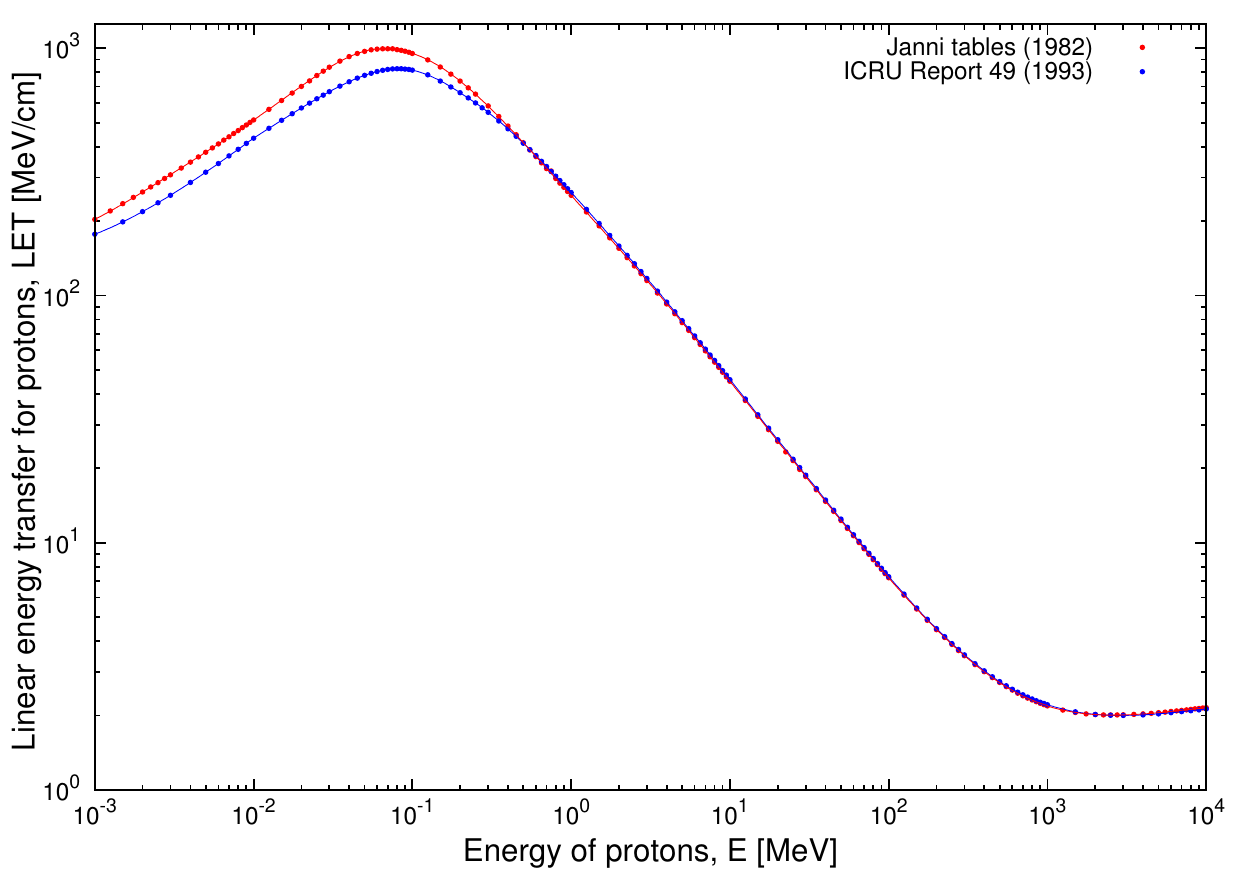}
\end{center}
\caption{The stopping power as a function of proton energy. Data, for liquid water, from tables of \cite{janni1982} are compared with recent data of \cite{icru1993}.}
\label{fig.let1}
\end{figure}

In calculations of the Track Structure model the values of stopping power for heavy ions, LET$_i$, of kinetic energy per nucleon $\frac{E}{A}$ are derived from the proton LET values using the expression given by \cite{barkas1964}:
\begin{equation}
\textrm{LET}_i(Z,\frac{E}{A}) = \textrm{LET}_p(E)\left( \frac{z^*(\beta)}{z^*_{p}(\beta)} \right)^2,
\label{eq.letion}
\end{equation}
here $z^*$ and $z^*_{p}$ are the effective charges (as defined below) of the ion and proton, respectively and LET$_p(E)$ is the stopping power of the proton of kinetic energy $E$. Calculations of LET$_p(E)$ as a function proton energy, originally made by Robert Katz and his co-workers, were based on the tables of stopping power published by \cite{janni1982}. In this work we based our calculations of LET$_p(E)$ on the more recent tables of proton stopping power published in 1993 by the International Commission on Radiation Units and Measurements (\cite{icru1993}). Comparison between these two data sources (points), together with respective parametrization (lines) of the data for protons is presented in Fig. \ref{fig.let1}. For parametrization details see Appendix \ref{ch. appendixa}. The algorithm for calculating ion stopping power values from proton stopping power values, as given by Janni, was described by \cite{waligorski1988}. As shown in Fig. \ref{fig.let1}, the main discrepancy between the \cite{janni1982} and \cite{icru1993} tables is observed over proton energies below $1\;MeV$.

\begin{figure}[!ht]
\begin{center}
\includegraphics[width=0.94\textwidth]{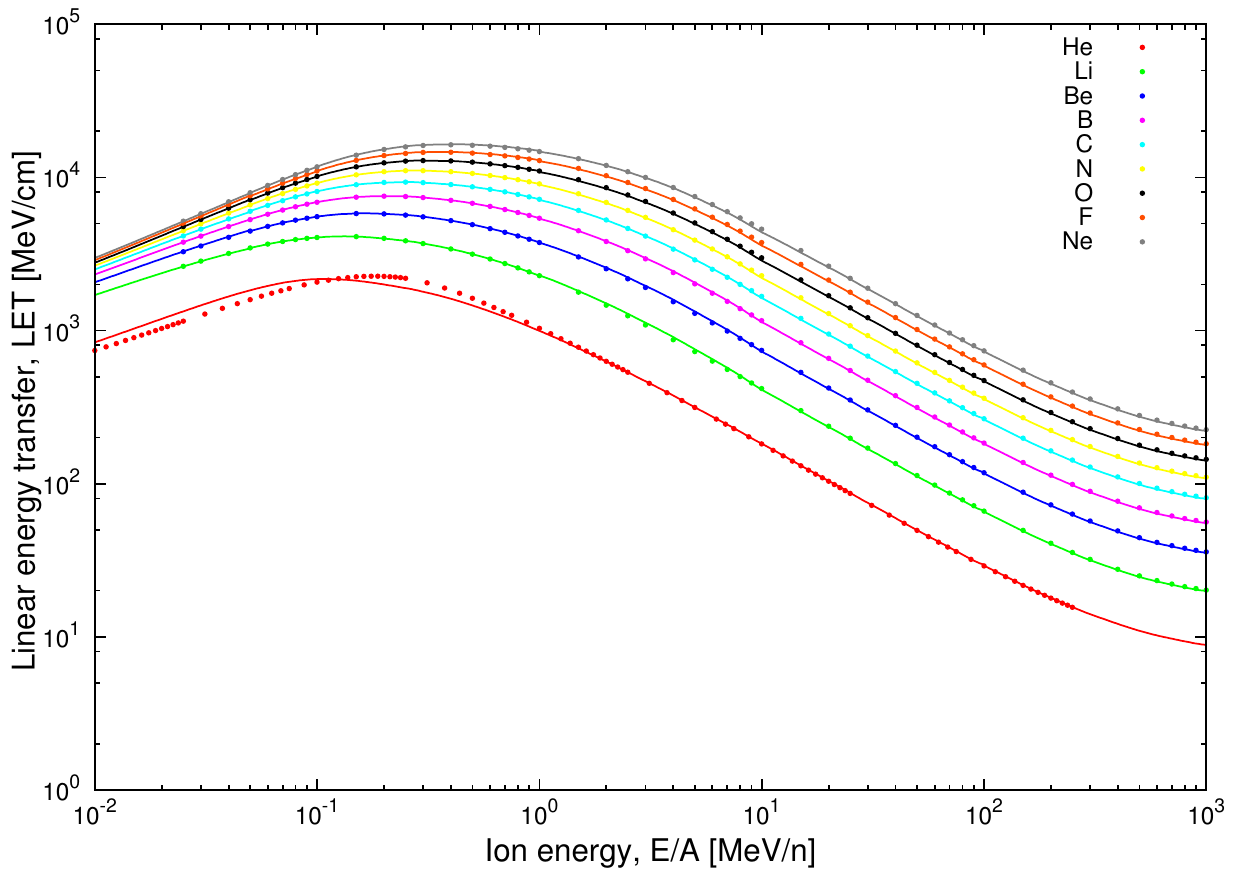}
\includegraphics[width=0.94\textwidth]{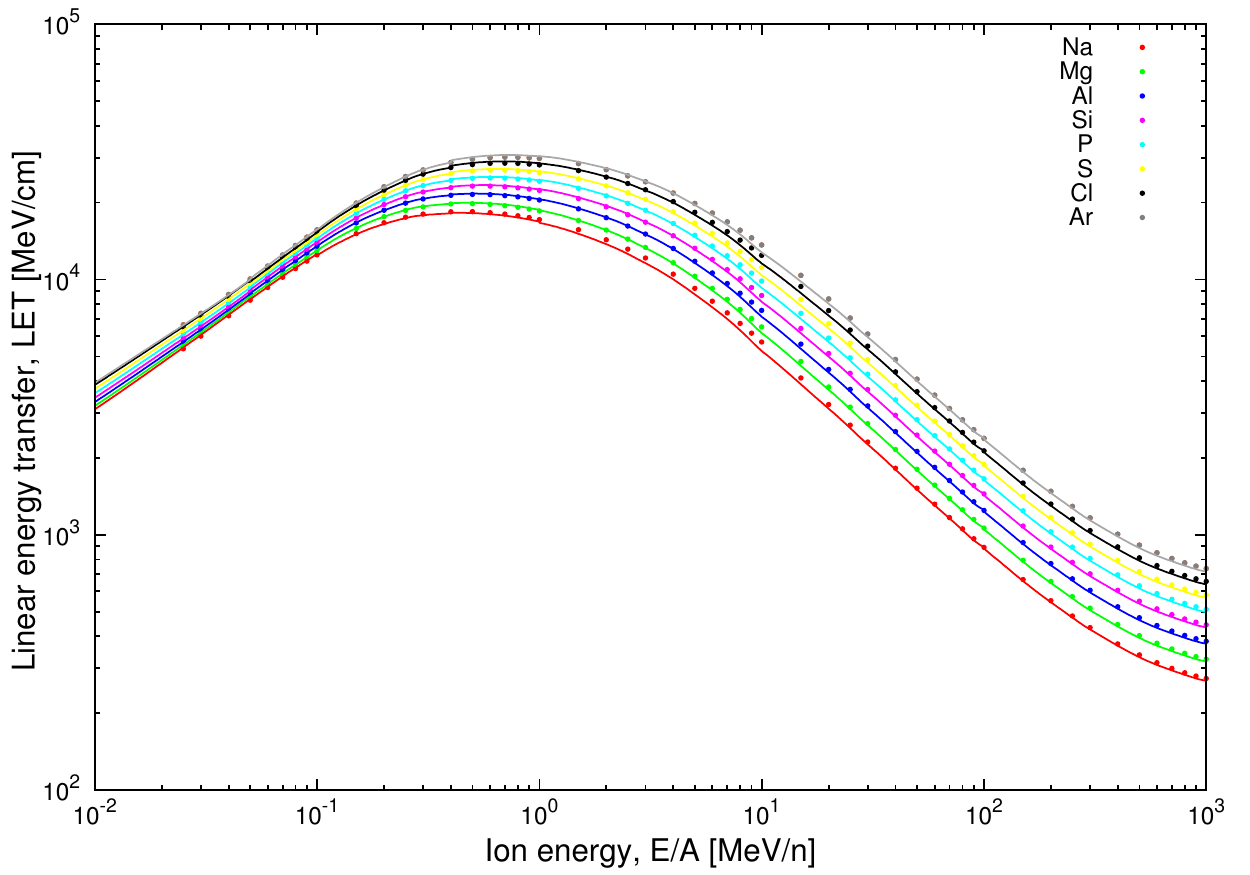}
\end{center}
\caption{Values of stopping power (LET$_{\infty}$) for He, Li, Be, B, C, N, O, F, Ne, Na, Mg, Al, Si, P, S, Cl and Ar ions. Points represent data for liquid water from \cite{icru1993} and \cite{icru2005}; Lines - values calculated using eq.(\ref{eq.letion}) multiplied by the $Z$-dependent correction factor where necessary (see Appendix \ref{ch. appendixa}).}
\label{fig.let2}
\end{figure}

In this work eq.(\ref{eq.letion}) was multiplied by an additional factor to obtain better agreement of the LET values calculated using this equation and those listed in (\cite{icru2005}) for ions of $3 \leq Z \leq 18$ and in the energy range below $10\;MeV/n$. The atomic number, $Z$-dependent factor was introduced for ions of atomic number $3 \leq Z \leq 18$ and of energies $ 1.5 \cdot 10^{-2} MeV/n \leq \frac{E}{A} \leq 10\; MeV/n$, as given in Appendix \ref{ch. appendixa}. Values of stopping power for ions heavier than protons, calculated using eq.(\ref{eq.letion}), together with the respective data from \cite{icru2005} are presented in Fig. \ref{fig.let2}. 

\begin{figure}[!ht]
\begin{center}
\includegraphics[width=1\textwidth]{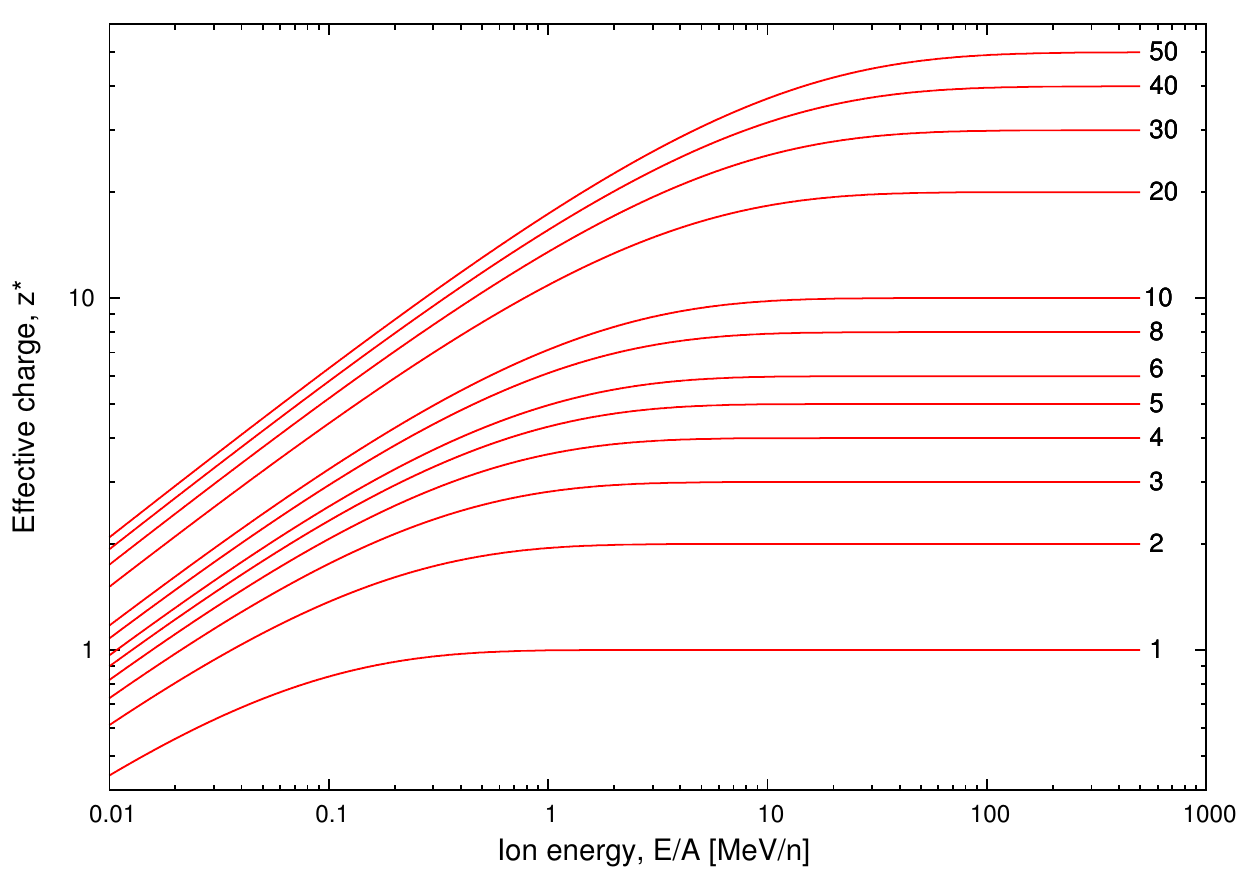}
\end{center}
\caption{'Effective charge' as a function of ion energy calculated by eq.(\ref{eq.zeff}) for ions of atomic numbers in the range from $1$ to $50$.}
\label{fig.zeff}
\end{figure}

In many equations the 'effective charge' of an ion of charge $Z$ and relative speed $\beta$, is given by the formula of \cite{barkas1963}: 
\begin{equation}
z^* = Z \left[ 1-e^{\left(-125 \beta Z^{-2/3}\right)}\right],
\label{eq.zeff}
\end{equation} 
and replaces the ion charge $Z$. The 'effective charge' represents the effect of partial screening of the charge of the ion at energies below a few $MeV/n$. For protons eq.(\ref{eq.zeff}) takes the form:
\begin{equation}
z^*_p = 1-e^{\left(-125 \beta\right)}.
\label{eq.zeffproton}
\end{equation}
In Fig. \ref{fig.zeff} effective charge, $z^*$, calculated for ions of different atomic numbers ($Z=1,2,3,4,5,6,8,10,20,30,40,50$) is plotted against the energy of the ion.

\newpage
\newpage
\section{Interaction of ionizing radiation with biological systems}
\label{ch. biologicalinteraction}

Effects caused by ionizing radiation may occur at any level of organization of the living species, ranging from single molecules within individual cell to its tissues and organs. At the basic molecular level, ionization of atoms within a particular biomolecule may result in a measurable biological effect which depends on a number of factors. The number of copies available and the importance of the molecule in the cell structure are crucial factors which determine the outcome of exposure to radiation. Because DNA, the basic element of cell replication, is present only as a single, double-stranded copy, changes in its structure due to exposure to ionizing radiation may lead to major consequences. Unrepaired or incorrectly repaired damage of the DNA may lead to potentially malignant cell transformation or to cell death. 

\begin{figure}[!ht]
\begin{center}
\includegraphics[width=0.45\textwidth]{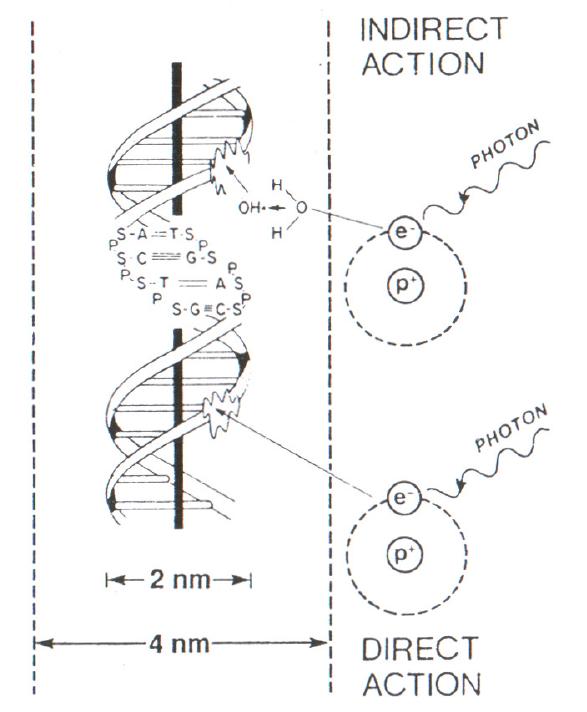}
\end{center}
\caption{Scheme of direct and indirect actions of radiation on the DNA. Figure reprinted from \cite{hall2006}. }
\label{fig.indirectdirect}
\end{figure}

Both indirect and direct radiation action may contribute to the DNA damage. Direct action of radiation occurs when particle ionizes molecules of the single DNA strand, or both strands, directly. Indirect action of radiation involves water molecules which are the most abundant molecules in the cell ($75-85\%$ of the cell mass). Radiation interacts with water to produce highly reactive free radicals that are able to migrate far enough to reach and damage the DNA. The scheme of direct and indirect actions of radiation on the DNA is shown in Fig. \ref{fig.indirectdirect}. The indirect process goes through the following stages: radiation ionizes the molecules of water found in the cell's nucleus:

\begin{displaymath}
H_2O \;\; ^{\underrightarrow{\small\textrm{radiation}\normalsize}} \;\; H_2O^+ + e^-
\end{displaymath}
 $H_2O^+$ is an ion radical. The primary ion radicals are short-lived (with a lifetime of about $10^{-10}$ second). The ionized water molecule reacts with another non-ionized water molecule to form aqueous hydrogen (hydrogen captured by a molecule of water) and a hydroxyl radical
\begin{displaymath}
H_2O^+ + H_2O \;\; \longrightarrow \;\; H_3O^+ + OH^{\bullet}
\end{displaymath} 
or interacts with a free electron to produce an excited water molecule 
\begin{displaymath}
H_2O^+ + e \;\; \longrightarrow \;\; H_2O^*
\end{displaymath}
which dissociates to hydrogen and hydroxyl radicals
\begin{displaymath}
H_2O^* \;\; \longrightarrow \;\; H^{\bullet} + OH^{\bullet}.
\end{displaymath}
The hydroxyl radical has a lifetime about $10^{-9}$ seconds and this free radical is highly reactive. It may diffuse a short distance to reach a critical target in the DNA within the cell nucleus.

Organic radicals of the longest lifetime (about $10^{-5}$ seconds) are formed either by direct ionization or by reaction of $HO^{\bullet}$ radicals with biologically important molecules, for example with the DNA (\cite{hall2006}). As a result of the action of the hydroxyl radical, a hydrogen atom is removed from an organic compound to produce water and an alkyl radical 
\begin{displaymath}
RH + OH^{\bullet} \;\; \longrightarrow \;\; H_2O + R^{\bullet}.
\end{displaymath}

It was \cite{mottram1936} who first discovered that the presence of oxygen enhances the biological effect of radiation. Nowadays we know that the oxygen reacts with the free hydrogen radical which had formed during water radiolysis, to produce the hydroperoxyl radical:
\begin{displaymath}
H^{\bullet} + O_2 \;\; \longrightarrow \;\; HO_2^{\bullet}.
\end{displaymath}
The resulting radical in the presence of another such radical or a hydrogen radical can form hydrogen peroxide, which is a highly oxidative molecule.
\begin{displaymath}
2HO_2^{\bullet} \;\; \longrightarrow \;\; H_2O_2^{\bullet} + O_2\\
\end{displaymath}
\begin{displaymath}
HO_2^{\bullet} + H^{\bullet} \;\; \longrightarrow \;\; H_2O_2^{\bullet} \\
\end{displaymath}
The alkyl radical also reacts promptly with oxygen to form the dangerous peroxy radical:
\begin{displaymath}
R^{\bullet} +O_2 \;\; \longrightarrow \;\;RO_2 \\
\end{displaymath}
From the point of view of cell death, the presence of these radicals is the most dangerous because they cause intensive damage to the DNA.

\begin{figure}[!ht]
\begin{center}
\includegraphics[width=1.0\textwidth]{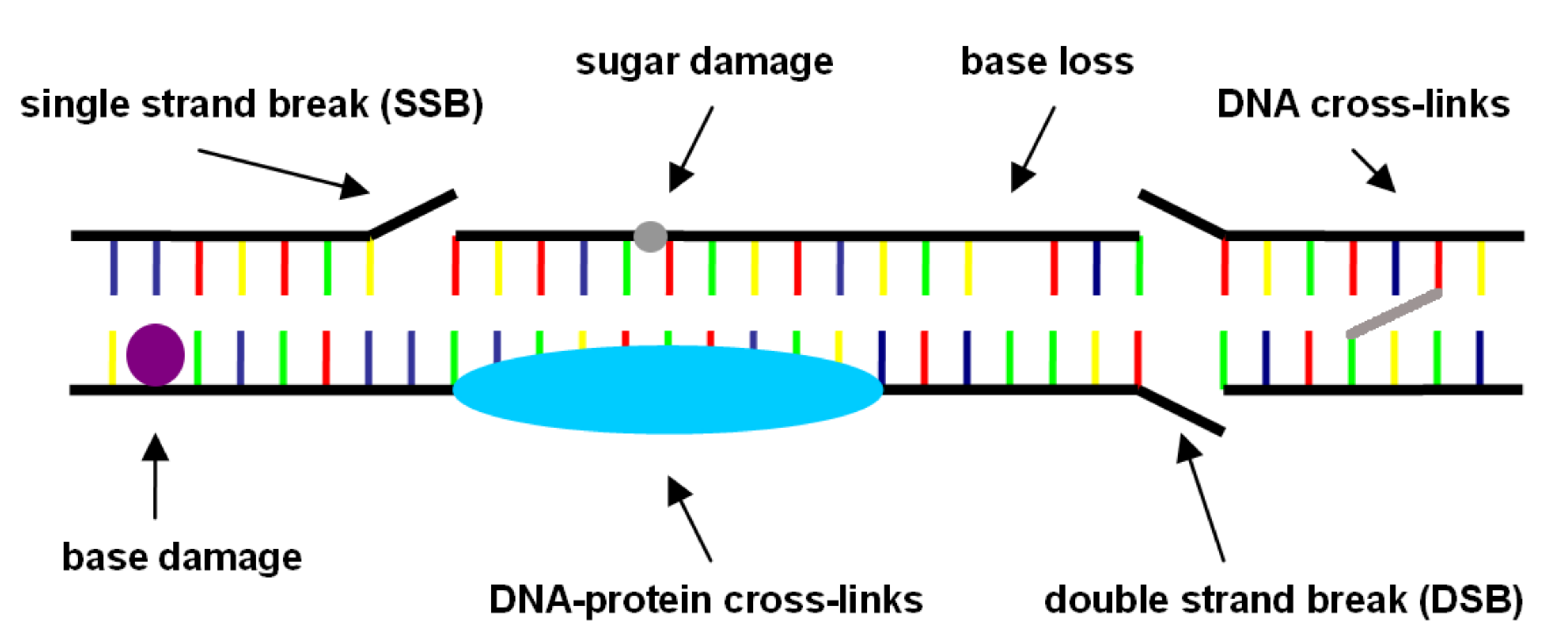}
\end{center}
\caption{Scheme of the DNA molecule with different types of modifications.}
\label{fig.dnabreaks}
\end{figure}

Indirect action constitutes about $70\%$ of the total damage produced in DNA after low-LET radiation, such as X-rays, whereas direct interaction is the dominant process when high-LET radiation interacts with living organisms (\cite{hall2006}). The damage to DNA resulting from the indirect and direct action of radiation is in principle similar. The type and frequency of the induced damage depends on the geometrical distribution of ionization events, i.e. on the LET of radiation. Base modification, base loss, sugar damage, single strand breaks (SSB), double strand breaks (DBS), DNA cross-links, DNA-protein cross-links have been identified as a different types of radiation damage to the DNA. Some of these are schematically presented in Fig. \ref{fig.dnabreaks}. Single-strand breaks are of little biological consequence as they can be easily repaired using the opposite strand as a template. The presence of double-strand breaks (DSB) of the DNA often has the most catastrophic consequences, in terms of the cell's reproductive integrity. These are difficult to repair as DSB may snap the chromatin into two parts. As a consequence the specific genetic information may be irreversibly lost, leading to cell death, or carcinogenesis. 

\section{Survival of cells in culture after exposure to ionizing radiation}
\label{ch. xresponse}

In radiobiology, 'cell death' means that the cell loses its 'reproductive' or 'clonogenic' activity, or it is no longer able to continue its tissue-specific functions (\cite{gasinska2001}, \cite{gunderson2000}), though it may be still physically present, metabolically active, and may even be able to undergo one or two mitoses. In contrast, 'cell survival' denotes the ability of the cell to sustain proliferation indefinitely, in the case of proliferating cells (including those cultured \emph{in vitro}, stem cells of normal tissues and tumour clonogens), and in the case of nonproliferating cells (nerve cells or muscle cells) - the ability to sustain specific biological functions after their exposure to ionizing radiation.
\begin{figure}[!ht]
\begin{center}
\includegraphics[width=0.6\textwidth]{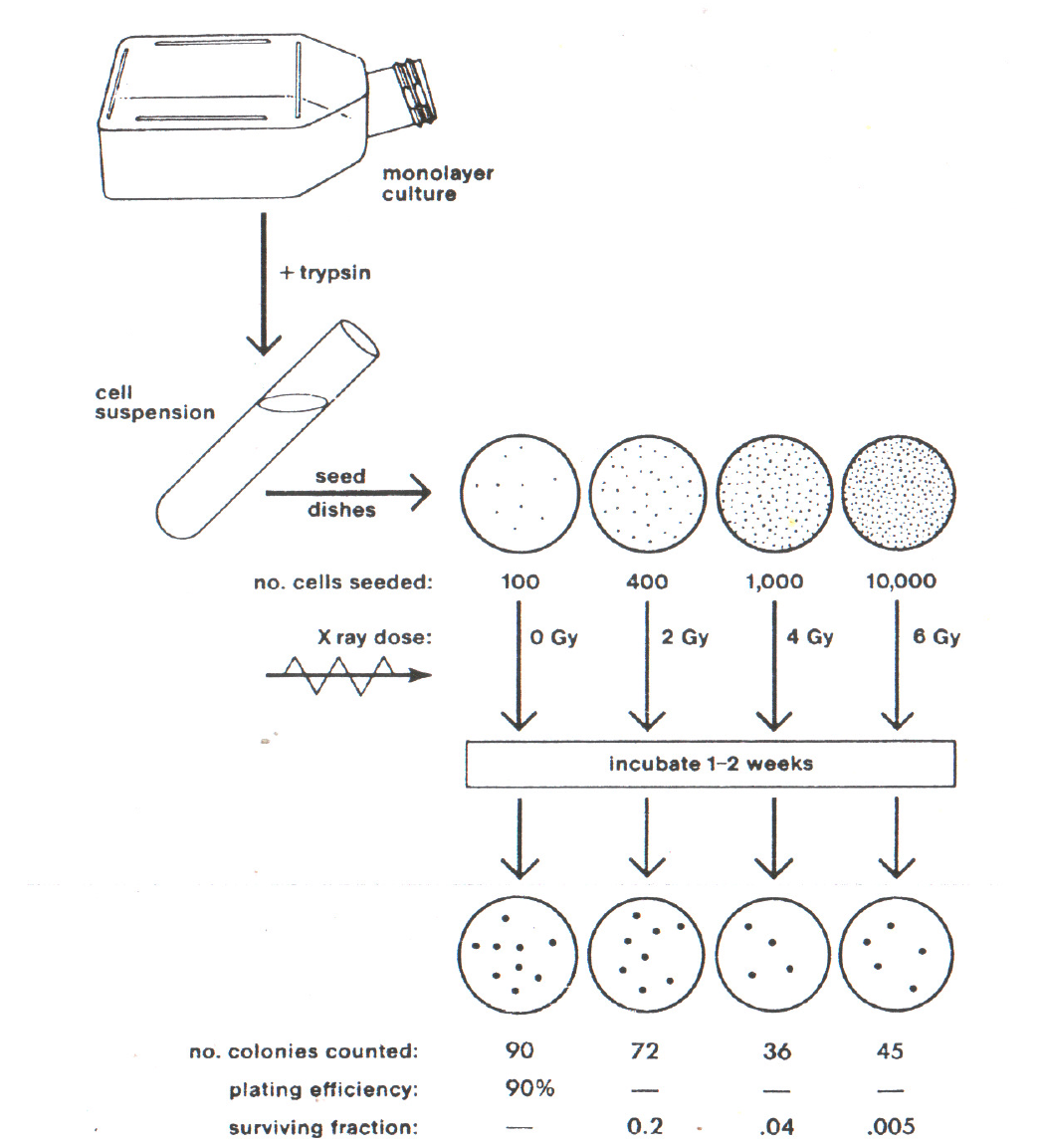}
\end{center}
\caption{Scheme of the cell culture technique used to generate a cell survival curve. Control - from 100 seeded cells 90 cells grow into a colony. Plating efficiency is equal $90\%$. For dose $2\;Gy$ - from 400 seeded cells 72 cells grow into a colony. Surviving fraction after exposure to $2\;Gy$ is equal: $72/400 \times 0.9 = 0.16$. Figure reprinted from \cite{hall2006}.}
\label{fig.sf}
\end{figure}

The capability of a single cell to grow into a colony of at least $50$ daughter cells after exposure to ionization radiation, is a proof that it has retained its reproductive capacity. The threshold of $50$ daughter cells is arbitrary (\cite{kausch2003}). A specimen of normal or tumour tissue is first mechanically cut into small pieces and next dissolved using the trypsin enzyme into a single-cell suspension. It is then seeded into a Petri dish, covered with an appropriate complex growth medium and maintained under specific conditions to grow and divide (\cite{hall2006}). Every few days the cells are removed from the surface of the dish and diluted with trypsin, which allows a small but known number of cells to be re-seeded in culture flasks after some time, to be used in radiobiological experiments. Cell samples are then irradiated with known doses of ionizing radiation. Following these exposures, cells are incubated and after 1-2 weeks are fixed and stained. Unfortunately, even if a cell sample is not irradiated, for variety of reasons that affect the ability of cells to reproduce, not all seeded cells will form a colony. The factor indicating the percentage of cells seeded, which grow into colonies is called 'plating efficiency' and is given by the formula:
\begin{equation}
\textrm{PE}=\frac{\textrm{Number} \;\;\textrm{of}\;\; \textrm{colonies}\;\; \textrm{counted}}{\textrm{Number}\;\; \textrm{of}\;\; \textrm{cells}\;\; \textrm{seeded}} \times 100\;.
\label{eq.platingefficiency}
\end{equation}
For example, if there are $60$ colonies counted on the dish, the plating efficiency is $60\%$, per $100$ seeded cells. For cells seeded in a parallel dish, exposed to a dose of ionization radiation, fixed and stained after 1-2 weeks, the surviving fraction of these cells is calculated as follows: 
\begin{equation}
\textrm{Surviving}\;\; \textrm{Fraction}=\frac{\textrm{Colonies}\;\;\textrm{counted}}{\textrm{Cells}\;\;\textrm{seeded}\; \times\; (\textrm{PE}/100)}\;\;.
\label{eq.survivalfraction}
\end{equation}
The number of cells seeded per dish is adjusted to the expected survival following the exposure to a dose of radiation. The surviving fraction can be evaluated for a set of cell samples irradiated separately to a certain set of doses (see also Fig. \ref{fig.sf}). In such a way one can obtain the dose-survival dependence - the survival curve for cells in culture (or '\emph{in vitro'}). The cell survival curve is usually plotted on semi-logarithmic scale, with dose values plotted on the linear x-axis. The cell survival curve provides a relationship between the absorbed dose of the radiation and the portion of the cells that survive (or retain their reproductive integrity) after that dose. The type of the cells, their oxygen status, the phase in the cell cycle they are irradiated at, and type (LET) of radiation are factors which affect the shape of the cell survival curve. Depending on these factors one may observe a variety of survival curve shapes, form purely exponential (linear on a semi-logarithmic scale) to shouldered ones (with a linear, or exponential, initial part, curved in the intermediate dose region and again linear-exponential at higher doses).

\begin{figure}[!ht]
\begin{center}
\includegraphics[width=1.0\textwidth]{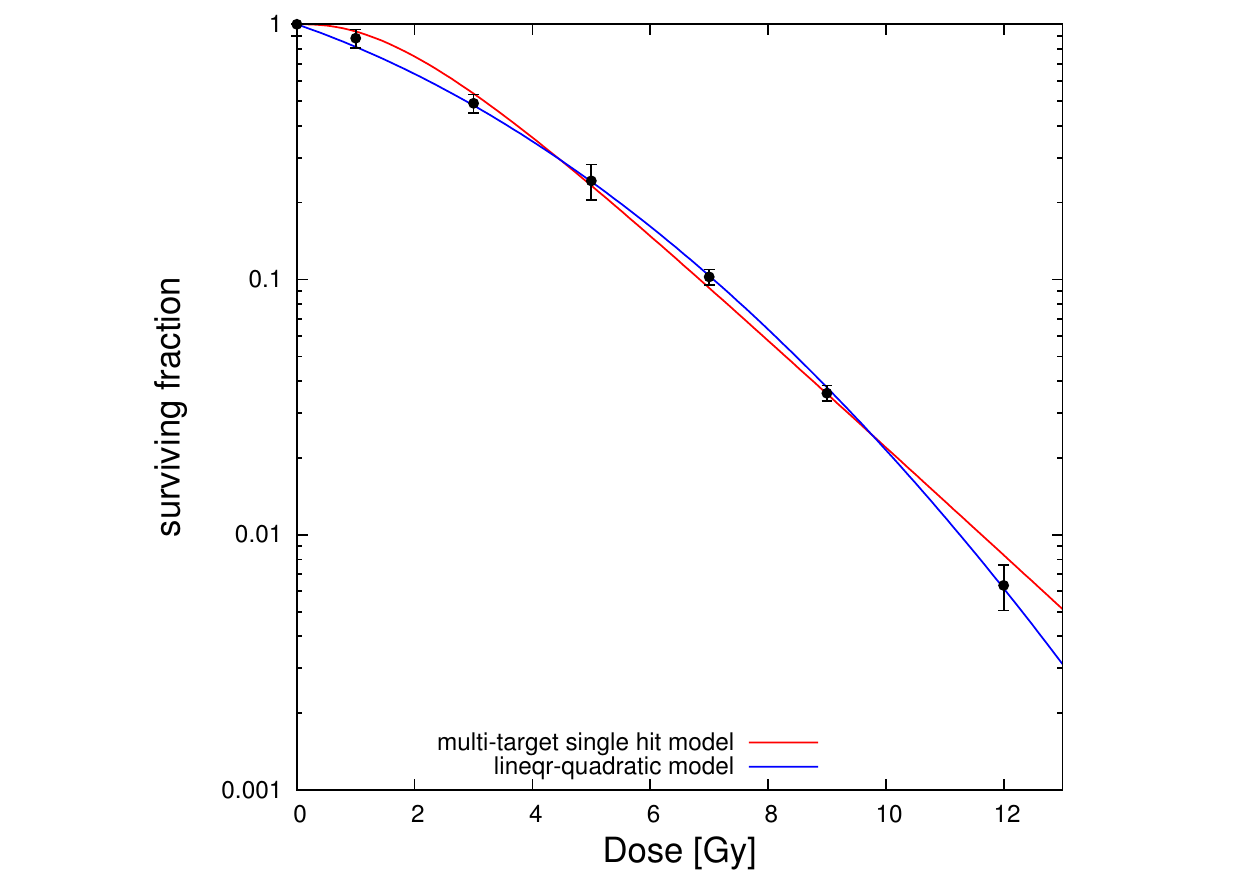}
\end{center}
\caption{Survival of V79 Chinese hamster cells after irradiation by $200 \;kV$ X-rays. Full lines show the multi-target or linear-quadratic representations best-fitted to the measured data points. Red line: multi-target model with $m=2.91$, and $D_0=2.05\;Gy$ (for parameter values, see Section \ref{ch. furusawa}). Blue line: linear-quadratic model with $\alpha = 0.184\;Gy^{-1}$ and $\beta=0.02\;Gy^{-2}$. Data points ($\bullet$) and their errors together with the values of $\alpha$ and $\beta$ are from \cite{furusawa2000}. }
\label{fig.sfexample}
\end{figure}

The cell survival curve can be represented by many different mathematical descriptions. For the purpose of this work two such descriptions will be discussed: linear-quadratic and multi-target. Details about other descriptions may be found elsewhere (\cite{gasinska2001}). As an example, the survival curve of V79 Chinese hamster cells, represented by the linear-quadratic and multi-target descriptions, both best-fitted to the survival data points, are presented in Fig. \ref{fig.sfexample}.

\subsection{Survival curves - multi-target single-hit description}
\label{ch. mhitmodel}
The general multi-target single-hit formula to describe the survival curve is as follows:
\begin{equation}
S(D) = 1 - \left(1-e^{-\frac{D}{D_{0}}}\right)^m.
\label{eq.mtarget}
\end{equation}
The two parameters used in this description are $D_0$, the 'characteristic' dose related to the radiosensitivity of the cell, and $m$ - the 'number of targets' parameter, which enables the 'curvature' of the survival curve to be generated. If $m=1$, eq.(\ref{eq.mtarget}) reduces to $1-e^{-D/D_0}$, and represents a purely exponential (linear) survival curve, with $D_0$ representing the dose that reduces cell survival to $37\%$ $(1/e)$ of the initial population.

One may interpret the last expression by assuming a single target in a cell to be inactivated after receiving a given portion of dose, i.e. a 'hit'. If the cell were to contain $m$ such '1-hits' targets, all $m$ of which must receive a 'hit' for the cell as a whole to be inactivated, then eq.(\ref{eq.mtarget}) obtains. In the case of $m$ '1-hit' targets in each cell, in a population of cells exposed to a low dose, only some of these targets will receive 'hits', so most cells will survive. As the dose increases, the number of targets 'hit' in each cell cumulatively increases, and finally, as the dose increases even further, the dose response of the cell population becomes exponential, as the remaining targets in each cell receive their 'hits'. Thus, initially, at low doses, the slope of the survival curve represented by eq.(\ref{eq.mtarget}) is zero, then there is a curvature at intermediate doses of the order of $D_0$, and exponential decrease at higher doses - the steeper, the higher the value of $m$.

In the above interpretation, the number $m$ of '1-hit' targets in each cell may acquire only integer values. In practice, as shown in Fig. \ref{fig.sfexample}, real numbers representing the $m$-parameter may be best-fitted to experimentally measured survival curves. The zero initial slope postulated by the '1-hit' $m$-target representation of the survival curve has often been contested as being inconsistent with results observed in survival curves measured for mammalian cells (\cite{gunderson2000}).

\subsection{Survival curves - linear-quadratic description}
\label{ch. lqmodel}
The linear-quadratic representation of measured survival curves is extensively used in radiation biology. The survival curve is described by an expression containing a linear and a quadratic component in the exponent, as follows:
\begin{equation}
S(D) = e^{\left(-\alpha \cdot D - \beta \cdot D^2 \right)}.
\label{eq.lq}
\end{equation}
An often quoted interpretation of this equation assumes that a double strand break in the DNA helix is the critical damage which can lead to cell death. At low doses, where the linear (purely exponential) dependence of survival on dose dominates, a double-strand break may arise from a single energy deposition event involving both strands of the DNA. At high doses, where the quadratic dependence on the dose dominates, two separate events, each involving a single strand may result in double strand beak of the DNA (\cite{chadwick1973}). 

Representation of the survival curve by eq.(\ref{eq.lq}) gives a linear (purely exponential) dependence at low doses which implies (in contrast to the multi-target single-hit representation) that even the smallest dose of radiation results in a finite chance of killing a cell. At high doses, the linear-quadratic representation gives a continuous curvature of the survival curve, which disagrees with much of the radiobiological data (\cite{hall2006}).

Whether represented by multi-target single-hit or linear-quadratic formulae, cell populations are far more complex. None of these oversimplified representations are able to account for low dose hypersensitivity (\cite{marples1993}) or non-target effects, such as bystander effects (\cite{mothersill1997}) or genomic instability (\cite{kadhim1992}).

\subsection{Relative Biological Effectiveness}
\label{ch. rbe}
The biological effect (for instance cell survival or number of chromosomal aberration) caused by the ionizing radiation depends on the pattern of energy deposition at the microscopic level. Equal doses of different types of radiation do not result in equal responses of the biological system. As high-LET radiation (charged particles) is more densely ionizing than low-LET radiation, it deposits much more energy in a particular 'micro' target volume (i.e. cell), which presumably leads to a larger number of double-strand breaks (\cite{brenner1992}) and more complex and severe damage of the DNA (\cite{anderson2002}). The fraction of cells killed is linked to the number of sites of irreparable DNA damage. Thus, high-LET radiation is more biologically effective than low-LET radiation. The factor which describes differences in the response of cells to doses of radiation of different quality is called relative biological effectiveness (RBE) and is defined as the ratio of the dose of reference, low-LET radiation (usually $X$- or $\gamma$-rays) to that of high-LET tested radiation required to achieve the same level of a given biological endpoint. In this work the comparison will be made mainly between $X$-rays and ion beams, hence the RBE at a given isoeffect level can be represented as follows:
\begin{equation}
\textrm{RBE} = \frac{D_{X-\textrm{rays}}}{D_{\textrm{ion}\; \textrm{beam}}} \Bigg|_{\textrm{isoeffect}},
\label{eq.rbe}
\end{equation}
where $D_{X-rays}$ and $D_{ion\; beam}$ is the dose of reference radiation and the dose of test radiation respectively, that yield the same level of biological effect (isoeffect). If the survival curves are represented by the linear-quadratic formula, eq.(\ref{eq.lq}) then an alternative definition of RBE may be used, namely RBE$_{\alpha}$, which represents the 'maximum RBE' at the 'zero-dose' limit. It is calculated as the ratio of the ${\alpha}$ coefficients representing the initial slopes of the linear-quadratic equation describing the survival curve after doses of heavy ions, $\alpha_i$, and of the reference radiation, $\alpha_{\gamma}$:
\begin{equation}
\textrm{RBE}_{\alpha} = \frac{\alpha_i}{\alpha_{\gamma}}.
\label{eq.rbealpha}
\end{equation}
The RBE value depends on the biological endpoint under consideration. The RBE cannot be uniquely defined for a given radiation, since it depends on many different factors. It depends on the linear energy transfer (LET) and also on the kind of particle. Different kinds of particles but of the same LET may lead to different values of RBE in the same cell system, because of the different track structure of these ions. Moreover, RBE varies with the dose, dose per fraction, degree of oxygenation, cell or type of tissue (\cite{iaea2008}).

\subsection{Oxygen Enhancement Ratio}
\label{ch. oer}
Oxygen is probably the best-known chemical agent that modifies the biological effect of ionizing radiation. Presence of oxygen during irradiation intensifies the action of free radicals and promotes the production of more stable and more toxic peroxides (see Section \ref{ch. biologicalinteraction}). Oxygen sensitizes cells and increases the radiation damage. Cells exposed to ionizing radiation in the presence of oxygen are more radiosensitive, and are more radioresistant in its absence, which is reflected in the shapes of the respective survival curves. For a given cell line, survival of cells is lower in aerobic than that in hypoxic conditions, after exposure to the same type of ionizing radiation. This is of importance in radiotherapy, as fast-growing tumour cells are usually hypoxic for lack of sufficient blood supply, and are therefore more radioresistant than the neighbouring healthy and well-oxygenated cells. The oxygen enhancement ratio (OER) describes the difference between the response of hypoxic and aerobic cells at a given level of survival, and is given by the formula: 
\begin{equation}
\textrm{OER} = \frac{D_{\textrm{hypoxic}}}{D_{\textrm{aerobic}}} \Bigg|_{\textrm{isoeffect}},
\label{eq.oer}
\end{equation}
where $D_{hypoxic}$ is the dose given under hypoxic and $D_{aerobic}$ is the dose given under aerobic condition, both resulting in the same level of biological effect. The oxygen enhancement ratio depends on the LET of radiation and usually decreases with increasing LET of ions, which is advantageous in ion beam radiotherapy (see Section \ref{ch. ionradiotherapy}).

\section{Ion beam radiotherapy}
\label{ch. ionradiotherapy}
The general aim of radiotherapy, regardless of its type, is to deposit the prescribed dose to the tumour volume in order to fully inactivate the tumour cells in that volume, while sparing to the extent possible the neighbouring healthy tissues.

In conventional external beam radiotherapy, doses of ionizing radiation are delivered by high-energy $X$-ray or electron beams of energy range $4-20\;MeV$, generated by medical electron accelerators.

In ion beam radiotherapy, inactivation of cells in the tumour volume is achieved by energetic ions (typically protons or carbon ions) accelerated to several hundred $MeV/n$ by dedicated accelerators, such as a synchrotron or cyclotron.

\subsection{Ion beam versus conventional radiotherapy}
\label{ch. ionrationale}
From the physical and clinical points of view, features of ion beam radiotherapy with proton beams and with beams of carbon ions are different. The distinction basically follows from the widely different stopping power (LET) of these ions: proton LET values are much lower then those of ions heavier than helium. With respect to photon and electron beams applied in conventional radiotherapy, the advantage of proton beam therapy stems mainly from physical considerations due to which a better dose distribution with depth in the patient may be obtained. On the other hand, the clinical advantages of heavier ions are due not only to these physical aspects, but also to biological considerations, such as the enhanced RBE or OER, characterising such ion beams (\cite{schulz2006}, \cite{schulz2007}).

\begin{figure}[!ht]
\begin{center}
\includegraphics[width=0.75\textwidth]{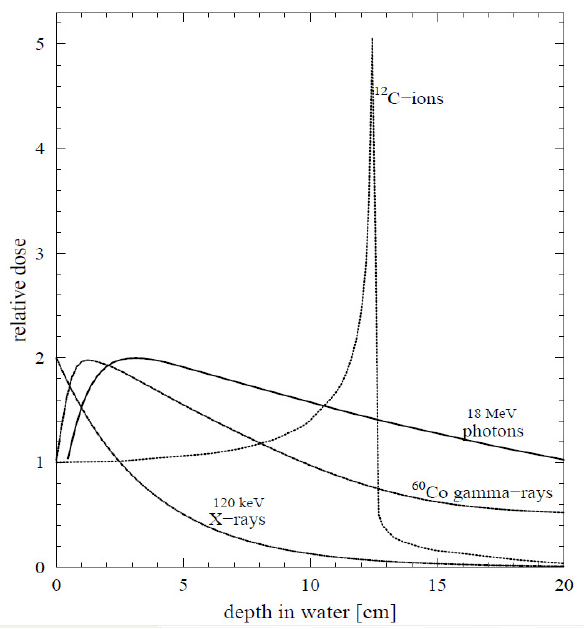}
\end{center}
\caption{A comparison of the relative depth-dose distributions from conventional external radiotherapy beams of $18\;MeV$ photons, $120\;keV$ X-rays, $^{60}$Co $\gamma$-rays and from a monoenergetic carbon ion beam of initial energy $250\;MeV/n$. Figure reprinted from \cite{kraft2001}. Note that the relative depth-dose distributions for conventional beams are normalized to a value of $2$ at dose maximum, while the carbon beam dose distribution is normalized to $1$ at beam entrance.}
\label{fig.physicaldose}
\end{figure}

The relative depth-dose distribution of low voltage $X$-ray machines, a $Co$-gamma radiotherapy beam, and a $18\;MeV$ $X$-ray beam from a medical linear accelerator, all used in conventional radiotherapy, and from a monoenergetic carbon beam of initial energy $250\;MeV$, are compared in Fig. \ref{fig.physicaldose}.

\begin{figure}[!ht]
\begin{center}
\includegraphics[width=0.95\textwidth]{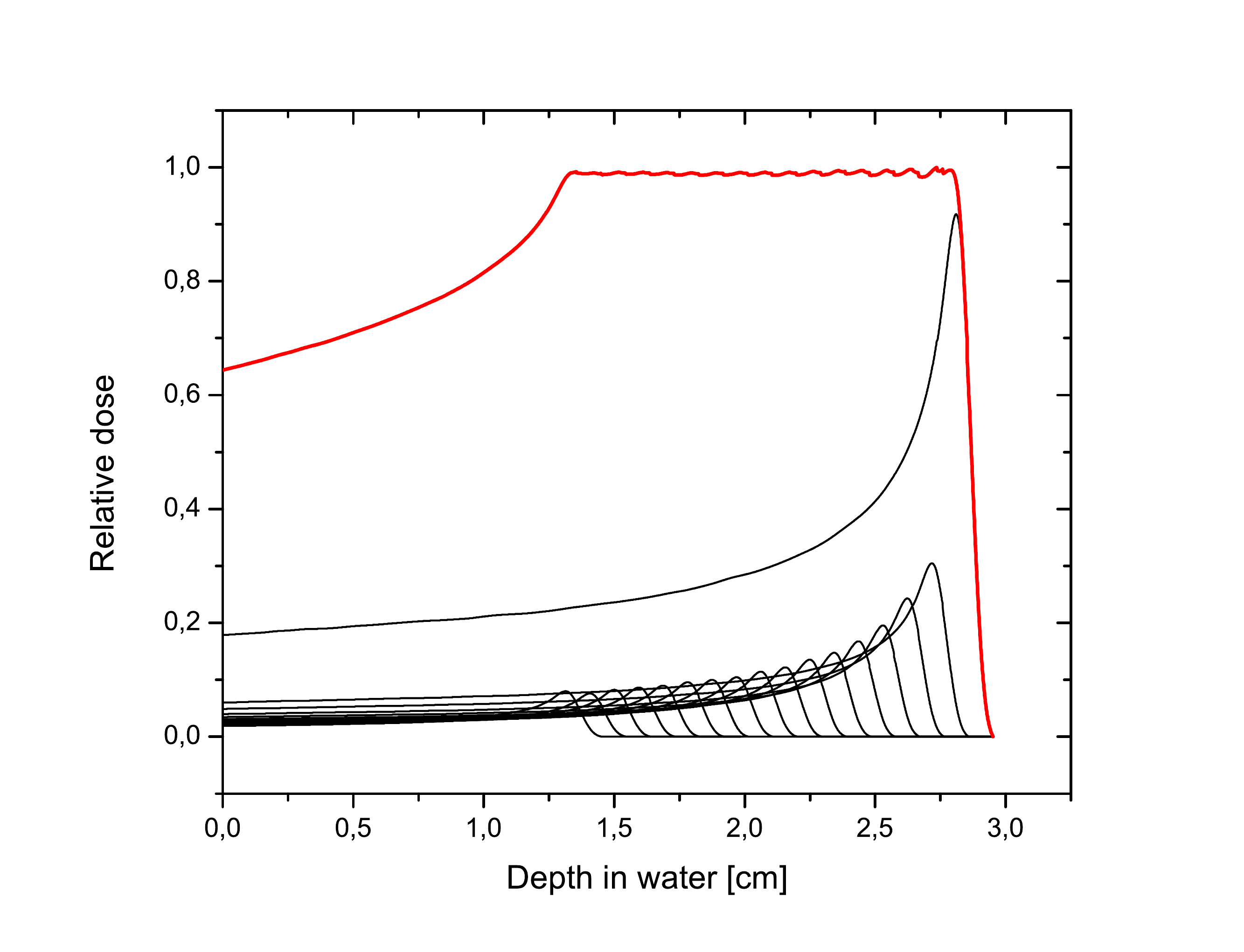}
\end{center}
\caption{The principle of the spread-out Bragg Peak. Beams of protons of different energies are superimposed in order to cover the tumour dimension, here extending between $1.25$ and $2.75\;cm$ depth with the desired dose. For a carbon beam, adjustment of this 'physical dose' profile to correct for the effective RBE of such a beam superposition would be necessary (see Section \ref{ch. biologicaldose} and Fig. \ref{fig.biologicaldose}).}
\label{fig.sobp}
\end{figure}

The photon beams deposit their highest dose rates at depths below $5\;cm$ in water, the dose rate next gradually decreasing at larger depths. This is not very convenient for treating deeply seated tumours, as tissues in front of such tumours receive an unnecessarily high dose. In the case of ion beams the relative dose-depth distributions are quite different - the entrance dose rate is quite low, while with increasing depth of penetration the ions lose their energy and their energy loss per unit path length (LET) increases. As a consequence, the relative dose rate will gradually rise until the track end, where ions deposit an extremely large amount of energy over a very narrow range, an effect called the Bragg peak. Next, for protons, the dose rate falls rapidly to zero as the charged particles finally stop at their full range. In the case of carbon ions, apart from the sudden decrease of dose rate beyond the Bragg peak, a characteristic 'tail' extends, representing the dose deposited by secondary ions, lighter than the primary ions of the beam. These secondary particles are produced by nuclear reactions of the primary ions with nuclei of the absorber atoms. The speed of these secondary particles is only slightly less than that of the primary particles, therefore, the newly created fragments, of charges lower than that of the beam particles, travel to distances exceeding the range of the main beam, thus adding unnecessary exposure of tissues close to the target volume. However, ion beams will deposit most of their dose at greater depths than photons or electron beams, the optimum depth being adjustable by varying the entrance energy of the ion beam.

The width od the Bragg peak in a beam of monoenergetic particles is usually too small to fully cover the treatment volume. Therefore, beams of different energies have to be superimposed, as shown in Fig. \ref{fig.sobp}, to produce a spread-out Bragg peak (SOBP) and to deliver the dose to the whole tumour.

Due to their higher LET, charged particles are biologically more effective than photons. The factor describing this difference is the relative biological effectiveness, RBE (see Section \ref{ch. rbe}). Typical values of RBE for carbon radiotherapy beams range between $2$ and $4$, depending on the treatment procedure and type of tumour treated (\cite{tsujii2008}).

Another advantage when applying carbon beams in radiotherapy is the possibility of reducing the oxygen enhancement ratio, i.e. enhancing the radiosensitivity of anoxic tumour cells (see Section \ref{ch. oer}). Also, since the cell-cycle dependence of radiation sensitivity of the cells and their repair ability are reduced with increasing LET (observed as a reduction of the curvature of their survival curves), the possibility arises of reducing the number of fractions in the patient's treatment course (hypofractionation). Since the lateral scattering of the beam of ions decreases with increasing charge $Z$ of the ion, better coverage of the tumour volume allows higher doses to be delivered to the irradiated volume(dose escalation).

Despite the many advantages of ion beam radiotherapy, clinical applications of this relatively new modality are limited to selected tumour types and localisations. The high cost of the ion accelerator technology, when faced with advances in the much cheaper and more available photon beam delivery techniques using medical accelerators (such as IMRT - Intensity Modulated Radiotherapy; IGRT - Image Guided Radiotherapy, etc.) supported by the rapidly developing medical imaging technology, will make ion beam radiotherapy the clinical choice only for very selective cases. Proton beam radiotherapy is presently the clinically recommended choice for treating uveal melanoma and an option in treating paediatric tumours, skull base tumours and head-and-neck tumours, or inoperable early stage lung cancer. Carbon beams have been used with success to treat skull base chordomas and chondrosarcomas, and other intracranial tumours, as well as in paraspinal and sacral bone tumours, but also at several other localizations (\cite{schulz2007}). The potential superiority of ion beam radiotherapy over other modalities is yet to be demonstrated by systematic clinical trials, but the clinical advantages of proton radiotherapy in treating ocular melanoma and paediatric tumours, as well as those of carbon beams in treating skull base chardomas and chondrosarcomas are quite clear, soon to be followed by other sites as ion radiotherapy matures and becomes more available worldwide. Several very exhaustive review articles an books are available, where the historical development, clinical advantages, and rationale of patient selection for ion radiotherapy are described (e.g., \cite{durante2010}, \cite{schulz2006}, \cite{tsujii2008}, \cite{linz1995}).

\subsection{Biologically weighted dose}
\label{ch. biologicaldose}
Due to the higher biological effectiveness of heavy ion beams, the basic question that has to be answered during the treatment planning routine for ion radiotherapy is how to adjust the dose profile of the ion beam over the tumour volume to achieve uniform cell inactivation in that volume. The tumour cell survival level should be the same as that achieved by prescribing the appropriate dose in conventional external photon beam radiotherapy. This is the how the concept arose of the biologically weighted dose referred to as 'biological dose', $D_{\textrm{RBE}}$, which is the product of the physical (absorbed) dose, $D_{Phys}$, multiplied by the value of RBE:
\begin{equation}
D_{\textrm{RBE}} = D_{Phys} \cdot \textrm{RBE},
\label{eq.biologicaldose}
\end{equation}
\begin{figure}[!ht]
\begin{center}
\includegraphics[width=0.65\textwidth]{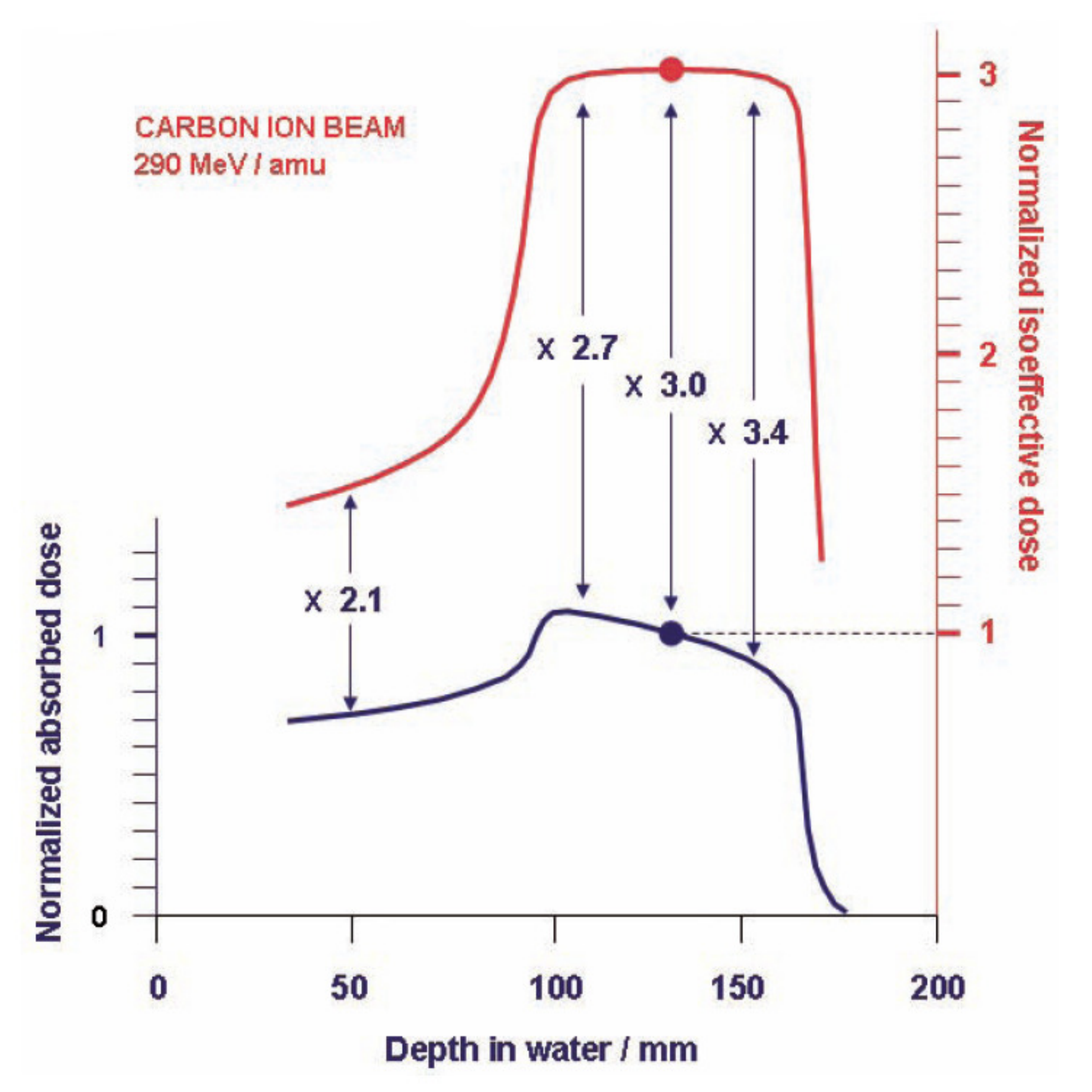}
\end{center}
\caption{Illustration of the idea of biological dose. The full line in black represents the 'physical dose' deposited by carbon ions within the SOBP, the line in red represents the isoeffective biologically weighted dose. Figure reprinted from \cite{iaea2008}.}
\label{fig.biologicaldose}
\end{figure}
The concept of biological dose is illustrated in Fig. \ref{fig.biologicaldose}, where a comparison is made between the dose vs. depth distributions of absorbed (or 'physical') dose (black line), and of isoeffective biologically weighted ('biological') dose (red line). The case considered concerns a tumour located between $100$ and $160 \;mm$ depth, irradiated by a SOBP of carbon ions of initial energy $290\;MeV/n$. The RBE of carbon ions as a function of LET significantly increases with depth. In order to obtain a uniform distribution of biological dose within the tumour volume the distribution of absorbed dose has to be properly adjusted. To compensate for the higher RBE at the distal end of the carbon ion beam range, the physical dose has to decrease with depth (\cite{iaea2008}). 

In ion radiotherapy the RBE plays the role of a weighting factor to account for changes in radiation quality along the beam range. Although the RBE appears to be a simple concept, its clinical application is complex because local values of RBE vary along the beam range and depend on many factors, such as particle type, energy, dose, dose per fraction and cell or tissue type (see Section \ref{ch. rbe}), in a manner difficult to predict or represent quantitatively. Only in the case of protons is this the situation somewhat simpler. The experimentally established degree of RBE variation in many \emph{in vitro} systems irradiated with protons of energies ranging between $60$ and $250 \; MeV$ is very low. The RBE values for protons are consistent with a mean RBE of $1.1$ (\cite{icru2007}). Therefore at all proton radiotherapy establishments only a single value of RBE$=1.1$ is employed in treatment planning systems, independently of dose, fractionation scheme, position in the SOBP, tissue type, \emph{etc}.
For heavy ion radiotherapy beams, determination of the RBE for clinical use is much more complicated, as discussed above. In addition, a further complication is the presence in the primary beam of lighter secondary fragments which provide an additional significant contribution to the final biological effect.

Experimental evaluation of the RBE values in a heavy ion beam for all clinically relevant conditions is impossible. Thus, application of relevant biophysical models is necessary in heavy ion beam therapy planning.

\section{Treatment planning systems for ion beam radiotherapy}
\label{ch. planningsystems}

A number of proton therapy planning systems (TPS) are presently in use, such as XiO (produced by Elekta), Eclipse (produced by Varian Medical Systems), RayStation (produced by RaySearch Laboratories). Currently, only two such therapy planning systems are used clinically for carbon beam radiotherapy: the Treatment Planning for Particles (TRiP) - developed at the Gesellschaft f\"ur Schwerionenforschung (GSI) in Darmstadt by the group of Prof. Gerhard Kraft (\cite{scholz1997}) and the carbon beam treatment planning system developed at the at National Institute of Radiological Sciences (NIRS) at Chiba, Japan (\cite{kanai1997}). These treatment planning systems were tailored to the specific needs of the respective facilities. The GSI-Darmastadt (and now HIT-Heidelberg) groups use active beam scanning, while the Japanese group at NIRS has up to now used passive spreading of the Bragg peak. Two further projects concerning treatment planning software for carbon ion scanning beams are under development, one by a group at the Istituto Nazionalle di Fisica Nucleare (INFN) in Italy, in collaboration with the Ion Beam Applications (IBA) company (\cite{russo2011}). The other scanning carbon beam TPS is being developed by the Japanese group at NIRS (\cite{inaniwa2008}), to be applied in a new treatment facility which will use the raster scan method, added to the existing Heavy-Ion Medical Accelerator (HIMAC) facility in Chiba.

It is generally accepted that in order to give a complete description of the biological dose distribution inside the patient's body, originating from carbon beams, the treatment planning systems should contain two distinct components: the physical part, or beam transport (to deliver the physical dose, or sets of energy-fluence spectra at different depths, as an output) and the biophysical part (to deliver the appropriate RBE values).

\subsection{Calculation of physical dose distribution}
\label{ch. physicaldose}

The existing physical dose calculation algorithms for ion beams have been usually based on the pencil beam approximation. Several Monte Carlo (MC) codes exist that are capable of computing dose distributions for light ion radiation therapy. MC codes such as FLUKA, GEANT 4, PHITS and SHIELD-HIT have also been adapted to calculate carbon beam transport through different media (\cite{hollmark2008}). Because the MC techniques are too machine time-consuming, an analytical pencil beam model of depth dose distributions for a range of ion species was proposed by \cite{hollmark2004}. This analytical method uses a pencil beam algorithm model in which the analytical approach is combined with the MC code SHIELD-HIT07 to derive physical dose distributions of light ions transported in tissue-equivalent media. Multiple scattering of primary and secondary ions was considered. The contribution to the dose from fragmentation processes has so far not been included. Another semi-analytical model was developed by \cite{kundrat2007}, using energy-loss and range tables generated by the SRIM code (version SRIM-2003.26) for calculating depth-dose distributions, and an analytical method to describe energy loss by straggling and nuclear reactions. However, reaction products were not included in the calculation and their contribution to dose depositions from primary ions was neglected. Such analytical approaches are much faster to compute but are not yet as accurate as MC simulation, so considerable further development in this area is necessary.

The solutions for the physical dose calculation developed within the particular treatment planning systems already applied for clinical use are discussed together with the respective TPS project within which they are integrated.

\subsection{Biophysical modelling for carbon beams}
\label{ch. biophysical models}
To be applicable in the treatment planning system for carbon ions, the biophysical model has to fulfil several basic conditions: it has to provide calculations of the RBE dependences on such factors as particle type, energy, dose and cell or tissue type; it has to provide the calculation of the RBE in a mixed radiation field of therapeutic ion beams, consisting of particles of different energies and atomic numbers, with accuracy required in clinical radiotherapy; its analytical formulation should be fast and robust in order to be applicable to massive calculations required for ion beam radiotherapy planning. 

Many different biophysical models have been developed to calculate and predict the response of cells \emph{in vitro} after their irradiation, but only a few are able to fulfil the above requirements. The models to be considered are: the Local Effect Model - LEM (\cite{scholz1997}), the modified Microdosimetric Kinetic Model - MKM (\cite{inaniwa2010}), the Probabilistic Two-Stage Model (\cite{kundrat2007}) and, as we shall demonstrate in this work, the cellular Track Structure Theory - TST, developed much earlier by Robert Katz and co-workers (\cite{katz1978}).

The two biophysical approaches currently applied in clinical treatment planning systems (TPS) for carbon radiotherapy beams, will now be briefly outlined.

\subsubsection[The GSI-Darmstadt/HIT Heidelberg TPS \\ project]{The GSI-Darmstadt/HIT Heidelberg TPS project}
\label{ch. lem}
The Treatment Planning for Particles (TRiP) code developed at Gesellschaft f\"ur Schwerionenforschung (GSI) Darmstadt by Prof. Gerhard Kraft's group is the most advanced TPS, dedicated to the active beam scanning technique. The TRiP software includes a physical beam model and a radiobiological model - the Local Effect Model (LEM). Additionally, inverse planning techniques are implemented in order to obtain a uniform distribution of biologically equivalent dose in the target volume.

The physical beam model developed by \cite{kramer2000} allows depth dose profiles for ion beams of various initial energies to be generated. This numerical transport code is based on the tabulated values of energy loss, and includes the most important basic interactions, such as energy loss, straggling and generation of secondary fragments. Calculations of the depth dose distribution made for homogeneous media (water) are pre-calculated for a set of initial ion energies and stored as a reference data set. Since the TRiP's transport code does not exploit time-consuming Monte Carlo methods, physical dose profiles can be evaluated efficiently, within a few minutes.

As for the biophysical part of the TRiP, the Local Effect Model (\cite{scholz1997}) is implemented. LEM and Track Structure Theory have the common feature of relating the biological effectiveness of charged particle radiations to the radial distributions of dose around the ion's path. Since there are differences in the formalisms of LEM and TST which link the radial dose distribution to cell survival, their framework may lead to different predictions of the clinical outcome of carbon therapy. A comprehensive study of the differences in the principles of the two track structure approaches as well as in the model predictions of cell survival of V79 cells after proton beam irradiation was performed by \cite{paganetti2001}. Here we only recapitulate the main features of the LEM model. 

The number of surviving cells is equal to the fraction of cells carrying no lethal event. If $\overline{N_{X}}$ denotes the average number of lethal events per cell, according to the Poisson distribution, the surviving fraction of cells after their exposure to a dose of photon radiation is:
\begin{equation}
S_{X}(D)=e^{-\overline{N_{X}(D)}},
\label{eq.lemsflethalref}
\end{equation}
and therefore:
\begin{equation}
\overline{N_{X}(D)} = -ln S_{X}(D).
\label{eq.lemnolethalref}
\end{equation}
From this number, the dose-dependent event density, $v_X(D)$ for photon radiation can be introduced:
\begin{equation}
v_{X}(D)=\frac{\overline{N_{X}}}{V_{\textrm{nucleus}}}=\frac{-ln S_{X}(D)}{V_{\textrm{nucleus}}},
\label{eq.lemdenstylethal}
\end{equation}
where $V_{\textrm{nucleus}}$ is the volume of the cell nucleus and $D$ is the photon dose.

The principal assumption of LEM model is that the biological effect is determined by the local spatial energy deposition in small sub-volumes (i.e. the cell nucleus), but is independent of the particular type of radiation leading to that energy deposition. Thus, the differences in biological action of ions are attributed to the energy deposition pattern of charged particles, as compared with photon irradiation. The energy deposition pattern after irradiation by charged particles is determined essentially by secondary electrons liberated by the passing ion, and within the LEM model this pattern is described as a function of distance $r$ from the ion's trajectory, as follows:
\begin{equation}
D(r) = \left\{ \begin{array}{ll}
\lambda \; \textrm{LET}/r^2_{min} \quad \textrm{if} \quad r<r_{min}\\
\lambda \; \textrm{LET}/r^2 \quad \quad \textrm{if} \quad r_{min} \leq r \leq r_{max}\\
0 \quad \quad \quad \quad \quad \quad \textrm{if} \quad r>r_{max}
\end{array} \right.
\label{eq.lemrdd}
\end{equation}
where LET denotes the linear energy transfer and $\lambda$ is a normalization constant to ensure that the radial integral reproduces the value of LET. The parameter $r_{min}$ describes the transition from the inner part of the track, where a constant local dose is assumed, to the $1/r^2$ behaviour, and $r_{max}$ is the maximum radius determined by the $\delta$-electrons with the highest energy. Given the local dose distribution, eq.(\ref{eq.lemrdd}), the average number of lethal events induced per cell by heavy ion irradiation, can be obtained (\cite{elsasser2008}):
\begin{equation}
\overline{N_{\textrm{Ion}}}= \int v_{\textrm{Ion}}(D(x,y,z)) \; dV_{\textrm{nucleus}},
\label{eq.lemnolethal1}
\end{equation}
where $v_{ion}$ denotes the lethal event density after ion irradiation. According to the main assumption of the LEM model, the local dose effect is independent of the radiation quality and thus $v_{\textrm{Ion}}(D)=v_{X}(D)$ is assumed. Inserting eq.(\ref{eq.lemdenstylethal}) into eq.(\ref{eq.lemnolethal1}) one obtains:
\begin{equation}
\overline{N_{\textrm{Ion}}}= \int v_{\textrm{Ion}}(D(x,y,z)) \; dV_{\textrm{nucleus}} = \int \frac{-logS_X(D(x,y,z))}{V_{\textrm{nucleus}}} \; dV_{\textrm{nucleus}},
\label{eq.lemnolethal2}
\end{equation}
where $S_X(d)$ denotes the photon dose-response curve given by a modified linear-quadratic model:
\begin{equation}
S_X(d) = \left\{ \begin{array}{ll}
e^{-(\alpha D + \beta D^2)} \quad \quad \quad \quad \quad \quad \textrm{if} \quad D \leq D_t\\
e^{-(\alpha D_t + \beta D^2_t + s_{max}(D-D_t))}  \quad \textrm{if}  \quad D>D_t
\end{array} \right.
\label{lemsfx}
\end{equation}
The survival curve for $X$-rays is assumed to be shouldered with a purely exponential tail, of slope $s_{max} = \alpha+2\beta D_t$, at doses greater than the threshold dose, $D_t$. Finally, the number of surviving cells is given by the fraction of cells carrying no lethal event. According to the Poisson distribution, for a given pattern of particle traversals, the ion survival probability $S_{\textrm{Ion}}$ for a cell is given by 
\begin{equation}
S_{\textrm{Ion}}=e^{-\overline{N_{\textrm{Ion}}}}.
\label{lemsf}
\end{equation}
In the LEM, the parameters necessary to predict the cell survival fraction are: $\alpha$, $\beta$, $D_t$ and $r_{min}$. In order to obtain the outcome of the LEM model according to eq.(\ref{eq.lemnolethal2}), calculations using Monte Carlo methods are necessary. This makes the computing times unacceptable when model predictions are implemented to the treatment planing system. To make the TRiP code usable in therapy planning, certain approximations have been introduced to speed up the computation (\cite{scholz1997}). \cite{kramer2006} proposed a fast calculation method using a low-dose approximation, allowing for more sophisticated treatment planning in ion radiotherapy. The LEM model was twice improved in order to obtain better agreement with experimental data obtained using beam dosimetry and cell cultures \emph{in vitro}. The original version of this model (LEM I) overestimated the effective (biologically-corrected) dose in the entrance channel and underestimated the effect in the Bragg peak region. To make more accurate calculations the effects of clustered damage in the DNA were included into the cell survival curve after photon irradiation (LEM II). Nevertheless, this model still showed a tendency similar to that of the original version. The next step was to introduce a parametrization of the $r_{min}$ which in the previous version was maintained constant. The new velocity-dependent radius $r_{min}$ of the inner part of the track, introduced in LEM III, almost completely compensated the systematic deviations in RBE predictions (\cite{elsasser2008}). Since in 2009 all the know-how achieved by the GSI - Darmstadt group was taken over by Siemens, no further information on consecutive improvements of this model is available. 

The TRiP TPS code software has been in clinical use at GSI since the beginning of the ion radiotherapy pilot project in 1997 until it ended in 2009. Over this period, $440$ patients were planned and treated. Currently TRiP is used in clinical therapy planning at the Heidelberg Ion Therapy Center (HIT) which began its operation in November 2009. Further development of the carbon TRiP for the HIT clinical facility is now handled on a commercial basis by Siemens.

\subsubsection{The NIRS-HIMAC TPS project}
\label{ch. nirs}

A different approach in designing the SOBP for carbon ion beam radiotherapy is applied at the Heavy Ion Medical Accelerator (HIMAC). In contrast to TRiP, the treatment planning system used at the HIMAC is designed for a beam-delivery system based on passive shaping techniques. For clinical trials at HIMAC the carbon beam was chosen, because it possesses similar LET characteristics as the fast neutron beam which had been used for radiotherapy at NIRS for over 20 years. Basing on clinical equivalence between carbon and fast neutron radiotherapy, best use was made of the clinical experience acquired so far with fast neutrons. The design of the SOBP at HIMAC is a three-step process. The first step consists in calculating the physical dose distribution. Next, the biological dose is designed in order to achieve the prescribed uniform survival level for tumour cells within the SOBP region. At the end of this process, the clinical dose is calculated to achieve a biological response similar to that which would be achieved from a fast neutron beam irradiation.

The depth-dose distribution and LET distributions of monoenergetic carbon beams are calculated using the HIBRAC code, developed by \cite{sihver1996}. This code includes ion fragmentation. A 'dose-averaged' LET value is deduced at each depth from the calculated LET spectra. The patient's body is assumed to be water-equivalent.

In order to design a uniform distribution of biologically equivalent dose within the SOBP, the dose-survival relationships of HSG (Human salivary gland) cells were chosen, as their response after carbon ion irradiation was found to be representative of typical tumour response and representative of patient outcome after fast neutron irradiation. Survival curves of HSG cells irradiated by carbon ions of various incident energies and incident LET of monoenergetic carbon beams, were characterized by best-fitted $\alpha$ and $\beta$ coefficients of the linear-quadratic description. It is next assumed that the survival curves describing the response of the cells after a mixed radiation field can be expressed by the linear-quadratic description with dose-averaged coefficients $\alpha$ and $\sqrt{\beta}$ for monoenergetic beams over the spectrum of the SOBP beam. The survival curve for a mixed irradiation field can then be described by the equation (\cite{kanai1997}):
\begin{equation}
S_{mix}(D) = e^{(- \alpha_{mix}D - \beta_{mix} D^2)},
\label{himacsfmixed}
\end{equation}
in which
\begin{displaymath}
\alpha_{mix} = \sum f_i \alpha_i,
\end{displaymath}
\begin{displaymath}
\sqrt{\beta_{mix}} = \sum f_i \sqrt{\beta_{mix}},
\label{himacalphabeta}
\end{displaymath}
where $f_i=d_i/D$ is the fraction of the dose of the \emph{i}th monoenergetic beam, $d_i$, and $D=\sum d_i$ is the total dose of the mixed beam. The thus-calculated $\alpha_{mix}$ and $\beta_{mix}$ coefficients, the values of which are based on 'dose-averaged' LET are now used for survival calculations at each depth. The biological dose over the SOPB region is designed to achieve a constant survival probability of $10 \%$ for HSG cells over the entire SOBP.

The last step in the designing the SOBP at HIMAC is to calculate the 'clinical dose' in order to achieve a biological response equivalent to that after a fast neutron beams. It is assumed that the carbon beam is clinically equivalent to the fast neutron beam at the point where the 'dose-averaged' LET value in $80\;keV/\mu m$ - the 'neutron-equivalent point'. From the NIRS's neutron therapy experience, the clinical RBE of neutron beam was found to be $3.0$. Therefore, the clinical RBE value of the carbon spread-out Bragg peak is determined to be $3.0$ at the neutron equivalent point, where the average LET value is $80\;keV/ \mu m$ (\cite{kanai1999}). Next, the entire SOBP is normalized by multiplying it by the factor equal to the ratio of the clinical RBE to the biological RBE determined at the neutron equivalent point.

Clinical carbon beam radiotherapy began at HIMAC in 1994. By now, the medico-physical group at NIRS has gathered considerable clinical experience by treating several tumour types at many locations in over $5000$ patients (\cite{schulz2007}, \cite{tsujii2008}).


\chapter[Formulation of Track Structure Theory]{Formulation of Track Structure Theory}
\label{ch. results1}

The track structure theory (TST), developed by Robert Katz and co-workers over 50 years ago (\cite{katz1971}), is a parametric phenomenological model able to quantitatively describe and predict the response of physical and biological detectors after ion irradiation. The TST calculations provided a quantitative description of the response to heavy ions of many physical detectors such as thermoluminescence detectors (\cite{waligorski1980}), alanine (\cite{waligorski1989}) or the Fricke dosimeter (\cite{katz1986}. TST has been also applied to \emph{in vitro} survival (\cite{katz1978}), and mutation induction and transformation endpoints (\cite{cucinotta1997}, \cite{waligorski1987a}) in a number of mammalian cell lines. Robert Katz was the first to show the importance of track structure in the analysis of the response of physical detectors and biological systems after irradiation with energetic ions, or 'high-LET radiation' (\cite{butts1967}).

The two main assumptions of TST are that the radiation effect of an energetic heavy ion is due to $\delta$-rays surrounding the ion's path and that, per average dose, the biological effect of those $\delta$-rays is the same as that of the reference radiation (e.g., $^{60}$Co $\gamma$-rays or $X$-rays). A general formula describing the radial distribution of $\delta$-ray dose around the path of an ion of a given charge and speed is applied and the response of a physical or biological system is calculated by folding into this formula the response of this system after a uniformly distributed dose of reference radiation.

The track structure theory has two variants. The first one is the 'full' track structure theory which gives the predictions for both physical and biological systems. Three model parameters are necessary to predict the response of any system after heavy ion bombardment: $c$ (or $m$ depending on the system), $D_0$ and $a_0$. The response of the detector, via the activation cross-section, is calculated here by numerically integrating the radial distribution od probability, calculated in turn by folding the probability of inactivation in a uniform field of reference radiation and the radial distribution of $\delta$-ray dose of the given ion, averaged over the volume of the sensitive site.

The second 'approximated' variant of TST is a four-parameter model - the cellular Track Structure Theory, which in principle refers only to cellular (or 'm-target') systems. Four model parameters are necessary to predict the response of the cells after heavy ion bombardment: $m$, $D_0$, $\sigma_0$ and $\kappa$ (definitions of these model parameters will be given in what follows).

In this section we focus only on the 'full' three-parameter track structure theory, while the details of the cellular Track Structure Theory are given in the next section. Here, we give the full description of the track structure theory formalism. In particular, we analyse the formulae which describe the radial dose distribution (RDD) around the ion's path. These formulae have been successively developed within the track structure theory. Next, we study the effect of different RDD on the predictions of track structure theory. Among the four RDD formulae analysed, we seek one which, by fulfilling certain scaling conditions, is the most suitable for application to the four parameter cellular Track Structure Theory. For an ion of specified charge and energy, the radial distribution of $\delta$-ray dose, $D_{\delta}(r)$ is required: i) to reproduce experimentally measured radial distributions of dose; ii) when integrated over all radii, to yield the correct value of LET of the ion; iii) to be represented by a relatively simple analytical formula; and iv) to exhibit appropriate scaling, permitting model calculations to be rapidly performed over a wide range of ions of different charges and energies.

To calculate the response of 'm-target' or 'c-hit' detectors (see Section \ref{ch. referenceradiation}) which contain sensitive sites of a given dimension, the average radial dose distribution over these sites is evaluated (see Section \ref{ch. averagedose}). To avoid confusion, the radial distribution of $\delta$-ray dose formulae, RDD or $D_{\delta}(r)$, discussed in Sections \ref{ch. radialdose} and \ref{ch. doseenergycomparison}, are called 'point-target' radial dose distributions in the Katz model jargon.

Scaling is an important feature of Katz's Track Structure Theory. The existence of specific scaling made it possible for Katz to propose the transition between the 'full' tree-parameter track structure, and his 'approximate' four-parameter cellular track structure models (\cite{katz1971}). We investigate in more detail to what degree are the various RDD formulations able to fulfil the above conditions, with emphasis on their scalability, as required by Katz's approach to amorphous track structure modelling.

To test the RDD formulae applied in Katz' TST, we choose measured endpoints in two systems described as '1-hit' detectors: inactivation cross-sections for \emph{E. coli} B$_{s-1}$ spores, and average relative effectiveness of generating ESR-measured free-radicals in alanine. The experimental ion irradiation data considered were, for \emph{E. coli} B$_{s-1}$ spores: neon, argon, krypton, gold, lead and uranium ions, and, for alanine irradiation: protons, and helium, lithium, oxygen, fluorine, neon, silicon, sulphur and argon ions. These two 'one-hit' ($m = c = 1$) detectors demonstrate widely differing radiosensitivities, as represented by values of $D_0$ (or $a_0$), allowing us to test the scaling factors used in Katz's TST for '1-hit' detectors over widely ranging target sizes and detector radiosensitivities.

\section{Detector response after exposure to reference radiation}
\label{ch. referenceradiation}

Track Structure Theory assumes that each detector consists of small identical, uniformly distributed radiosensitive elements. A sensitive element can be activated by absorbing a single, quantized value of the radiation dose, called a 'hit'. In physical detectors, the sensitive site could be represented by a grain of nuclear emulsion, in biological detectors - by the cell nucleus and elements within (e.g., chromosomes). The response (signal normalized to its saturation value) of a physical detector after the exposure to a uniformly distributed dose, $D_{\gamma}$ of reference radiation (e.g., $X$-rays or $\gamma$-rays) is described by the multi-hit (or c-hit) formula (\cite{katz1978}):
\begin{equation}
P\left(c,A\right) =  \sum_{x=c}^{\infty} \frac{\left(A\right)^x \;\cdot \; e^{-\left(A\right)}}{x!}  = 1 - \sum_{x=0}^{c-1} \frac{\left(A\right)^x \;\cdot \; e^{-\left(A\right)}}{x!},
\label{eq.chit}
\end{equation}
where $A = D_{\gamma}/D_{0}$, and $D_0$ is the characteristic dose representing the radiosensitivity of the target of a radius $a_0$ \footnote{the size of the target, $a_0$, does not appear explicitly in eq.(\ref{eq.chit}) - we shall later discuss the relation between $a_0$ and $D_0$ in TST.}. If activation of the sensitive sites in the detector results from one or more 'hits', then such a detector is called a '1-hit' detector. If it takes at least c or more 'hits' to activate the sensitive sites, the detector is called a 'c-hit' detector, and its dose response is given by eq.(\ref{eq.chit}).
\begin{figure}[!ht]
\begin{center}
\includegraphics[width=0.875\textwidth]{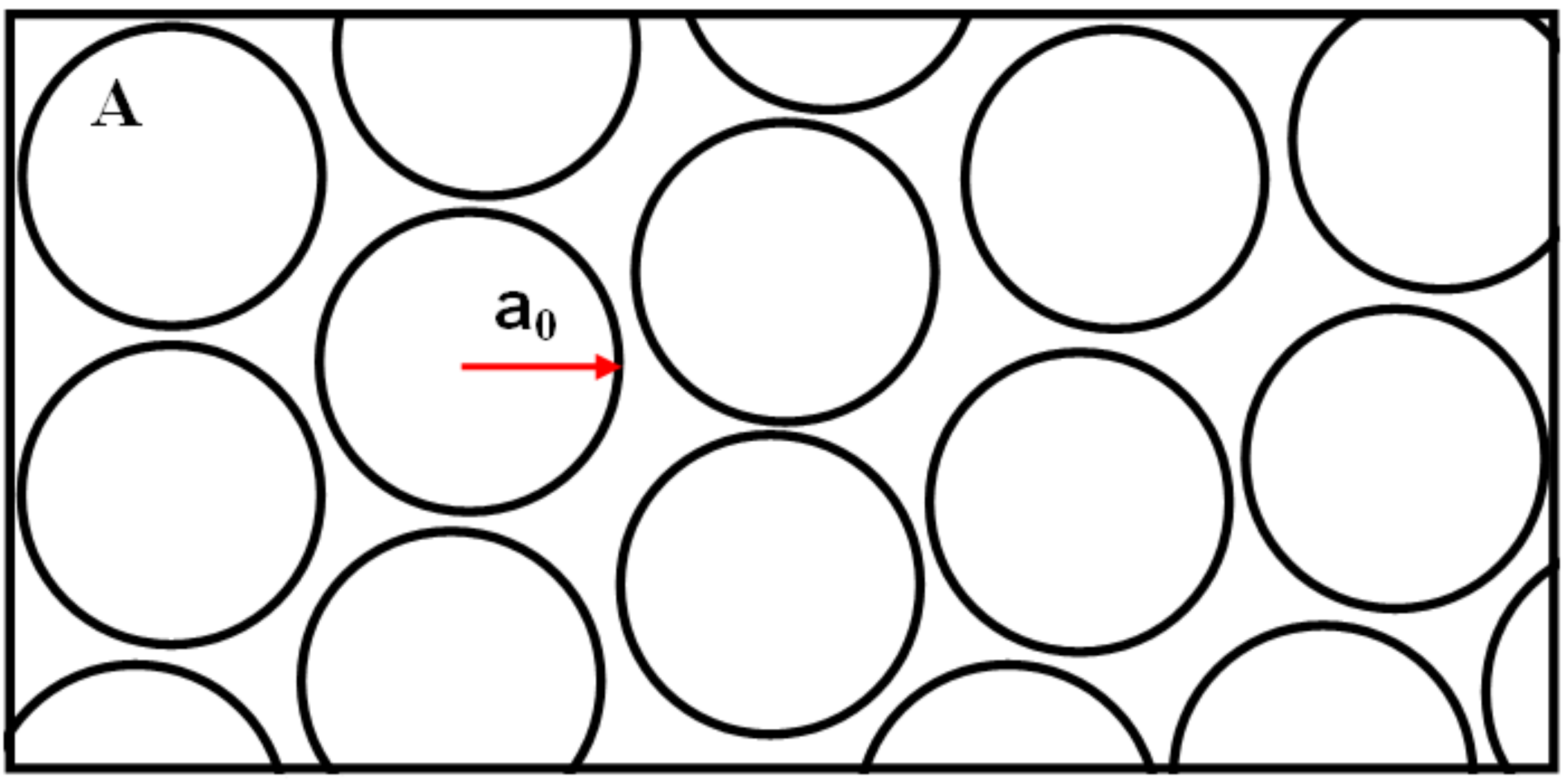}
\includegraphics[width=0.875\textwidth]{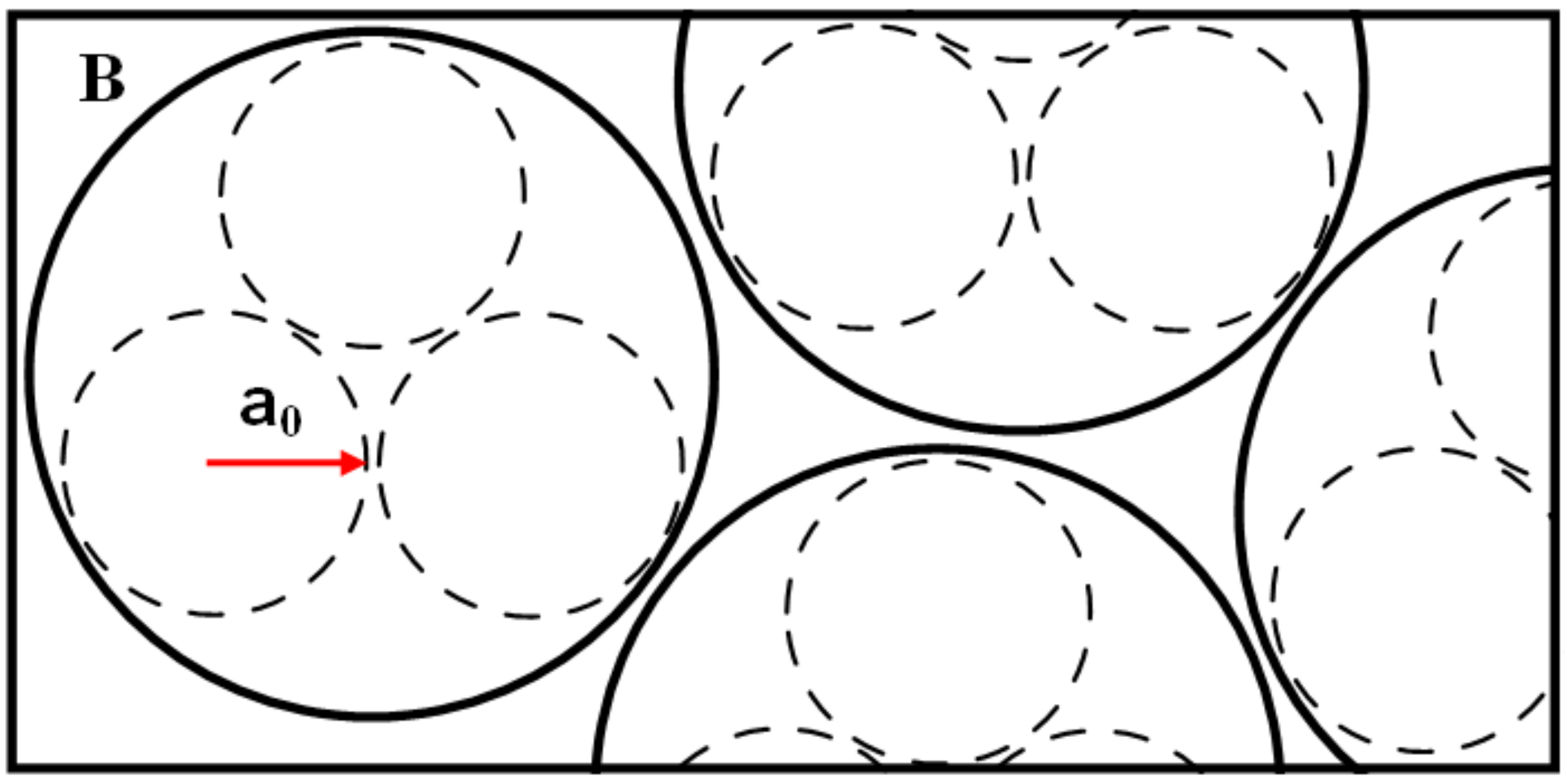}
\end{center}
\caption{ The concept of sensitive sites present in 'physical', or 'c-hit' (panel A) and 'biological', or 'm-target' (panel B) detectors.}
\label{fig.cmmodels}
\end{figure}
A somewhat different description of the response after reference radiation is used for biological detectors, where the multi-target single hit formula is applied:
\begin{equation}
P\left(m,A\right) =  \left(1-e^{-A}\right)^m.
\label{eq.singlehitmultipletarget}
\end{equation}
This '1-hit m-target' model (colloquially called the 'bean-bag' model) assumes that a sensitive site in a cell (e.g., the cell nucleus), incorporates a certain number, $m$, of 1-hit sub-targets of radius $a_0$ each. All of them need to be inactivated, each by a single 'hit' (or more 'hits') of radiation in order to achieve the observed end-point, such as inactivation of the cell. Because in radiobiology cellular survival is typically evaluated, rather than 'cell killing', eq.(\ref{eq.singlehitmultipletarget}) is usually represented by its complement to $1$ - the multi-target model, eq.(\ref{eq.mtarget}). The concept of sensitive sites in 'physical' and 'biological' detectors is shown in Fig. \ref{fig.cmmodels}.

\section{Energy-range relationship for $\delta$-electrons and Radial Distribution of Dose (RDD)}
\label{ch. radialdose}

To proceed with the calculation of the response of physical and biological detectors after ion irradiation, the 'point-target' radial distribution of dose (RDD) around the ion's path is needed. It is assumed that the ionization, due to electron ejection from the atoms of the target material, is the dominant mode of energy loss by the charged particle passing through the absorber. Based on this assumption the radial dose, $D_{\delta}(r)$, as a function of the radial distance, $r$, from the path of an ion of atomic number $Z$ and relative velocity $\beta$, can be derived.
 
We shall analyse four formulations of $D_{\delta}(r)$, developed by \cite{butts1967}, \cite{zhang1985}, \cite{waligorski1986} and by \cite{cucinotta1997}. The formulations differ mainly with respect to the incorporated energy-range and angular dependence for $\delta$-ray electron ejection. The original formula of Butts and \cite{butts1967} was derived using the Rutherford equation for $\delta$-ray production by the heavy ion. The $\delta$-rays were assumed to be ejected perpendicularly to the ion's path, and a linear electron energy-range relation was assumed. In the formula proposed by \cite{zhang1985} a more accurate power law approximation of the range of $\delta$-electrons and a fixed value of ionization potential were assumed while perpendicular ejection of $\delta$-rays was maintained. When integrated radially, Zhang's formula was found to yield about $50\%$ of the total value of LET of the respective ion. To correct for this discrepancy, \cite{waligorski1986} introduced a multiplicative correction factor to Zhang's formula, valid at radial distances below $10\;nm$. The last formula, developed by \cite{cucinotta1997}, is based on a phenomenological model describing the energy distribution of secondary electrons which includes the distribution of their ejection angles and uses a semi-empirical formula to describe the electron energy-range relationship developed by \cite{tabata1972}. Accurate reconstruction of the total LET value in Cucinotta's formula is achieved by adding a second 'excitation' term which is calculated via radial integration of the RDD and suitably normalised. 

A more detailed derivation of the radial distribution of dose formula given by \cite{butts1967} and also of the RDD formula proposed by \cite{zhang1985} has been presented elsewhere (\cite{waligorski1988}).

\subsection{The RDD formula of \cite{butts1967}}
\label{ch. rddbutts}
Derivation of the formula for $\delta$-ray dose, $D(r)$, as a function of the radial distance, $r$, starts with the equation describing the energy spectrum of $\delta$-ray production by the passing ion. \cite{butts1967} used the modified Rutherford formula for delta-ray production from a medium containing $N$ electrons per unit volume, as follows:
\begin{equation}
\frac{dn}{d\omega} = \frac{2 \pi N e^{4} z^{*2}}{m_{e}c^2 \beta^2} \frac{1}{\omega^2},
\label{eq.rutherford1}
\end{equation}
where $dn$ is the number of delta-rays per unit pathlength of energy between $\omega$ and $\omega + d\omega$, produced by an ion of effective charge $z^{*2}$ moving with the relative speed $\beta$, $m_{e}$ and $e$ are the electron mass and charge. To simplify the calculations it was assumed that all electrons are ejected normally to the ion's path. To calculate $D_{\delta}(r)$ the following linear dependence between the range and energy of $\delta$-electrons, $\omega$ was assumed: 
\begin{equation}
R_{max}=k_{1} \cdot \omega_{max} \quad \left[kg\;m^{-2}\right],
\label{eq.linearrange}
\end{equation}
where 
\begin{displaymath}
k_{1} = 0.1 \quad [kg \; m^{-2} MeV^{-1}].
\end{displaymath}
$R_{max}$ is the maximum range of $\delta$-electrons if the maximum energy $\omega_{max}$, eq.(\ref{eq.maximum.energy}), is applied. The value of the $k$ coefficient results from fitting eq.(\ref{eq.linearrange}) to experimentally evaluated extrapolated ranges in aluminium of electrons of energies between $0.7 \;keV$ and $5 \; keV$, because only these limited data were available at that time. 
Considering that the dose $D_{\delta}(r)$ at a distance $r$ is defined as the energy lost by electrons passing the volume of cylindrical shell of radius $r$ and thickness $dr$ coaxial with the ion's path, and taking into account the above assumptions, the following RDD formula obtains:

\begin{equation}
D_{\delta}(r) = C\;\frac{z^{*2}}{\beta^2}\;\frac{1}{\rho}\;\frac{1}{r^{2}}\left(1-\frac{r}{R_{max}}\right) \quad [MeV \; kg^{-1}],
\label{eq.rddbutts}
\end{equation}
where $r$ is the distance from the ion's path $[m]$, $\rho$ is the density of the medium, $z^{*2}$ and $\beta$ are the ion's effective charge and relative velocity, $R_{max}$ is the maximum range of $\delta$-electrons. Water as absorber is assumed to represent tissue-equivalent medium, so in all TST model calculations it is assumed that for biological material $\rho = 1000 \;[kg\cdot m^{-3}]$. The constant $C$ is related to the electron density of the medium of the absorber, i.e. water:
\begin{equation}
C = \frac{Ne^4}{m_{e}c^2(4\pi \epsilon _{0})^2} \quad  = 1.355 \; [MeV \;m^{-1}],
\label{eq.constantc}
\end{equation}
where $N$ is the electron density (for water $N = 3.341 \cdot 10^{29} m^{-3}$), $m_{e}$ and $e$ are the electron mass and charge and $\epsilon_{0}$ is the permittivity of vacuum. The SI system of units is used throughout this work (in the original papers of Katz, it was the CGS system), so all $\delta$-ray dose formulae are expressed in units of $MeV/kg$. Applying the conversion: $1\;eV\;=\;1.602\times10^{-19}\;J$ and $1\;Gy\;=\;1\;J/kg$; the radial distribution of dose can be expressed in $Gy$. 

Although the RDD of \cite{butts1967} was oversimplified it was successfully used to show the importance of track structure in analyzing the response of physical and biological detectors after high-LET radiation, and led Katz to an explanation of the 'thindown effect' known in radiobiology and in studies of ion tracks in nuclear emulsion (\cite{katz1985}).

\subsection{The RDD formula of \cite{zhang1985}}
\label{ch. rddzhang}

The $D_{\delta}(r)$ equation given by \cite{zhang1985} was based on the same Rutherford formula, eq.(\ref{eq.rutherford1}) but here the electrons in the absorber atoms were assumed to be bound with an ionization potential $I = 10 \; eV$:
\begin{equation}
\frac{dn}{d\omega} = \frac{2 \pi N e^{4} z^{*2}}{m_{e}c^2 \beta^2} \frac{1}{(\omega + I)^2}.
\label{eq.rutherford2}
\end{equation}
To derive the radial distribution of the dose, perpendicular ejection of $\delta$-rays was maintained, but a more accurate power law representation of the electron energy-range relationship was used:
\begin{equation}
R_{max}=k_{2} \cdot \omega_{max}^{\alpha} \quad \left[kg\;m^{-2}\right],
\label{eq.powelowrange}
\end{equation}
Equation (\ref{eq.powelowrange}) was fitted separately to the data representing the extrapolated electron range in aluminium for electrons of energy below and above $1\;keV$ - the choice of $\alpha$ depending on the value of $\omega_{max}$. For $\omega_{max}$ $<$ $1\; keV$ $\alpha= 1.079$ and for $\omega_{max}$ $>$ $1 \;keV$ $\alpha = 1.667$. Solving eq.(\ref{eq.maximum.energy}), the corresponding relative ion speeds are:
\begin{displaymath}
\alpha = \left\{ \begin{array}{ll}
1.079 \quad \textrm{if} \quad\beta \;<\;0.031265\\
1.667 \quad \textrm{if} \quad\beta \;\geq\;0.031265
\end{array} \right.
\end{displaymath}
In terms of ion energy, expressed in $MeV$, the $k$ coefficient takes the values:
\begin{displaymath}
k_{2} = \left\{ \begin{array}{ll}
1.0355\cdot 10^{-1} \quad [kg \; m^{-2} MeV^{-\alpha}]\quad \; \; \textrm{for} \quad  \alpha = 1.079\\
6.0138 \quad \quad \quad \;\;\; [kg \; m^{-2} MeV^{-\alpha}]  \; \quad \; \textrm{for} \quad \alpha = 1.667
\end{array} \right.
\end{displaymath}
The corresponding formula describing the radial distribution of dose, derived by \cite{zhang1985} is:
\begin{equation}
D_{\delta}(r) = C\;\frac{z^{*2}}{\beta^2}\;\frac{1}{\rho}\;\frac{1}{r^{2}}\;\frac{1}{\alpha}\;\left(\frac{\left(1- \frac{r+\theta}{R_{\max}+\theta}\right)^{\frac{1}{\alpha}}}{1+\frac{\theta}{r}}\right) \quad [MeV \; kg^{-1}],
\label{eq.rddzhang1}
\end{equation}
where the symbol denotations are the same as in eq.(\ref{eq.rddbutts}). Additionally, $\theta$ is the 'range' of an electron of energy equal to its binding potential $I = 10\;eV$: 
\begin{equation}
\theta = k_{2} \cdot I^{\alpha} \quad \left[kg\;m^{-2}\right].
\end{equation}
In our investigation to select the most appropriate phenomenological RDD formula for TST calculations, we also considered the formula of \cite{zhang1985}, but without the assumption of bound electrons, i.e. for the case $I = 0\;eV$. Then the following formula obtains:
\begin{equation}
D_{\delta}(r) = C\;\frac{z^{*2}}{\beta^2}\;\frac{1}{\rho}\;\frac{1}{r^{2}}\;\frac{1}{\alpha}\;\left(1- \frac{r}{R_{max}}\right)^{\frac{1}{\alpha}} \quad [MeV \; kg^{-1}],
\label{eq.rddzhang2}
\end{equation}

\subsection{The RDD formula of \cite{waligorski1986}}
\label{ch. rddwaligorski}
When integrated radially, Zhang's formula, eq.(\ref{eq.rddzhang1}), was found to yield about 50\% of the total value of linear energy transfer (LET) of the respective ion (see Fig. \ref{fig.LETnormal}). To correct for this discrepancy, \cite{waligorski1986} introduced a multiplicative correction factor to Zhang's formula, valid at radial distances below $10\; nm$. In developing this correction factor, tabulated values of proton LET in water and results of MC calculations of the radial distribution of dose in water around protons of different energy, were exploited. The following equation was then developed:
\begin{equation}
D_{\delta}(r) = C\;\frac{z^{*2}}{\beta^2}\;\frac{1}{\rho}\;\frac{1}{r^{2}}\;\frac{1}{\alpha}\;\left(\frac{\left(1- \frac{r+\theta}{R_{max}+\theta}\right)^{\frac{1}{\alpha}}}{1+\frac{\theta}{r}}\right)\left(1+K\left(r\right) \right) \quad [MeV \; kg^{-1}],
\label{eq.rddwaligorski}
\end{equation}
The correction factor $K(r)$ for radial distances $r < 0.1\; nm$
\begin{displaymath}
K\left(r\right) = 0, 
\end{displaymath}
and for radial distances $r \geq 0.1 \;nm$ from the ion's path
\begin{displaymath}
K\left(r\right)= L \cdot \left(\frac{r-M}{N}\right)\cdot \exp\left(-\frac{r-M}{N}\right),
\end{displaymath}
where
\begin{displaymath}
L = \left\{ \begin{array}{ll}
8 \cdot \beta^{\frac{1}{3}} \quad \quad \;\; \textrm{for} \quad \beta \leq 0.031265\\
19 \cdot \beta^{\frac{1}{3}} \quad \quad \textrm{for} \quad \beta > 0.031265\\
\end{array} \right.
\end{displaymath}

\begin{displaymath}
M = 1 \cdot 10^{-10}m, 
\end{displaymath}
and 
\begin{displaymath}
N = 1.5 \cdot 10^{-9}m + \beta \cdot 5 \cdot 10^{-9}m.
\end{displaymath}

\subsection{The RDD formula of \cite{cucinotta1997}}
\label{ch. rddcucinotta}
The radial distribution of dose formula proposed by \cite{cucinotta1997} was based on the \cite{bradt1948} formula to describe the energy spectrum of $\delta$-electrons. The number of $\delta$-electrons produced per unit pathlength by an ion of energy between $\omega$ and $\omega+d\omega$ is given by:
\begin{equation}
\frac{dn}{d\omega}=\frac{2\pi Nz^{*2}e^4}{mc^2\beta^2}\frac{1}{\omega^2} \left[1-\frac{\beta^2 \omega}{\omega_{max}}+\frac{\pi \beta z^{*2}}{137} \sqrt{\frac{\omega}{\omega_{max}}} \left(1-\frac{\omega}{\omega_{max}} \right) \right],
\end{equation}
where $\omega_{max}$ denotes the maximum energy that an ion can transfer to a free electron given by eq.(\ref{eq.maximum.energy}). \cite{cucinotta1997} added to their RDD formula the angular distribution of the secondary electrons together with a more sophisticated formula to describe the energy-range relationship for $\delta$-electrons. All formulae describing the range of $\delta$-electrons, $R$, mentioned earlier, were restricted only to the case of alanine absorber.The semi-empirical equation describing the electron range developed by \cite{tabata1972}, used by Cucinotta, allows one to use it also for other absorbers of atomic numbers ranging between $Z=6$ to $Z=92$ in the $\delta$-ray energy range $0.7 \;keV$ to $30 \;MeV$. The corresponding formula is as follows:
\begin{equation}
R_{max} = a_1\left[ \frac{1}{a_2}ln(1+a_2\frac{\omega_{max}}{m_ec^2})-\frac{a_3 \frac{\omega_{max}}{m_ec^2}}{1+a_4 \left( \frac{\omega_{max}}{m_ec^2}\right) ^{a_5}}\right] \quad \left[kg\;m^{-2}\right],
\label{eq.tabatarange}
\end{equation}
where the parameters $a_i$ ($i=1,2,...,5$) are given by simple functions of atomic number $Z$ and mass number $A$ of the absorber:
\begin{displaymath}
\begin{array}{ll}
a_1 \; = \; b_1 A\ / \bar{Z}^{b_2} \quad \left[kg\;m^{-2}\right],\\
a_2 \; = \; b_3 Z,\\
a_3 \; = \; b_4 - b_5 Z,\\
a_4 \; = \; b_6 - b_7 Z,\\
a_5 \; = \; b_8 / Z^{b_9},\\
\end{array}
\end{displaymath}
where the symbols $b_i$ ($i=1,2,...,9$) denote constants independent of absorber material. Values of these nine constants expressing $a_i$ have been determined by the group of Tabata by least-squares fitting to a total of 232 experimental points representing the extrapolated range of electrons measured in absorbers of atomic number from $6$ to $92$. The values of the constants $b_{i}$ are listed in Table \ref{tab.1}. 
\begin{table}[!ht]
\begin{center}
\caption{Values of the constants $b_i$. Table reprinted from \cite{tabata1972}.}
\begin{tabular}{c l}
\hline
$i$ & $b_i$    \\
\hline
$1$  & 0.2335\\
$2$  & 1.209\\
$3$  & 1.78 $\cdot 10^{-4}$ \\
$4$  & 0.9891 \\
$5$  & 3.01 $\cdot 10^{-4}$ \\
$6$  & 1.468 \\
$7$  & 1.180 $\cdot 10^{-2}$ \\
$8$  & 1.232 \\
$9$  & 0.109 \\
\hline
\end{tabular}
\label{tab.1}
\end{center}
\end{table}
\\
In the case where eq.(\ref{eq.tabatarange}) is applied to the absorbers which are mixtures or compounds, the atomic number and mass number should be replaced by respective average values:
\begin{displaymath}
Z_{avg}= \sum_{i}\;f_{i}Z_{i},
\end{displaymath}
\begin{displaymath}
A_{avg} = Z_{avg}\left(Z/A\right)^{-1}_{avg}, 
\end{displaymath}
where
\begin{displaymath}
\left(Z/A\right)_{avg} = \sum_{i}\;f_{i}Z_{i}/A_{i},
\end{displaymath}
and $f_{i}$ is the fraction by weight of the constituent element with atomic number $Z_{i}$ and atomic weight $A_{i}$. For water medium $Z_{avg}=7.22$ and $A_{avg}=13.0$.\\

The RDD formula developed by \cite{cucinotta1997} takes into account the radial $\delta$-ray dose, $D_{\delta -rays}(r)$, and the radial dose component arising from excitation effects, $D_{exc}(r)$:
\begin{equation}
D_{\delta}(r) = D_{\delta -rays}(r) + D_{exc}(r) \quad [MeV \; kg^{-1}].
\label{eq.rddcucinotta}
\end{equation}
The radial dose $D_{\delta-rays}(r)$ presents an inverse square dependence on radial distance $r$ from the ion's path, modified by two functions: $f_S(r)$ - at small, and $f_L(r)$ - at large distances
\begin{equation}
D_{\delta -rays}(r)=C\;\frac{z^{*2}}{\beta^2}\;\frac{1}{\rho}\;\frac{1}{r^2}\;f_S(r)\;f_L(r),
\end{equation}
where $r$ is the distance from the ion's path $[m]$, $z^*$ is the effective charge, eq.(\ref{eq.zeff}), $\beta$ is the relative speed of the ion, $C$ is the constant given by eq.(\ref{eq.constantc}), $\rho$ is the density of the medium (for water, which is assumed to be tissue-equivalent medium, $\rho = 1000 \;kg\cdot m^{-3}$). \\
The function $f_S(r)$ modifies the short-distance behaviour and is represented by:
\begin{displaymath}
f_S(r) =  \left( \frac{10^{-9}\;[m]}{ r } + (0.6 + 1.7\beta+1.1\beta^2)\right) ^{-1}.
\end{displaymath}
The function $f_L(r)$ modifies the long-distance behaviour and is represented by:
\begin{displaymath}
f_L(r) = exp \left [-\left( \frac{r}{0.37 \cdot R_{max}}\right)^2 \right].
\end{displaymath}
$D_{exc}(r)$ describes the energy transferred in excitation processes and its contribution to the total dose $D_{\delta}(r)$
is limited to low radii of less than $10\;nm$ (\cite{ponomarev2006}):
\begin{equation}
D_{exc}(r) = S(\textrm{LET})\frac{1}{\rho} \frac{\;exp\left(- \frac{r}{2d} \right)}{r^2},
\end{equation}
and
\begin{displaymath}
d = \left( \frac{\beta}{2}\right)\left( \frac{hc}{2\pi\omega_r}\right),
\end{displaymath}
where $c$ is the speed of light, $h$ is Plank's constant, $\omega_r$ is the constant for water $(\omega_r = 13\;eV)$, and $\rho$ is the density of the material $[kg/m^3]$. 

In order to yield the correct value of the ion LET, integration of eq.(\ref{eq.rddcucinotta}) is performed
\begin{equation}
\textrm{LET} = 2 \pi \int_{R_{min}}^{R_{max}}\;[D_{\delta}(r)\;+D_{exc}(r)]\;r\;dr,
\label{eq.normdose}
\end{equation}
where the cut-off value is set to $R_{min}=10^{-10}m$ (\cite{krauter1977}). $S(\textrm{LET})$ is then determined by normalizing eq.(\ref{eq.rddcucinotta}) to the total LET value through eq.(\ref{eq.normdose}):
\begin{displaymath}
S(\textrm{LET}) = \frac{\textrm{LET} - 2 \pi \int_{R_{min}}^{R_{max}}\;D_{\delta}(r)\;r\;dr}{2 \pi \int_{R_{min}}^{R_{max}}\;D_{exc}(r)\;r\;dr}.
\end{displaymath}
Thus, finally, eq.(\ref{eq.rddcucinotta}) takes the form:
\begin{equation}
D(r) = C\frac{z^*}{\beta^2}\;\frac{1}{\rho}\;f_S(r)\;f_L(r)\;\frac{1}{r^2} + S(\textrm{LET})\frac{1}{\rho} \frac{\;exp\left(- \frac{r}{2d} \right)}{r^2}.
\label{eq.rddcucinotta2}
\end{equation}

\section{Comparison of electron energy-range and RDD formulae with experimental data}
\label{ch. doseenergycomparison}
Different formulations of the energy-range relationship for $\delta$-electrons were used in the derivation of 'point-target' RDD formulae of \cite{butts1967}, \cite{zhang1985}, \cite{waligorski1986} and by \cite{cucinotta1997}. All considered RDD formulas together with their appropriate representations of the electron energy-range relationship for $\delta$-rays are listed in Table \ref{tab.radialdose}. \\

\begin{table}[!ht]
\begin{center}
\caption{RDD formulae, $D_{\delta}(r)$, considered in this work, together with respective energy-range relationships for $\delta$-electrons, $R$.}
\begin{tabular}{l c @{} }
\hline \hline 
\multicolumn{2}{c}{\rule{0pt}{3ex} \cite{butts1967}} \\
\rule{0pt}{3ex} eq.(\ref{eq.linearrange}) & $R_{max}=k_{1} \cdot \omega_{max}$  \\
\rule{0pt}{3ex} eq.(\ref{eq.rddbutts}) & $D_{\delta}(r) = C\;\frac{z^{*2}}{\beta^2}\;\frac{1}{\rho}\;\frac{1}{r^{2}}\left(1-\frac{1}{R_{max}}\right)$ \\ \\ \hline
\multicolumn{2}{c}{\rule{0pt}{3ex} \cite{zhang1985} with $I = 10\; eV$} \\
\rule{0pt}{3ex} eq.(\ref{eq.powelowrange}) & $R_{max}=k_{2} \cdot \omega^{\alpha}_{max}$  \\
\rule{0pt}{3ex} eq.(\ref{eq.rddzhang1}) & $D_{\delta}(r) = C\;\frac{z^{*2}}{\beta^2}\;\frac{1}{\rho}\;\frac{1}{r^{2}}\;\frac{1}{\alpha}\;\left(\frac{\left(1- \frac{r+\theta}{R_{max}+\theta}\right)^{\frac{1}{\alpha}}}{1+\frac{\theta}{r}}\right)$  \\ \hline
\multicolumn{2}{c}{\rule{0pt}{3ex} \cite{zhang1985} with $I = 0\; eV$} \\
\rule{0pt}{3ex} eq.(\ref{eq.powelowrange}) & $R_{max}=k_{2} \cdot \omega^{\alpha}_{max}$  \\
\rule{0pt}{3ex} eq.(\ref{eq.rddzhang2}) & $D_{\delta}(r) = C\;\frac{z^{*2}}{\beta^2}\;\frac{1}{\rho}\;\frac{1}{r^{2}}\;\frac{1}{\alpha}\;\left(1- \frac{r}{R_{max}}\right)^{\frac{1}{\alpha}}$ \\ \\ \hline
\multicolumn{2}{c}{\rule{0pt}{3ex} \cite{waligorski1986}} \\
\rule{0pt}{3ex} eq.(\ref{eq.powelowrange}) & $R_{max}=k_{2} \cdot \omega^{\alpha}_{max}$  \\
\rule{0pt}{3ex} eq.(\ref{eq.rddwaligorski}) & $D_{\delta}(r) = C\;\frac{z^{*2}}{\beta^2}\;\frac{1}{\rho}\;\frac{1}{r^{2}}\;\frac{1}{\alpha}\;\left(\frac{\left(1- \frac{r+\theta}{R_{max}+\theta}\right)^{\frac{1}{\alpha}}}{1+\frac{\theta}{r}}\right)\left(1+K\left(r\right) \right)$  \\ \hline
\multicolumn{2}{c}{\rule{0pt}{3ex} \cite{cucinotta1997}} \\
\rule{0pt}{3ex} eq.(\ref{eq.tabatarange}) & $R_{max} = a_1\left[ \frac{1}{a_2}ln(1+a_2\frac{\omega_{max}}{m_ec^2})-\frac{a_3 \frac{\omega_{max}}{m_ec^2}}{1+a_4 \left( \frac{\omega_{max}}{m_ec^2}\right) ^{a_5}}\right] $  \\
\rule{0pt}{4ex} eq.(\ref{eq.rddcucinotta2}) & $D_{\delta}(r) = C\frac{z^{*2}}{\beta^2}\;\frac{1}{\rho}\;f_S(r)\;f_L(r)\;\frac{1}{r^2} + S(\textrm{LET})\frac{1}{\rho} \frac{\;exp\left(- \frac{r}{2d} \right)}{r^2}$  \\ \\ \hline \hline
\end{tabular}
\label{tab.radialdose}
\end{center}
\end{table}

\begin{figure}[!ht]
\begin{center}
\includegraphics[width=1.0\textwidth]{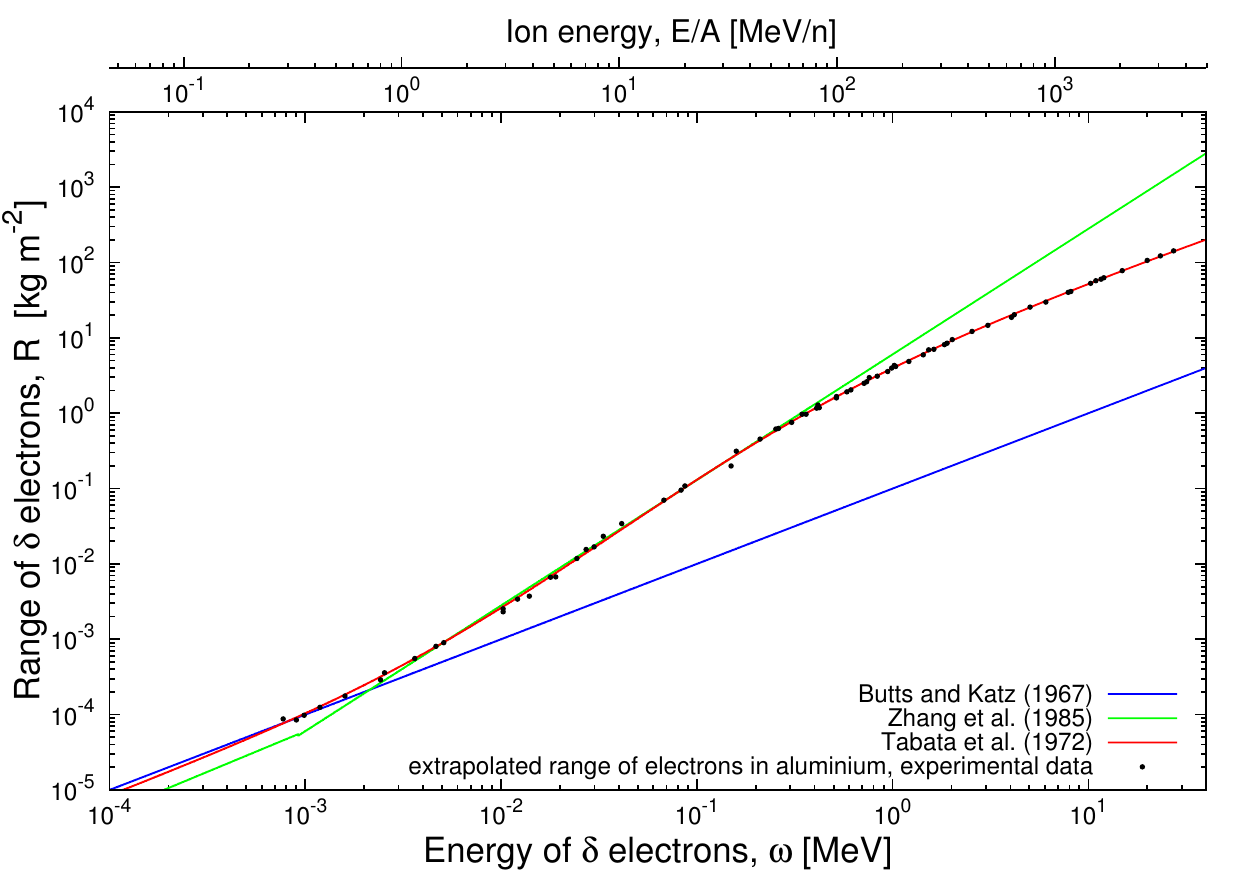}
\end{center}
\caption{Electron energy $(\omega)$ vs. range $(R)$ for electrons in aluminium. Experimental data are represented by full symbols (for sources of data, see main text). Blue line - linear approximation, green line - power law, red line - equation of \cite{tabata1972}. The upper abscissa represents the energy of the ion $(MeV/n)$ which limits the maximum $\delta$-ray range, $R_{max}$ via the kinematical constraint - eq.(\ref{eq.maximum.energy}). }
\label{fig.electronrange}
\end{figure}

\begin{figure}[!ht]
\begin{center}
\includegraphics[width=1.0\textwidth]{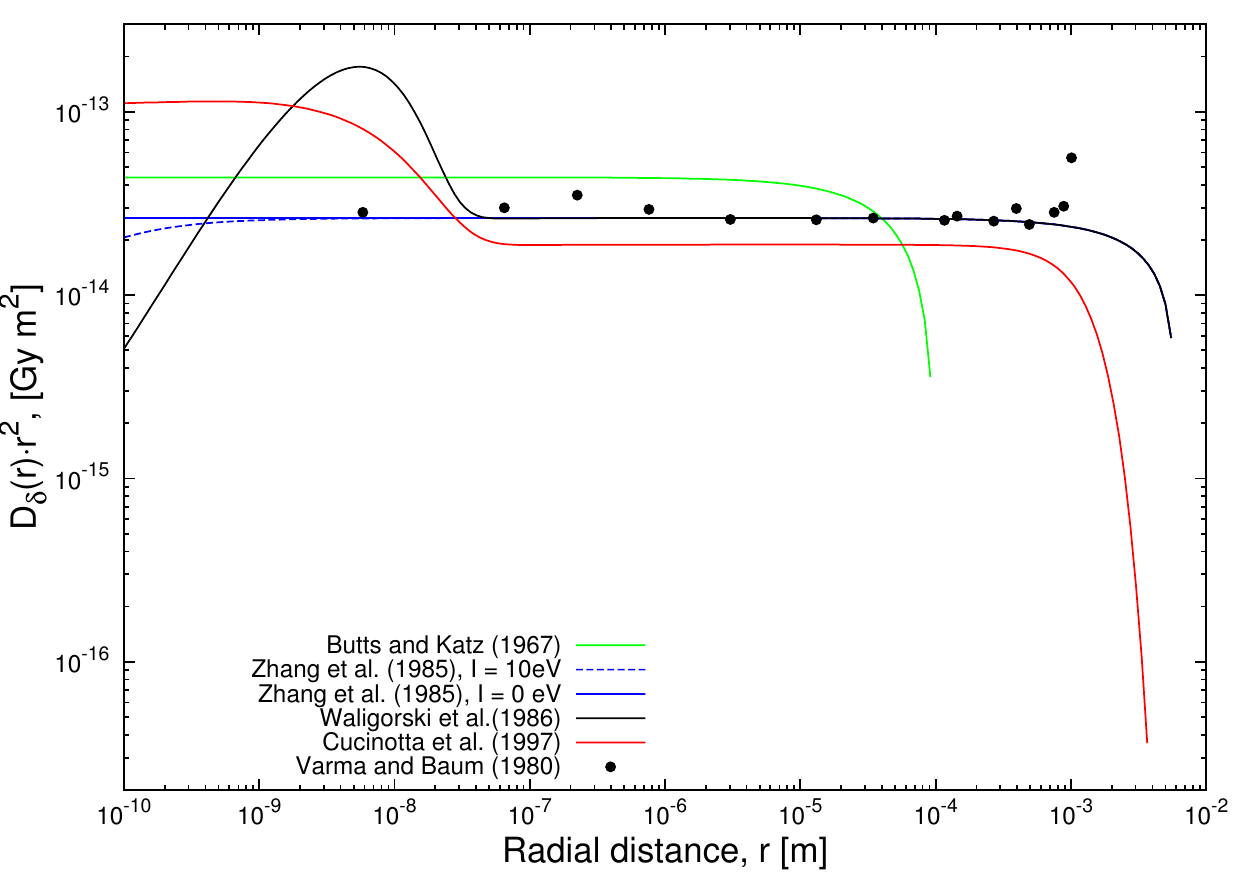}
\end{center}
\caption{Comparison between radial distributions of dose for $377 MeV/n$ neon ions measured in tissue-equivalent gas (\cite{varma1980}) and calculated in water, using the formulae by \cite{butts1967} - green line, by \cite{zhang1985}, with $I = 10\; eV$ - dashed blue line, and with $I = 0 \;eV$ - solid blue line, by \cite{waligorski1986} - black line, and by \cite{cucinotta1997} - black line. All radial dose values are multiplied by $r^2$. No error values were given in the paper of \cite{varma1980}.}
\label{fig.rddose}
\end{figure}

\begin{figure}[!ht]
\begin{center}
\includegraphics[width=1.0\textwidth]{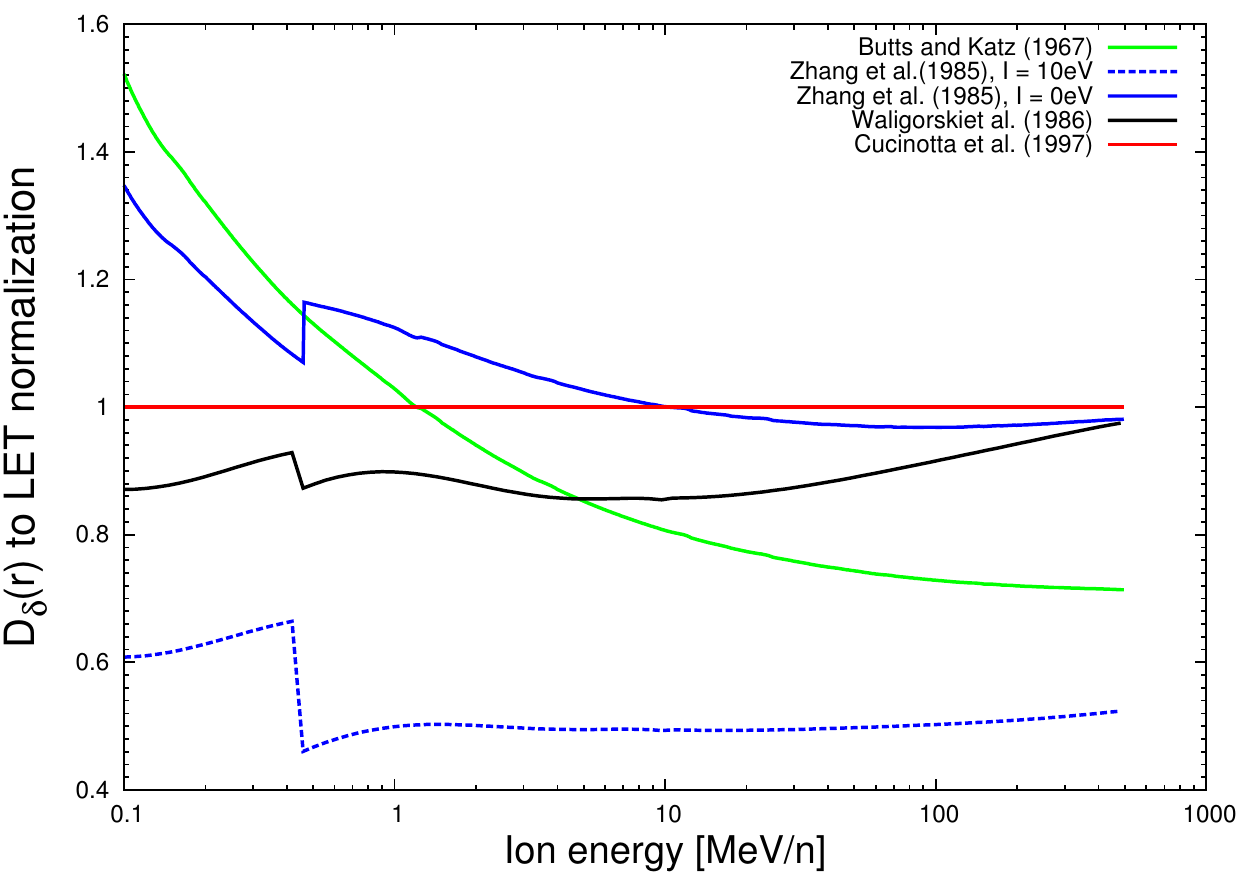}
\end{center}
\caption{Quotients of radially integrated RDD formulas, $D_{\delta}(r)$, over respective LET values from eq.(\ref{eq.letion}) versus proton energy, in $MeV/n$. Green line - formula of \cite{butts1967}; dashed blue line - formula of \cite{zhang1985} with ionization potential $I=10\;eV$; solid blue line - formula of \cite{zhang1985} with ionization potential $I=0\;eV$, eq.(\ref{eq.rddzhang2}); black line - formula of \cite{waligorski1986}; red line formula of \cite{cucinotta1997}. }
\label{fig.LETnormal}
\end{figure}

\subsection{Electron energy-range relationship}
\label{ch. energyrangecomparison}
In the derivation of the analytic formula describing the 'point-target' radial distribution of dose, $D_{\delta}(r)$, use is made of the analytic representation of the electron energy-range relationships in calculating the average dose from $\delta$-rays deposited within cylindrical shells around the path of the ion. Assumption of a linear energy-range relationship made it possible to derive analytically the first formulation of the RDD in the paper of \cite{butts1967}. \cite{zhang1985} made use of the power-law representation of the electron energy-range relationship in deriving their analytic RDD formula. The RDD formula introduced by \cite{cucinotta1997} uses a more complicated analytical expression of \cite{tabata1972} to describe the electron energy-range relationship and requires numerical radial integration to establish its 'excitation' term (cf. Table \ref{tab.radialdose}). 
In Fig. \ref{fig.electronrange} we show the experimentally measured energy-range dependence in aluminium over the electron energy range $10^{-4} - 10^{2}\;MeV$, where data of \cite{kanter1962}, \cite{lane1954}, \cite{schonland1925}, \cite{seliger1955}, \cite{agu1957}, \cite{miller1970}, \cite{grimshaw1966}, \cite{miller1968}, \cite{ebert1969}, \cite{tabata1971} and \cite{harder1967} are superimposed. By applying the ion energy-dependent kinematic constraint, eq.(\ref{eq.maximum.energy}), we also represent this data via the maximum range of the $\delta$-rays, $R_{max}$, in aluminium versus ion energy (upper abscissa in Fig. \ref{fig.electronrange}) and plot the linear and power-law expressions used to derive the RDD formulae of \cite{butts1967}, of \cite{zhang1985}, of \cite{cucinotta1997}, and the expression used by Tabata et al. The energy-range formula of \cite{butts1967} is valid only for ions of low energy ($0.35\; MeV/n - 0.7\; MeV/n$). The two-step power-law approximation applied by \cite{zhang1985} formula is satisfactory for ions of energies ranging between about $2\; MeV/n$ and $100 \;MeV/n$. The semi-empirical formula of Tabata et al.(1972) describes the $\delta$-ray range quite well over the ion energies shown ($0.1 - 1000\; MeV/n$). \\

\subsection{Radial distribution of dose (RDD)}
\label{ch. dosecomparison}
The rather scarce measurements of radial distributions of dose around accelerated ions, performed in hydrogen, neon or tissue equivalent gas, have been gathered by \cite{katz1991}. In Fig. \ref{fig.rddose}, as an example, we compare the radial distribution of dose formulae of \cite{butts1967}, \cite{zhang1985}, \cite{waligorski1986}, and of \cite{cucinotta1997}, calculated for $377\;MeV/n$ neon ions with the radial distribution of dose measured by \cite{varma1980} for this ion, performed in tissue equivalent gas. Results of measurements close to the ion's path and at the maximum range of $\delta$-rays are uncertain. To facilitate this comparison, the formulae and results of measurements have been shown in Fig. \ref{fig.rddose} multiplied by $r^2$ (thus a strict $1/r^2$ dependence is seen as a horizontal line). The observed differences in the calculated maximum ranges of $\delta$-electrons follow from differences in the assumed energy-range relationships, as demonstrated in Fig. \ref{fig.electronrange}.\\ \\

\subsubsection{Radial integration of the RDD and the value of ion LET}
\label{ch. letnormalization}
For consistency of track structure modelling, integration of the RDD over all radial distances from the cut-off value, $R_{min}$, to the maximum range of $\delta$-electrons, $R_{max}$, should yield the correct value of the ion's LET$_{\infty}$. Quotients of radially integrated RDD formulae, over respective ion LET values, calculated from eq.(\ref{eq.letion}), as a function of ion energy, are shown in Fig. \ref{fig.LETnormal}. Radial integration of the \cite{butts1967} formula overestimates of the respective LET values by up to $50\%$ at low ion energies and increasingly underestimates it, by some $30\%$, at higher ion energies. Considerable underestimation of LET by up to $50\%$ is seen in the case of Zhang's formula with $I=10\;eV$ (cf. Table \ref{tab.radialdose}). However, assuming $I=0\;eV$ in that formula leads to much better agreement. The RDD formula of \cite{waligorski1986} generally somewhat underestimates the LET values, and the formula introduced by \cite{cucinotta1997} correctly reproduces the respective LET values by its designed addition of a LET-dependent 'excitation' term, $S(\textrm{LET})$ (cf. Table \ref{tab.radialdose}). For the first three RDD formulations the calculated ratio strongly depends on the choice of the lower limit of integration (due to the $1/r^2$ dependence of the RDD, its integral tends to infinity as $r$ tends to zero). Throughout this work the lower limit of integration was always set to $R_{min}=10^{-10}m$ (\cite{krauter1977}).

\section{The average radial distribution of dose}
\label{ch. averagedose}

In order to consider the effect after particle irradiation, the track structure model needs, as an input, the radial dose distribution, $D_{\delta}(r)$, around the ion's path, i.e. any one of the four formulas presented in Table \ref{tab.radialdose}. Next, this 'point-target' RDD is transformed into the radial distribution of dose (RDD$_{avg}$) averaged over the sensitive target - a short cylinder of radius $a_{0}$ oriented along the direction of the ion, the centre of which is at a distance $t$ from the ion's path, as follows:

\begin{equation}
\overline{D}(t) = \frac{1}{\pi a_{0}^{2}} \int_{t-a_{0}}^{t+a_{0}} D(r)\; \phi (r, t, a_{0})\;dr \quad \quad [Gy],
\label{eq.averagedose}
\end{equation}
where $\phi(r,t,a_{0})$ is the length of an arc over which integration occurs within the volume of the cylinder at distance $t$ (\cite{waligorski1988}). A detailed description concerning the calculation of $\phi(r,t,a_{0})$ can be found in Appendix \ref{ch. appendixb}. The shape of the average radial dose distribution (the 'extended-target' RDD, $\overline{D}(t)$) will depend on the choice of the 'point-target' formula, $D_{\delta}(r)$, as given in Table \ref{tab.radialdose}, and on the value of $a_{0}$. In order to distinguish the 'extended-target' dose from 'point-target' dose in equations we use the symbol $t$ instead of $r$ to denote the radial distance of the centre of the cylinder representing the sensitive target, from the ion's path. \\

\begin{figure}[!ht]
\begin{center}
\includegraphics[width=0.9\textwidth]{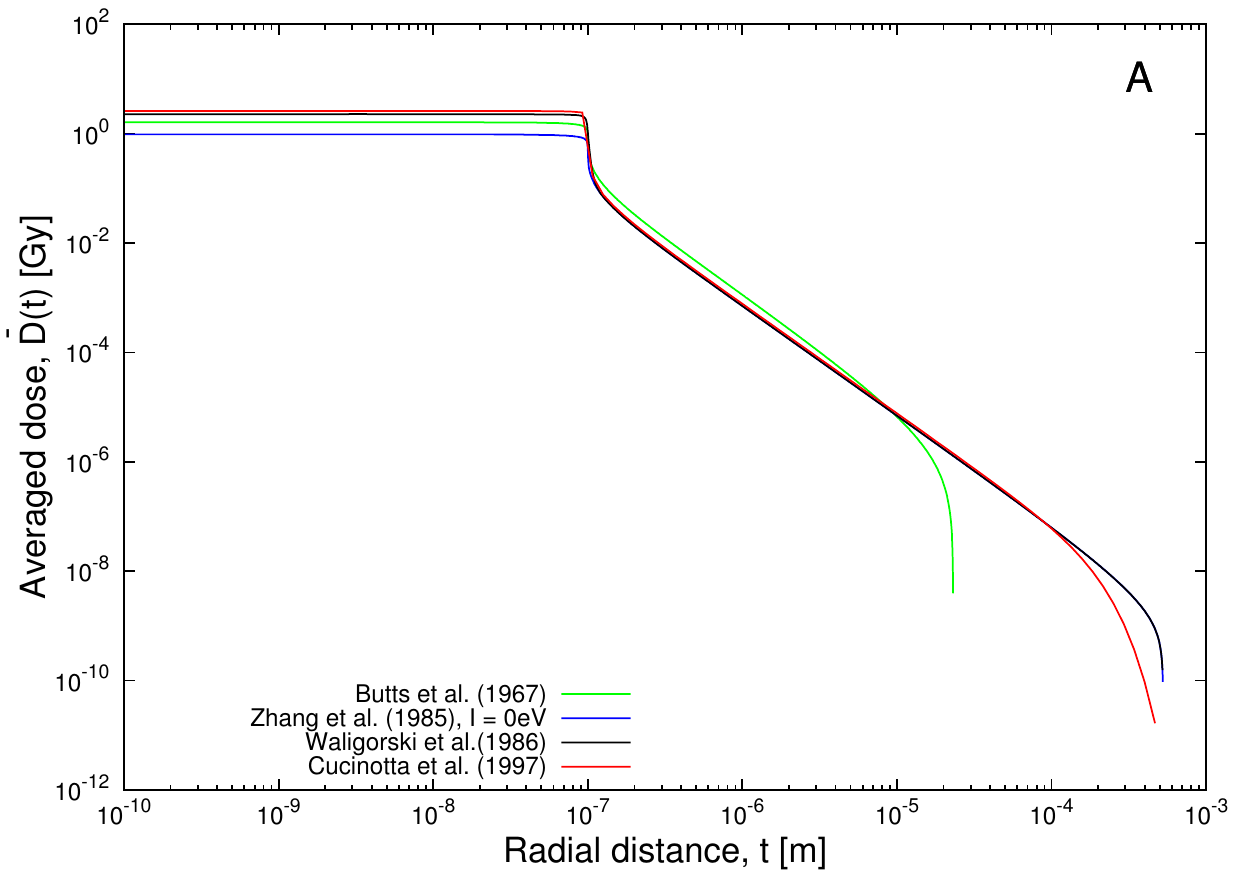}
\includegraphics[width=0.9\textwidth]{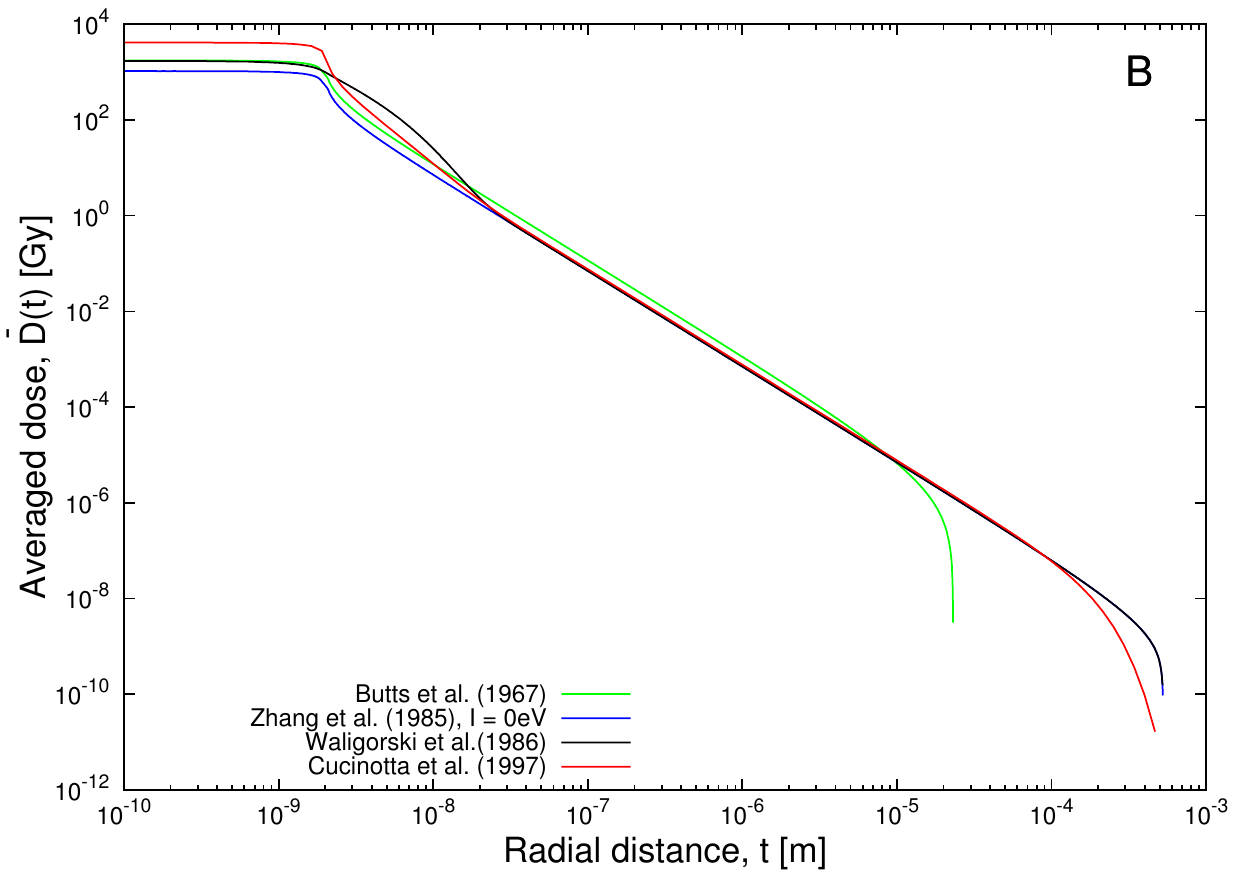}
\end{center}
\caption{Radial distributions of dose averaged over cylindrical targets of diameter $a_0 = 1.0\cdot10^{-7}m$ (panel A) and $a_0 = 2.0\cdot10^{-9}m$ (panel B), calculated for $100\;MeV/n$ protons in water. Green lines - RDD formula of \cite{butts1967}; blue lines - formula of \cite{zhang1985}, with $I=0\;eV$, eq.(\ref{eq.rddzhang2}); black lines - formula of \cite{waligorski1986}; red lines - formula of \cite{cucinotta1997}. }
\label{fig.esr}
\end{figure}

\begin{figure}[!ht]
\begin{center}
\includegraphics[width=1.0\textwidth]{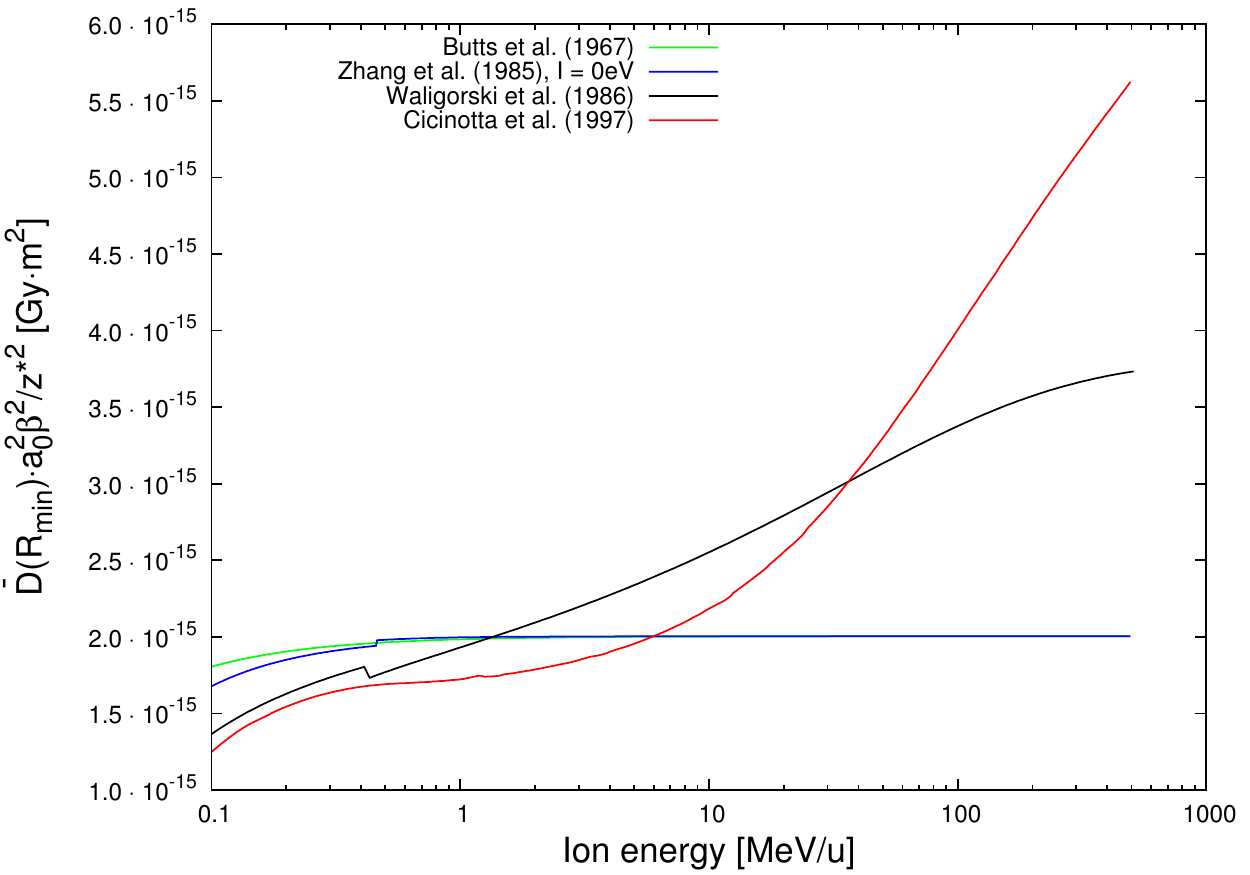}
\end{center}
\caption{'Scaling' of the RDD$_{avg}$, as represented by the almost constant value of the 'plateau' part of the average dose distribution (cf. Fig. \ref{fig.esr} and main text), irrespective of ion energy or target size, if scaling of average dose by $z^{*2}/ a_0^2\beta^2$ is applied. Green line: formula of \cite{butts1967}; blue line - formula of \cite{zhang1985}, with $I=0\;eV$, eq.(\ref{eq.rddzhang2}); black line - formula of \cite{waligorski1986}; red line - formula of \cite{cucinotta1997}. Calculations prepared for $a_0 = 1.0\cdot10^{-6}\;m$, hence $10^{-10}m<R_{min}<10^{-6}m$, cf. Fig \ref{fig.esr}.}
\label{fig.plateauaverdose}
\end{figure}

\subsection{Averaging the RDD over sensitive sites of different sizes}
\label{ch. averagingdose}
To illustrate the dependence of RDD$_{avg}$ on the radius of the sensitive site $a_0$, we calculated radial distributions of dose around protons of energy $100\; MeV/n$, averaged over $a_0=1.0\cdot10^{-7}m$ (Fig. \ref{fig.esr}, panel A) and $a_0=2.0\cdot10^{-9}m$ (Fig. \ref{fig.esr}, panel B), according to eq.(\ref{eq.averagedose}). These values of $a_0$ are reported by \cite{cucinotta1997} and \cite{waligorski1989} and may represent the 'sensitive sites' in \emph{E. coli B$_{s-1}$} spores and alanine, respectively. The average radial distributions have been calculated in water. Characteristically, a 'plateau' in average dose is observed at radial distances $t$ below $a_0$ on a logarithmic plot, followed at larger distances by a $1/t^2$-dependence, similar to that of the 'point-target' RDD, until the maximum $\delta$-ray range occurs. As may be expected, the choice of a given RDD formula is better reflected when averaging over a smaller sensitive site. Thus, effect of the different RDD formulae should be better distinguished when analysing experimental data on the response of alanine ($a_0=2.0\cdot10^{-9}\;m$) than that concerning bacterial spore survival ($a_0=1.0\cdot10^{-7}m$) after ion irradiation. Differences in the values of the 'plateau' region, due to the choice of the RDD formulae, are also more pronounced in the case of the smaller value of $a_0$ (in panel B of Fig. \ref{fig.esr}). The averaged radial dose distributions are not normalized to the respective values of the ion LET. \\

\subsection{Scaling of averaged radial distributions of dose}
\label{ch. averageddosescaling}
Katz was the first to notice that if average dose distributions, $\overline{D}(t)$, calculated for various ion species of different energies, are multiplied by $a_0^2\beta^2/z^{*2}$ and plotted against $t/a_0$, the RDD$_{avg}$ curves lie atop of one another (\cite{katz1972}). In his analysis the original radial dose distribution of \cite{butts1967} was used and the common 'plateau' values of averaged dose multiplied by $a_0^2\beta^2/z^{*2}$ occurred around $2\cdot10^{-15}\;Gy\;m^2$. This observation allowed Katz to introduce \begin{equation}
\kappa = \frac{D_{0}a_{0}^{2}}{2 \cdot 10^{-15} \;\; Gy \cdot m^{2}},
\label{eq.kappa1}
\end{equation}
as a parameter in his cellular Track Structure Theory (\cite{katz1978}). We performed similar calculations for the studied $D_{\delta}(r)$ formulae, averaging them over a site of $a_0 = 1.0\cdot10^{-6}\;m$ (as this value corresponds to the size of the mammalian cell) and plotting the 'plateau' value of the respective RDD$_{avg}$, multiplied by $a_0^2\beta^2/z^{*2}$, versus ion energy, in Fig. \ref{fig.plateauaverdose}. Above about $0.5 \;MeV/n$ the formulae of \cite{butts1967} and of \cite{zhang1985} (with ionization potential $I=0\;eV$) yield a constant 'plateau' value of $2\cdot10^{-15}\;Gy\;m^2$. The formulae of \cite{waligorski1986} and of \cite{cucinotta1997} do not show such scaling, due to the presence of the correction factor $K(r)$ and to the presence of an additive second term, respectively. The contributions of either of these corrections to the radial distribution of dose depend on the energy of the ion and on its charge.\\ \\

\begin{figure}[!ht]
\begin{center}
\includegraphics[width=0.85\textwidth]{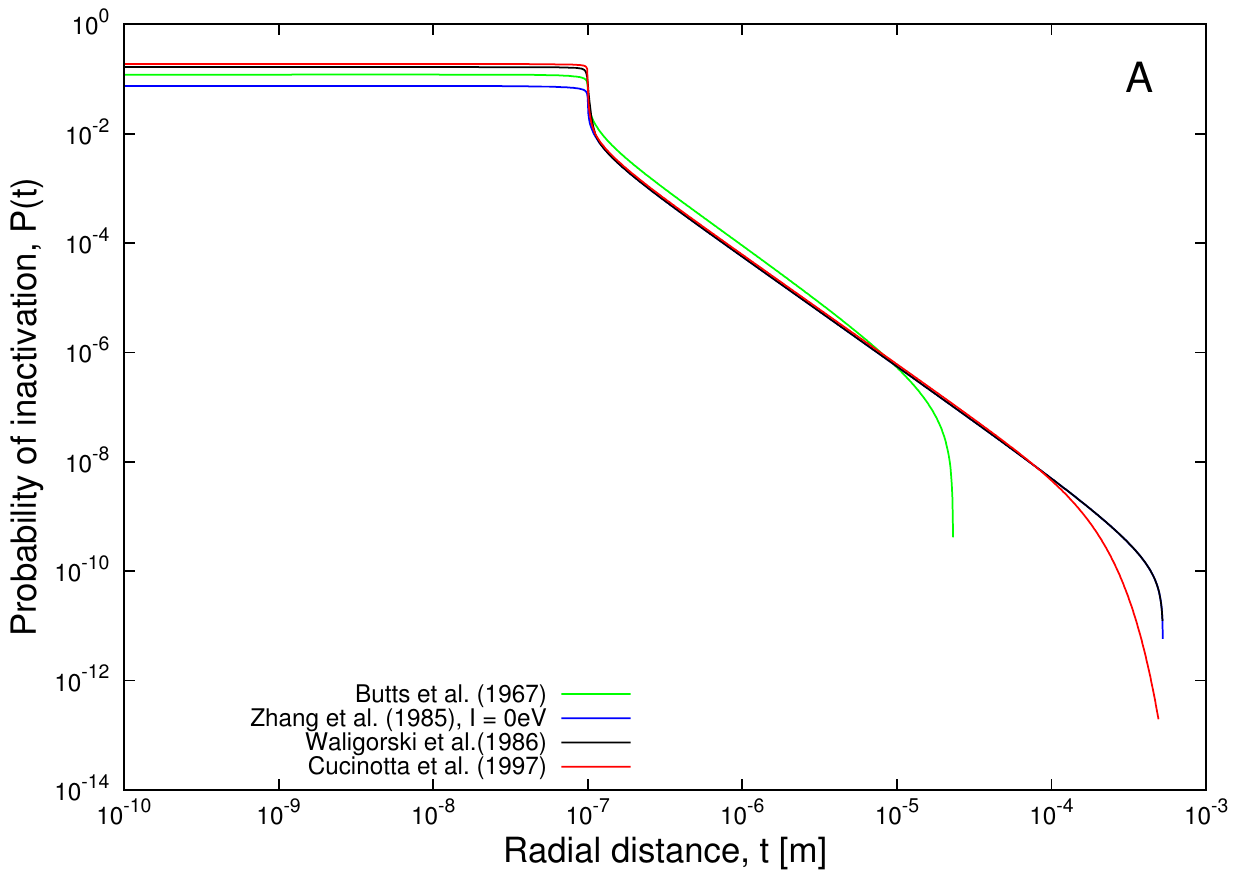}
\includegraphics[width=0.85\textwidth]{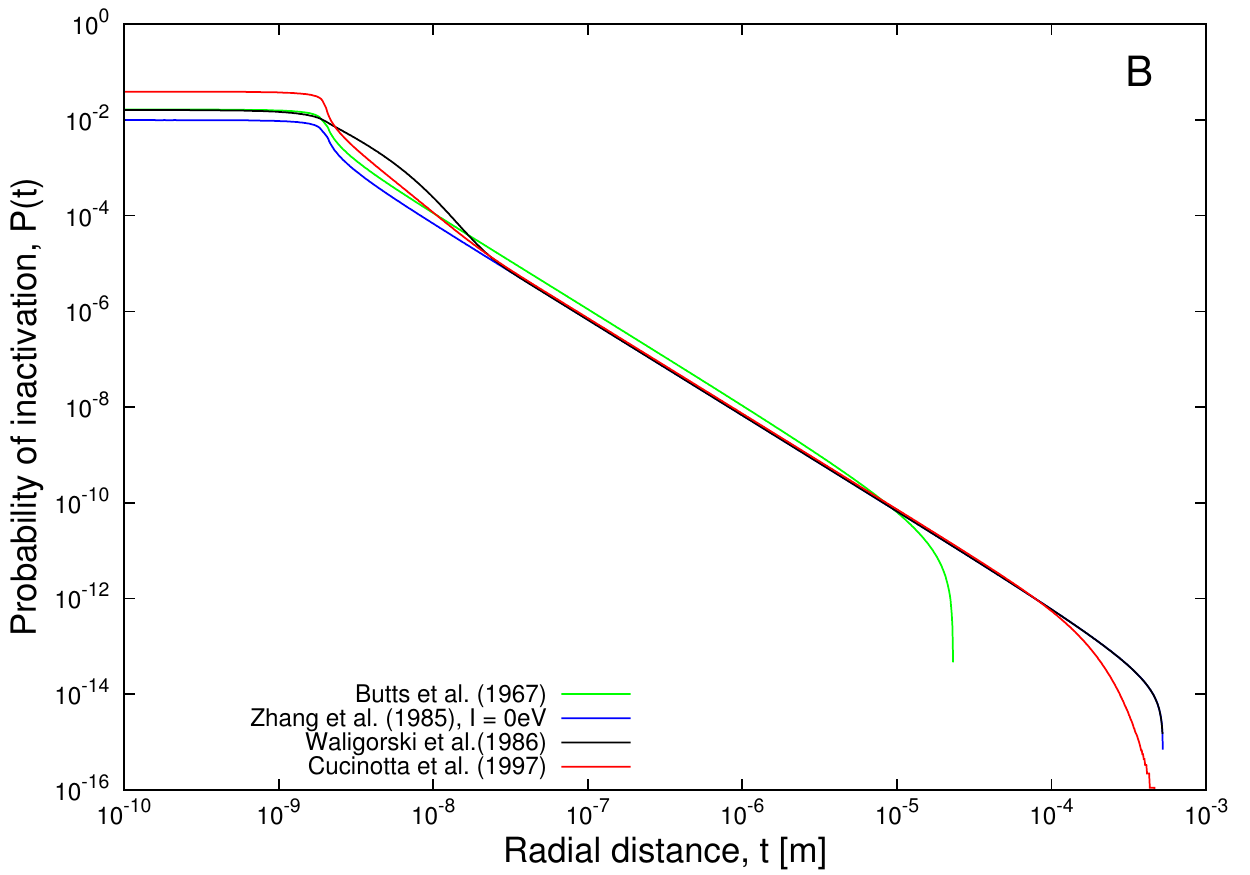}
\end{center}
\caption{Radial distributions of activation probability calculated using results from Fig. \ref{fig.esr} and the $P(t)$ formula given by eq.(\ref{eq.probability}). Calculations were performed for $100\;MeV/n$ protons in water. Model parameters are: $m=1$, $D_{0}=12.6\;Gy$, $a_0=1.0\cdot10^{-7}m$ (panel A) and $m=1$, $D_{0}=10.5\cdot10^5\;Gy$, $a_0=2.0\cdot10^{-9}m$ (panel B). Green lines - RDD formula of \cite{butts1967}; blue lines - formula of \cite{zhang1985}, with $I=0\;eV$, eq.(\ref{eq.rddzhang2}); black lines - formula of \cite{waligorski1986}; red lines - formula of \cite{cucinotta1997}.}
\label{fig.radialprob}
\end{figure}

\section{The radial distribution of activation probability}
\label{ch. radialprobability}

The basic assumption of track structure theory is that the effect arising from a dose of secondary electrons, averaged over the target volume, is equivalent to that of the same average dose of reference radiation. The radial distribution of averaged dose is converted into a radial distribution of activation probability through the application of the dose-response function after reference radiation, given by eq.(\ref{eq.singlehitmultipletarget}) or eq.(\ref{eq.chit}) (depending on the type of the considered system):
\begin{align}
P(t) & = P\left(c,\overline{D}(t)/D_0\right), \quad \textrm{or}  \nonumber \\
& = P\left(m,\overline{D}(t)/D_0\right).
\label{eq.probability}
\end{align}
Fig. \ref{fig.radialprob} illustrates the basic assumption of track structure theory: the biological effect arising from a dose of $\delta$-electrons, averaged over the target volume, is equivalent to that of the same average dose of reference radiation. This implies that, for $m > 1$, the radial dose distribution around the ion path is responsible for any enhancement of effect against reference radiation. Simulated radial distributions of activation probability presented in Fig. \ref{fig.radialprob} for four different RDDs were calculated for $100\;MeV/n$ protons in water.\\

\newpage
\section{Calculation of activation cross-section in one-hit systems}
\label{ch. crosssection}
The action cross-section, which may describe formation of radicals in alanine detectors, and in case of bacteria spores, their inactivation, can be obtained by integration of the radial distribution of activation probability over all radial distances from ion's path till the maximum range of $\delta$-electrons:
\begin{equation}
\sigma = \int_{R_{min}}^{R_{max}}\;2\pi P(t)\; t\; dt \quad \quad [m^2],
\label{eq.crossection}
\end{equation}
where $R_{min}$ is the cut-off value which, in our calculations, is taken to be $1.0\cdot10^{-10}m$ (\cite{krauter1977}). Cross-sections calculated in such a manner are representative of track segment conditions of irradiation with heavy ions. Here, 'track segment conditions' imply that the thickness of the detector is much smaller than the range of the ion, or that the relative speed $\beta$ of the ion remains constant.\\

\begin{figure}[!ht]
\begin{center}
 \begin{tabular}{cc}
\includegraphics[width=0.5\textwidth]{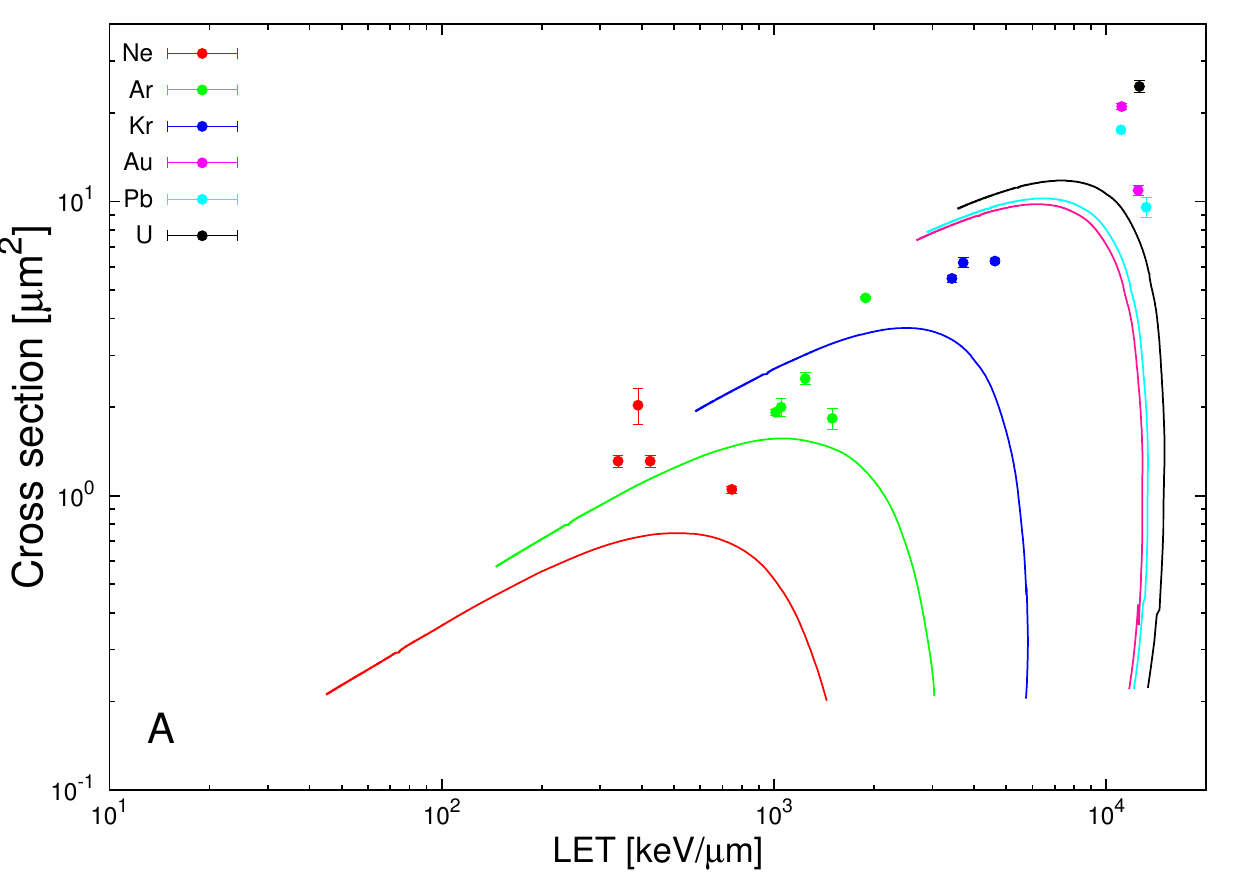} &
\includegraphics[width=0.5\textwidth]{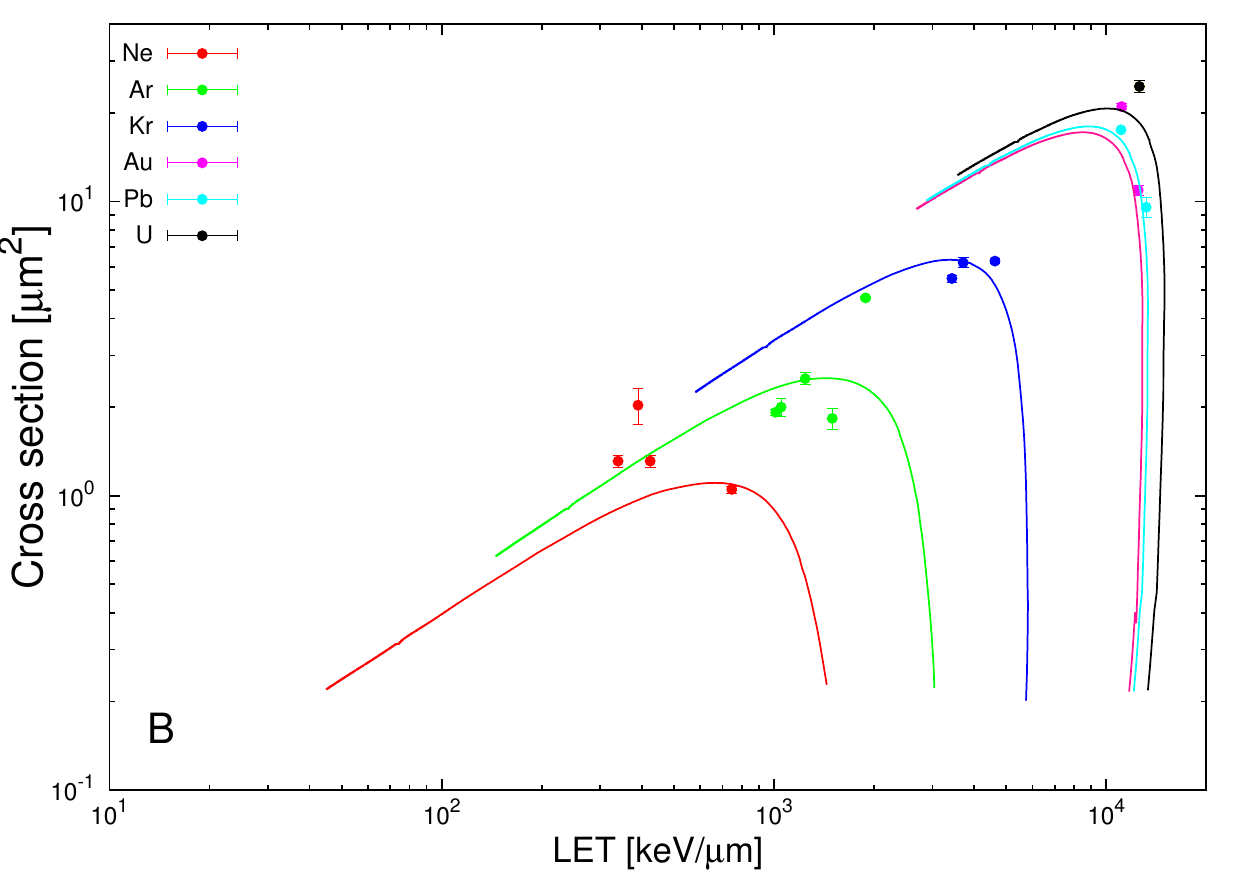} \\
\includegraphics[width=0.5\textwidth]{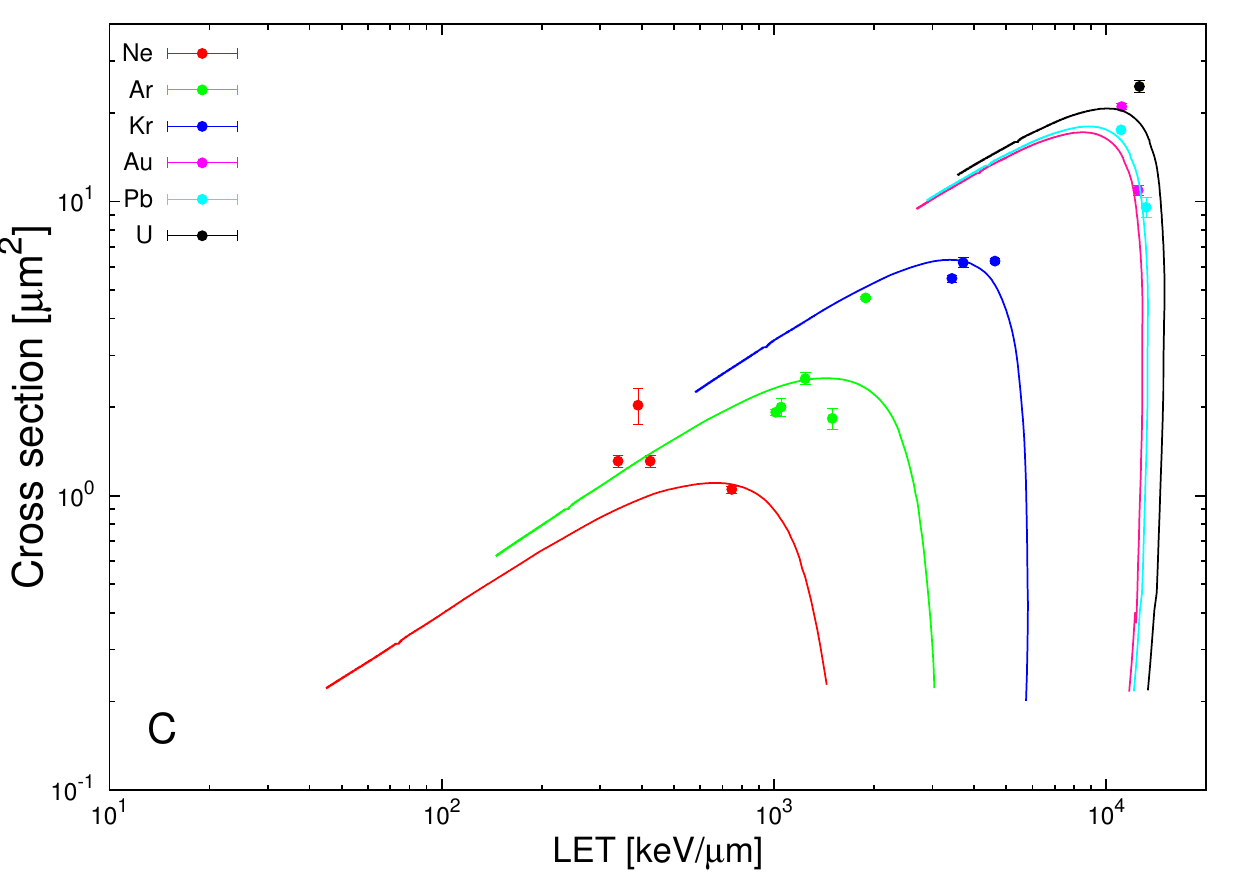} &
\includegraphics[width=0.5\textwidth]{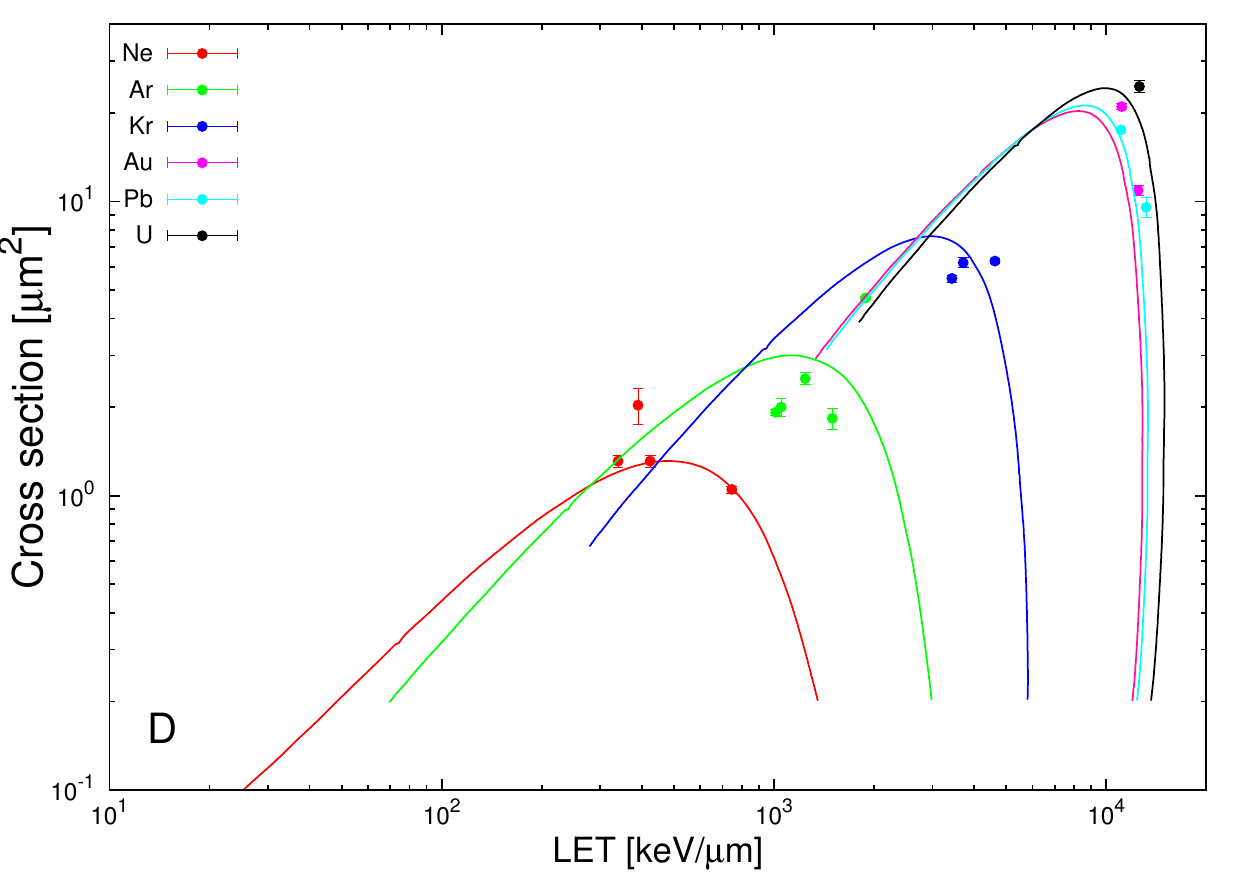} \\
 \end{tabular}
\end{center}
\caption{Calculated dependences of inactivation cross-sections of \emph{E. coli} B$_{s-1}$ spores versus LET of different ions, using model parameters: $c = 1$, $D_0 = 12.6\;Gy$, $a_0 = 1.0\cdot10^{-7}m$ (\cite{cucinotta1997}, panel D). Experimental data are taken from \cite{schafer1994}. The 'point-target' RDD formulae, $D_{\delta}(r)$,used in these calculations are those of \cite{butts1967} - panel A; of \cite{zhang1985}, with $I=0\;eV$, eq.(\ref{eq.rddzhang2}) - panel B; of \cite{waligorski1986} - panel C; and of \cite{cucinotta1997} - panel D.}
\label{fig.sigma}
\end{figure}

\subsection{Bacteria \emph{E. coli} B$_{s-1}$}
\label{ch. bacteria}
To calculate the cross-section for inactivation of \emph{E. coli} B$_{s-1}$ spores versus ion LET using the different $a_0$-averaged $D_{\delta}(r)$ formulae, we used eq.(\ref{eq.crossection}) with calculated values of LET for different ions using eq.(\ref{eq.letion}). Experimental data obtained by irradiation with Ne, Ar, Kr, Au, Pb and U ions, are taken from \cite{schafer1994}. The choice of radiosensitivity parameters to represent \emph{E. coli} B$_{s-1}$ spores ($m=1$, $D_0=12.6\;Gy$, and $a_0=1.0\cdot10^{-7}m$) is that of \cite{cucinotta1997} who used their $D_{\delta}(r)$ formula and obtained results shown in Fig. \ref{fig.sigma}, panel D. Results shown in the remaining panels of this figure have been obtained from calculations performed using the $D_{\delta}(r)$ formulae of \cite{butts1967} (panel A), of \cite{zhang1985}, with ionization potential $I=0\;eV$, i.e. eq.(\ref{eq.rddzhang2}) (panel B), and of \cite{waligorski1986} (panel C), for the same set of radiosensitivity parameters. As may be seen in Fig. \ref{fig.sigma} and in Table \ref{tab.chi2}, all formulae, except that of \cite{butts1967}, appear to represent the experimental data to a similar degree. Due to the relatively large size of the sensitive site, $a_0$, representing \emph{E. coli} B$_{s-1}$ spores, no significant differences are apparent between results of calculations and the experimental data for the remaining $D_{\delta}(r)$ formulae, except for calculations using the original formula of \cite{butts1967} which could perhaps be improved by a different choice of radiosensitivity parameters (results not shown).  \\

\begin{figure}[!ht]
\begin{center}
 \begin{tabular}{cc}
\includegraphics[width=0.5\textwidth]{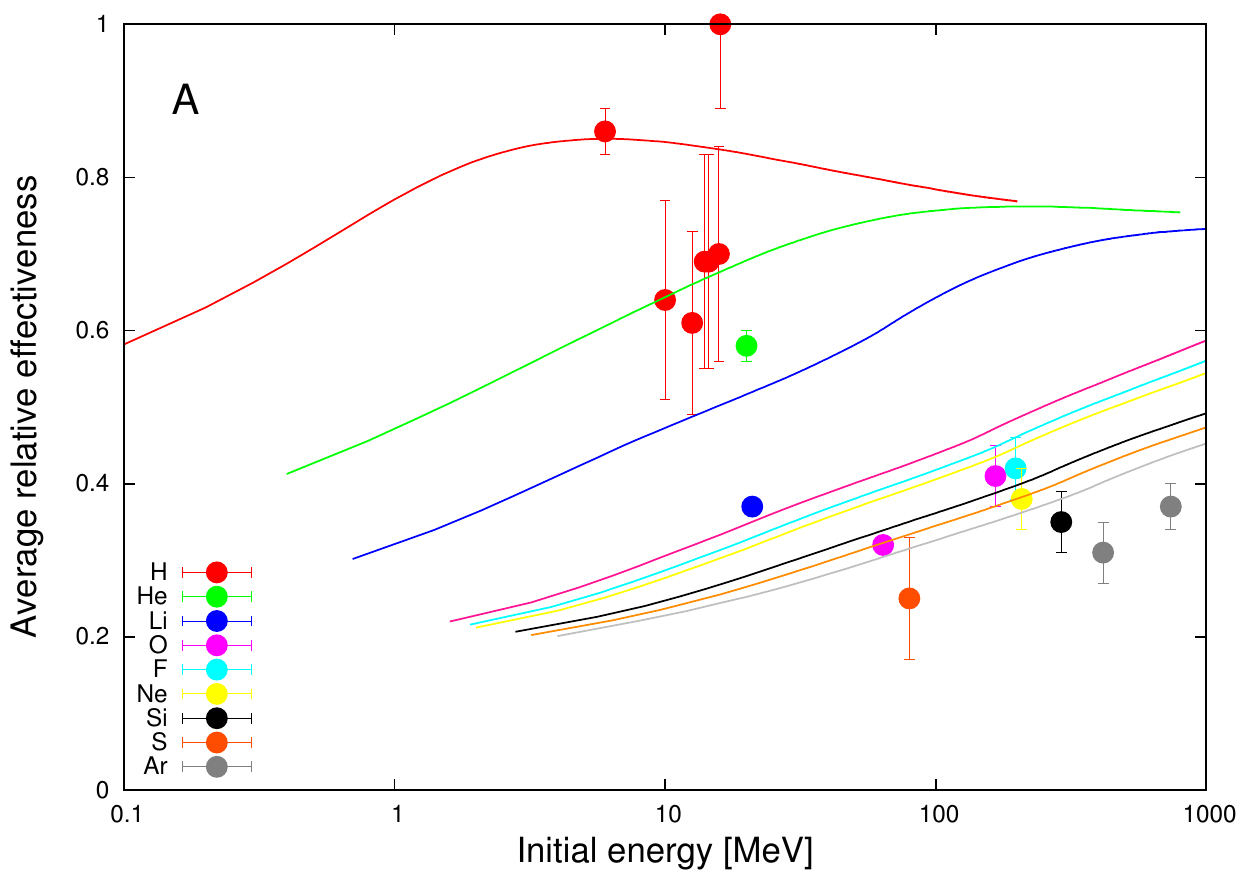} &
\includegraphics[width=0.5\textwidth]{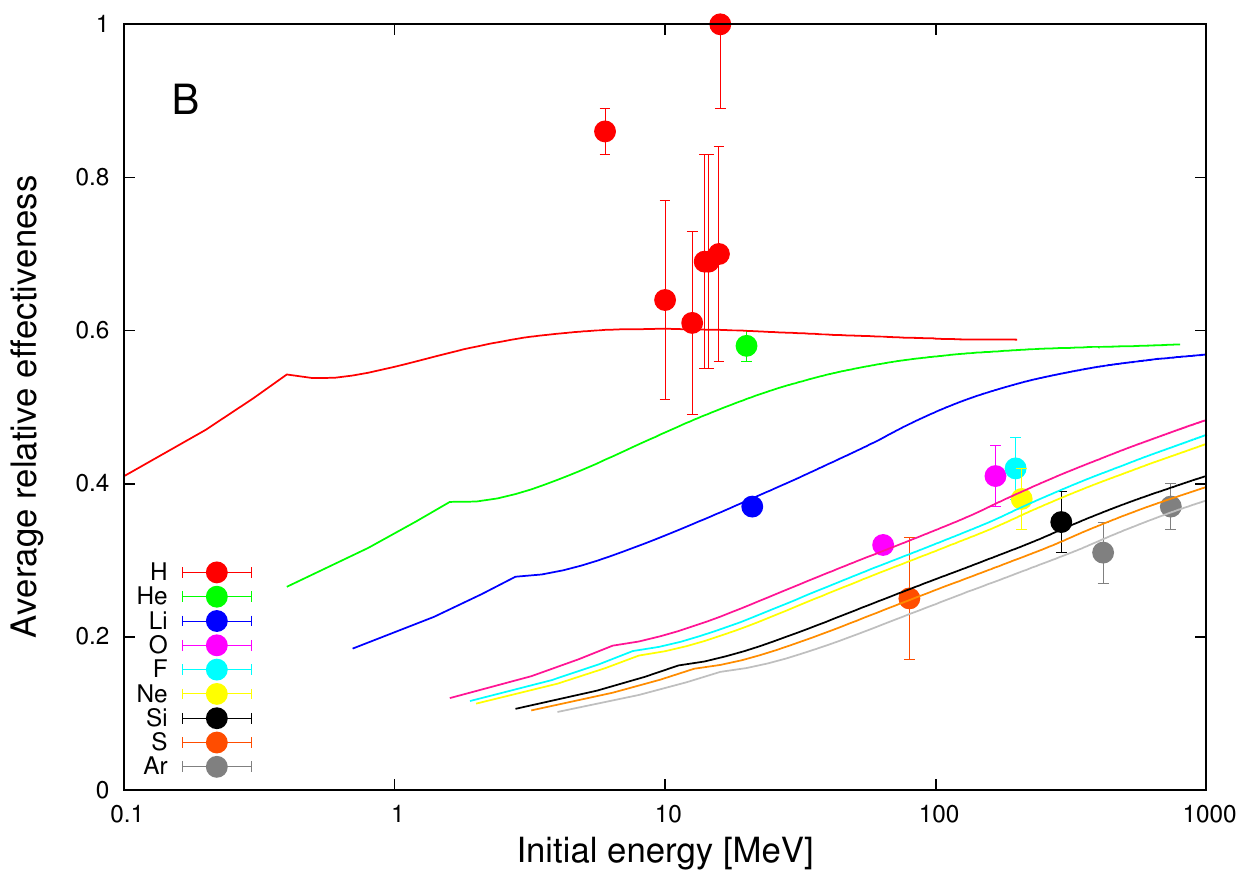} \\
\includegraphics[width=0.5\textwidth]{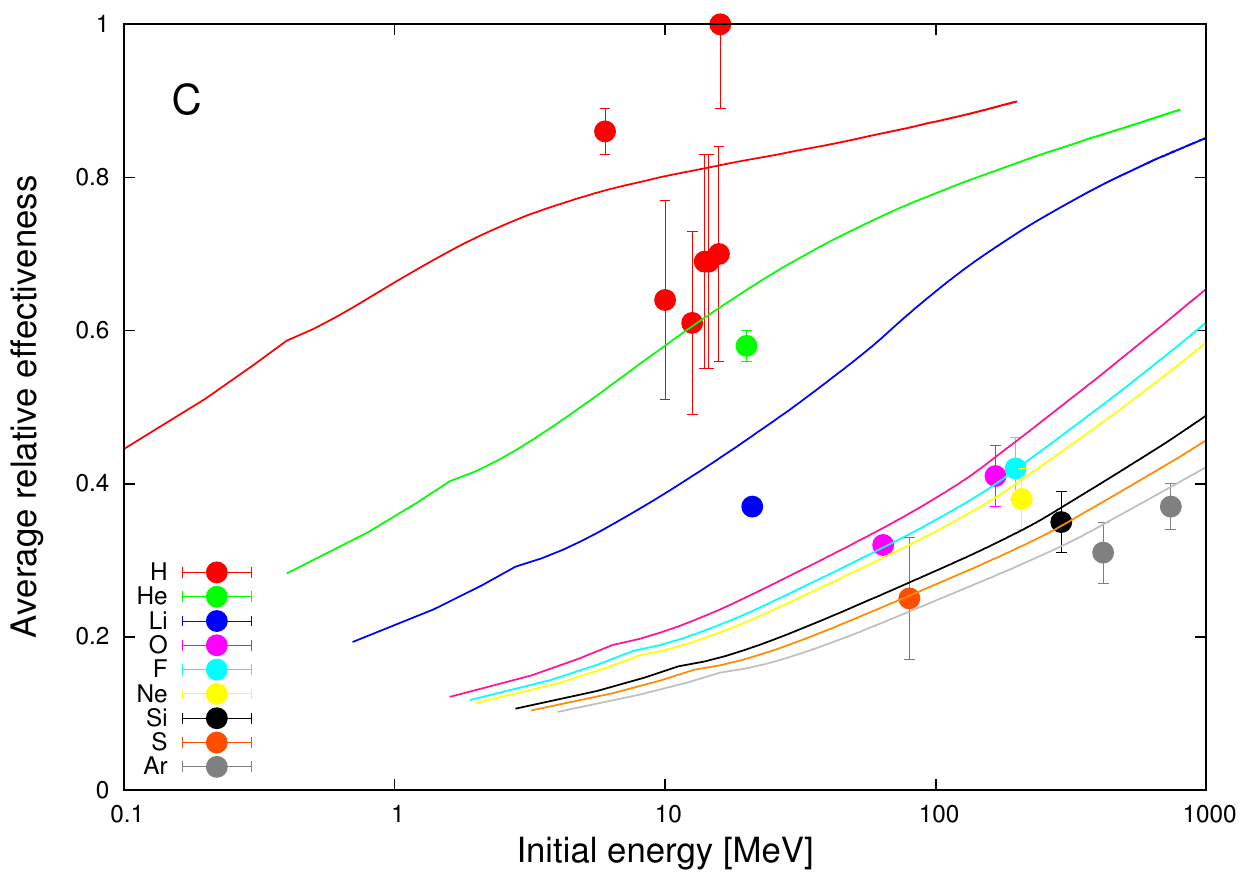} &
\includegraphics[width=0.5\textwidth]{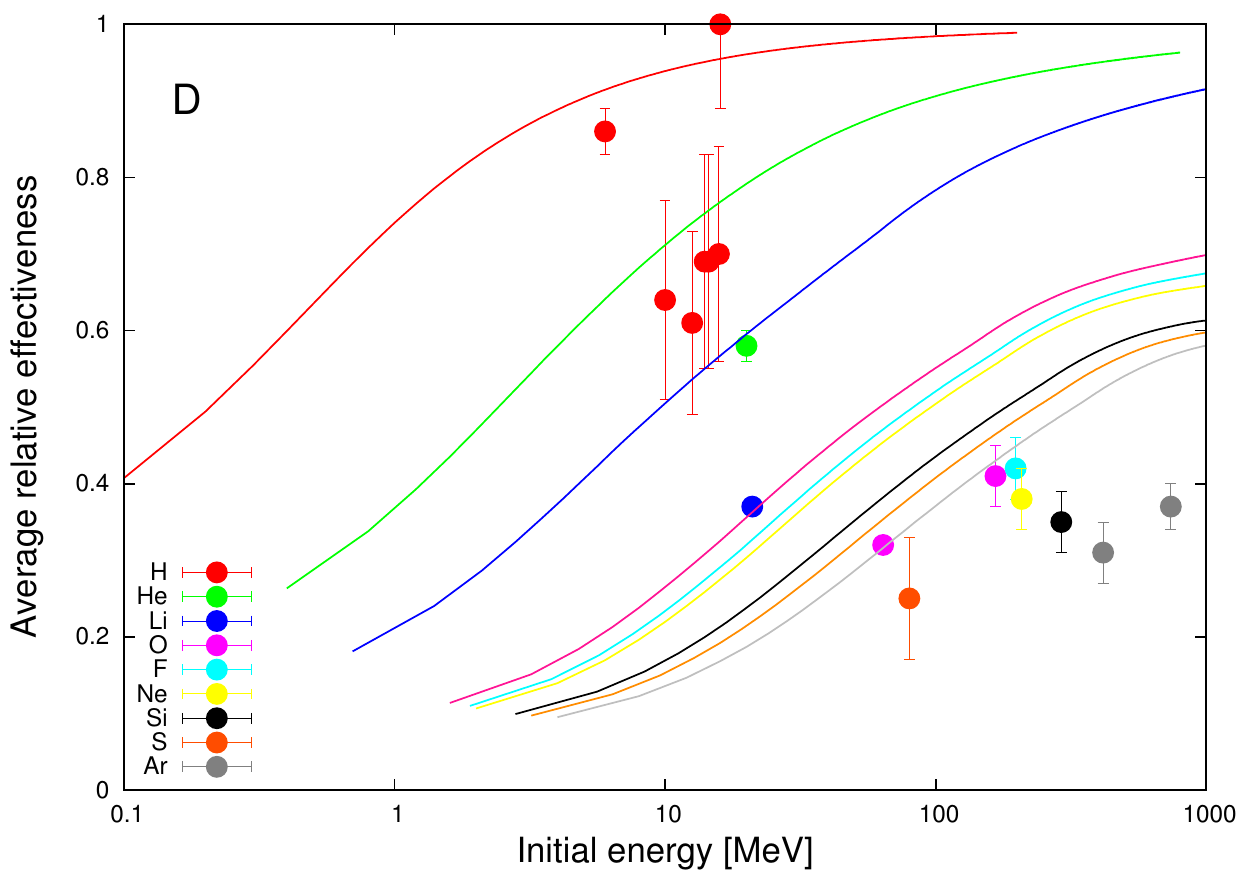} \\
 \end{tabular}
\end{center}
\caption{Calculated dependences of average relative effectiveness of the alanine detector irradiated by different ions, versus their initial energy, using model parameters: $c = 1$, $D_0 = 1.05\cdot10^5\;Gy$, $a_0 = 2.0 \cdot10^{-9}m$ (\cite{waligorski1986}). Experimental data is also quoted from \cite{waligorski1986}. The 'point-target' RDD formulae, $D_{\delta}(r)$,used in these calculations are those of \cite{butts1967} - panel A; of \cite{zhang1985}, with $I=0\;eV$, eq.(\ref{eq.rddzhang2}) - panel B; of \cite{waligorski1986} - panel C; and of \cite{cucinotta1997} - panel D.}
\label{fig.alanina}
\end{figure}

\subsection{The alanine detector}
\label{ch. alaninedetectors}
The signal, $S_{\gamma}(D_{\gamma})$, of the '1-hit' detector after its exposure to uniformly distributed dose, $D_{\gamma}$, of reference radiation according to eq.(\ref{eq.chit}) takes the form:
\begin{equation}
S_{\gamma}(D_{\gamma}) = 1-e^{-\frac{D_{\gamma}}{D_0}}.
\label{eq.1hitsignalreference}
\end{equation}
The signal, $S_i(D_i)$, observed in a thin segment of the '1-hit' detector (track segment irradiation) after bombardment by a heavy ion of fixed values of energy (or $\beta$), charge and fluence, can be calculated as follows (\cite{waligorski1989}) :
\begin{equation}
S_i(D_i) = 1-e^{-\frac{\sigma\cdot D_i}{\textrm{LET}}},
\label{eq.1hitsignal}
\end{equation}
where $D_i$ denotes the ion dose and $\sigma$ is the action cross-section calculated with the aid of the average radial distribution of dose,RDD$_{avg}$, and is given by eq.(\ref{eq.crossection}). The relative efficiency in a track segment is defined as the ratio of the detector signal after ion dose, $S_i(D_i=D)$, and the signal after the same value of dose of reference radiation ($X$-rays or $\gamma$-rays), $S_{\gamma}(D_{\gamma}=D)$:
\begin{equation}
\textrm{RE} = \frac{S_i(D)}{S_{\gamma}(D)}.
\label{eq.relativeefficiency1}
\end{equation}
After substituting eq.(\ref{eq.1hitsignalreference}) and eq.(\ref{eq.1hitsignal}) into eq.(\ref{eq.relativeefficiency1}), expanding the exponential function as a Taylor series and considering only the linear parts of the series, we obtain:
\begin{equation}
\textrm{RE} = \frac{\sigma \cdot D_0}{\textrm{LET}}.
\label{eq.relativeefficiency2}
\end{equation}
In order to reconstruct the experimental conditions where the thickness of the detector exceeds the ion range, we evaluate the average relative efficiency, RE$_{avg}$, by calculating the average value of $\sigma$ over the total path length of the ion, i.e. over consecutive track segments corresponding to gradually decreasing energy (or relative speed $\beta$) of the ion, in the continuous slowing-down approximation (CSDA) (\cite{waligorski1989}). In eq.(\ref{eq.relativeefficiency2}) both the value of LET and $\sigma$ depend on the relative speed $\beta$ of the ion.

Average relative effectiveness of the alanine detectors after heavy ion irradiation was calculated using eq.(\ref{eq.relativeefficiency2}) with the set of parameters: $c = 1$, $D_0 = 1.05\cdot10^5\;Gy$, and $a_0 = 2.0\cdot10^{-9}m$, proposed by \cite{waligorski1986}. Sources of experimental data: irradiation with H, He, Li, O, F. Ne, Si and Ar ions, have also been quoted in their paper, and results of their calculations are shown in Fig. \ref{fig.alanina}, panel C. Calculations based on the same set of parameters and using the $D_{\delta}(r)$ formulae of \cite{butts1967}, of \cite{zhang1985}, with ionization potential $I=0\;eV$, i.e. eq.(\ref{eq.rddzhang2}), and of \cite{cucinotta1997}, are presented in panels A, B and D, respectively. As may be seen in Fig. \ref{fig.alanina} and Table \ref{tab.chi2}, the best overall fit, both to the lighter ion and heavier ion data appears to be provided by the expressions of \cite{waligorski1986} and of \cite{zhang1985}, however with possible underestimation of measured relative effectiveness for protons and more energetic heavier ions. Calculations based on radial dose distributions by \cite{cucinotta1997} and by \cite{butts1967} appear to better represent the measured relative efficiency after irradiation with protons of higher energies.

Due to its small radius of the sensitive site, $a_0$, analysis of model calculations representing the relative effectiveness of the alanine detector offers better insight into the choice of the most appropriate analytical formula to represent the 'point-target' RDD, $D_{\delta}(r)$, as the impact of the radial dose distribution formula on the TST results is more pronounced than in the case of inactivation of \emph{E. coli} B$_{s-1}$ spores (cf. Fig. \ref{fig.sigma} and Fig. \ref{fig.alanina}). On the other hand, averaging the $D_{\delta}(r)$ over the much larger radius $a_0$ representing the \emph{E. coli} B$_{s-1}$ bacterial spores tends to eliminate the effect of the multiplicative correction factor introduced by \cite{waligorski1986} which acts at distances below $10\;nm$, hence the similarity of calculated results displayed in panel B and panel C of Fig. \ref{fig.sigma}.

\section{Selection of RDD formula for Cellular Track Structure Theory calculations}
\label{ch.bestradialdose}
We performed our analysis of '1-hit' detectors ($c=1$), represented in track structure modelling by two other parameters: $D_0$ and $a_0$, to later apply of our results to Katz's four-parameter cellular track structure model. Our selection of the most appropriate RDD formula is intimately related to the principles of Katz's track structure theory, as we intend to apply this formula in calculations required in ion radiotherapy planning where the predictive capacity of Katz's cellular track structure model can be best exploited. Inherent in Katz's approach to track structure modelling is extensive application of various scaling features within his model: scaling of absorber density to that of water, application of 'effective charge' $z^*$ of the ion, eq.(\ref{eq.zeff}), representation of ion LET by proton LET, eq.(\ref{eq.letion}), representation of cross-section and RBE data versus $z^{*2}/\beta^2$ rather than LET. Far more subtle is Katz's observation (\cite{katz1971}) that if all the calculated doses averaged over the target size, RDD$_{avg}$, are scaled by $a_0^2\beta^2 /z^{*2}$, a fairly constant value (of about $ 2 \cdot 10^{-15} Gy\cdot m^2$, in water) over the 'plateau' parts of RDD$_{avg}$ is observed over wide ranges of $a_0$ and of ions of different charges $Z$, and speeds $\beta$ (or energies). The existence of such scaling made it possible for Katz to propose $\kappa = D_0\;a_0^2/(2 \cdot 10^{-15}\; Gy\cdot m^2$) as a parameter in his four-parameter cellular track structure model (\cite{katz1971}). Here, non-trivially, $\kappa$ implies a simple relation between the size of the sensitive site in a detector ($a_0$) and its radiosensitivity ($D_0$).

\begin{table}[!ht]
\begin{center}
\caption{Values of $\chi^2$ reflecting the difference between TST model calculations and experimental data. For \emph{E. coli} B$_{s-1}$ spores the definition of $\chi^2$ was based on the difference of logarithmic values of calculated and measured cross-sections for bacteria spores, and logarithmic values of errors of experimental data, $\chi^2=\sum_i \left( \frac{log(\sigma_{th})_i-log({\sigma_{exp}})_i}{log(\vartriangle  \sigma_{exp})_i} \right)^2 $. For the alanine detector the definition of $\chi^2$ was based on the difference between the calculated and measured values of relative efficiency of alanine detectors, and values of errors of experimental data, $\chi^2=\sum_i \left( \frac{{\sigma_{th}}_i-{\sigma_{exp}}_i}{\vartriangle  {\sigma_{exp}}_i } \right)^2 $. For each RDD formula the same number of experimental points within each system was considered.}
\begin{tabular}{l c c}
\hline
$\quad \quad \quad$ RDD $\quad \quad \quad \quad \quad \quad$ {\footnotesize Eq. No} & {\footnotesize \emph{E. coli} B$_{s-1}$ spores }& {\footnotesize  Alanine detector }\\
\hline \\
{\footnotesize \cite{butts1967} $\quad \quad \quad \quad \;\;\;$ eq.(\ref{eq.rddbutts})} & {\footnotesize 203.09}  & {\footnotesize  376.43}\\
{\footnotesize \cite{zhang1985}, $I=0\;eV$ $\quad$ eq.(\ref{eq.rddzhang2})}& {\footnotesize 11.42} & {\footnotesize 106.29}\\
{\footnotesize \cite{waligorski1986} $\quad \quad \quad \quad$eq.(\ref{eq.rddwaligorski})}& {\footnotesize  11.42 } & {\footnotesize 122.11}\\
{\footnotesize \cite{cucinotta1997} $\quad \quad \quad \quad$ eq.(\ref{eq.rddcucinotta2})}& {\footnotesize  25.16 } & {\footnotesize 1175.35}
 \\
\hline
\end{tabular}
\label{tab.chi2}
\end{center}
\end{table}

In our search for a most suitable phenomenological formulation of the 'point-target' radial distribution of dose, $D_{\delta}(r)$, we found that the analytically simple formula of Zhang, with $I= 0\;eV$, i.e. eq.(\ref{eq.rddzhang2}), best reproduces experimentally measured radial distributions of dose (Fig. \ref{fig.rddose}), reproduces accurately enough the ion LET (Fig. \ref{fig.LETnormal}) and fulfils the requirement of scaling with respect to the by $a_0^2\beta^2/z^{*2}$ parameter (Fig. \ref{fig.plateauaverdose}) by yielding the constant plateau values of averaged dose of about $2\cdot 10^{-15} Gy\cdot m^2$ over a wide range of ion energies. When tested (using the $\chi^2$ goodness of fit criterion, see Table \ref{tab.chi2}) against experimentally measured inactivation cross-sections of \emph{E. coli} B$_{s-1}$ spores (Fig. \ref{fig.sigma}) and of average cross-sections for stopping particles in alanine (Fig. \ref{fig.alanina}), this formula appears to best fit the heavier ion data (Fig. \ref{fig.alanina}, panel B), indicating, as expected, that a 'radioresistant' 1-hit detector with a small sensitive target ($D_0 = 1.05\cdot10^5\; Gy$, $a_0 = 2.0\cdot10^{-9}m$) is able to better distinguish between various formulations of $D_{\delta}(r)$. The difference with which various formulations of $D_{\delta}(r)$ appear to represent the response of '1-hit' detectors after irradiation by light ions (H and He in Fig. \ref{fig.alanina}) is a matter for further consideration.

While the modified RDD formula developed by \cite{zhang1985}, eq.(\ref{eq.rddzhang2}), like all other phenomenological formulations analyzed, does not fully meet our postulated requirements, we believe it to be most suitable in our further work as it is quite simple, appears to fulfil these requirements better than the other formulae and, above all, demonstrates the $a_0^2\beta^2/z^{*2}$ scaling which we believe to be essential in accurate calculations of the response of cells using the cellular track structure model with $\kappa$ as one of the four parameters of this model.


\chapter[Formulation of cellular Track Structure Theory]{Formulation of cellular Track Structure Theory}
\label{ch. results2}

In this section we shall discuss the transition from the three-parameter 'full' model discussed in Section \ref{ch. results1} to the four-parameter 'approximated' cellular Track Structure version of this model. The cellular Track Structure Theory (TST) was originally developed by Katz and co-workers (\cite{katz1971}) as a four-parameter analytical model able to predict and quantify the complex dependence of cellular survival and RBE on the properties of the detector, as specified by four detector parameters, and of the energetic ion, as specified by its charge and energy, via its relative speed $\beta$.
 
In the cellular Track Structure Theory one of the three model parameters used in the 'full' model, namely the size of the sensitive site $a_0$, is replaced by two other parameters. These are: $\sigma_0$ which represents the plateau (or saturation) value of the action cross-section, proportional to $\pi a_{0}^{2}$, and $\kappa$, which, as we shall see later, implies a simple relation between the size of the sensitive site in the cell nucleus, $a_0$, and its radiosensitivity, $D_0$. Two model parameters $m$ and $D_0$ remain from the 'full' version of TST. Introduction of $\sigma_0$ and $\kappa$ instead of $a_0$, enables one to approximate the cross-section given by eq.(\ref{eq.crossection}) (see previous section) by a simple formula. In the three-parameter track structure theory, calculation of the action cross-section requires time-consuming double integration in equations eq.(\ref{eq.averagedose}) and eq.(\ref{eq.crossection}). In the cellular TST, the applied approximation of the action cross-section does not require any integration, which makes the model robust and highly applicable to massive calculations required in radiotherapy planning.

In this section we focus only on the four-parameter cellular Track Structure Theory. We show the transition between the three-parameter TST and four-parameter cellular TST. We give a complete description of the cellular Track Structure Theory. In particular, we reconstruct the algorithm for approximation of the action cross-section over the 'track-width' regime, which has never been fully published. In these model calculations, we apply the radial dose distribution of \cite{zhang1985}, with $I=0\;eV$, eq.(\ref{eq.rddzhang2}), which we found to best fulfil the model's scaling requirements as shown in Section \ref{ch.bestradialdose} and in this chapter. Next, based on this new algorithm, we perform a model analysis of two extensive biological data sets published by the groups from the National Institute of Radiological Science in Japan (NIRS). Four model parameters ($m,\; D_0, \; \sigma_0$ and $\kappa$) of the cellular TST, representing the survival endpoint in normal human skin fibroblasts irradiated with heavy ions, are best-fitted to the complete set of the experimental data published by \cite{tsuruoka2005}. We also find two sets of model parameter describing the survival of Chinese hamster cells V79 irradiated under hypoxic or aerobic conditions, from data published by \cite{furusawa2000}. We compare TST-calculated survival curves for normal human skin fibroblasts with those evaluated experimentally. We analyze the influence of model parameters on cellular TST predictions of RBE. We show model predictions of the RBE-LET dependences for both cell lines and compare them with experimental results. We also calculate the OER-LET dependence for V79 cells.

\section{Approximation of the activation cross-section}
\label{ch.approximation}

\begin{figure}[!ht]
\begin{center}
\includegraphics[width=0.87\textwidth]{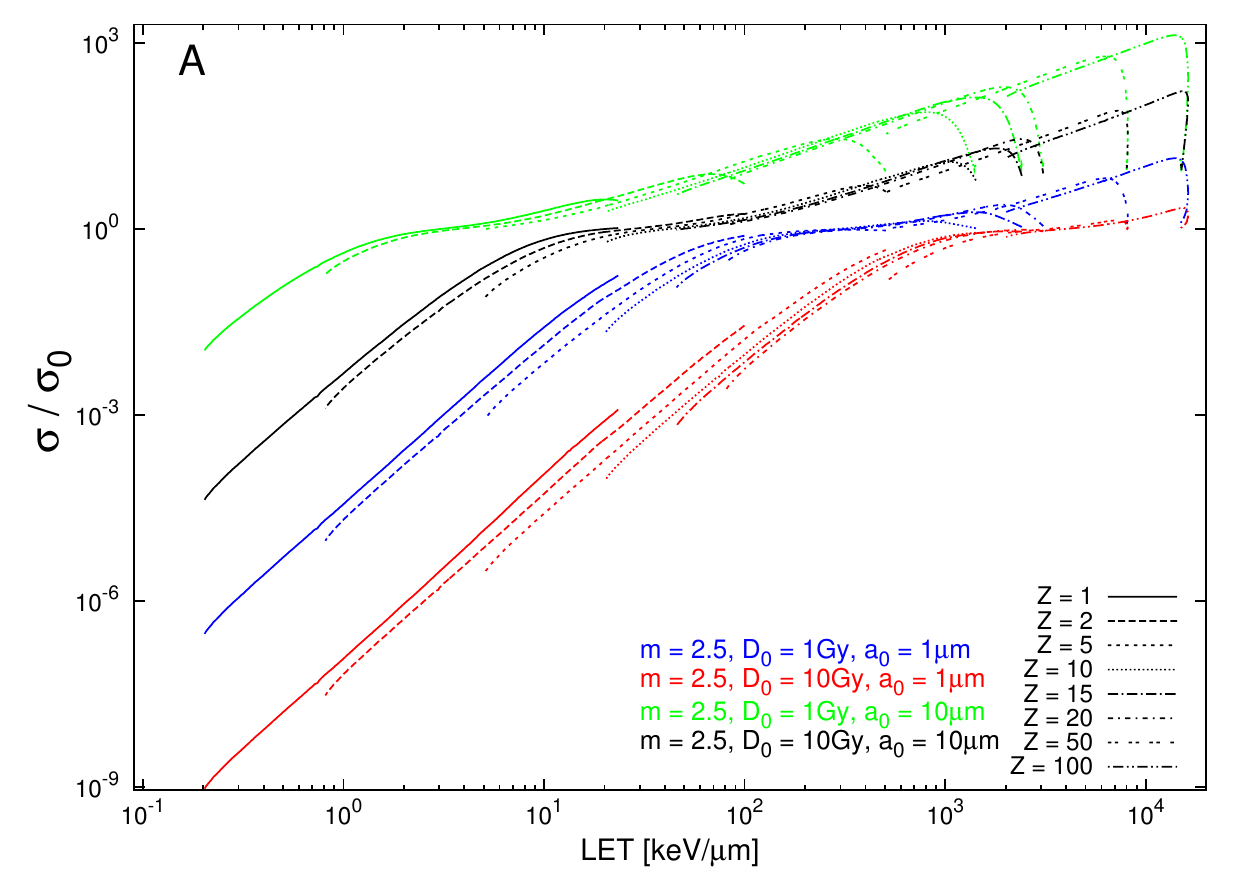}
\includegraphics[width=0.87\textwidth]{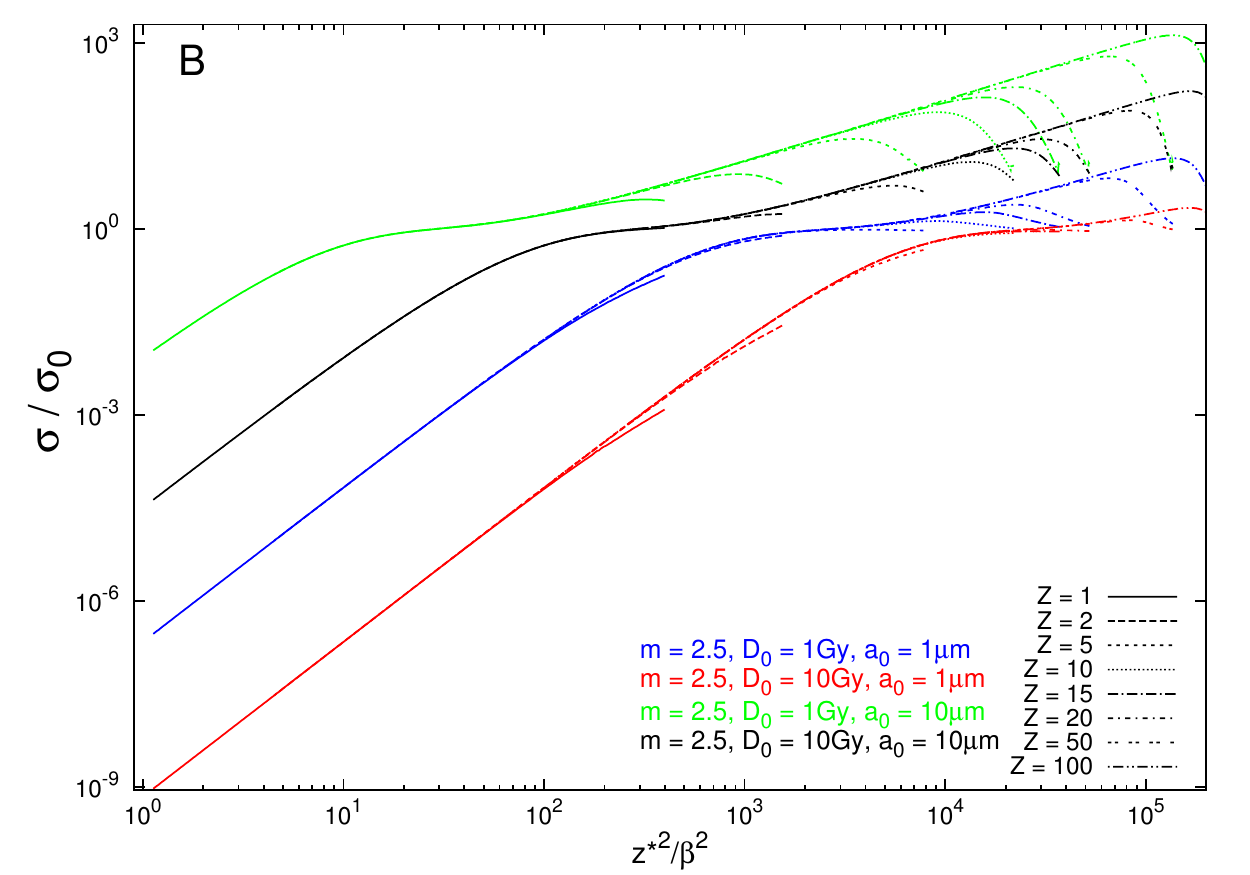}
\end{center}
\caption{Calculated values of the single-particle action cross-section, given by eq.(\ref{eq.crossection}), normalized to the saturation or 'plateau' value and plotted as a function of LET (panel A) or $z^{*2}/\beta^2$ (panel B). The families of curves were calculated for $m = 2$; $D_0 = 1\;Gy$, or $10\; Gy $; $a_0 = 1\;\mu m$, or $10\; \mu m$, integer values of ion charge $Z$ (1, 2, 5, 10, 20, 50, 100), and varying $\beta$ (between $0.05$ and $0.99$). Calculations were performed using the modified RDD formula of \cite{zhang1985}, with $I=0\;eV$, eq.(\ref{eq.rddzhang2}).}
\label{fig.krzywe1}
\end{figure}

In the early 70's last century, when the cellular Track Structure Theory was developed (\cite{katz1971}), there was hardly any information available about the mechanisms leading to inactivation of biological cells nor about the structure of any sensitive targets inside the cells, responsible for cell death, following exposure to ionizing radiation. Robert Katz and his co-workers constructed a model which did not require this detailed knowledge. TST assumes that the cell nucleus (or most of its volume) is the sensitive site which consists of a number, $m$, of 1-hit sensitive sub-targets. All these sub-targets need to be inactivated by radiation in order to achieve the observed end-point, i.e. inactivation of the sensitive site as a whole, resulting, e.g., in cell death or mutation. There are thus two relevant sizes to consider: one representing the sensitive sub-target, of radius $a_0$ (see also Fig. \ref{fig.cmmodels}), and another, roughly corresponding to the dimension of the cell nucleus within the volume of which the sub-targets are contained. In order to make model predictions for \emph{in vitro} cell survival, TST defines the fourth model parameter, $\sigma_0$, representing roughly the cross-section area of the cell nucleus (\cite{katz1988}). From now on we will speak about the cellular Track Structure Theory approach which applies four model parameters to describe the biological system: $m$ and $D_0$ attributed to the response of the cells after uniformly distributed doses of reference radiation ($X-$ or $\gamma-$rays) as given by the m-target formula, eq.(\ref{eq.mtarget}); $\sigma_0$ representing the plateau (or saturation) value of the inactivation cross-section, a multiple of $\pi a_{0}^{2}$ (see Fig. \ref{fig.krzywe1}), nominally corresponding the cross-sectional area of the cell nucleus; and $\kappa$ which replaces $a_0$, related to the linear dimension of the sensitive sub-target, $a_0$ and its radiosensitivity, $D_0$. The detailed definition of the last parameter will be given later.

In Fig. \ref{fig.krzywe1} we present results of calculations of the single-particle action cross-sections, given by eq.(\ref{eq.crossection}), after irradiating the cellular detector by a range of ions of varying speed, for a fixed number $m=2.5$ \footnote{note that in TST model calculations, a real value of $m$ may be applied, presumably representing some average over the numbers of sub-targets in a population of cell nuclei.} of sub-targets of different radii $a_0$, and their characteristic dose $D_0$. In these calculations, the radial dose distribution given by \cite{zhang1985} with $I=0\;eV$, eq.(\ref{eq.rddzhang2}), was applied. The family of curves were calculated for $m = 2.5$; $D_0 = 1\;Gy$, or $10\; Gy $; $a_0 = 1\;\mu m$, or $\;10\; \mu m $, and were plotted in a normalized way as a function of LET (panel A) or $z^{*2}/\beta^2$ (panel B). These values of parameters were chosen as being representative of many mammalian cells \cite{katz1994}. Cross-sections presented in Fig. \ref{fig.krzywe1} were calculated for increasing values of ion charge $Z$ (1, 2, 5, 10, 20, 50, 100), of increasing energy, in terms of their relative speed $\beta$ (ranging form $0.05$ to $0.99$).

The prominent features of these graphs is the rise in the cross-section values with LET (or $z^{*2}/\beta^2$), appearance of a transition region or the 'plateau' which lies in the neighbourhood of $\pi a_{0}^{2}$, and next, a further increase of $\sigma$, ending with a rapid decrease (a 'hook') which represents the thindown region at low ion velocities. This thindown region is due to the kinematic constraint on the maximum energy of $\delta$-rays, eq.(\ref{eq.maximum.energy}), and their corresponding range $R_{\delta}$ (\cite{katz1985}), reflecting the appearance of ion track thinning over the final range, seen in nuclear emulsions.

Indeed, if we consider an energetic ion passing through a nuclear photographic emulsion made up of radiosensitive grains dispersed in a gelatin matrix, the blackening of each developed grains is caused by the absorption of dose deposited by $\delta$-electrons generated by the passing ion. The probability of activation (i.e. blackening) of the grains nearest to the ion's path will be the highest since most of the radial $\delta$-ray dose is deposited there. TST differentiates between two geometric features, or 'regimes' of ion tracks:
\begin{itemize}
\item the 'grain-count' regime, when single and separated black grains of emulsion are visible along the ion's path; 
\item the 'track-width' regime, when black grains surround a wider area around the ion's path. 
\end{itemize}
The first, 'grain-count' regime, corresponds to a situation where the probability of grain activation by the passing ion is lower than or equal one \footnote{in the 'grain-count' regime, the value of cross-section (a probabilistic concept) less than its 'plateau' value implies that only a fraction of grains are activated along the ion's path; if the cross-section reaches its 'saturation' value, then every grain along the ion's path is activated.}. In the 'grain-count' regime the activation cross-section, $\sigma$, is lower or reaches its plateau (or saturation) value, $\sigma_0$, approximating the cross-sectional area of the sensitive site, here a grain of emulsion. The cross-section increases again in the 'track-width' regime indicating that the probability of grain activation is very high, even at short distances from the ion's path (\cite{waligorski1987b}). In cellular Track Structure Theory the cutoff value between the 'grain-count' and 'track-width' regimes has been set at $0.98$, i.e. the 'grain-count' regime is the region where $\sigma/\sigma_0 \leq 0.98$, and 'track-width' regime where $\sigma/\sigma_0 > 0.98$. Features of the ion's track in a detector are governed non-separably by the properties of the ion, as described by its energy and charge, and by the properties of the detector, as described by its radiosensitivity parameters, including characteristic dose $D_0$ and its statistical properties ($m$ or $c$ - parameters) (\cite{katz1971}). The above description of tracks in nuclear emulsion will also apply with regard to the number of inactivated cells or chromosomal changes in the cell nucleus, following the passage of a beam of energetic ions.

Calculations of the cross-sections performed for a range of model parameters, $m$, $D_0$ and $a_0$, illustrated in Fig. \ref{fig.krzywe1} panel B, demonstrate another important feature, namely, that if plotted as a function of $z^{*2}/\beta^2$, the families of calculated dependences of activation cross-sections for a given set of $D_0$ and $a_0$, and for ions of different charge $Z$ over a range of energies (as given by the ion's relative speed $\beta$), lie atop of one another. This is another demonstration of the scalability of the model via the use of an appropriate RDD formula and of $z^{*2}/\beta^2$ rather than LET, to characterize the radiobiological properties of ions.

\subsection{Cross-section in the 'grain-count' regime}
\label{ch.graincount}
\begin{figure}[!ht]
\begin{center}
\includegraphics[width=0.87\textwidth]{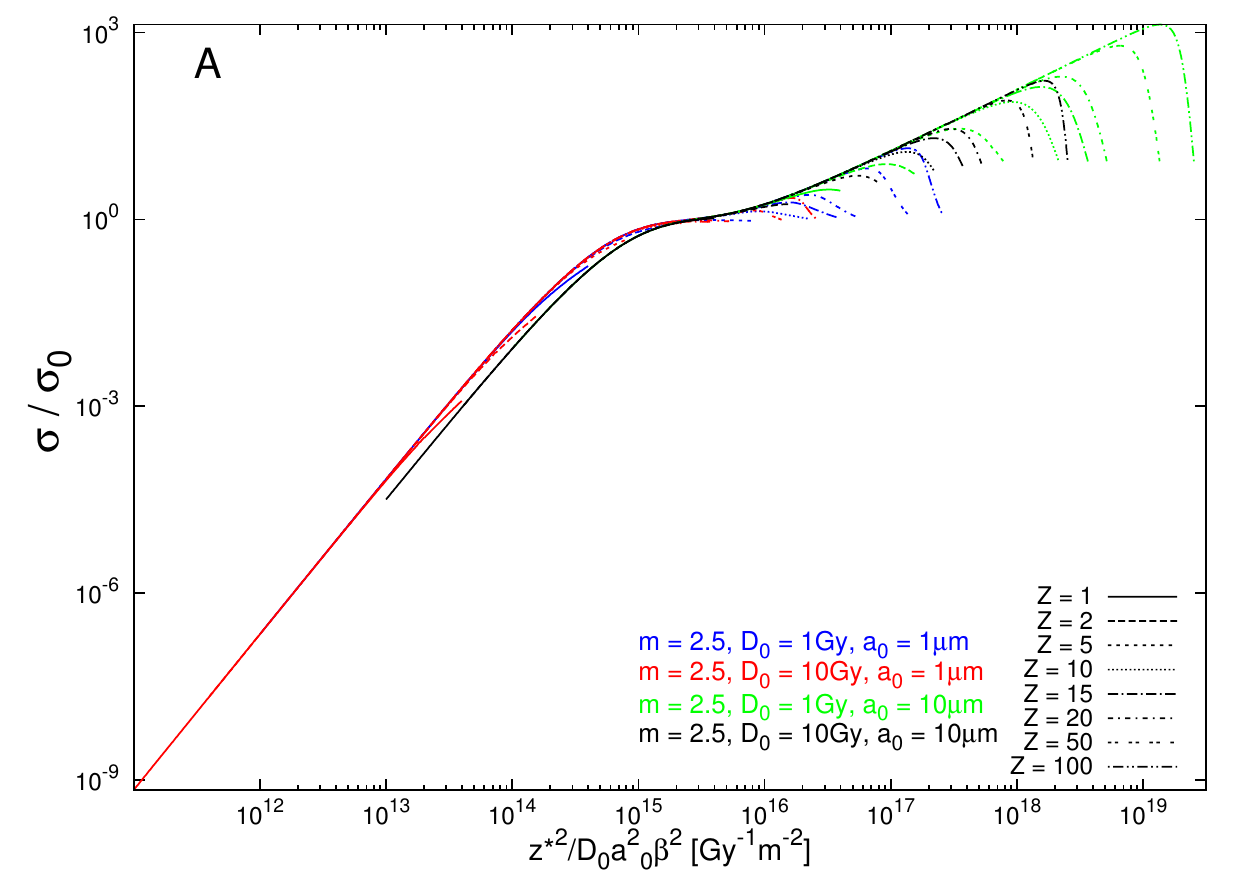}
\includegraphics[width=0.87\textwidth]{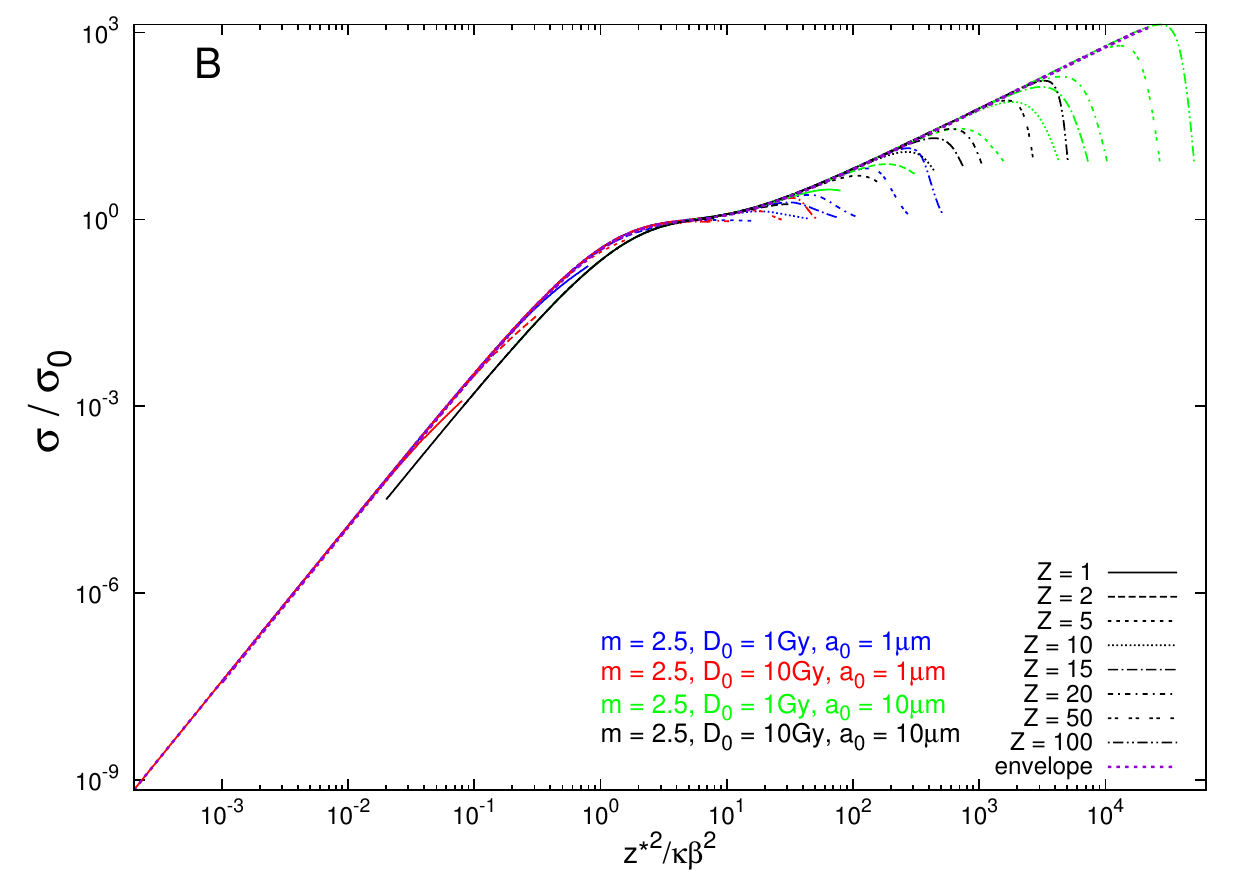}
\end{center}
\caption{Calculated values of the cross-section, given by eq.(\ref{eq.crossection}), normalized to the saturation or 'plateau' value, plotted as a function of $z^{*2}/D_{0}a_{0}\beta^{2}$ (panel A) and $z^{*2}/\kappa\beta^2$ (panel B). The family of curves were calculated for $m = 2$; $D_0 = 1\;Gy$, or $10\; Gy $; $a_0 = 1\;\mu m$, or $10\; \mu m$, constant values of ion charge $Z$ (1, 2, 5, 10, 20, 50, 100), and varying $\beta$ (form $0.05$ to $0.99$). Simulations were done using the modified RDD given by \cite{zhang1985}, with $I=0\;eV$, eq.(\ref{eq.rddzhang2}).}
\label{fig.krzywe2}
\end{figure}

Families of cross-section dependences similar to those of Fig. \ref{fig.krzywe1} are also presented in Fig. \ref{fig.krzywe2}. In panel A of this figure, calculated cross-sections are plotted as a function of $z^{*2}/D_{0}a_{0}\beta^{2}$, which reflects the properties of the radiation via $z^{*2}/\beta^2$ and the properties of the biological system described by two TST parameters, $D_0$ and $a_0$. All curves, calculated for $m = 2.5$; $D_0 = 1\;Gy$, or $10\; Gy $; $a_0 = 1\;\mu m$, or $10\; \mu m$, merge together if they are plotted versus $z^{*2}/D_{0}a_{0}\beta^{2}$. For values of $m$ greater than 1, the curves largely overlap, except at values of $\beta$ below about $0.10$, forming separate families of curves characteristic for a given value of $m$. For $m = 1$ no such overlap occurs (see Fig. \ref{fig.envelope} in later discussion). If $z^{*2}/\kappa\beta^{2}$ scaling is applied, for a particular value of $m$, curves representing the activation cross-section for the multi-target system can be approximated by a single envelope, if one neglects the 'hooks' in the thindown region, (Fig. \ref{fig.krzywe2} panel B). Robert Katz noticed that in the 'grain-count' regime (where $\sigma \leq \sigma_0$) the envelope of these curves varies as an exponent to the $m$'th power, next arriving at a 'plateau' if $m>1$, signalling the end of the 'grain-count' regime, and then increasing linearly with $z^{*2}/\beta^2$ in the 'track-width' regime (\cite{katz1988}).

\begin{figure}[!ht]
\begin{center}
\includegraphics[width=1\textwidth]{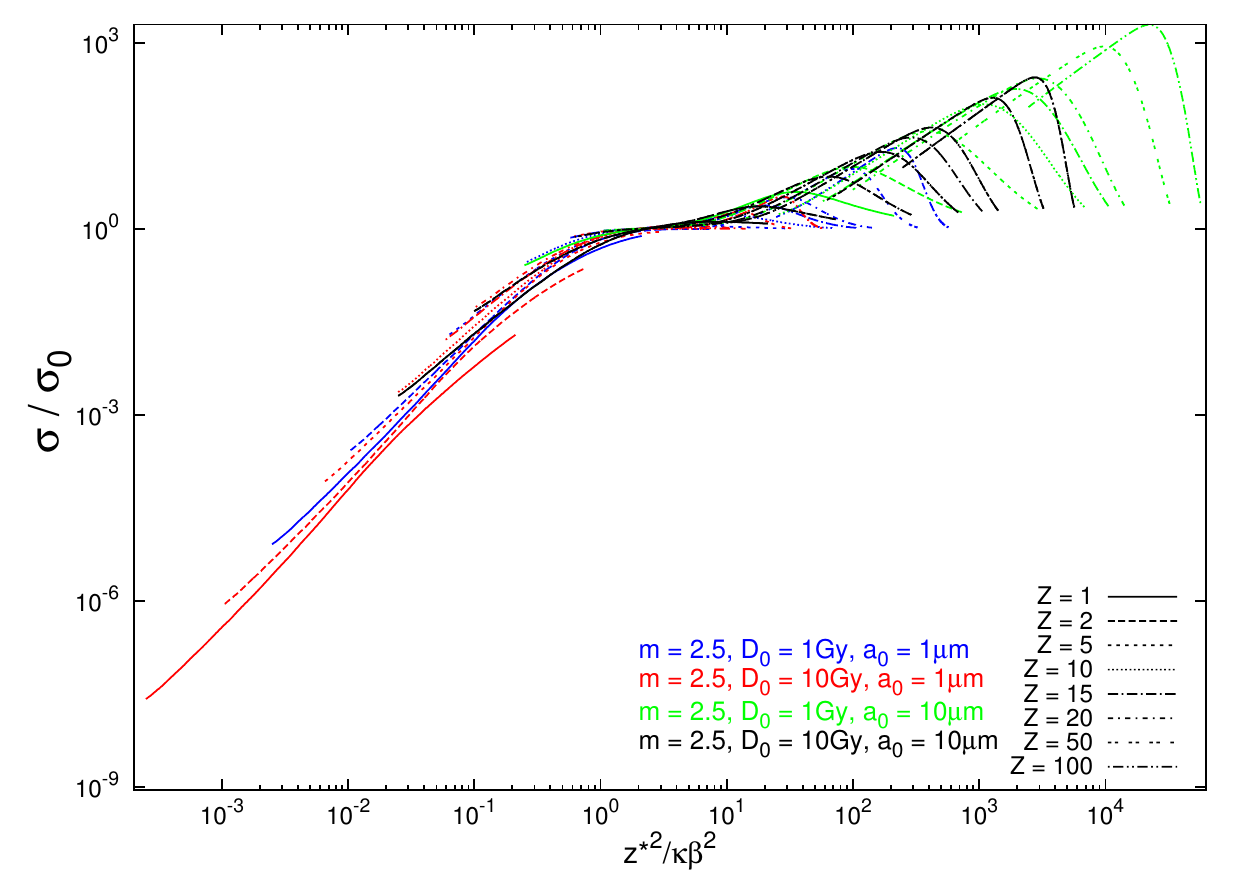}
\end{center}
\caption{Calculated values of the cross-section, given by eq.(\ref{eq.crossection}), normalized to the saturation or 'plateau' value, plotted as a function of $z^{*2}/\kappa\beta^2$. The family of curves were calculated for $m = 2.5$; $D_0 = 1\;Gy$, or $10\; Gy $; $a_0 = 1\;\mu m$, or $10\; \mu m$, increasing values of ion charge $Z$ (1, 2, 5, 10, 20, 50, 100), and varying $\beta$ (from $0.05$ to $0.99$). Calculations were performed using the RDD formula of \cite{cucinotta1997}, eq.(\ref{eq.rddcucinotta2}).}
\label{fig.krzywe3}
\end{figure}

In the grain-count regime (where $\sigma/\sigma_0 \leq 0.98$), the cross-section envelope can be approximated by the function, $P$, of the form: 
\begin{equation}
P = \frac{\sigma}{\sigma_{0}} = \left(1 - e^{-z^{*2}/\kappa\beta^{2}}\right)^m.
\label{eq.sigmaapprox1}
\end{equation}
If the transition between the 'grain-count' regime and the 'track-width' regime or the plateau value is set to be achieved at $z^{*2}/\kappa\beta^{2} \simeq 4$, then the $\kappa$ parameter is defined as follows (\cite{katz1971}):
\begin{equation}
\kappa = \frac{D_{0}a_{0}^{2}}{2 \cdot 10^{-15} \;\; [Gy \cdot m^{2}]}.
\label{eq.kappa2}
\end{equation}
The explanation of the transition from the $z^{*2}/\beta^2$ (panel A) variable to $z^{*2}/\kappa\beta^2$ (panel B) variable at the axis of abscissa in Fig. \ref{fig.krzywe2} is simple. If $\sigma/\sigma_{0}$ values from this figure are plotted against $z^{*2}/D_{0}a_{0}\beta^{2}$, saturation of the cross-section is achieved at about
\begin{equation}
z^{*2}/D_{0}a_{0}\beta^{2} \simeq 2 \cdot 10^{15} \quad [Gy^{-1} \cdot m^{-2}].
\label{eq.plateau}
\end{equation}
By applying the condition $z^{*2}/\kappa\beta^{2} \simeq 4 $ to the above identity, one arrives at Katz's definition of $\kappa$, as given by eq.(\ref{eq.kappa2}).

In Fig. \ref{fig.krzywe3} we present action cross-sections for inactivation given by eq.(\ref{eq.crossection}), calculated for the same sets of model parameters as those in Fig. \ref{fig.krzywe1} and Fig. \ref{fig.krzywe2}. In order to illustrate the importance of the scaling properties which have to be met by the RDD formula used for developing the cellular Track Structure Theory, these cross-sections were calculated using the radial dose distribution formula of \cite{cucinotta1997}, eq.(\ref{eq.rddcucinotta2}), which does not fulfil the condition of scaling. Here the particular curves representing the cross-section for a given value of $m$ only partly overlap, hence one cannot uniquely determine their envelope, especially over the 'track-width' regime.

\subsection{Saturation of the cross-section}
\label{ch.sigmao}

Before we specify the function which approximates the values of $P=\sigma/\sigma_0$ in the 'track-width'regime, we have to explain, in more detail the definition of $\sigma_0$. In the original construction of the TST model the action cross-section, $\sigma$, was rescaled by an empirical constant factor $\sigma_0=1.4\pi a_{0}^2$ identified as the cross-sectional area of the cell nucleus. When performing calculations of the envelope, we noticed that the plateau value is not constant for all values of $m$. It decreases with increasing $m$, but it is always greater than $\pi a_0^2$. In our approach, to achieve a better agreement of the function $P$ with the cross-section envelope we assumed an additional dependence of $\sigma_0$ on $m$. We substituted the constant value $1.4$ in the definition of $\sigma_0$ by an $m$-dependent coefficient $u(m)$ as follows:
\begin{equation}
\sigma_{0} = u(m) \cdot \pi \cdot a_{0}^{2},
\label{eq.sigmao}
\end{equation}
where $u(m)$ can be described by the function:
\begin{equation}
u(m) = \exp \left(a_{0} + \sum_{i=1}^{5}a_{i}\cdot ln(m)^{i} \right).
\label{eq.uapprox}
\end{equation}

\begin{figure}[!ht]
\begin{center}
\includegraphics[width=1.0\textwidth]{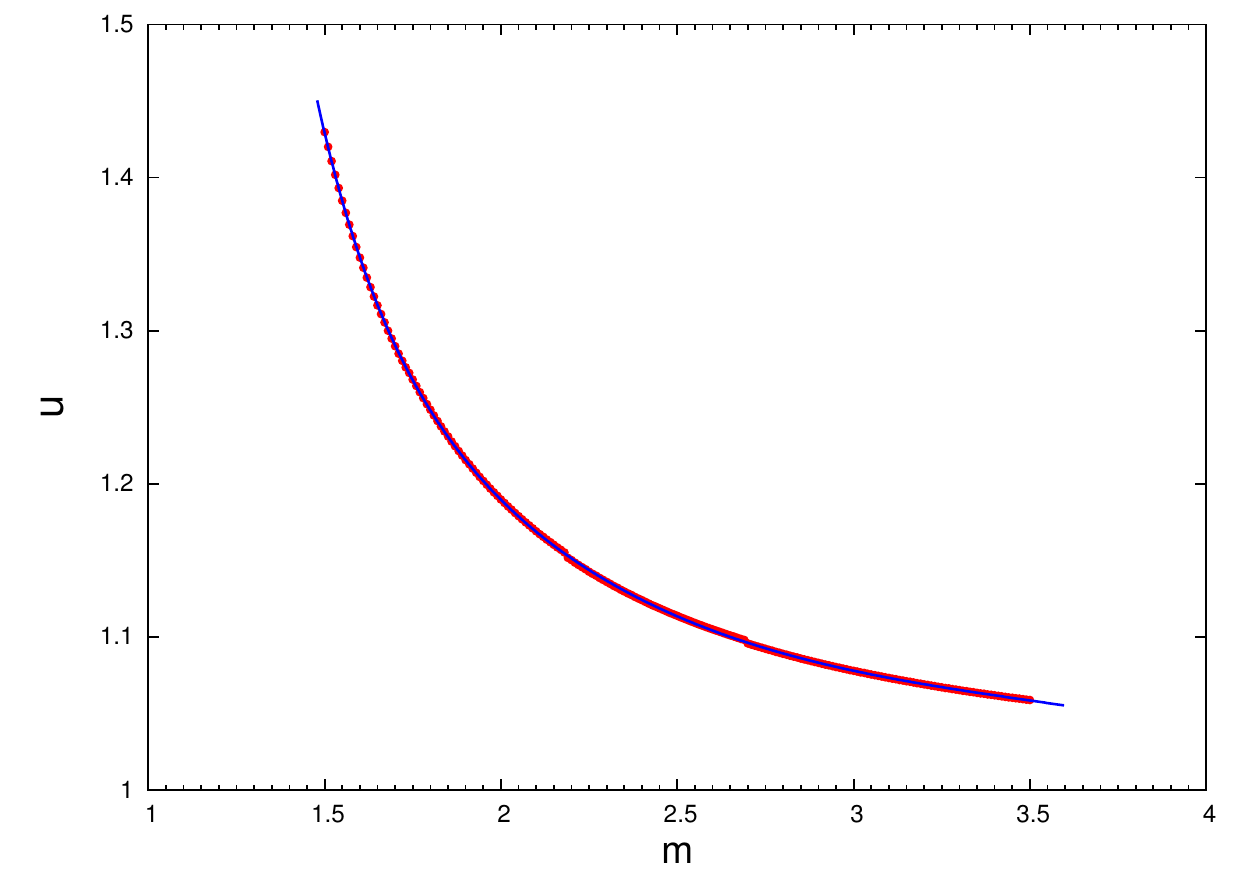}
\end{center}
\caption{ Calculated values of the $u$-factor (red points), as a function of m, and their approximation, as given by eq.(\ref{eq.uapprox}) (blue line).}
\label{fig.um}
\end{figure}

The values of power expansion coefficients, $a_j$ of the $u$-factor, are given in Table \ref{tab.3}. Values of $u(m)$ determined for $1.5 \leq m \leq 3.5$ are shown in Fig. \ref{fig.um}. In our approach $\sigma_0$ formally represents the value used to re-scale the envelope in order to ensure a good approximation of the envelope by the function $\left(1 - e^{-z^{*2}/\kappa\beta^{2}}\right)^m$ at the point of transition from the 'grain-count' to the 'track-width' regime occurring around $z^{*2}/\kappa\beta^{2} \simeq 4 $, which corresponds to the condition $\sigma/\sigma_0 = 0.98$.

\begin{table}[!ht]
\begin{center}
\caption{Power expansion coefficients of eq.(\ref{eq.uapprox}), $a_{i}$.}
\begin{tabular}{c c}
\hline
$i$ & $a_i$    \\
\hline
$0$  & 1.35719  \\
$1$  & -4.75757  \\
$2$  & 8.30616  \\
$3$  & -8.07658  \\
$4$  & 4.11203  \\
$5$  & -0.85068  \\
\hline
\end{tabular}
\label{tab.3}
\end{center}
\end{table}

After a closer inspection of model parameters we decided to restrict the approximation to values used most frequently to characterise mammalian cells (\cite{katz1994}). It must be emphasized that the algorithm for calculating the 'grain-count' cross-section envelope presented in this work is limited only to values of $m$ within the range $1.5 \leq m \leq 3.5$. 

\subsection{Cross-section in the 'track-width' regime}
\label{ch.trackwidth}
The functions which approximate the cross-section envelope over the 'track-width' regime, i.e. over the region where $\sigma/\sigma_0>0.98$, were never published or documented in the work of Prof. Robert Katz's group. However the codes of these functions were available in the FORTRAN library of computer codes used by that group, unfortunately without any derivation. Therefore, we had to reconstruct the approximation of the envelope, basing on functions used in the original TST computer program library.

\begin{figure}[!ht]
\begin{center}
\includegraphics[width=1\textwidth]{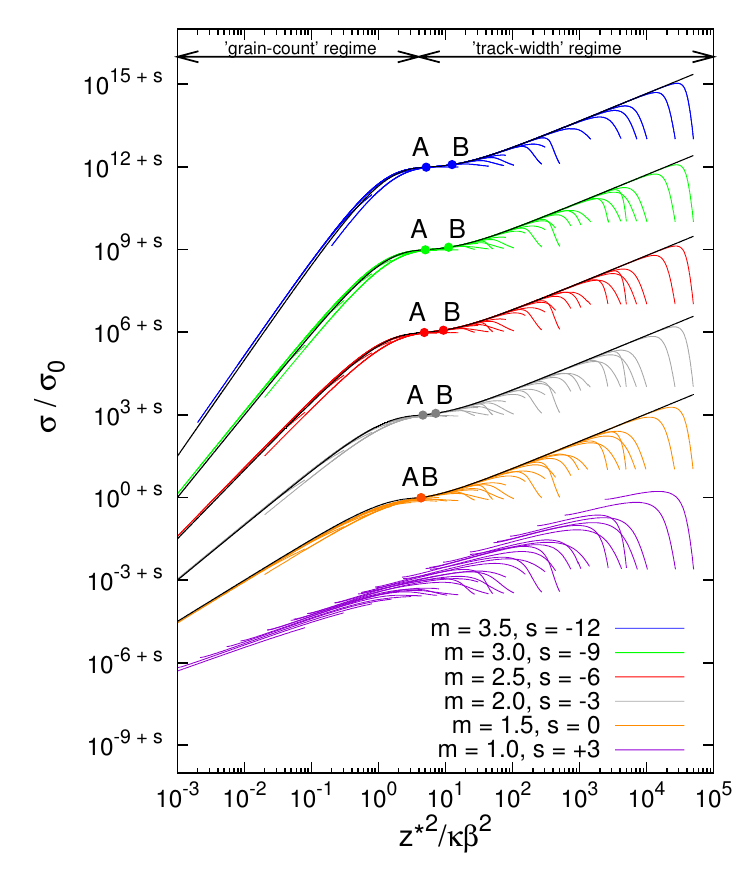}
\end{center}
\caption{$\sigma/\sigma_0$ plotted against $z^{*2}/\kappa\beta^{2}$ for integer values of $Z$ (1, 2, 5, 10, 20, 50, 100) and varying $\beta$. The family of curves were calculated for $m = 1.0,\;1.5,\;2.0,\;2.5,\;3.0$ and $3.5$; $D_0 = 1\;Gy$, or $10\; Gy $; $a_0 = 1\;\mu m$, or $10\; \mu m$. Calculations were performed using the modified RDD given by \cite{zhang1985}, eq.(\ref{eq.rddzhang2}). The points on the 'plateau' region of calculated cross-sections correspond to points $A$ and $B$, as given in eq.(\ref{eq.sigmaapprox2}) and eq.(\ref{eq.sigmaapprox3}) and determined as shown in Appendix \ref{ch. appendixc}.}
\label{fig.envelope}
\end{figure}

Thus, guided by formulae available from these codes, we determined a new set of coefficients to fit suitable envelopes for different values of $m$, for cross-sections in the 'track-width' regime. Over the range of the $\sigma/\sigma_0$ plateau, the envelope is approximated by the linear function:
\begin{equation}
P = \frac{\sigma}{\sigma_{0}} = \left( \frac{Y_A-Y_B}{X_A-X_B}\left(\frac{z^{*2}}{\kappa\beta^{2}}-X_A \right) \right) + Y_A,
\label{eq.sigmaapprox2}
\end{equation}
and for higher values of $z^{*2}/\kappa\beta^{2}$, by an exponential function
\begin{equation}
P = \frac{\sigma}{\sigma_{0}} = \frac{Y_B \cdot 0.8209}{1-e^{-\left( X_B \cdot 1.72/ \frac{z^{*2}}{\kappa\beta^{2}} \right) }},
\label{eq.sigmaapprox3}
\end{equation}
where $X_A$, $X_B$, $Y_A$ and $Y_B$ are coordinates of two points $A(X_A,Y_A)$ and $B(X_B,Y_B)$ lying directly on the envelope, and fixed for a given $m$ (see fig \ref{fig.envelope}). These points were chosen in a manner such that continuous transition is assured between functions approximating the envelope, given by eq.(\ref{eq.sigmaapprox1}), eq.(\ref{eq.sigmaapprox2}) and eq.(\ref{eq.sigmaapprox3}), respectively. Simultaneously, $X_A$ and $X_B$ determine the range of applicability of these particular equations. The envelope is approximated: for $z^{*2}/\kappa\beta^{2} \leq X_A$ by eq.(\ref{eq.sigmaapprox1}), for $X_A < z^{*2}/\kappa\beta^{2} \leq X_B$ by eq.(\ref{eq.sigmaapprox2}), and for $z^{*2}/\kappa\beta^{2} \geq X_B$ by eq.(\ref{eq.sigmaapprox1}). The value of the constant $1.72$ in eq.(\ref{eq.sigmaapprox3}) was determined by a $\chi^2 $ fitting of eq.(\ref{eq.sigmaapprox2}) and eq.(\ref{eq.sigmaapprox3}) to the envelope over the region of 'track-width' regime. The value of the constant $0.8209$ results form the continuity condition at the point of transition between these two formulae. The details of how the values of all of these coefficients were determined and all formulae used to determine the values of $X_A$, $X_B$, $Y_A$ and $Y_B$ are discussed in Appendix \ref{ch. appendixc}. \\

In Fig. \ref{fig.envelope} we plot the cross-sections normalized to their saturation value, $\sigma/\sigma_0$, against $z^{*2}/\kappa\beta^{2}$ for increasing values of $Z$ (1, 2, 5, 10, 20, 50, 100) and varying $\beta$ (from $0.05$ to $0.99$). The family of curves were calculated for $m = 1,1.5,2,2.5,3,3.5$; $D_0 = 1\;Gy$, or $10\; Gy $; $a_0 = 1\;\mu m$, or $10\; \mu m$. Calculations were performed using the modified RDD formula of \cite{zhang1985}, eq.(\ref{eq.rddzhang2}). At values of $m>1$ there is substantial overlap of curves calculated for different values of $D_0$ and $a_0$ except for values of $\beta$ below 0.1. For $m = 1$ no such overlap occurs. Additionally, for a given value of $m$ we plot the envelopes approximated by the $P$-function given by eq.(\ref{eq.sigmaapprox1}), eq.(\ref{eq.sigmaapprox2}), and eq.(\ref{eq.sigmaapprox3}). These envelopes neglect the 'hooks' in the thindown region. Regions of the 'grain-count' regime as well as 'track-width' regime are also specified in Fig. \ref{fig.envelope}. \\ \\

\newpage
\vspace*{\stretch{2}}

\section{Cellular Track Structure Theory - the four-parameter formalism}
\label{ch. survival}
The complexity of the response of cells after the exposure to heavy ions is displayed in the shape of the survival curve. Depending on the quality of the radiation we observe a range of shapes of the survival curves. Through the use of 'ion-kill' (or \emph{intra}-track) and 'gamma-kill' (or \emph{inter}-track) components of the (in)activation probability, dose-response curves of shapes ranging from purely exponential (such as those in the neighbourhood of maximum relative biological effectiveness, RBE) to shouldered ones (such as those after doses of $\gamma$-rays) can be generated. Cells can be inactivated directly by the passage of a single ion, as expressed by the 'ion-kill' mode of inactivation. Additionally, cell response will depend on its irradiation history, namely, a cell may 'remember', by cumulating sublethal damage, that an ion had passed trough. The idea of 'gamma-kill' mode was proposed by Katz (\cite{katz1973}) to account for the cumulative mode of inactivation by overlapping delta-rays produced by several ions. Here, the $P$-function plays the role of a 'mixing factor' introduced to combine the 'ion-kill' and 'gamma-kill' modes of cell inactivation.

In TST, 'ion-kill' and 'gamma-kill' modes of inactivation, represented by probabilities $\Pi_{i}$ and $\Pi_{\gamma}$ respectively, contribute independently to cell survival, S:
\begin{equation}
S = \Pi_{i} \cdot \Pi_{\gamma}.
\label{eq.survival}
\end{equation}
The combined probability of cell survival after irradiation by a beam of ions of charge $Z$, speed $\beta$ track-segment value of LET$(z,\beta)$ and fluence $F$, is calculated as follows: The 'ion-kill' component is: 
\begin{equation}
\Pi_{i} = e^{-\sigma \cdot F},
\end{equation}
where $\sigma$ is calculated utilizing its approximation described in Section (\ref{ch.approximation})
\begin{equation}
\sigma = \sigma_{0} \cdot P,
\end{equation}
and $P$ is described by eq.(\ref{eq.sigmaapprox1}), eq.(\ref{eq.sigmaapprox2}), and eq.(\ref{eq.sigmaapprox3}). 

The 'gamma-kill' component is the probability of cell survival after an 'ion dose', $D_{i}$
\begin{equation}
D_{i} = \frac{1}{\rho}\;F \cdot \textrm{LET}_i(z,\frac{E}{A}),
\end{equation}
where LET$_i(z,\frac{E}{A})$ is ion's LET given by eq.(\ref{eq.letion}), $F$ is the ion fluence (number of ions per $m^2$) and $\rho$ is the density of the medium (in $kg\cdot m^{-3}$). The survival in 'gamma-kill' mode follows the same functional form as that used for the description of cell response after gamma (reference) radiation, eq.(\ref{eq.mtarget}), but with the substitution:
\begin{equation}
D_{\gamma} = \left(1-P\right) D_{i}.
\end{equation}
Thus, the probability in the 'gamma-kill' mode takes the form 
\begin{equation}
\Pi_{\gamma} = 1-\left(1-e^{-\frac{(1-P)D_{i}}{D_{0}}}\right)^m,
\end{equation}
and is responsible for the presence of the shoulder in the shape of the survival curve.

As we have discussed earlier, 'track-width' and 'grain-count' modes of inactivation are distinguished in track structure calculations. In the 'track-width' regime, where no 'overlap' of $\delta$-rays between neighbouring ion tracks occurs, i.e. if $P>0.98$, where $\sigma>\sigma_{0}$, $P$ is set to $1$. Thus, in the 'track-width' mode of inactivation, cell survival is given by the first term of eq.(\ref{eq.survival}) only, as the second terms equals $1$ and cells are inactivated through 'ion-kill' only
\begin{equation}
\quad \quad \quad \quad \quad  S =  e^{-\sigma \cdot F } \quad \quad \quad \quad \quad \textrm{for} \quad P>0.98.
\label{eq.survivaltw}
\end{equation}
In the so-called 'grain-count' regime, where $\sigma<\sigma_{0}$, i.e. where $P$ attains values less than $0.98$, cell survival is calculated using both terms of eq.(\ref{eq.survival}):
\begin{equation}
S =  e^{-\sigma \cdot F } \cdot \left[1-\left(1-e^{-\frac{(1-P)D_{i}}{D_{0}}} \right)^m\right] \quad \textrm{for} \quad P\leqslant 0.98,
\label{eq.survivalgc}
\end{equation}
where $P$ is calculated, using eq.(\ref{eq.sigmaapprox1}), eq.(\ref{eq.sigmaapprox2}), and eq.(\ref{eq.sigmaapprox3}).

\begin{figure}[!ht]
\begin{center}
\includegraphics[width=1.0\textwidth]{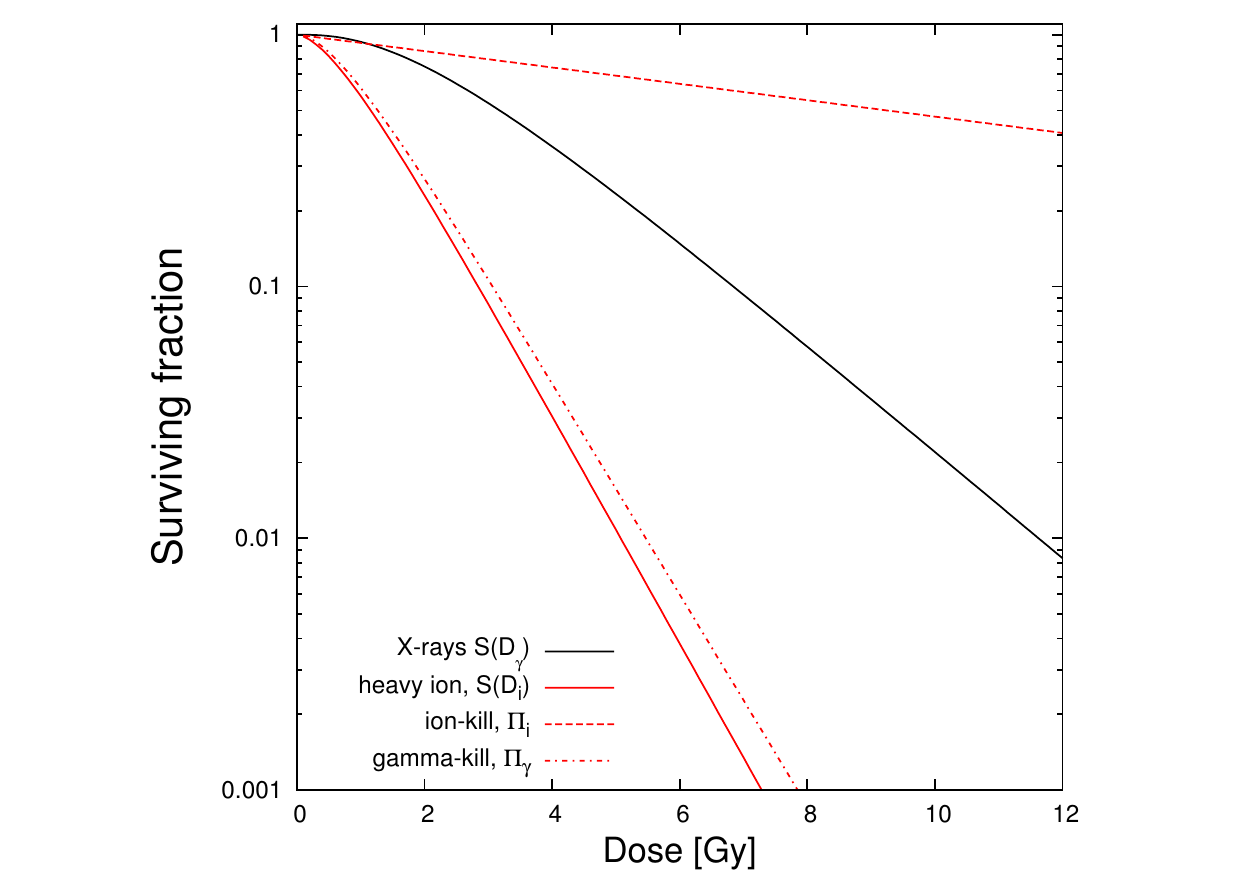}
\end{center}
\caption{Simulated survival curves for a biological system represented by arbitrary values of four model parameters: $m = 2.0, D_0=1.0 \; Gy, \sigma_0 = 10 \; \mu m^2, \kappa = 1000$. The black solid line represents the survival curve for the biological system after exposure to reference radiation. The red solid line represents the survival curve for the same system after irradiation with $100 \; MeV/n$ carbon ions. The red dashed line and red dash-dotted line represent the 'ion-kill' and 'gamma-kill' components for the last case, respectively. }
\label{fig.krzywerys}
\end{figure}

To illustrate the impact of the 'ion-kill' and 'gamma-kill' mode of inactivation on the shape of the survival curve, in Fig. \ref{fig.krzywerys} we present dose-response dependences calculated for arbitrary values of four model parameters: $m = 2.0,\; D_0=1.0 \; Gy,\; \sigma_0 = 10 \; \mu m^2, \;\kappa = 1000$. The black line represents the survival curve resulting from the multi-target single-hit formula, eq.(\ref{eq.mtarget}), and represents the response of the biological system after exposure to reference radiation. The red line represents the dose-survival dependence, eq.(\ref{eq.survival}), for the same system after irradiation with $100 \; MeV$ carbon ions. For this value of ion energy both modes of inactivation - 'ion-kill' and 'gamma-kill' - affect the shape of the survival curve. The 'ion-kill' and 'gamma-kill' components are represented by the red dashed line and red dash-dotted line respectively.

\section{Best-fitting of TST model parameters}
\label{ch. fittingprocedure}
The four model parameters $(m, D_0, \sigma_0$ and $\kappa)$ of the cellular TST representing the survival endpoint in cells are all fitted simultaneously, applying eq.(\ref{eq.survival}), to the complete set of the experimentally measured survival curves, including the cell response after exposure to reference radiation ($X$-rays or $^{60}$Co $\gamma$-rays). Each experimental point is given as a function of dose, ion type $Z$, LET, $\beta$, $z^{*2}/\beta^2$, and the four sought model parameters. During the fitting procedure the model calculations for the reference radiation may be approximated by, e.g., $100 \; MeV$ protons for which there is no significant difference in the response of the cells in comparison to the reference radiation. The parameter optimized is the $\chi^2$-difference between model-calculated values of surviving fraction, $SF_{th}$, and those measured experimentally, $SF_{exp}$, weighted by their respective experimental errors for all data points, $\vartriangle SF_{exp}$:
\begin{equation}
\chi^2 = \sum_{i=1}^{n} \left( \frac{{SF_{th}}_i-{SF_{exp}}_i}{\vartriangle {SF_{exp}}_i} \right)^2 ,
\end{equation}
where $n$ is the number of experimental points.

In the fitting procedure the MINUITS minimising code of \cite{james1977} was used. The MINOS subroutine of this code delivers parabolic errors of the best-fitted parameter values (which can be interpreted as one standard deviations and describe the width of the minimum of the function $\chi^2$ found for those values). Several consistency checks, including temporary elimination of the data subsets, are made prior to establishing the final best-fitted parameter values. For the cellular Track Structure Theory the fitting procedure utilize the approximation envelope defined for the cross-section. Thus the fit is independent of the radial dose distribution, even if the the envelope has been found applying a specific radial dose distribution. In this work, the envelope calculation was based on the modified RDD formula developed by \cite{zhang1985}, eq.(\ref{eq.rddzhang2}). We note that in the fitting procedure, all four parameters are best-fitted independently, i.e. no use is made of the relations between $\kappa$, $D_0$ and $a_0$, as given by eq.(\ref{eq.kappa2}), nor between $\sigma_0$and $a_0$, as given by eq.(\ref{eq.sigmao}). Having found the best-fitted values of $m, D_0, \sigma_0$ and $\kappa$, values of $a_0$ derived from eq.(\ref{eq.kappa2}) and eq.(\ref{eq.sigmao}) should agree but they do not. They are no more correlated through eq.(\ref{eq.kappa2}) and eq.(\ref{eq.sigmao}). Therefore, 'disentanglement' of $a_0$, and the direct transition from four-model parameter cellular TST ($m, D_0, \sigma_0$ and $\kappa$) to the three-model parameter TST $(m, D_0$ and $a_0)$ is impossible. More details about the fitting procedure can be found elsewhere (\cite{roth1976}, \cite{paganetti2001}).

\section{Survival curves after mixed radiation}
\label{ch. mixedfield}

Track Structure Theory also provides for estimation the effect of secondary charged particles in an ion beam. Cell survival resulting from irradiation by a mixed ion field consisting of $J$ different ions $(j=1,2...J)$, each of different track-segment energy $E_{j}$ (or respective relative speed $\beta_{j}$), charge $Z_{j}$ (or effective charge $z^{*}_{j}$, as calculated from eq.(\ref{eq.zeff})), and fluence $F_{j}$, each ion contributing its 'ion dose' $D_{j}=F_{j} \cdot \textrm{LET}_{j}$ (where the value of LET$_{j}$ is calculated on the basis of the energy $E_{j}$ and charge $Z_{j}$ of each ion), can be calculated from the resulting survival probability $S_{eff}$ versus total 'ion dose', $D_{ion-tot}=\sum_{j}D_{j}$:
\begin{equation}
S_{eff} =  \left( e^{-\sigma_{0} \sum_{j}P_{j}F_{j}}\right)\cdot\left( 1-\left( 1-e^{-\frac{\sum_{j}(1-P_{j})D_{j}}{D_{0}}}\right)^{m}\right),
\label{eq.survivaleff}
\end{equation}
where values of $P_{j}$ for each ion are calculated, using eq.(\ref{eq.sigmaapprox1}), eq.(\ref{eq.sigmaapprox2}), and eq.(\ref{eq.sigmaapprox3}), by suitably replacing $z^*$ and $\beta$ by $z^{*}_{j}$ and $\beta_j$, for each ion respectively. The same set of the four TST parameters is used throughout this calculation.

\section{Track Structure analysis of survival and RBE in normal human skin fibroblasts}
\label{ch. tsuruoka}
In this section we recapitulate fragments of a published article (\cite{korcyl2009}) to which we refer to for further details. The experimental data of \cite{tsuruoka2005} consists of $40$ survival data sets concerning normal human skin fibroblasts irradiated by five ion beams: carbon (of initial energies $135$ and $290\;MeV/n$), neon ($230$ and $400\;MeV/n$), silicon ($490\;MeV/n$) and iron ($500\;MeV/n$), generated by the Heavy ion Medical Accelerator in Chiba (HIMAC) at the National Institute of Radiological Science (NIRS) in Japan. To obtain a variety of ion LET values in track-segment irradiation conditions, various layers of polymethyl methacrylate (PMMA) absorbers of different thickness were applied, as listed in Table I of the publication of Tsuruoka et al. Reference radiation was provided by $200 \;kV$ X-rays, filtered with $0.5 \;mm$ aluminium and $0.5 \;mm$ copper.

\begin{figure}[!ht]
\begin{center}
\includegraphics[width=1.0\textwidth]{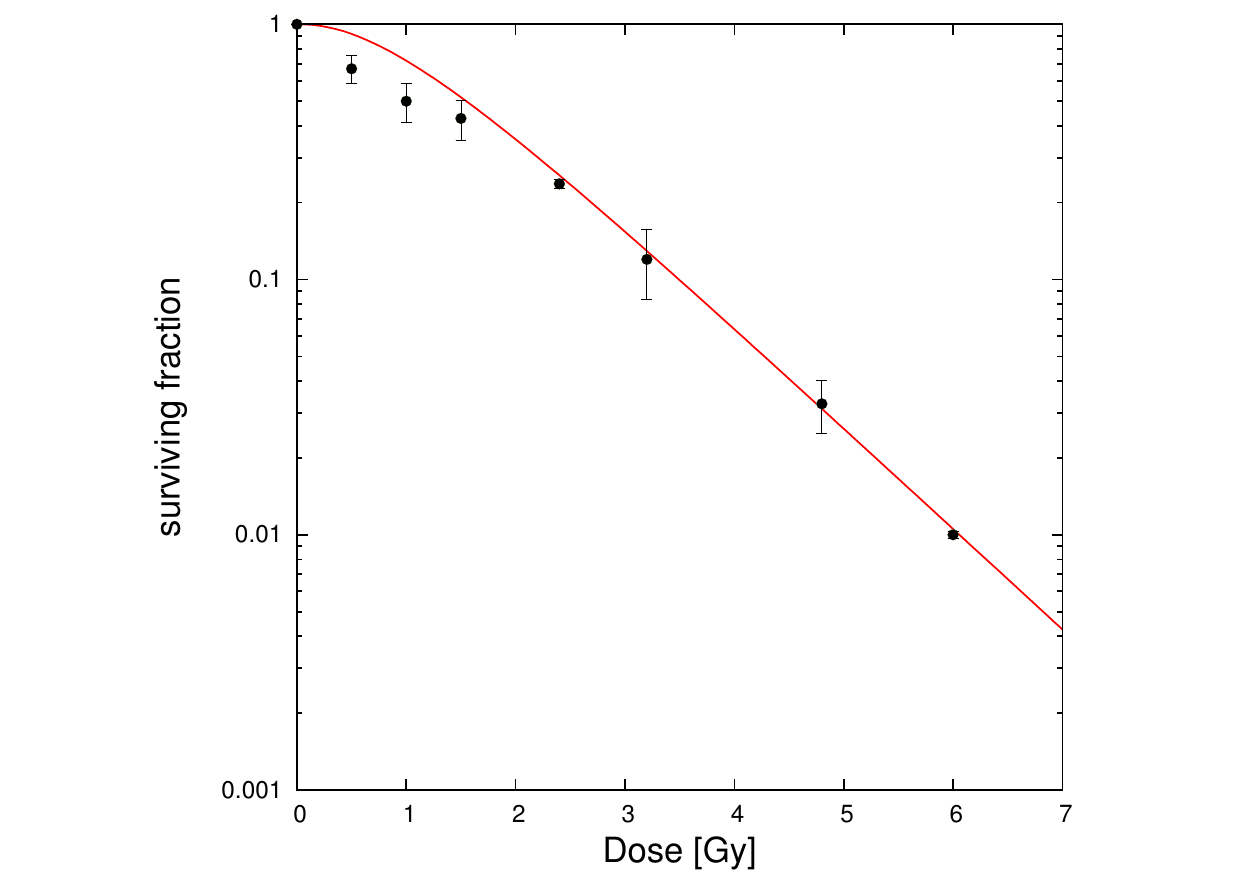}
\end{center}
\caption{Survival of normal human skin fibroblasts after irradiation by $200 \;kV$ X-rays. The full line represents the TST calculation, with $m=2.36$, and $D_0=1.10\;Gy$. Data points ($\bullet$) and their errors (one standard deviation for three independent experiments) are from \cite{tsuruoka2005}. }
\label{fig.xraystsuruoka}
\end{figure}

From the publication of \cite{tsuruoka2005} we read the values of doses and survival points (including error bars) for each of the survival curves presented in their Figure 1. Assuming that each beam consisted of only a single ion species, we next calculated the track-segment energy of each ion from its respective value of LET listed in their Table I. In this calculation, for each ion species we applied, in an inverse manner, eq.(\ref{eq.letion}) describing the ion energy-LET dependence. The corresponding values of $z^*$, $\beta^2$ and $z^{*2}/\beta^2$ required for TST calculations were then established from the thus calculated track-segment energy of each ion species. In Table \ref{tab.rbeNHSF} we list the ion source, source ion energy and LET values for ions, as given by Tsuruoka et al., followed by our LET-calculated track-segment ion energy and the respective values of $z^{*2}/\beta^2$ (in the paper of Tsuruoka et al. we noticed an error in their calculation of $z^{*2}/\beta^2$).
\\ \\
After a series of model fits to the complete experimental data set, the following best-fitting model parameters were found, given in Table \ref{tab.parametersNHSF}:

\begin{table}[!ht]
\begin{center}
\caption{The values of best-fitted TST parameters (together with their parabolic errors) representing \emph{in vitro} survival in normal human skin fibroblast cells in the experiment of \cite{tsuruoka2005}.}
\begin{tabular}{c c c c}
\hline
$\quad m \quad \quad$ & $D_0 \quad$ & $\sigma_0$ & $\quad \quad \kappa \quad$ \\
\hline \\
$\quad 2.36 \pm 0.03 \quad \quad$  & $1.1006 \pm 0.004 \;Gy \quad$ & $140.8 \pm 3.0\;\mu m^2$ & $\quad \quad 1204 \pm 34\quad$ \\ \\
\hline
\end{tabular}
\label{tab.parametersNHSF}
\end{center}
\end{table}
The best-fitting TST model parameters, presented in Table \ref{tab.parametersNHSF}, are given together with their parabolic errors according to the MINOS subroutine of MINUIT minimizing code. The values of our best-fitted TST parameters representing \emph{in vitro} survival in normal human skin fibroblast cells in the experiment of \cite{tsuruoka2005}, generally fall in line with values found for other mammalian cell lines, such as aerated V79, T-1 kidney cells (\cite{roth1976}), or other human skin fibroblasts (\cite{katz1994}). This set of four parameters was used in all TST model calculations for normal human skin fibroblasts shown in this work. In Fig. \ref{fig.xraystsuruoka} we present the TST-calculated response after $200 \;kV$ X-rays against the experimental data. Here, two of the four best-fitted TST parameters: $m=2.36$, and $D_0=1.10\;Gy$ were applied to eq.(\ref{eq.mtarget}). 

\begin{figure}[!ht]
\begin{center}
 \begin{tabular}{cc}
\includegraphics[width=0.55\textwidth]{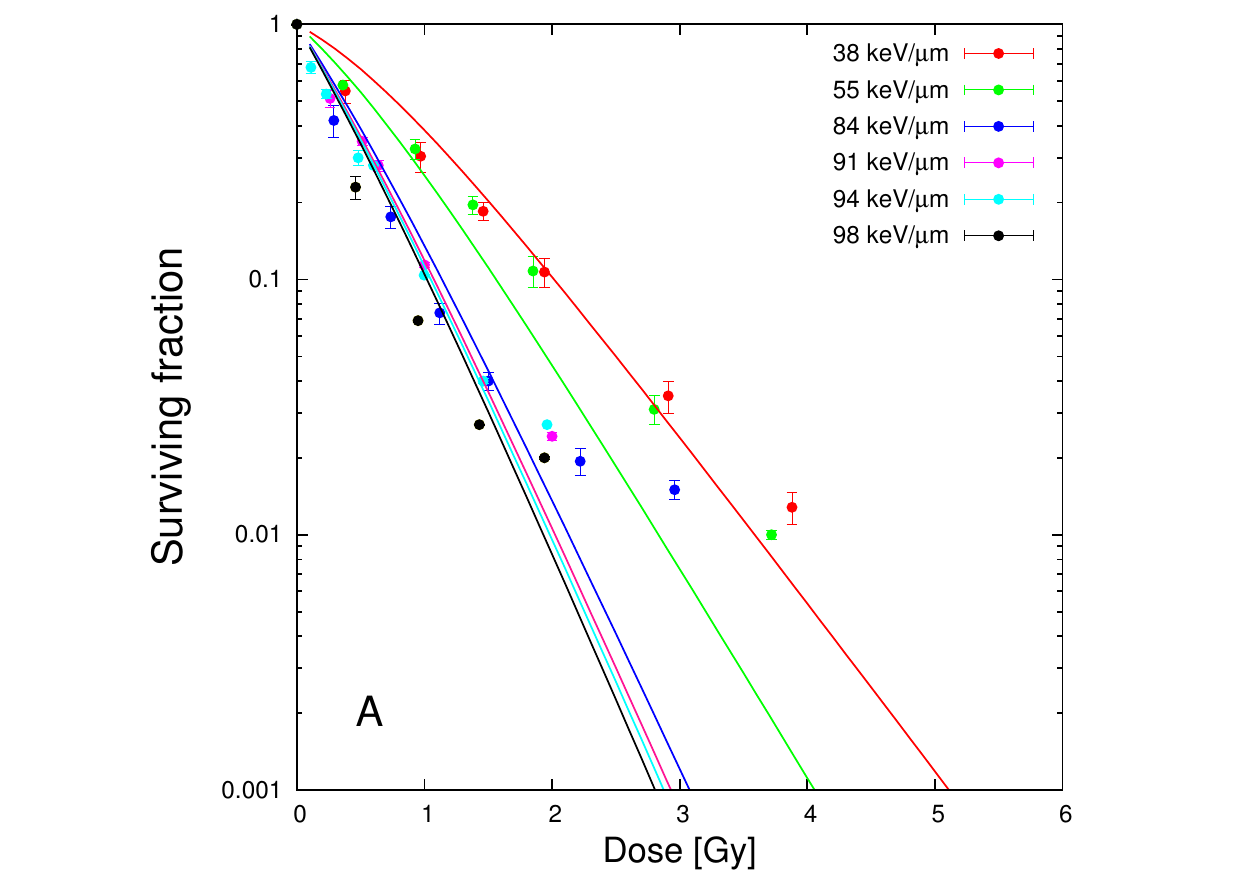} &
\includegraphics[width=0.55\textwidth]{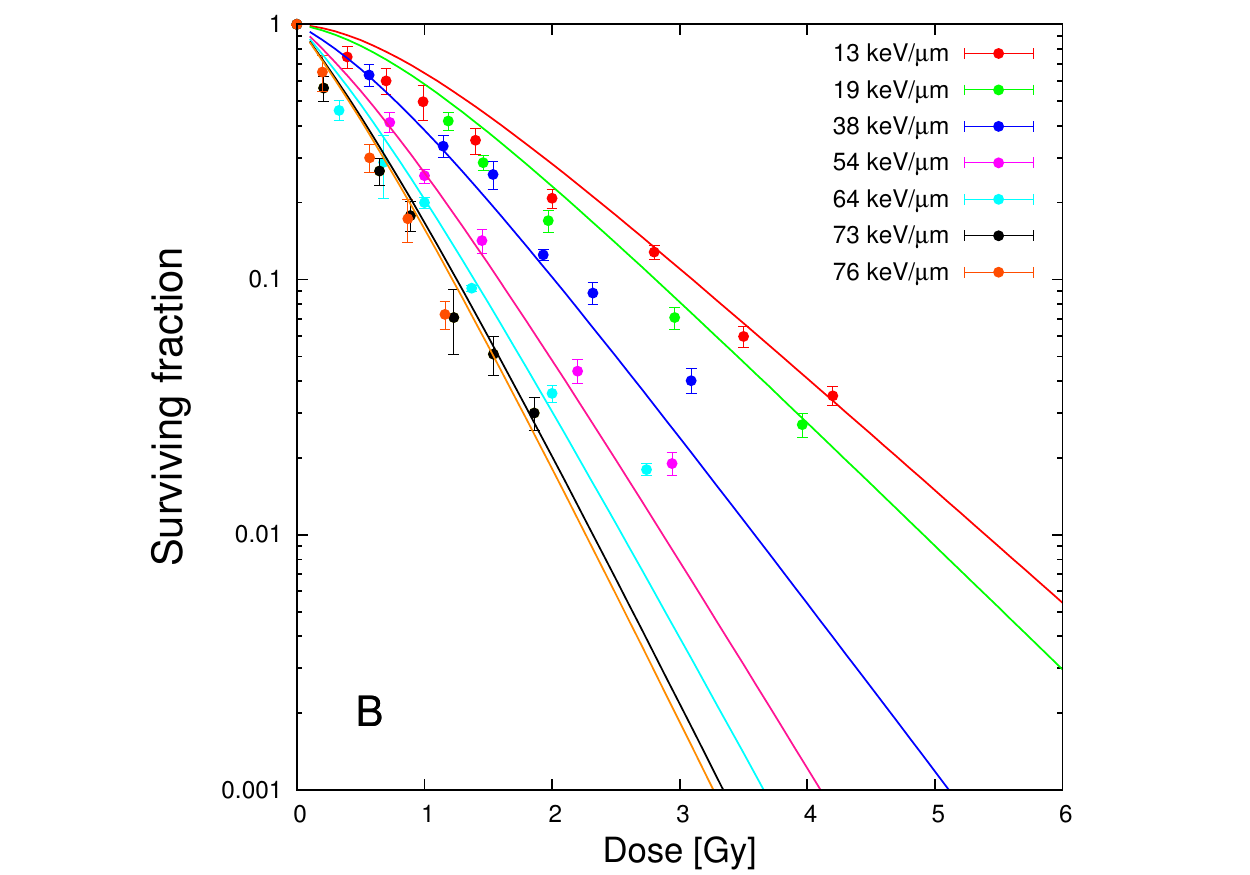} \\
\includegraphics[width=0.55\textwidth]{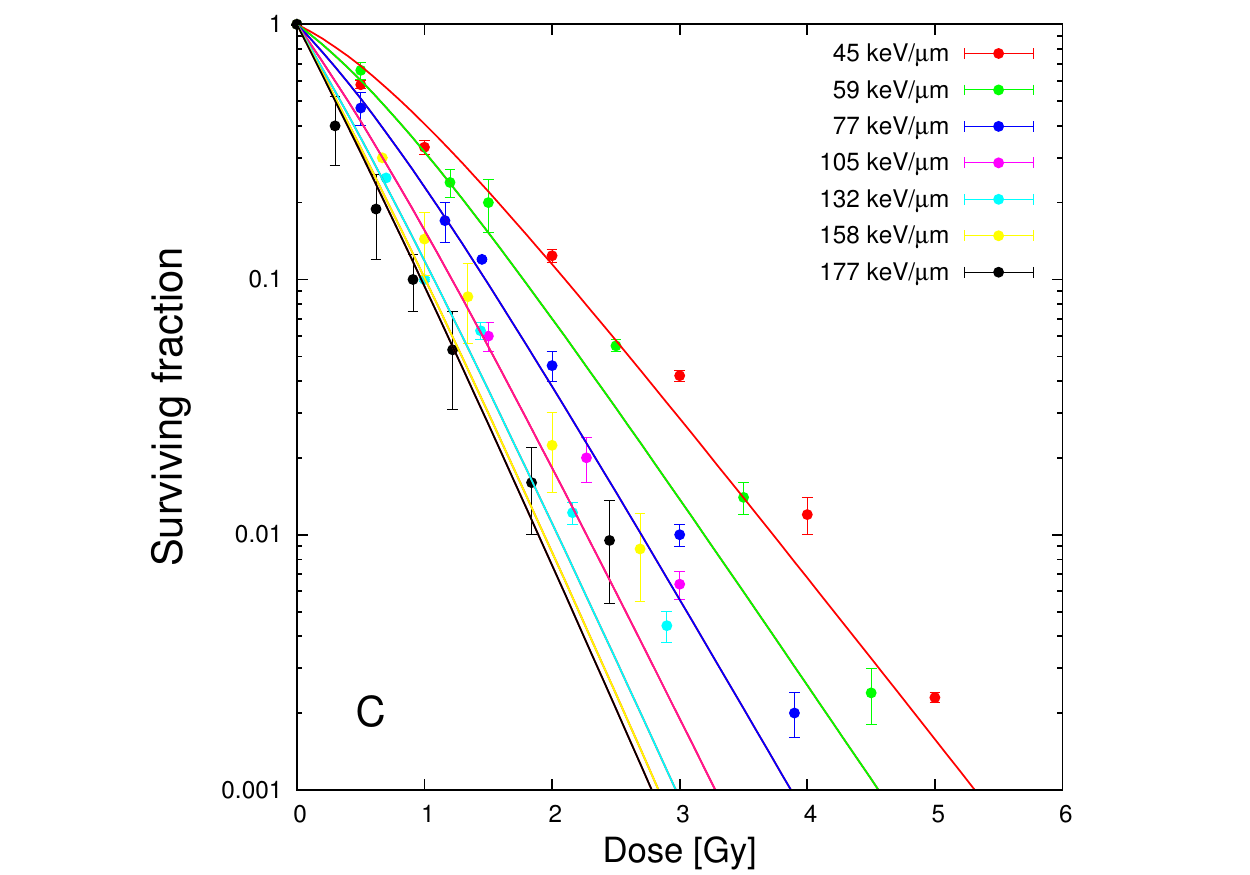} &
\includegraphics[width=0.55\textwidth]{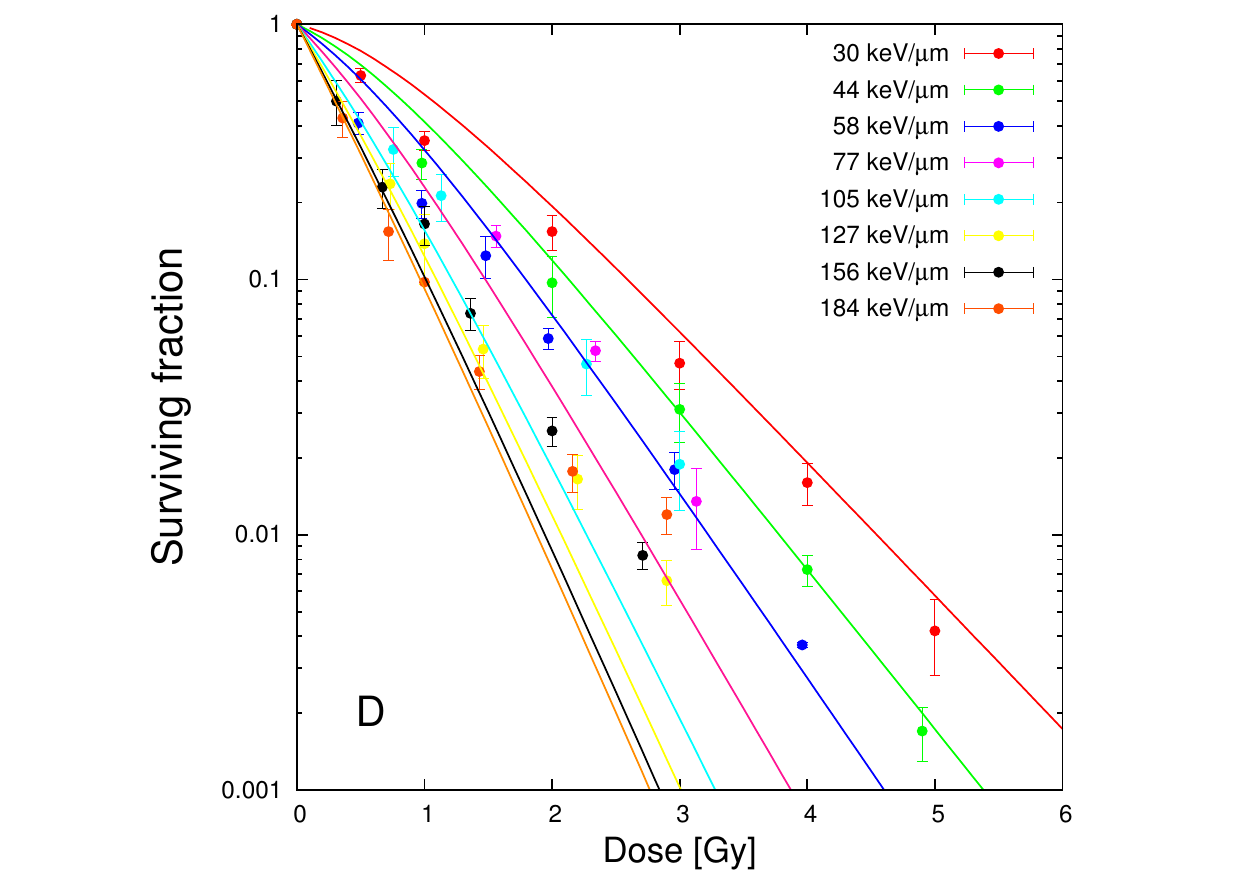} \\
\includegraphics[width=0.55\textwidth]{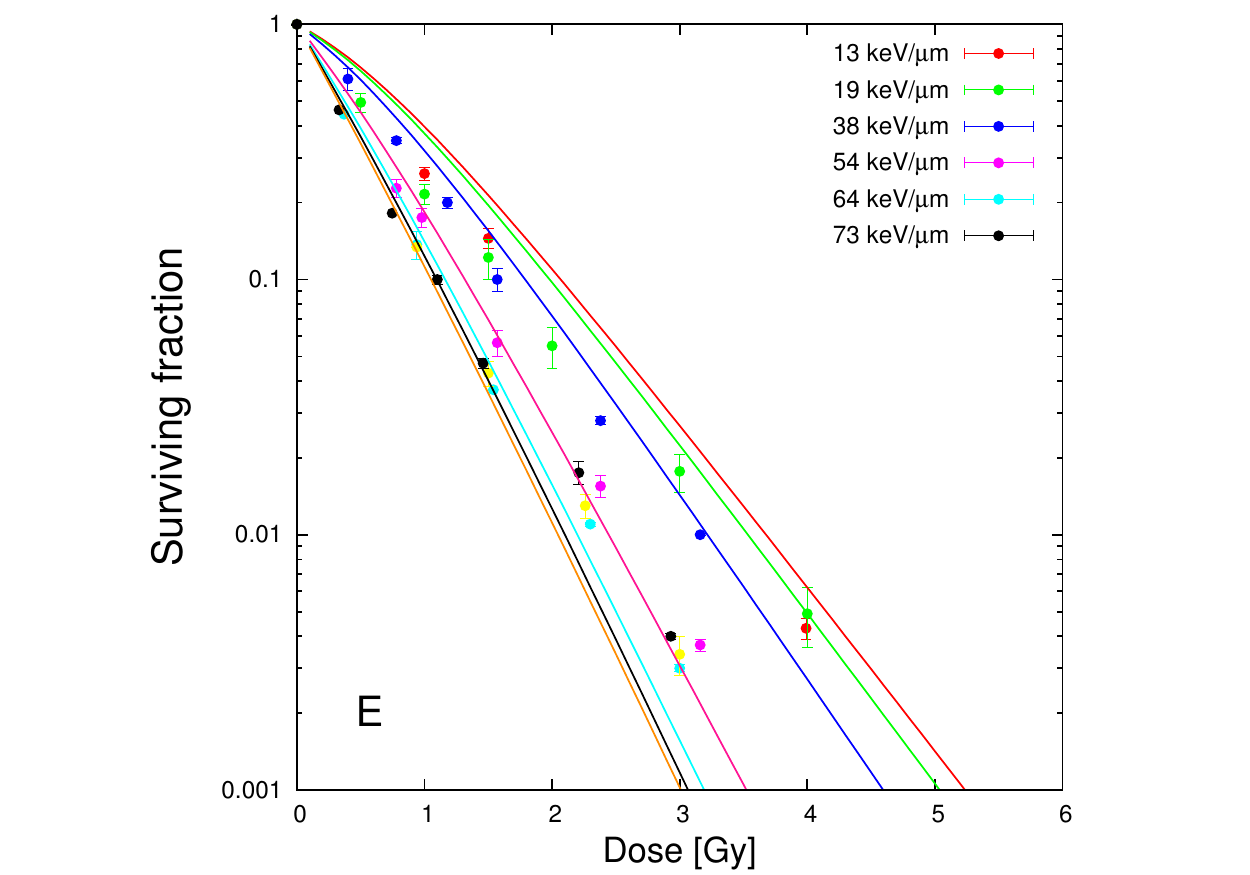} &
\includegraphics[width=0.55\textwidth]{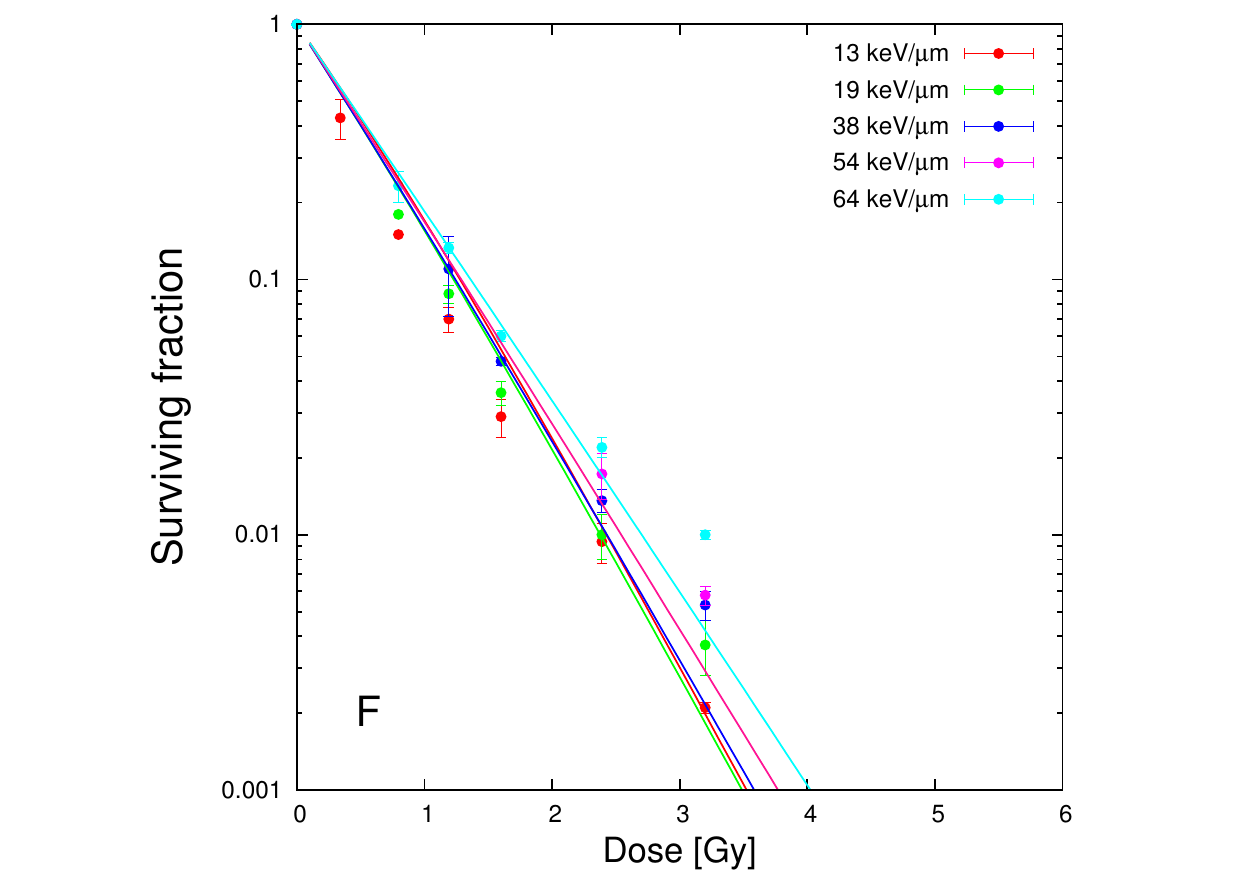} \\
 \end{tabular}
\end{center}
\caption{Survival of normal human skin fibroblasts after irradiation by ion beams of initial energy: $135\;MeV/n$ carbon ions (A); $290\;MeV/n$ carbon ions (B); $230\;MeV/n$ neon ions (C); $400\;MeV/n$ neon ions (D); $490\;MeV/n$ silicon ions (E); $500\;MeV/n$ iron ions (F). The track-segment values of LET for each ion species are given in each panel. Solid lines represent calculations using the best-fitted TST parameter values, see Table \ref{tab.parametersNHSF}. Data points, their errors (one standard deviation for three independent experiments) and values of LET are from \cite{tsuruoka2005}.}
\label{fig.sftsuruoka}
\end{figure}

In Fig. \ref{fig.sftsuruoka} the set of TST-calculated survival curves is compared with the respective experimental data points of \cite{tsuruoka2005}. Initial beam energies of the carbon, neon, silicon and iron ions and values of LET for these ions listed in this figure are those given by Tsuruoka et al. We wish to pay attention to 'hockey-stick' features present in some of the measured dose-survival dependences, especially observed in survival curves obtained at higher values of LET. Some survival curves do not decrease with the dose as fast as it is expected. The explanation given by \cite{tsuruoka2005}. for this behaviour is that the cell population has not been irradiated uniformly because of stopping effects of ion beams. We will refer to this phenomenon during our later discussion.

\begin{figure}[!ht]
\begin{center}
 \begin{tabular}{cc}
\includegraphics[width=0.95\textwidth]{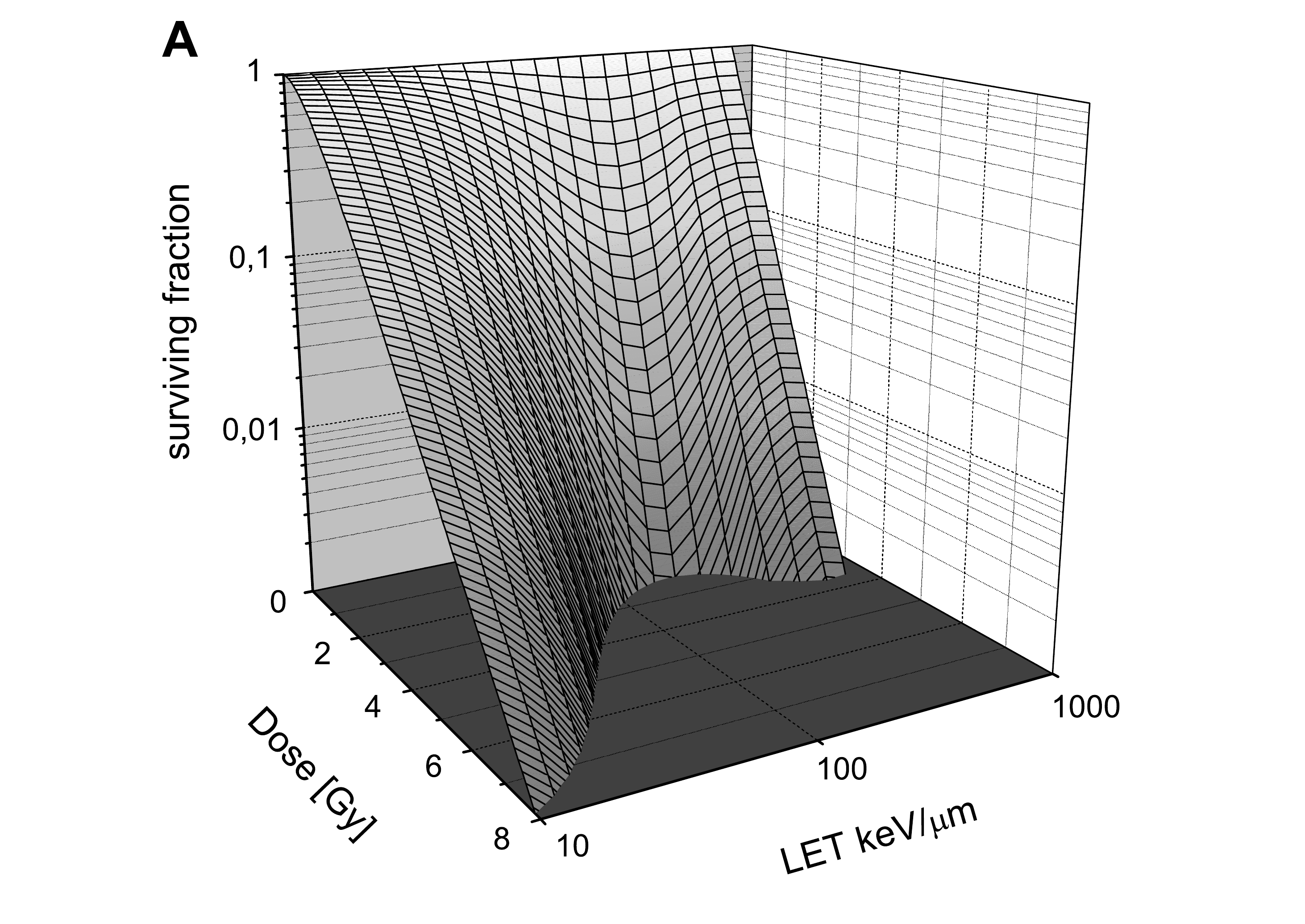} \\
\includegraphics[width=0.95\textwidth]{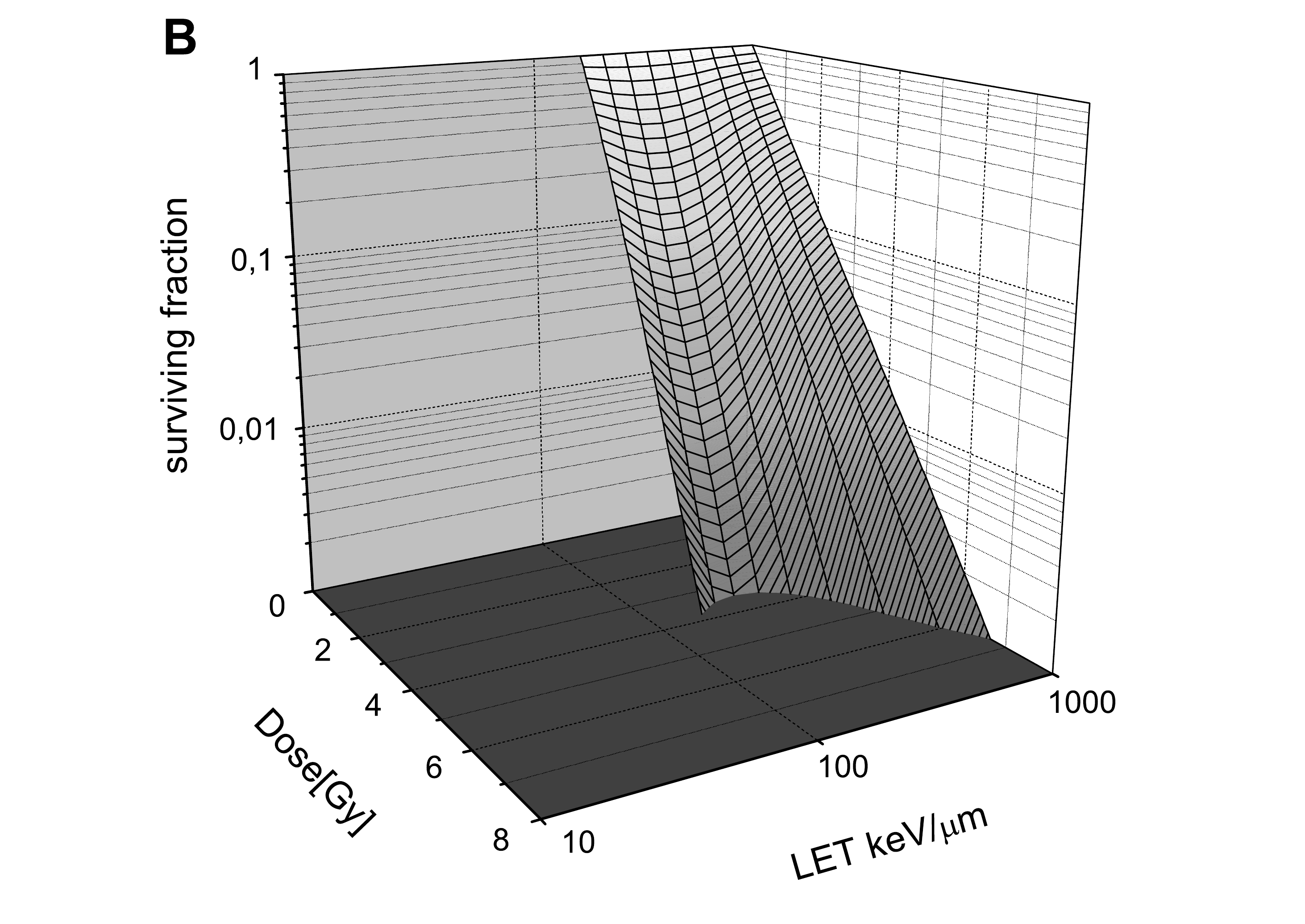} \\
 \end{tabular}
\end{center}
\caption{TST-predicted \emph{in vitro} survival of normal human skin fibroblasts irradiated by (A) carbon ions, and (B) iron ions, over the available ranges of LET values of these ions. For TST parameter values, see Table \ref{tab.parametersNHSF}.}
\label{fig.matrix}
\end{figure}

To illustrate the predictive capability of cellular Track Structure Theory, we show in Fig. \ref{fig.matrix} predictions of the shapes of the track-segment survival vs. dose curves, with LET (in logarithmic representation) as the third continuous parameter. These 3-D plots were calculated using the best-fitted parameter values derived in this work, representing \emph{in vitro} cell survival in normal human skin fibroblasts, for carbon and iron ions. Once the respective set of TST parameters have been established for a given endpoint, track structure calculations will yield such predictions for any ion species.

\begin{figure}[!ht]
\begin{center}
\includegraphics[width=1\textwidth]{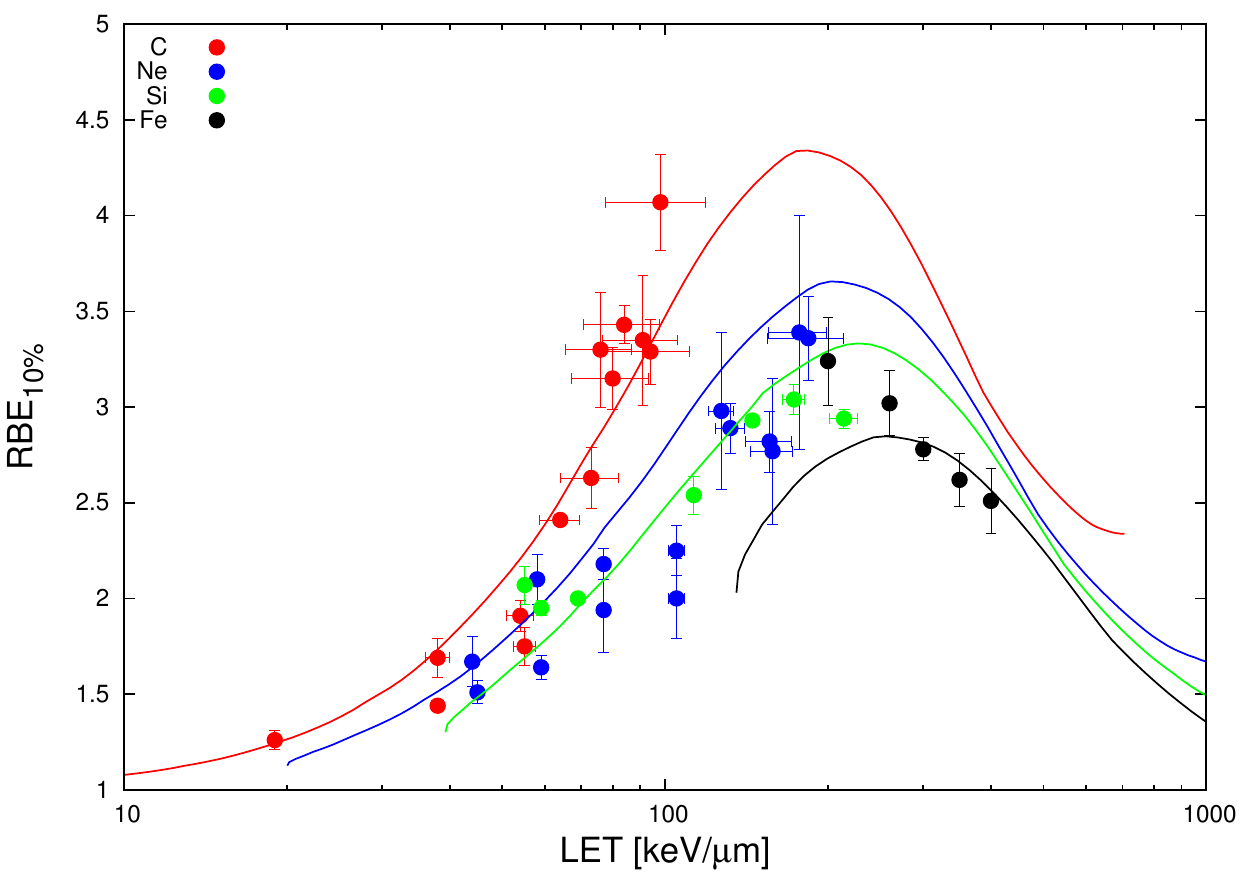}
\end{center}
\caption{RBE vs. LET for cell killing of normal human skin fibroblasts, at $10\%$ survival. Full lines represent TST calculations, for C (red line), Ne (blue line), Si (green line) and Fe (black line) ions. For TST parameter values, see Table \ref{tab.parametersNHSF}. Data points and their errors (one standard deviation for three independent experiments) are from \cite{tsuruoka2005}.}
\label{fig.rbetsuruoka}
\end{figure}

Model predictions allow for the comparison of absorbed dose of reference radiation and dose originating from the beam of ions resulting in the same biological effect. Thus, TST enables one to calculate the RBE-LET dependences at any level of survival. In Fig. \ref{fig.rbetsuruoka} we show the track-segment RBE-LET dependences at $10\%$ survival for normal human skin fibroblasts irradiated with carbon, neon, silicon and iron ions. In this figure, the experimental values of RBE and their error bars have been taken from Table II of the paper of \cite{tsuruoka2005}, while the LET error bars represent the range of values of LET as listed in Table I of their paper. In Table \ref{tab.rbeNHSF} we compare the TST-calculated values of RBE with experimental data listed in Table II of Tsuruoka et al. for the four ion species at their respective values of LET.

\begin{center}
\begin{longtable}{l c c c c c}
\hline

            &             &  {\footnotesize Track}          &                    &             &     \\
            & {\footnotesize LET}      &  {\footnotesize segment energy} &                    &             & {\footnotesize RBE} \\
{\footnotesize Ion species} & {\footnotesize $keV/\mu m$} & {\footnotesize ($MeV/n$)}       & {\footnotesize ($z^{*2}/\beta^2)$} & {\footnotesize RBE$_{exp}$} & {\footnotesize (this work)} \\
\hline 
\endhead

\hline \multicolumn{5}{c}{\emph{Continued on next page}} \\ \hline
\endfoot

\endlastfoot

{\footnotesize Carbon beams,}   &{\footnotesize 13 } &{\footnotesize 286.0 }& {\footnotesize 86.8 } &{\footnotesize 1.11  $\pm$ 0.01 }& {\footnotesize 1.12 } \\
{\footnotesize initial energy:} &{\footnotesize 19 }&{\footnotesize 156.9  }& {\footnotesize 134.5 } &{\footnotesize 1.26 $\pm$ 0.05 }& {\footnotesize 1.24 } \\
{\footnotesize $135\; MeV/n$ or} &{\footnotesize 38 }&{\footnotesize 61.60  }& {\footnotesize 299.2 } &{\footnotesize 1.44 $\pm$ 0.01 }& {\footnotesize 1.73 } \\
{\footnotesize $290\;MeV/n$}     &{\footnotesize 38 }&{\footnotesize 61.60  }&  {\footnotesize 299.2 }&{\footnotesize 1.69 $\pm$ 0.10 }& {\footnotesize 1.73 } \\
 &{\footnotesize 54 }&{\footnotesize 39.66 }& {\footnotesize 449.9 }&{\footnotesize 1.91 $\pm$ 0.08 }& {\footnotesize 2.20 }\\
 &{\footnotesize 55 }&{\footnotesize 38.74 }& {\footnotesize 459.9 }&{\footnotesize 1.75 $\pm$ 0.10 }& {\footnotesize 2.24 }\\
 &{\footnotesize 64 }&{\footnotesize 32.09 }& {\footnotesize 549.4 }&{\footnotesize 2.41 $\pm$ 0.02 }& {\footnotesize 2.51 }\\
 &{\footnotesize 73 }&{\footnotesize 27.20 }& {\footnotesize 643.4 }&{\footnotesize 2.63 $\pm$ 0.16 }& {\footnotesize 2.78 }\\
 &{\footnotesize 76 }&{\footnotesize 25.92 }& {\footnotesize 673.6 }&{\footnotesize 3.30 $\pm$ 0.30 }& {\footnotesize 2.86 }\\
 &{\footnotesize 80 }&{\footnotesize 24.41 }& {\footnotesize 713.7 }&{\footnotesize 3.15 $\pm$ 0.16 }& {\footnotesize 2.96 }\\
 &{\footnotesize 84 }&{\footnotesize 22.98 }& {\footnotesize 756.2 }&{\footnotesize 3.43 $\pm$ 0.10 }& {\footnotesize 3.07 }\\
 &{\footnotesize 91 }&{\footnotesize 20.76 }& {\footnotesize 833.9 }&{\footnotesize 3.35 $\pm$ 0.34 }& {\footnotesize 3.25 }\\
 &{\footnotesize 94 }&{\footnotesize 19.96 }& {\footnotesize 866.0 }&{\footnotesize 3.29 $\pm$ 0.17 }& {\footnotesize 3.32 }\\
 &{\footnotesize 98 }&{\footnotesize 18.92 }& {\footnotesize 912.1 }&{\footnotesize 4.07 $\pm$ 0.25 }& {\footnotesize 3.42 }\\
\hline 
{\footnotesize Neon beams,}   &{\footnotesize 30 } &{\footnotesize 409.6 }& {\footnotesize 193.2  }&{\footnotesize 1.43  $\pm$ 0.09 }& {\footnotesize 1.34 } \\
{\footnotesize initial energy:} &{\footnotesize 44 }&{\footnotesize 206.6  }& {\footnotesize 302.9  }&{\footnotesize 1.67 $\pm$ 0.13 }& {\footnotesize 1.64 } \\
{\footnotesize $230\; MeV/n$ or} &{\footnotesize 45 }&{\footnotesize 199.4  }& {\footnotesize 310.9 } &{\footnotesize 1.51 $\pm$ 0.06 }& {\footnotesize 1.66 } \\
{\footnotesize $400\;MeV/n$}     &{\footnotesize 58 }&{\footnotesize 137.1  }& {\footnotesize 416.3  }&{\footnotesize 2.10 $\pm$ 0.13 }& {\footnotesize 1.94 } \\
 &{\footnotesize 59 }&{\footnotesize 133.8 }& {\footnotesize 424.7 }&{\footnotesize 1.64 $\pm$ 0.06 }& {\footnotesize 1.96 }\\
 &{\footnotesize 77 }&{\footnotesize 92.86 }& {\footnotesize 577.7 }&{\footnotesize 1.94 $\pm$ 0.22 }& {\footnotesize 2.35 }\\
 &{\footnotesize 77 }&{\footnotesize 92.86 }& {\footnotesize 577.7 }&{\footnotesize 2.18 $\pm$ 0.08 }& {\footnotesize 2.35 }\\
 &{\footnotesize 105 }&{\footnotesize 62.06 }& {\footnotesize 826.2 }&{\footnotesize 2.00 $\pm$ 0.21 }& {\footnotesize 2.87 }\\
 &{\footnotesize 105 }&{\footnotesize 62.06 }& {\footnotesize 826.2 }&{\footnotesize 2.25 $\pm$ 0.13 }& {\footnotesize 2.87 }\\
 &{\footnotesize 127 }&{\footnotesize 48.61 }& {\footnotesize 1033.3 }&{\footnotesize 2.98 $\pm$ 0.41 }& {\footnotesize 3.19 }\\
 &{\footnotesize 132 }&{\footnotesize 46.36 }& {\footnotesize 1079.6 }&{\footnotesize 2.89 $\pm$ 0.13 }& {\footnotesize 3.25 }\\
 &{\footnotesize 156 }&{\footnotesize 37.76 }& {\footnotesize 1307.2 }&{\footnotesize 2.82 $\pm$ 0.16 }& {\footnotesize 3.46 }\\
 &{\footnotesize 158 }&{\footnotesize 37.13 }& {\footnotesize 1328.2 }&{\footnotesize 2.77 $\pm$ 0.38 }& {\footnotesize 3.47 }\\
 &{\footnotesize 177 }&{\footnotesize 32.16 }& {\footnotesize 1520.4 }&{\footnotesize 3.39 $\pm$ 0.61 }& {\footnotesize 3.58 }\\
 &{\footnotesize 184 }&{\footnotesize 30.58 }& {\footnotesize 1594.4 }&{\footnotesize 3.36 $\pm$ 0.22 }& {\footnotesize 3.61 }\\
\hline 
{\footnotesize Silicon beam,}   &{\footnotesize 55 } &{\footnotesize 472.0 }& {\footnotesize 350.2  }&{\footnotesize 2.07 $\pm$ 0.10 }& {\footnotesize 1.69 } \\
{\footnotesize initial energy:} &{\footnotesize 59 }&{\footnotesize 406.7 }& {\footnotesize 380.2  }&{\footnotesize 1.95 $\pm$ 0.04 }& {\footnotesize 1.76 } \\
{\footnotesize $490\; MeV/n$}   &{\footnotesize 69 }&{\footnotesize 299.3 }& {\footnotesize 458.8  }&{\footnotesize 2.00 $\pm$ 0.01 }& {\footnotesize 1.95 } \\
 &{\footnotesize 113 }&{\footnotesize 138.2 }& {\footnotesize 811.0  }&{\footnotesize 2.54 $\pm$ 0.10 }& {\footnotesize 2.65 } \\
 &{\footnotesize 145 }&{\footnotesize 98.56 }& {\footnotesize 1063.2 }&{\footnotesize 2.93 $\pm$ 0.01 }& {\footnotesize 2.94 }\\
 &{\footnotesize 173 }&{\footnotesize 77.72 }& {\footnotesize 1322.9 }&{\footnotesize 3.04 $\pm$ 0.08 }& {\footnotesize 3.19 }\\
 &{\footnotesize 214 }&{\footnotesize 58.87 }& {\footnotesize 1697.0 }&{\footnotesize 2.94 $\pm$ 0.05 }& {\footnotesize 3.32 }\\
\hline 
{\footnotesize Iron beam,}   &{\footnotesize 200 } &{\footnotesize 421.4 }& {\footnotesize 1285.3  }&{\footnotesize 3.24 $\pm$ 0.23 }& {\footnotesize 2.73 }\\
{\footnotesize initial energy:} &{\footnotesize 260 }&{\footnotesize 256.3 }& {\footnotesize 1755.2  }&{\footnotesize 3.02 $\pm$ 0.17 }& {\footnotesize 2.84 }\\
{\footnotesize $500\; MeV/n$}   &{\footnotesize 300 }&{\footnotesize 203.8 }& {\footnotesize 2067.2  }&{\footnotesize 2.78 $\pm$ 0.06 }& {\footnotesize 2.81 }\\
 &{\footnotesize 350 }&{\footnotesize 160.5 }& {\footnotesize 2478.2  }&{\footnotesize 2.62 $\pm$ 0.14 }& {\footnotesize 2.71 }\\
 &{\footnotesize 400 }&{\footnotesize 132.8 }& {\footnotesize 2882.4} &{\footnotesize 2.51 $\pm$ 0.17 }& {\footnotesize 2.56 }\\
\hline
\caption{Values of track-segment energy, $z^{*2}/\beta^2$ and TST-calculated RBE. Experimental values of LET, RBE$_{exp}$ are from Tsuruoka et al. 2005.}
\label{tab.rbeNHSF}
\end{longtable}
\end{center}

Our interpretation of the measured RBE-LET dependences is somewhat different from that offered by Tsuruoka et al. For each ion species (C, Ne, Si and Fe) the TST-calculated RBE-LET dependences gradually increase with increasing LET until they reach a maximum value and next slowly decrease. The calculated RBE maxima and respective values of LET at which they occur are as follows: RBE$= 4.3$ at $183\;keV/\mu m$ for carbon ions, $3.6$ at $202\;keV/\mu m$ for neon ions, $3.3$ at $227\;keV/\mu m$ for silicon ions and $2.8$ at $255\;keV/\mu m$ for iron ions. In general, RBE values tend to become higher for lighter ions over the whole range of LET values of the respective ions. Also, LET values at which RBE maxima occur tend to move upwards with increasing charge $z$ of the ion.

The factor affecting our assessment of the best-fitted values of TST parameters was the presence of 'hockey-stick' features in some of the measured dose-survival dependences above $1.5\;Gy$ (e.g., at $64,\;91\; 94$ and $98\; keV/\mu m$ for C ions, at $177$ and $184\;keV/\mu m$ for Ne ions, or in all measured data for Fe ions), cf. Fig \ref{fig.sftsuruoka}. In our parameter fitting procedure we initially eliminated these outlying data points and observed the stability of the fitted values in our $\chi^2$ optimisation by successively adding some of the omitted data.

\begin{figure}[!ht]
\begin{center}
\includegraphics[width=0.9\textwidth]{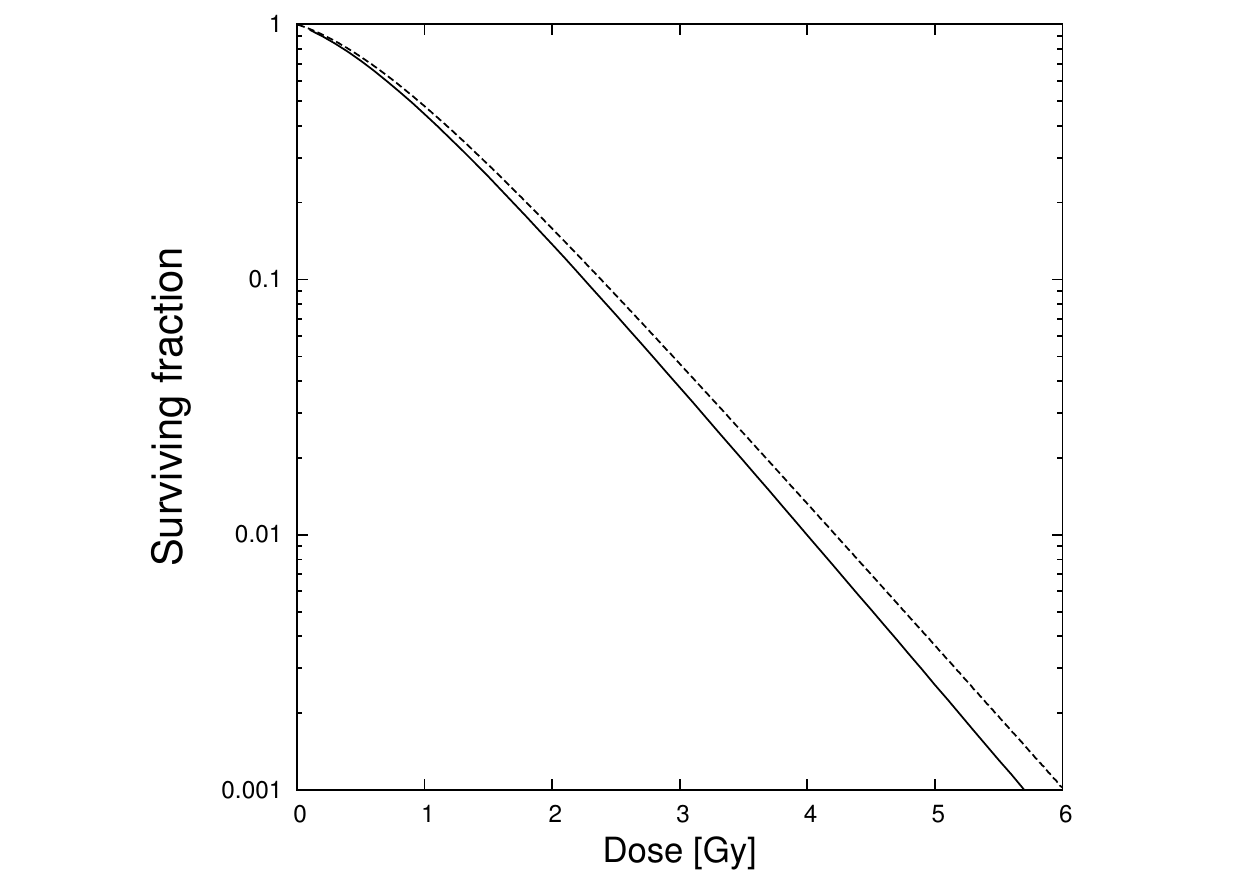}
\end{center}
\caption{TST-calculated survival for normal human skin fibroblasts after irradiation by C beams of initial energy $290\;MeV/n$ at depth $130\;mm$ in water. Full line represents survival after C ions of LET$=30\;keV/\mu m$ only, dashed line - survival after a mixture of carbon (primary) and secondary particles produced in PMMA, at the same depth.}
\label{fig.mixingNHSF}
\end{figure}

Since we based our analysis of the data of Tsuruoka et al. on the track-segment ion energy values re-calculated from the values of LET provided by these authors, we implied that only ions of a given species (C, Ne, Si or Fe) contribute to the LET values listed by these authors. In fact in the experiments of Tsuruoka et al., secondary particles generated in the PMMA absorbers of different thickness accompanied the primary ions along the beam depth and have additionally contributed to the LET values measured by these authors at cell sample positions. To evaluate by a TST calculation the likely effect of such secondary beam particles on the cell survival-dose dependences, we applied the data quoted in the paper of \cite{matsufuji2003}. We estimated the contribution to the total dose from their Figure 12, and the mean LET values of the primary and secondary particles in a carbon beam of initial energy $290\;MeV/n$ irradiating a PMMA layer of $130\;mm$ water equivalent thickness - from their Figure 10. We note that since \cite{matsufuji2003} estimate the residual range of this beam at $153\;mm$ in water (cf. their Table I), this comparison of secondary particles is not representative of the Bragg peak region. Using our eq.(\ref{eq.survivaleff}), we calculated the surviving fraction of normal human skin fibroblasts for this carbon beam taking into account the presence of lighter fragments, as estimated by \cite{matsufuji2003}. The relative contribution of these ions to the total dose, the mean LET value, track segment energy (estimated from the mean LET of the ion by our inverse energy-LET calculation, fluence (estimated at $1 \;Gy$ of total dose), effective charge of the ion species present in carbon beam, are shown in Table \ref{tab.mixingNHSF}. From the data of \cite{matsufuji2003} we estimated the 'dose-averaged' LET of a $290\;MeV$ carbon beam irradiating a PMMA layer of $130\; mm$ water equivalent thickness to be about $30\;keV/\mu m$ (including primary and secondary ions). The result of our TST calculations shown in Fig. \ref{fig.mixingNHSF} demonstrate that one may expect higher survival of normal human skin fibroblasts if the effect of secondary particles is included in such calculations, than if only carbon ions of $30\;keV/\mu m$ are assumed. This appears to correctly represent the trend observed for carbon ions (red line) around $30\;keV/\mu m$ in Fig. \ref{fig.sftsuruoka}. 

\begin{table}[!ht]
\begin{center}
\caption{Relative contribution to the total dose, mean LET value, back-calculated mean track-segment energy, fluence (calculated at $1\;Gy$ of total dose), effective charge for ion species contributing to a beam of carbon ions with initial energy $290\;MeV/n$ after its passage through PMMA of $130\;mm$ water equivalent thickness, as estimated from the data of \cite{matsufuji2003}.}
\begin{tabular}{l c c c c c}
\hline
 & {\footnotesize Relative} &  &{\footnotesize Fluence} & \\
{\footnotesize Atomic} & {\footnotesize contribution to }& {\footnotesize Mean LET }& {\footnotesize Mean energy }& {\footnotesize ($particles/cm^2$) }& \\
{\footnotesize number} & {\footnotesize dose, $\%$} & {\footnotesize ($keV/\mu m$)} & {\footnotesize ($MeV/n$) }& {\footnotesize per $1\;Gy$ total dose} &  {\footnotesize $z^{*}$ }\\
\hline
{\footnotesize 6 }&{\footnotesize 78.9 }&{\footnotesize 30.0 }&{\footnotesize 83.5 }&{\footnotesize 1.6 $\cdot$ 10$^7$ }&{\footnotesize 6.00 }\\
{\footnotesize 5 }&{\footnotesize 6.5  }&{\footnotesize 20.0 }&{\footnotesize 88.0 }&{\footnotesize 2.0 $\cdot$ 10$^6$ }&{\footnotesize 5.00 }\\
{\footnotesize 4 }&{\footnotesize 1.3  }&{\footnotesize 10.0 }&{\footnotesize 123.5 }&{\footnotesize 9.1 $\cdot$ 10$^5$ }&{\footnotesize 4.00 }\\
{\footnotesize 3 }&{\footnotesize 1.6  }&{\footnotesize 6.0 }&{\footnotesize 113.0 }&{\footnotesize 1.6 $\cdot$ 10$^6$ }&{\footnotesize 3.00 }\\
{\footnotesize 2 }&{\footnotesize 6.9  }&{\footnotesize 4.5 }&{\footnotesize 56.7 }&{\footnotesize 9.6 $\cdot$ 10$^6$ }&{\footnotesize 2.00 }\\
{\footnotesize 1 }&{\footnotesize 4.8  }&{\footnotesize 2.0 }&{\footnotesize 27.6 }&{\footnotesize 1.5 $\cdot$ 10$^7$ }&{\footnotesize 1.00 }\\
\hline
\end{tabular}
\label{tab.mixingNHSF}
\end{center}
\end{table}

\begin{figure}[!ht]
\begin{center}
\includegraphics[width=0.92\textwidth]{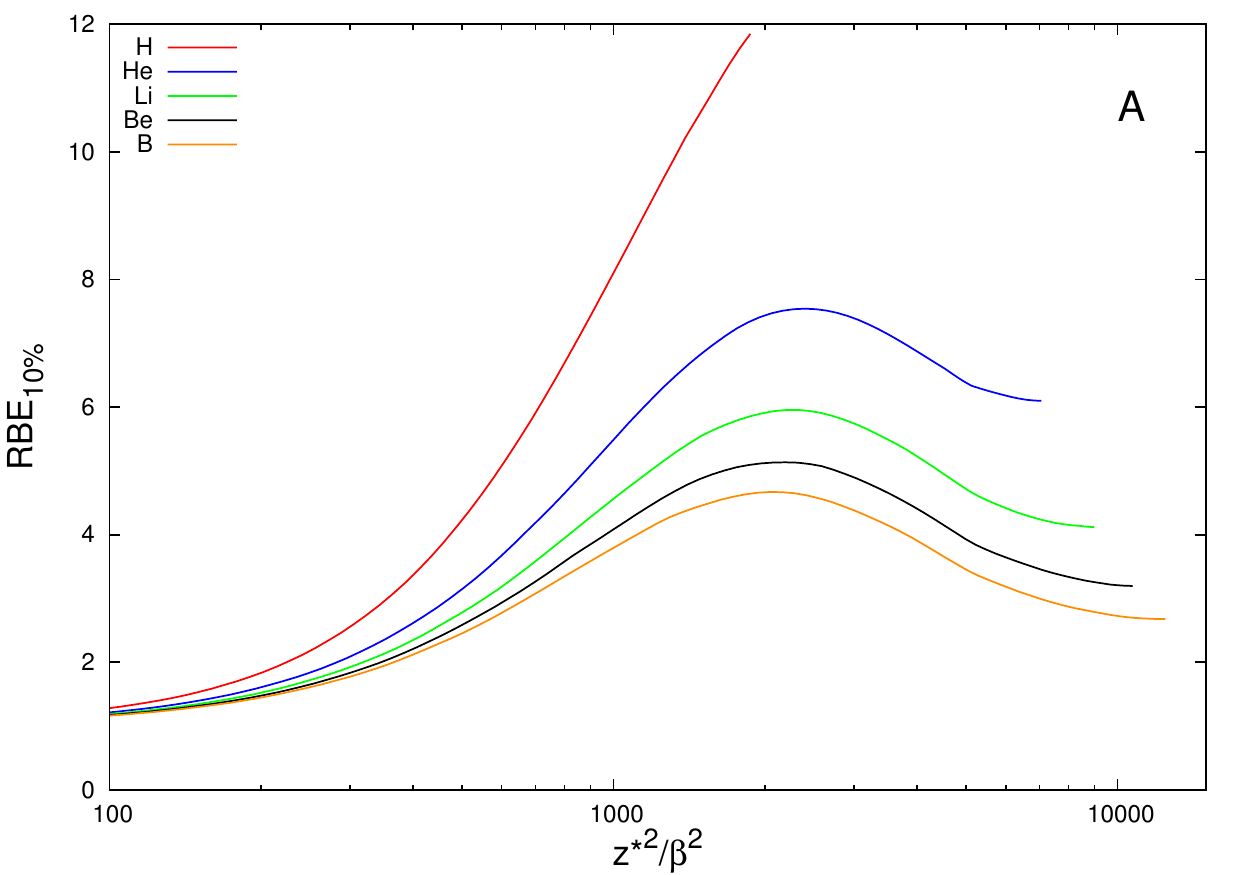}
\includegraphics[width=0.92\textwidth]{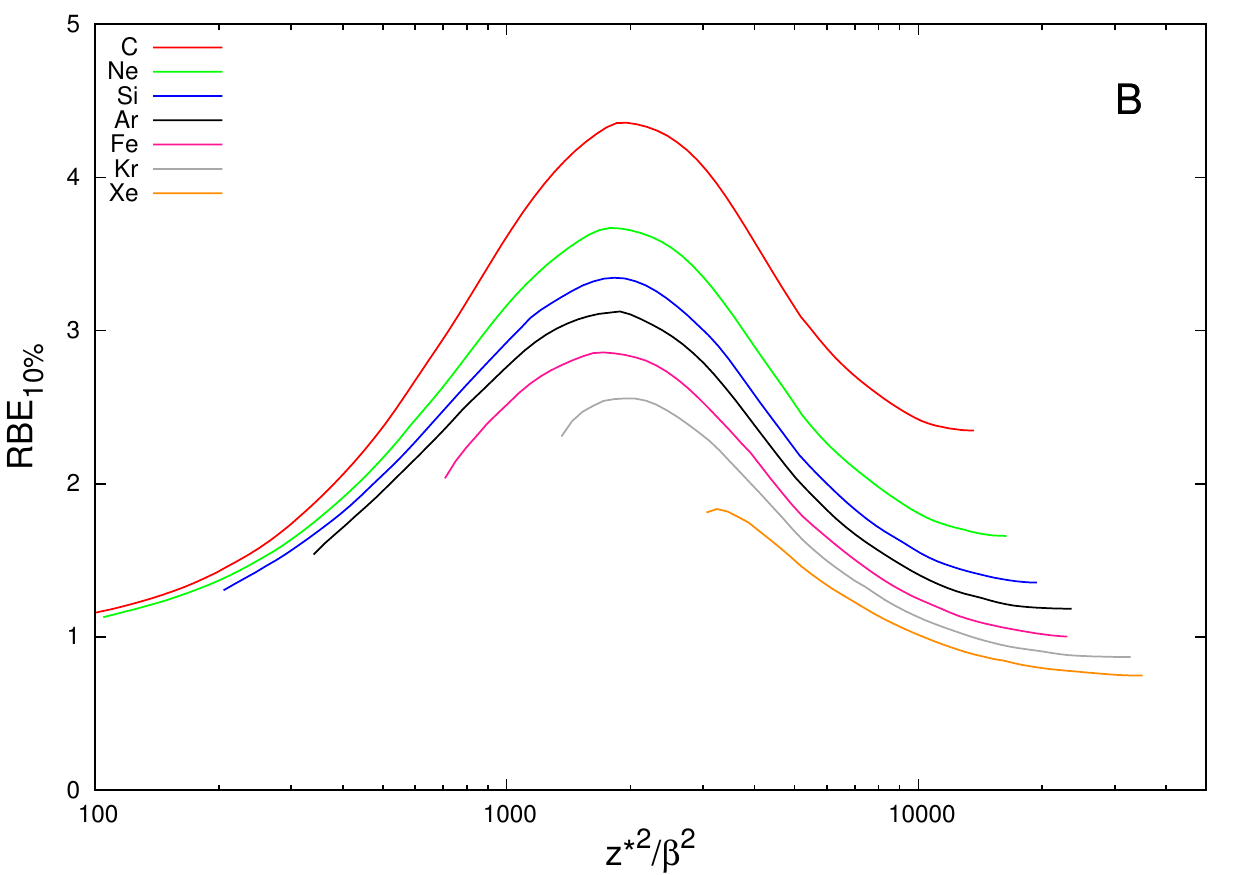}
\end{center}
\caption{TST model predicted RBE versus $z^{*2}/\beta^2$ dependences for cell survival in normal human skin fibroblasts, at $10\%$ survival, for a range on ions. Lines of different colours represent results of TST calculations for H, He, Li, Be and B ions in panel A, and for C, Ne, Si, Ar, Fe, Kr and Xe ions in panel B. For TST parameter values, see Table \ref{tab.parametersNHSF}. }
\label{fig.tsuruokarbeletlekkiejony}
\end{figure}

On our closer inspection of the experimental results and their model representation shown in this work, we may generally conclude that our inability (for lack of sufficient input data) to account for the effect of secondary charged particles generated in ion beams attenuated by PMMA absorbers of varied thickness is the main source of the discrepancy between model calculations and experiment. These discrepancies are most apparent, e.g., for carbon data at their highest LET values (LET$= 98\;keV/\mu m$), for Ne ions at LET values above $100\;keV/\mu m$, for Si ions at LET$= 214\;keV/\mu m$ or for Fe ions of LET of $200$ or $260\;keV/\mu m$, where 'hockey-stick' features are seen in the dose-survival curves and where model calculations overestimate the observed survival (Fig. \ref{fig.sftsuruoka}) thus generally underestimating the RBE values measured in these experiments (Fig. \ref{fig.rbetsuruoka}). Another example is the slight difference observed in the RBE values from two carbon ion bombardments, both of LET$= 38\;keV/\mu m$, obtained from carbon beams of initial energies $290\;MeV/n$ and $135\;MeV/n$. Here, PMMA absorbers of different thickness were used to attenuate the ion energy and arrive at similar values of their LET.

Once the representative set of TST parameters have been established for a given endpoint, track structure calculations will yield such predictions for any ion species. Thus based on model parameters describing the survival of normal human skin fibroblasts we calculated the values of RBE for ions of charge lighter than those used in experiment, in particular for hydrogen, helium, lithium, beryllium and boron ions. TST-predicted values of RBE versus $z^{*2}/\beta^2$ are plotted in Fig. \ref{fig.tsuruokarbeletlekkiejony} (panel A). The unique dependence of RBE on the charge $Z$ of the ion remains, with RBE maxima occurring around a single value of $z^{*2}/\beta^2$ of about $2000$, or $z^{*2}/\kappa\beta^2 = 1.8$. To better illustrate the systematic pattern offered by the track structure approach, TST-predicted dependences of RBE versus $z^{*2}/\beta^2$ for a wider range of ion species are also presented in this figure (panel B). While the shift of the positions of RBE maxima with $z^{*2}/\beta^2$ is reduced compared to RBE-LET plots (cf. Fig. \ref{fig.rbetsuruoka}), the overall pattern of separation with respect to ion charge ($Z$) remains. The shape of RBE-LET dependences simulated for ions of heavier charges seems to be correct. In the case of protons the track-segment RBE achieves values rising up to about $12$, which is inconsistent with the well-established value of RBE$=1.1$ used in proton radiotherapy. It is unfeasible to compare model predictions for protons in the high-LET region with experiment, because in reality the fulfillment of the track-segment conditions for such a light ion in this region of LET is difficult. There is a discrepancy between LET plotted in Fig. \ref{fig.tsuruokarbeletlekkiejony} which represents 'track-segment' values of LET corresponding to a single value of particle's energy and dose-averaged values of LET reported in many publications concerning experimental data. This discrepancy is more apparent in the case of a stopping light ion, than for heavier ions that suffer less energy loss straggling and multiple scattering. Without a suitable averaging approach, track structure model calculations will generally overestimate the experimentally measured RBE of light ions, such as protons or He, especially at their distal range. Unfortunately, the lack of any appropriate averaging procedure does not fully explain the appearance of the very inflated values of RBE calculated for light ions from TST. This deficiency of the TST model will be also noticed in our next model analysis of the survival of V79 Chinese hamster cells (see Section \ref{ch. furusawa}).

For other details of the cellular TST analysis of the survival of normal human skin fibroblasts, the paper by \cite{korcyl2009} should be consulted.

\section{Track Structure analysis of RBE and OER in V79 Chinese hamster cells}
\label{ch. furusawa}
The second set of TST-analysed biological data, published by \cite{furusawa2000}, concerns cell survival of V79 Chinese hamster cells \emph{in vitro} after their exposure to accelerated ion beams: helium (of initial energy $12\;MeV/n$), carbon ($12$ and $135\;MeV/n$) and neon ($135\;MeV/n$). The experiment was carried out at the NIRS medical cyclotron (NIRS-MC) and at the RIKEN ring cyclotron (RRC) facilities in Japan. V79 cells were irradiated in track segment conditions under aerobic or hypoxic conditions. The respective track-segment values of LET of different ions were obtained by changing the thickness of range shifters made of PMMA disks or of aluminium foils inserted upstream of the samples. For exposures under aerobic conditions, the inside of the irradiation chamber was flushed with air containing $5\%$ CO$_2$ or kept in atmospheric air and then exposed to the ion beams. For exposure under hypoxic conditions, the cells in their Petri dishes were flushed with $1000\;ml/min$ of pure nitrogen gas containing $5\%$ CO$_2$ gas for over $1\;h$ immediately before exposure. During radiation exposures in the chambers the flow rate of this gas mixture was $200\;ml/min$. Reference radiation was the same as in previous case - $200 \;kV$ X-rays, filtered with $0.5 \;mm$ aluminium and $0.5 \;mm$ copper (\cite{furusawa2000}).

We analysed a set of $67$ survival curves of V79 Chinese hamster cells measured under aerobic conditions and $59$ survival curves measured in hypoxic conditions, published by \cite{furusawa2000} and later corrected (\cite{furusawa2012}). All these survival curves were reported in the form of set of best-fitted $\alpha$ and $\beta$ parameters of the linear-quadratic model, eq.(\ref{eq.lq}), obtained from the measured survival data points. The survival curves for V79 were reconstructed from these values of $\alpha$ and $\beta$ parameters (\cite{sowa2005}). Next, 'data points' were calculated for the reconstructed survival curves at regular dose intervals, and a $10\%$ a 'experimental uncertainty' arbitrarily assigned to these 'data points'. 

We next best-fitted the TST parameters for V79 cells irradiated in aerobic or hypoxic conditions, in a a manner similar to that applied in the case of human skin fibroblasts, as described n the preceding section. The best-fitted values of TST parameters are presented in Table \ref{tab.parametersV79}, together with their parabolic errors, according to MINOS subroutine of MINUIT minimizing code.

\begin{table}[!ht]
\begin{center}
\caption{Values of best-fitted TST parameters (together with their parabolic errors) representing \emph{in vitro} survival in Chinese hamster cells V79 in aerobic or anoxic conditions, from the experimental data of \cite{furusawa2000}.}
\begin{tabular}{c c c c c}
\hline
& $\quad m \quad \quad$ & $D_0 \quad$ & $\sigma_0$ & $\quad \quad \kappa \quad$ \\
\hline \\
\footnotesize{aerobic V79} & \footnotesize{$\quad$ $2.91  \pm 0.12 $ $\quad \quad$}  & \footnotesize{$2.0504 \pm 0.0009\;Gy \quad$} & \footnotesize{$50.6 \pm 0.2\; \mu m^2$} & \footnotesize{$\quad \quad 689  \pm  2\quad$} \\
\footnotesize{hypoxic V79} & \footnotesize{$\quad$ $3.22  \pm 0.09$ $\quad \quad$}  & \footnotesize{$5.26  \pm 0.13 \;Gy \quad$} & \footnotesize{$55.29  \pm 0.03\;\mu m^2$} & \footnotesize{$\quad \quad 1002.2  \pm 0.1 \quad$} \\ \\
\hline
\end{tabular}
\label{tab.parametersV79}
\end{center}
\end{table}

\begin{figure}[!ht]
\begin{center}
\includegraphics[width=1.0\textwidth]{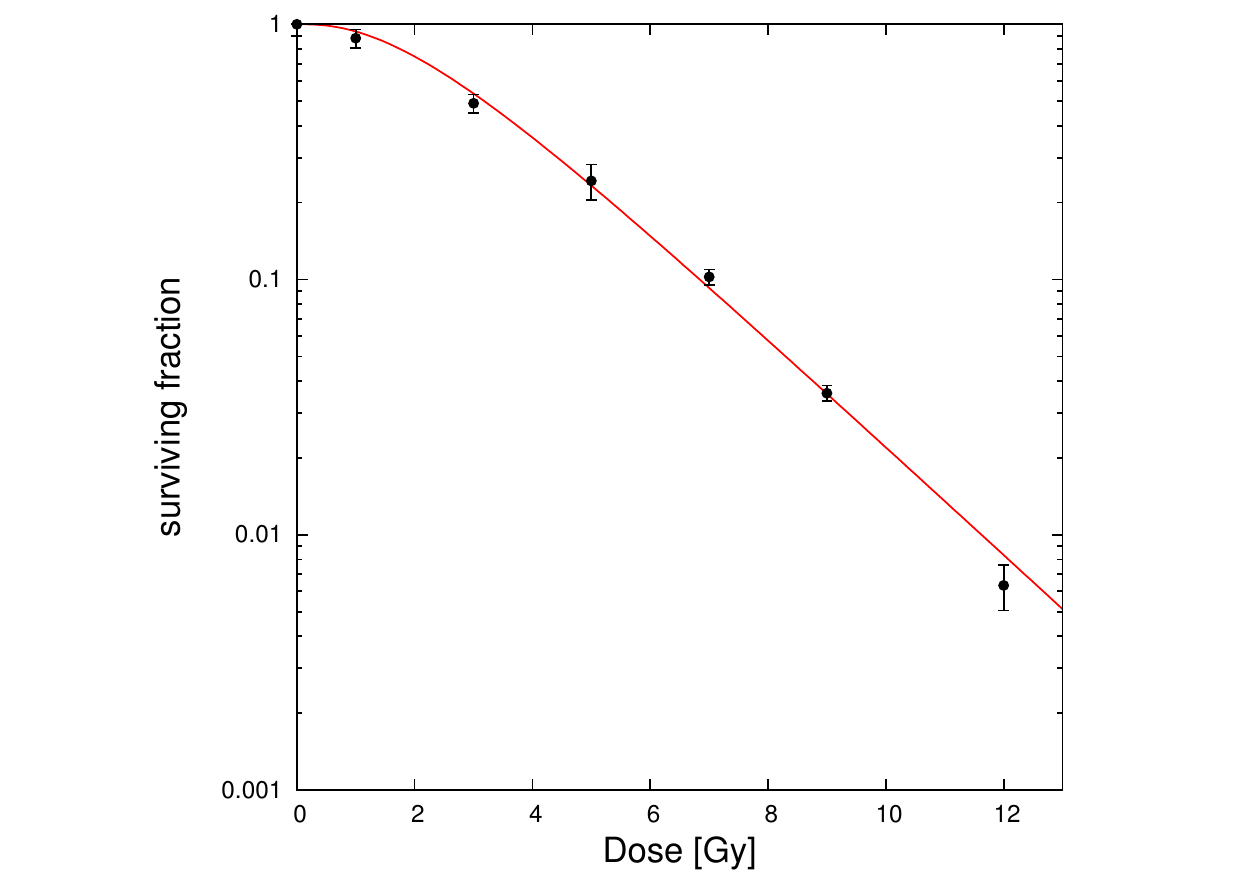}
\end{center}
\caption{Survival of Chinese hamster cells V79 after irradiation by $200 \;kV$ X-rays. The full line represents the TST calculation, with $m=2.91$, and $D_0=2.05\;Gy$. Data points ($\bullet$) and their errors (one standard deviation for five independent experiments) are from \cite{furusawa2000}. }
\label{fig.xraysfurusawa}
\end{figure}

\begin{figure}[!ht]
\begin{center}
\includegraphics[width=0.9\textwidth]{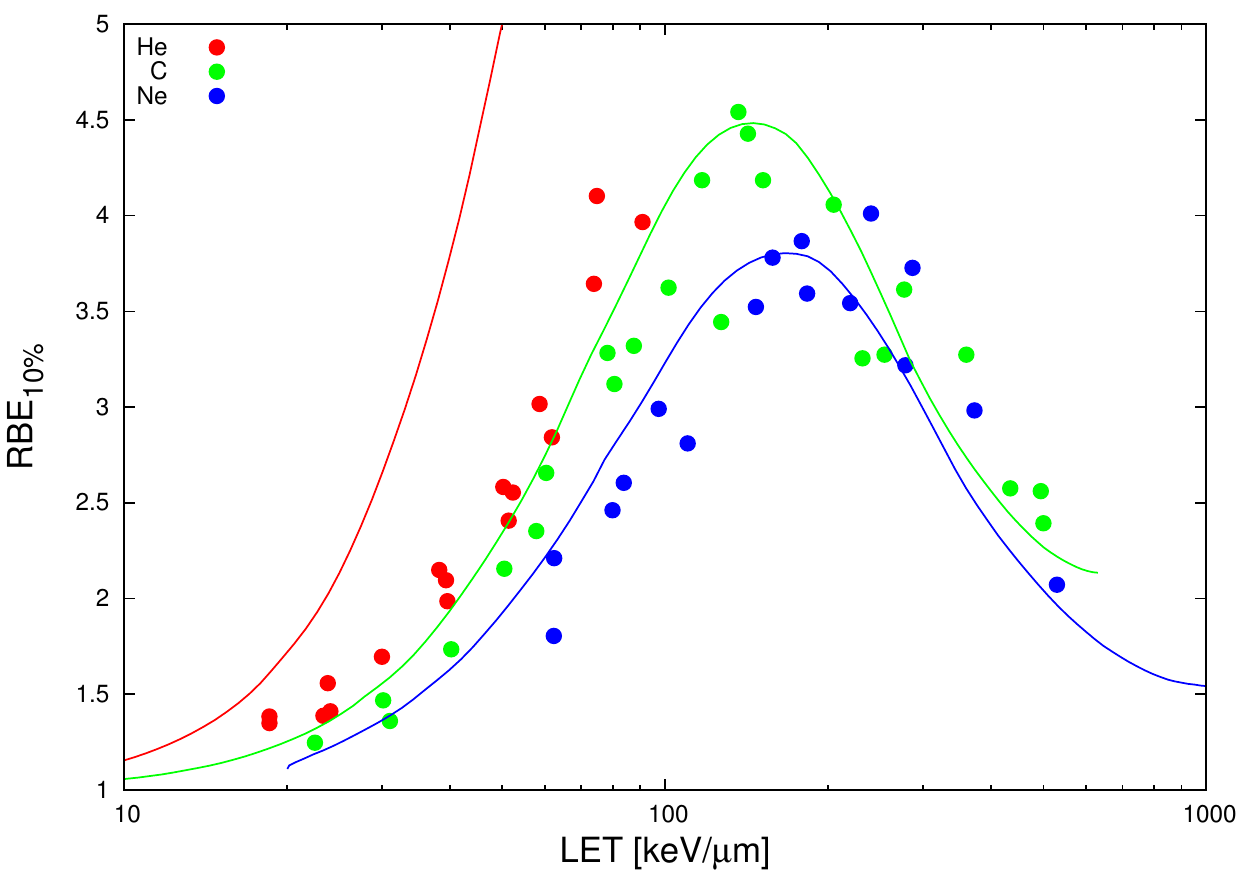}
\end{center}
\caption{RBE versus LET for cell killing of V79 irradiated in aerobic conditions, at $10\%$ survival. Full lines represent TST calculations, for He (red line), C (green line) and Ne (blue line) ions. For TST parameter values, see Table \ref{tab.parametersV79}. Data points are taken from Figure 3 from the publication by \cite{furusawa2000} and are shown without their error bars, as the respective experimental points on the survival curves are derived from their linear-quadratic representations.}
\label{fig.rbev79ox}
\end{figure}

The TST-calculated V79 $X$-ray survival curve under aerobic conditions, is compared in Fig. \ref{fig.xraysfurusawa} with experimental data points measured by \cite{furusawa2000}. Two of the four model parameters: $m=2.91$ and $D_0=2.05\;Gy$ best-fitted for the complete set of $67$ aerobic survival curves, were used to reconstruct this reference survival curve, as given by the m-target formula, eq.(\ref{eq.mtarget}). From Figure 3 of the paper by Furusawa et al. we read the values of LET and RBE at $10\%$ survival estimated by these authors for V79 irradiated with He, C and Ne ions under aerobic conditions and compared them in Fig. \ref{fig.rbev79ox} with TST predictions. Unfortunately, we were unable to find a set of model parameters to correctly reproduce the experimental data measured for all the ions studied - by attaching more weight to the heavier ion (carbon and neon) data in the fitting procedure. We find strongly diverging model predictions for the lighter helium ions, for either of irradiation conditions (aerobic or hypoxic). This divergence is also seen between measured and calculated RBE values, with model predictions highly overestimating the measured RBE values (see Table \ref{tab.rbeV79} and Fig. \ref{fig.rbev79ox}). We observe a similar pattern in TST model representation of survival of normal human skin fibroblasts after proton irradiation (see Fig. \ref{fig.tsuruokarbeletlekkiejony} panel A in section \ref{ch. tsuruoka}). The reasons for this discrepancy may be twofold: in the incorrect model representation of the radial distribution of dose around lighter ions, or in the incorrectness of the assumption that for a light ion passing through a cell, track segment irradiation holds (i.e. of a constant value of ion LET over the thickness of the cell nucleus). The degree of overestimation of TST predicted track-segment values of RBE for V79 cells irradiated with helium ion increases with increasing LET, which may suggest the lack of suitable LET averaging. For heavier ions, such as carbon or neon, the calculated RBE-LET dependences agree well with experiment. A clear separation of the RBE-LET dependences calculated for ion of different charges is seen in Fig. \ref{fig.rbev79ox}. This implies that the value of RBE cannot be uniquely predicted by the ion's LET. While the value of LET of ions of different charge may be the same, their track structure is not. Like in the case of normal human skin fibroblasts, TST-calculated RBE-LET dependences gradually increase with increasing LET until they reach a maximum value and next slowly decrease. The calculated RBE maxima and respective values of LET at which they occur are as follows: RBE$= 4.48$ at $145\;keV/\mu m$ for carbon ions, $3.8$ at $166\;keV/\mu m$ for neon ions.

\begin{center}
\begin{longtable}{l c c c c c}
\hline
            &             &  {\footnotesize Track}          &                    &             &     \\
            & {\footnotesize LET}      &  {\footnotesize segment energy} &                    &             & {\footnotesize RBE} \\
{\footnotesize Ion species} & {\footnotesize $keV/\mu m$} & {\footnotesize ($MeV/n$)}       & {\footnotesize ($z^{*2}/\beta^2)$} & {\footnotesize RBE$_{exp}$} & {\footnotesize (this work)} \\
\hline
\endhead

\hline \multicolumn{5}{c}{\emph{Continued on next page}} \\ \hline
\endfoot

\endlastfoot

{\footnotesize Helium beams,}   &{\footnotesize 18.6 } &{\footnotesize 9.74 }& {\footnotesize 194.0 } &{\footnotesize 1.35 }& {\footnotesize 1.61 } \\
{\footnotesize initial energy:} &{\footnotesize 18.6 }&{\footnotesize 9.74  }& {\footnotesize 194.0 } &{\footnotesize 1.38 }& {\footnotesize 1.61 } \\
{\footnotesize $12\; MeV/n$ } &{\footnotesize 23.0 }&{\footnotesize 7.53  }& {\footnotesize 250.1 } &{\footnotesize 1.39 }& {\footnotesize 1.94 } \\
 &{\footnotesize 23.8 }&{\footnotesize 7.18 }& {\footnotesize 262.4 }&{\footnotesize 1.56 }& {\footnotesize 2.02 } \\
 &{\footnotesize 24.0 }&{\footnotesize 7.13 }& {\footnotesize 264.0 }&{\footnotesize 1.41 }& {\footnotesize 2.03 }\\
 &{\footnotesize 29.9 }&{\footnotesize 5.37 }& {\footnotesize 349.7 }&{\footnotesize 1.70 }& {\footnotesize 2.63 }\\
 &{\footnotesize 38.1 }&{\footnotesize 3.92 }& {\footnotesize 477.1 }&{\footnotesize 2.15 }& {\footnotesize 3.57 }\\
 &{\footnotesize 39.2 }&{\footnotesize 3.78 }& {\footnotesize 494.4 }&{\footnotesize 2.09 }& {\footnotesize 3.70 }\\
 &{\footnotesize 39.4 }&{\footnotesize 3.76 }& {\footnotesize 497.4 }&{\footnotesize 1.98 }& {\footnotesize 3.72 }\\
 &{\footnotesize 50.0 }&{\footnotesize 2.73 }& {\footnotesize 681.2 }&{\footnotesize 2.58 }& {\footnotesize 5.00 }\\
 &{\footnotesize 51.9 }&{\footnotesize 2.59 }& {\footnotesize 718.3 }&{\footnotesize 2.40 }& {\footnotesize 5.22 }\\
 &{\footnotesize 52.3 }&{\footnotesize 2.57 }& {\footnotesize 722.5 }&{\footnotesize 2.55 }& {\footnotesize 5.24 }\\
 &{\footnotesize 58.9 }&{\footnotesize 2.18 }& {\footnotesize 846.6 }&{\footnotesize 3.02 }& {\footnotesize 5.90 }\\
 &{\footnotesize 61.9 }&{\footnotesize 2.04 }& {\footnotesize 907.9 }&{\footnotesize 2.84 }& {\footnotesize 6.18 }\\
 &{\footnotesize 73.9 }&{\footnotesize 1.57 }& {\footnotesize 1163.3 }&{\footnotesize 3.64 }& {\footnotesize 7.02 }\\
 &{\footnotesize 74.6 }&{\footnotesize 1.55 }& {\footnotesize 1176.7 }&{\footnotesize 4.10 }& {\footnotesize 7.05 }\\
 &{\footnotesize 90.8 }&{\footnotesize 1.17 }& {\footnotesize 1523.6 }&{\footnotesize 3.97 }& {\footnotesize 7.46 }\\
\hline 
{\footnotesize Carbon beams,}   &{\footnotesize 22.5 } &{\footnotesize 123.56 }& {\footnotesize 163.2 } &{\footnotesize 1.25 }& {\footnotesize 1.31 } \\
{\footnotesize initial energy:} &{\footnotesize 30.0 }&{\footnotesize 83.43  }& {\footnotesize 228.3 } &{\footnotesize 1.47 }& {\footnotesize 1.55 } \\
{\footnotesize $12\; MeV/n$ or} &{\footnotesize 31.0 }&{\footnotesize 80.10  }& {\footnotesize 236.7 } &{\footnotesize 1.36 }& {\footnotesize 1.58 } \\
{\footnotesize $135\; MeV/n$ or} &{\footnotesize 40.1 }&{\footnotesize 57.40  }&  {\footnotesize 319.3 }&{\footnotesize 1.73 }& {\footnotesize 1.93 } \\
 &{\footnotesize 50.3 }&{\footnotesize 43.16 }& {\footnotesize 415.7 }&{\footnotesize 2.15 }& {\footnotesize 2.35 }\\
 &{\footnotesize 57.8 }&{\footnotesize 36.56 }& {\footnotesize 485.7 }&{\footnotesize 2.35 }& {\footnotesize 2.65 }\\
 &{\footnotesize 60.0 }&{\footnotesize 34.80 }& {\footnotesize 509.5 }&{\footnotesize 2.65 }& {\footnotesize 2.74 }\\
 &{\footnotesize 78.5 }&{\footnotesize 24.91 }& {\footnotesize 699.8 }&{\footnotesize 3.28 }& {\footnotesize 3.45 }\\
 &{\footnotesize 80.6 }&{\footnotesize 24.15 }& {\footnotesize 721.0 }&{\footnotesize 3.12 }& {\footnotesize 3.51 }\\
 &{\footnotesize 88.0 }&{\footnotesize 21.60 }& {\footnotesize 798.2 }&{\footnotesize 3.32 }& {\footnotesize 3.73 }\\
 &{\footnotesize 102.0 }&{\footnotesize 17.94 }& {\footnotesize 960.4 }&{\footnotesize 3.62 }& {\footnotesize 4.10 }\\
 &{\footnotesize 117.0 }&{\footnotesize 15.15 }& {\footnotesize 1127.9 }&{\footnotesize 4.18 }& {\footnotesize 4.34 }\\
 &{\footnotesize 127.0 }&{\footnotesize 13.69 }& {\footnotesize 1247.7 }&{\footnotesize 3.44 }& {\footnotesize 4.43 }\\
 &{\footnotesize 137.0 }&{\footnotesize 12.49 }& {\footnotesize 1363.6 }&{\footnotesize 4.54 }& {\footnotesize 4.47 }\\
 &{\footnotesize 142.0 }&{\footnotesize 11.90 }& {\footnotesize 1421.1 }&{\footnotesize 4.42 }& {\footnotesize 4.48 }\\
 &{\footnotesize 206.0 }&{\footnotesize 7.40 }& {\footnotesize 2252.8 }&{\footnotesize 4.06 }& {\footnotesize 4.07 }\\
 &{\footnotesize 232.0 }&{\footnotesize 6.28 }& {\footnotesize 2627.7 }&{\footnotesize 3.25 }& {\footnotesize 3.79 }\\
 &{\footnotesize 255.0 }&{\footnotesize 5.50 }& {\footnotesize 2973.3 }&{\footnotesize 3.27 }& {\footnotesize 3.54 }\\
 &{\footnotesize 276.0 }&{\footnotesize 4.93 }& {\footnotesize 3285.1 }&{\footnotesize 3.61 }& {\footnotesize 3.32 }\\
 &{\footnotesize 360.0 }&{\footnotesize 3.31 }& {\footnotesize 4673.7 }&{\footnotesize 3.27 }& {\footnotesize 2.73 }\\
 &{\footnotesize 432.0 }&{\footnotesize 2.43 }& {\footnotesize 6035.5 }&{\footnotesize 2.57 }& {\footnotesize 2.44 }\\
 &{\footnotesize 493.0 }&{\footnotesize 1.89 }& {\footnotesize 7356.2 }&{\footnotesize 2.56 }& {\footnotesize 2.28 }\\
 &{\footnotesize 502.0 }&{\footnotesize 1.81 }& {\footnotesize 7593.9 }&{\footnotesize 2.39 }& {\footnotesize 2.26 }\\
\hline 
{\footnotesize Neon beams,}   &{\footnotesize 62.1 } &{\footnotesize 124.78 }& {\footnotesize 450.3 } &{\footnotesize 1.80 }& {\footnotesize 2.27 } \\
{\footnotesize initial energy:} &{\footnotesize 62.2 }&{\footnotesize 124.46  }& {\footnotesize 450.8 } &{\footnotesize 2.21 }& {\footnotesize 2.27 } \\
{\footnotesize $135\; MeV/n$ or} &{\footnotesize 80.0 }&{\footnotesize 88.13  }& {\footnotesize 604.6 } &{\footnotesize 2.46 }& {\footnotesize 2.79 } \\
 &{\footnotesize 84.6 }&{\footnotesize 82.31  }&  {\footnotesize 641.9 }&{\footnotesize 2.60 }& {\footnotesize 2.89 } \\
 &{\footnotesize 96.9 }&{\footnotesize 69.08 }& {\footnotesize 750.0 }&{\footnotesize 2.99 }& {\footnotesize 3.16 }\\
 &{\footnotesize 110.0 }&{\footnotesize 58.16 }& {\footnotesize 876.4 }&{\footnotesize 2.81 }& {\footnotesize 3.42 }\\
 &{\footnotesize 146.0 }&{\footnotesize 41.00 }& {\footnotesize 1210.4 }&{\footnotesize 3.52 }& {\footnotesize 3.76 }\\
 &{\footnotesize 158.0 }&{\footnotesize 37.26 }& {\footnotesize 1323.7 }&{\footnotesize 3.78 }& {\footnotesize 3.79 }\\
 &{\footnotesize 178.0 }&{\footnotesize 32.01 }& {\footnotesize 1527.1 }&{\footnotesize 3.86 }& {\footnotesize 3.79 }\\
 &{\footnotesize 182.0 }&{\footnotesize 31.02 }& {\footnotesize 1573.2 }&{\footnotesize 3.59 }& {\footnotesize 3.78 }\\
 &{\footnotesize 219.0 }&{\footnotesize 24.76 }& {\footnotesize 1947.5 }&{\footnotesize 3.54 }& {\footnotesize 3.60 }\\
 &{\footnotesize 239.0 }&{\footnotesize 22.14 }& {\footnotesize 2165.5 }&{\footnotesize 4.01 }& {\footnotesize 3.45 }\\
 &{\footnotesize 277.0 }&{\footnotesize 18.27 }& {\footnotesize 2596.6 }&{\footnotesize 3.21 }& {\footnotesize 3.16 }\\
 &{\footnotesize 287.0 }&{\footnotesize 17.50 }& {\footnotesize 2704.5 }&{\footnotesize 3.72 }& {\footnotesize 3.09 }\\
 &{\footnotesize 373.0 }&{\footnotesize 12.49 }& {\footnotesize 3708.2 }&{\footnotesize 2.98 }& {\footnotesize 2.51 }\\
 &{\footnotesize 528.0 }&{\footnotesize 7.72 }& {\footnotesize 5723.7 }&{\footnotesize 2.07 }& {\footnotesize 1.97 }\\
\hline 
\caption{Values of track-segment energy, $z^{*2}/\beta^2$ and TST-calculated RBE. Experimental values of LET are taken from Table 2, and RBE$_{exp}$ are read from Figure 3 in the paper of \cite{furusawa2000}.}
\label{tab.rbeV79}
\end{longtable}
\end{center}

Additionally, comparison between experimental values of RBE and those predicted by TST is shown in Table \ref{tab.rbeV79}, where the ion source, LET values for ions as given by Furusawa et al. followed by our LET-calculated track-segment ion energy and respective values of $z^{*2}/\beta^2$ are also listed.

\begin{figure}[!ht]
\begin{center}
\includegraphics[width=1\textwidth]{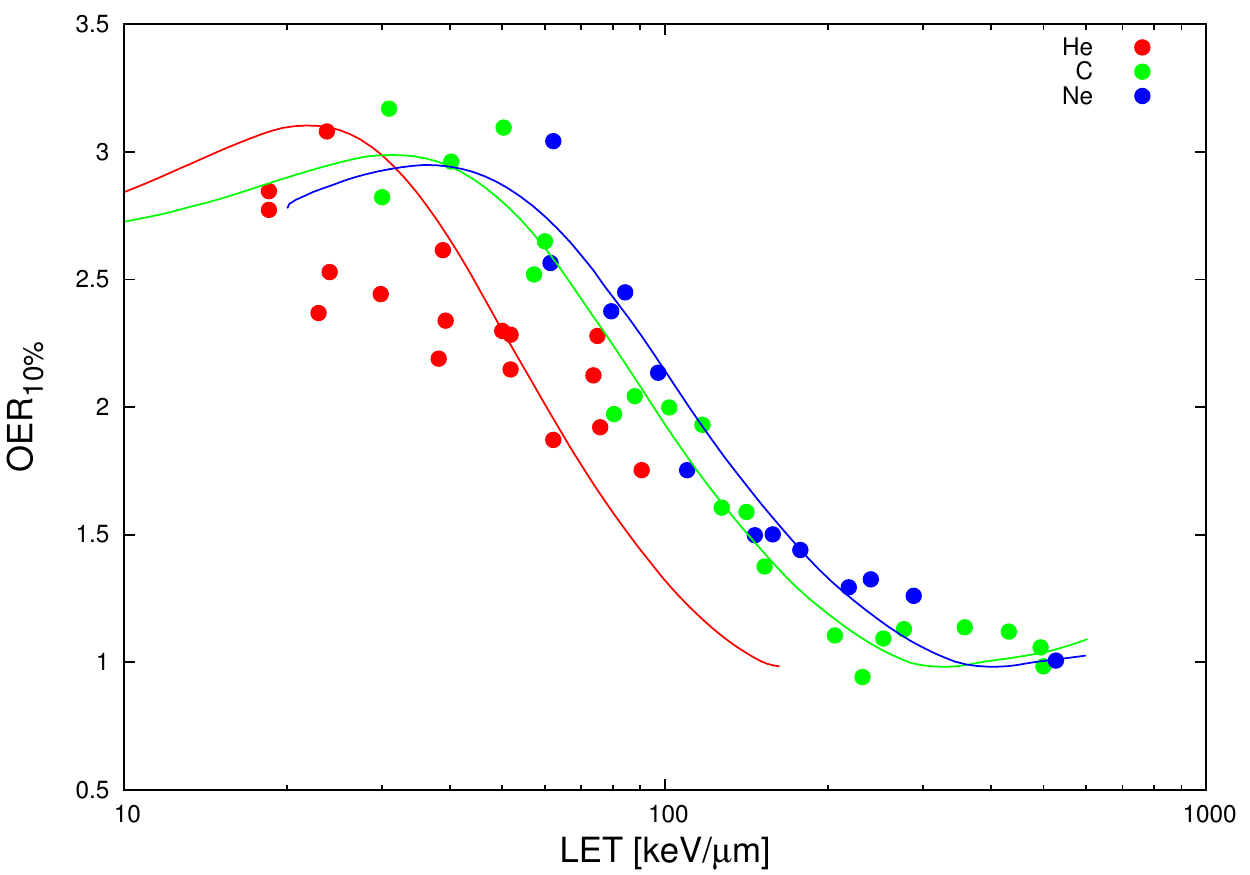}
\end{center}
\caption{OER vs. LET for cell killing of V79, at $10\%$ survival. Full lines represent TST calculations, for He (red line), C (green line), Ne (blue line) ions. For TST parameter values, see Table \ref{tab.parametersV79}. Data points are from \cite{furusawa2000}.}
\label{fig.oer}
\end{figure}

Since we established model parameters representing survival of Chinese hamster cells V79 irradiated in aerobic as well as in hypoxic conditions (see Table \ref{tab.parametersV79}), we are able to calculate the Oxygen Enhancement Ratio, which describes the difference in the survival of cells irradiated under these two conditions. For a specific ion LET value we calculated the OER at $10\%$ survival by comparing the doses delivered under hypoxic cell conditions to the dose delivered under oxygenated conditions, both resulting in $10\%$ survival, eq.(\ref{eq.oer}). Model-predicted values of OER and those estimated by \cite{furusawa2000} are presented in Fig. \ref{fig.oer}. For carbon and neon ion beams, good agreement of model predictions with experiment was found. TST-calculated values of OER slightly increase with increasing LET until they reach a maximum value and next gradually decrease around $50 \;keV/\mu m$, reaching a value of less than $2$ around $100\; keV/\mu m$, and then of approximately $1$ in the very high-LET region. We note that although the calculations of RBE made for V79 cells irradiated with helium ions are overestimated, model predictions of OER values do not demonstrate this tendency. The explanation for this behaviour is that while the TST-calculated survival of Chinese hamster cells after helium irradiation is underestimated (because TST overestimates the values of RBE) in aerobic as well in hypoxic conditions, the ratio of these two overestimated values may still be valid.

\chapter[Summary and Conclusions]{Summary and Conclusions}

Let us first briefly discuss the main results obtained in this work. Its principal objective was to recapitulate Katz's cellular Track Structure Theory (TST) in order to apply it as the 'radiobiological engine' in a future radiotherapy planning system for carbon radiotherapy, with potential inverse-planning capability. The main advantages of Katz's TST are its predictive power with regard to \emph{in vitro} cellular survival (or RBE) after ion irradiation and its simple analytic formulation which make it extremely efficient in terms of computer coding, compared with calculations based on Monte-Carlo simulations. The Katz model relies extensively on scaling principles, so particular attention was paid in this work to analyse the scaling properties of the key elements of this model - the 'point-target' radial distribution of dose formulae (in Section \ref{ch. radialdose}) or the action cross-section (in Section \ref{ch.approximation}) and their effect on the validity of the four phenomenological track structure parameters characterising a cell line in culture. Based on fulfillment of conditions which were specified in this work (in Section \ref{ch. doseenergycomparison}), on its scaling properties (as discussed in Section \ref{ch. averagedose}), and on analysis of experimental data concerning '1-hit' systems (in Section \ref{ch. crosssection}), the modified formula of \cite{zhang1985}, eq.(\ref{eq.rddzhang2}), was selected as the basic 'point-target' dose distribution formula to be used in all further model calculations. Analytic approximation of the cross-section in the 'track-width regime' (discussed in Section \ref{ch.trackwidth} and in Appendix \ref{ch. appendixc}) was completely reconstructed, as it was never published in the original work of Robert Katz and his collaborators. The final formulation of Katz's updated cellular Track Structure Theory is given in Section \ref{ch. survival}. This newly re-formulated Track Structure model was then applied to published sets of experimental data concerning \emph{in vitro} survival of normal human skin fibroblasts (Section \ref{ch. tsuruoka}) and V79 Chinese hamster cells in aerated and anoxic conditions (Section \ref{ch. furusawa}), after ion irradiation in order to obtain sets of model parameters representing these cell lines. Once established for a given cell line in given conditions, application of the four cellular parameters in model calculations enables quantitative predictions to be made of survival curves of these cells after irradiation by any ion of any energy, as shown, e.g., in Fig. \ref{fig.matrix} for normal human skin fibroblasts irradiated by carbon and iron ions of widely ranging energies. Moreover, TST predictions of the survival curve of a given cell line (in aerated or anoxic conditions) can be made after their irradiation by a mixed field of primary ions, secondary charged particles and associated $\gamma$- or $X$-rays, as encountered in ion radiotherapy conditions. However, the physical input for such calculations must consist of depth distributions of energy-fluence spectra of all these ions along the beam range, rather than distributions of averaged dose vs. depth, which are usually calculated. 

Some issues requiring further study have been raised by this work. In its present form, TST appears to underestimate the response (or overestimate the RBE) after light ion irradiation (protons or He ions), as shown in Fig. \ref{fig.alanina}, panel B, for the '1-hit' alanine detector, and in Fig. \ref{fig.tsuruokarbeletlekkiejony} for normal skin fibroblasts, or in Table \ref{tab.rbeV79} and Fig. \ref{fig.rbev79ox} for V79 cells. Two possible reasons for this discrepancy may be suggested: one is the likely change in the excitation and ionization contributions within the radial distributions of dose for light ions, as opposed to heavier ion species, the other the likely incorrect use of a 'dose-averaged' LET value (such as given by eq.(\ref{eq.doselet})) to represent 'effective' LET values for particles stopping in the detector. Another interesting issue raised by this work is the stated independence of $\sigma_0$ and $\kappa$ as fitting parameters of the cellular TST, despite their formal correlation through eq.(\ref{eq.kappa2}) and eq.(\ref{eq.sigmao}). This precludes the possibility of reducing the four parameters of the cellular model to three (see p. 77 in text), as could be expected from such the presence of such a correlation. Possibly, this may be connected with the manner in which the average radial dose distribution is calculated by averaging the 'point-target' dose distribution over the sensitive site of size $a_0$, whereby on radial integration this averaged RDD does not yield the correct value of ion LET. 

In the TST approach the concepts of the single-particle action cross-section and of the 'ion-kill' and 'gamma-kill' components of the survival probability (Section \ref{ch. survival}), intimately related to the multi-target description of the survival curve, eq.(\ref{eq.mtarget}), are applied to yield directly the predicted shape of the survival curve, thus obviating the need to calculate RBE-corrected 'biological dose' distributions. However, as stressed earlier, the required physical input consists of the ion energy-fluence spectra, and the sets of cellular parameters need to be pre-established from best-fits to survival curves measured after irradiation of this cell line by a sufficiently wide range of different ions of different energies.

The requirement of Katz's Track Structure Theory (TST) that energy-fluence ion spectra rather than dose distributions are needed as input from beam transport calculations illustrates the profound difference between TST and other approaches to radiobiological modelling of ion irradiation effects in cells. The most common approach to such modelling is to apply the concept of biologically weighted dose (Section \ref{ch. biologicaldose}) whereby this 'biological dose' is obtained as a product of 'physical' (i.e. absorbed) dose and the respective value of the radiobiological effectiveness (RBE). However, RBE - a deceptively simple concept - is a complicated function of several variables, especially for heavier ions, such as carbon. In particular, due to track structure effects, RBE is not a unique function of the ion's LET. The difficulties in correctly evaluating the local values of RBE along the beam depth are yet to be overcome by the LEM approach (Section \ref{ch. lem}) in which the survival curves are described by the linear-quadratic formula, eq.(\ref{eq.lq}). While, in principle, the $\alpha$-term can be quantitatively evaluated by LEM calculations, the prediction of the $\beta$-term relies on complex and rather unclear approximations. Fundamental questions concerning the LEM approach have also been raised, by \cite{bueve2009} and \cite{katz2003}, to which authors of the LEM have responded (\cite{scholz2004}).

In the modelling approach applied by the Japanese groups (Section \ref{ch. nirs}) the clinical experience from earlier fast neutron radiotherapy provided the basis for introducing a 'clinical RBE' factor (of about $3$) for the carbon beam, by which the independently calculated physical depth-dose distribution is divided. The linear-quadratic representation of survival curves is applied, whereby the survival curve for mixed - field radiation is calculated by suitably averaging alpha and beta terms found from irradiation of human salivary gland (HSG) cells \emph{in vitro} by mono-energetic ion beams, eq.(\ref{himacsfmixed}). The values of 'clinical RBE' are suitably modified by careful observation of the clinical results obtained. In principle, once the values of cellular parameters for HSG cells are established, track structure calculations performed retroactively both for the fast neutron beam and for therapy plans used by the Japanese groups, could clinically validate the proposed therapy planning system based on Katz's TST. 

Ion beam therapy is a rapidly developing field where the benefits of high-LET radiation (Bragg peak, enhancement of relative biological effectiveness - RBE, hypofractionation, oxygen effect) are weighted against the high cost of therapy and lack of generally established clinical treatment protocols. Development of a therapy planning system compatible with that presently under way at Heidelberg Ion Therapy Center (HIT) would allow cross-checking of some controversial issues, such as reporting ion radiotherapy procedures, application of appropriate RBE values or of representative Local Effect Model parameters for specific treatment localities, reporting late effects, etc. It is not yet quite clear whether dose-RBE or fluence-based approaches are more appropriate in ion beam radiotherapy. Application of Track Structure Theory-based biophysical modelling which requires the knowledge of energy-fluence spectra distributions along the beam depth rather than dose-depth distribution only, may lead to the emergence of novel concepts of ion therapy planning and optimization. Possibly, due to the possibility that 'biologically weighted dose' may not be additive, 'biological dose'-volume histograms (DVH) in ion therapy may need to be replaced by 'effect-volume histograms' or 'kill-volume histograms' (KVH), where 'effect' could represent the probability of cell killing over the target volume. As a result, the local distribution of cell survival or 'cell killing' could become the means of transferring the experience of conventional radiotherapy using photon beams to heavy ion beam therapy. This could lead to better understanding and accelerated development of ion radiotherapy, also using ions lighter than carbon, thanks to the predictive capability of Katz's TST. 

Let us now discuss the results of this work more systematically: in its first, introductory part (Chapter \ref{ch. introduction}) we discussed ion interactions, updating the calculation of ion LET to conform with present ICRU data (Section \ref{ch. ioninteraction} and Appendix \ref{ch. appendixa}), introduced the basic radiobiological concepts for later use (Section \ref{ch. xresponse}) and briefly discussed the present approaches to carbon beam therapy planning used by German and Japanese groups (Section \ref{ch. planningsystems}). 

In Chapter \ref{ch. results1} we recapitulated the three-parameter Track Structure Theory in order to analyse in more detail the scaling properties of several formulae of radial distribution of dose (RDD) successively developed within Katz's Track Structure Theory. We re-formulated all model expressions in terms of SI units instead of the CGS units used so far by the group of Robert Katz. Next, we studied how do different RRD formulations affect the integrity of the Katz model. Of the four RDD formulae studied we wished to select one that best fulfilled scaling conditions, to be later used in the re-formulated four-parameter cellular Track Structure Theory. For an ion of specified charge and energy, the 'point-target' radial distribution of $\delta$-ray dose, $D_{\delta}(r)$ was required: i) to reproduce experimentally measured radial distributions of dose; ii) when integrated over all radii, to yield the value of LET of the ion; iii) to be represented by a relatively simple analytical formula; and iv) to exhibit appropriate scaling, permitting model calculations to be rapidly performed over a wide range of ions of different charges and energies. We found the somewhat modified RDD formula of \cite{zhang1985}, eq.(\ref{eq.rddzhang2}), to exhibit the best scaling properties, enabling us later (in Chapter \ref{ch. results2}) to consistently use it in defining $\kappa$ as the fourth model fitting parameter for cellular m-target detectors. Finally, in Section \ref{ch. crosssection}, we compared predictions of the different RDD formulae against published values of inactivation cross sections of \emph{E. coli} B$_{s-1}$ spores and of average relative effectiveness of alanine after heavy ion bombardment. The \emph{E. coli} B$_{s-1}$ spores and alanine system, described as 1-hit detectors in TST, had widely differing values of model parameters: radiosensitivity, $D_0$ $(12.6\; Gy$ and $1.05 \cdot 10^5 Gy)$ and $a_0$ $(1 \cdot 10^{-7}m$ and $2 \cdot 10^{-9}m)$, respectively, offering a good test of the scaling concepts studied and confirming the choice of eq.(\ref{eq.rddzhang2}) as the 'point-target' RDD formula to be used exclusively in the revised cellular Track Structure model. 

In Chapter \ref{ch. results2}, we discussed the manner in which the factors $z^{*2}/\beta^2$, $z^{*2}/\kappa \beta^2$ and $a_0^2 \beta^2/z^{*2}$, rather than LET, may serve as scaling factors in the Track Structure model. Application of the selected radial dose distribution allowed us to eliminate repetitive calculations of double integrations by approximating the envelopes of values of action cross section using simple analytic formulae. Of particular significance is the re-capitulation of the cross section approximation algorithm in the 'track-width' regime, developed in this work and presented in detail in Appendix \ref{ch. appendixc}. Up to now, this approximation was never published, though a similar one was used in earlier publications of Katz's group. Based on our newly-developed algorithms, we performed model analysis of two extensive biological data sets representing the survival curves of cells irradiated by a range of heavy charged particles, published by the National Institute of Radiological Science (NIRS) groups in Japan. We evaluated a set of track structure parameters ($m$, $D_0$, $\sigma_0$ and $\kappa$) representing \emph{in vitro} survival of normal human skin fibroblasts (data by \cite{tsuruoka2005}) and two sets of model parameters representing the survival of Chinese hamster cells (data by \cite{furusawa2000}) irradiated under aerobic or hypoxic conditions. We compared TST - calculated survival curves and RBE-LET dependences with those evaluated experimentally for both cell lines. Additionally, in the case of Chinese hamster cells, we showed the TST-predicted OER-LET dependences for different ions and compared them with experiment. Based on model parameters evaluated for normal human skin fibroblasts we analysed the influence of model parameters on cellular TST predictions of RBE. We compared model predictions of survival curves for mono-energetic carbon beams and for mixed beams, consisting of the primary carbon beam and secondary lighter ions. The overall results of our TST analysis of the data set published by \cite{tsuruoka2005} and \cite{furusawa2000} have confirmed the basic track structure features generally observed in such data.

The biophysical model, to be applicable in the treatment planning system for carbon ions, has to fulfil some basic conditions: i) it has to provide calculation algorithms of cell survival or RBE dependences on factors such as particle type, energy, dose and cell or tissue type; ii) it has to provide the calculation of RBE in mixed radiation field of therapeutic ion beams, consisting of particles of different energies and atomic numbers with the accuracy necessary for therapy; and iii) its analytical formulation should be fast and robust in order to be applicable to massive calculations required for ion beam radiotherapy planning. Our analysis of Katz's Track Structure Theory indicates that this model allows the dependence of RBE to be calculated at any level of survival, on LET (or ion energy), particle type, cell type and degree of oxygenation. Unfortunately, model predictions of RBE for light ions are not yet satisfactory and require further work. Track Structure Theory provides the mathematical structure to deal with mixed radiation fields of therapeutic ion beams. The existence of scaling properties in Track Structure Theory made it possible to greatly simplify the algorithms of this model and greatly enhances the speed of model calculations, compared, e.g., with Monte Carlo simulations. Therefore, Katz's Track Structure Theory fulfils all the required conditions for being applicable in treatment planning systems, making it a strong competitor to the Local Effect Model. All studies presented in this work have been provided with future investigations in mind. The algorithm of the four-parameter cellular Track Structure Theory presented in this work can be integrated with the physical beam model to be yet developed, and should become an integral part of our proposed development version of a treatment planning system dedicated to carbon ion radiotherapy. Additionally, three sets of TST model parameters describing \emph{in vitro} survival of normal human skin fibroblasts and Chinese hamster cells irradiated by heavy charged particles under aerobic and hypoxic conditions will serve as input parameters in our future preliminary tests of this treatment planning system.

\appendix
\chapter[Algorithm for calculating ion LET]{Algorithm for calculating ion LET}
\label{ch. appendixa}
To calculate the values of LET$(Z,\frac{E}{A})$ for an ion, heavier than the proton, of mass number $A$, atomic number $Z$, relative speed $\beta$, and kinetic energy per nucleon $\frac{E}{A}$ we use the expression given by \cite{barkas1964}
\begin{equation}
\textrm{LET}(Z,\frac{E}{A}) = \textrm{LET}_{p}(E)\left( \frac{z^*(\beta)}{z^*_{p}(\beta)} \right)^2,
\label{eq.letionappendixa}
\end{equation}
where $z^*$ and $z^*_{p}$ are the effective charge numbers of the ion and proton given by eq.(\ref{eq.zeff}) and eq.(\ref{eq.zeffproton}) respectively, and LET$_p(E)$ is the stopping power of the proton of kinetic energy $E$. In this work, the calculation of the LET$_p(E)$ is based on \cite{icru1993} tables of the proton stopping power. Least square polynomials were fitted to the published values of stopping powers. LET$_{p}\;[MeV/cm]$ for protons in water presented in tables in \cite{icru1993} can then be parametrized as follows:
\begin{equation}
\textrm{LET}_{p}(E) = a_{0} + \sum_{j=1}^{4}a_{j}(E)^j,
\label{eq.letprotonappendixa}
\end{equation}
where the values of energy ranges $E$ are in $[MeV]$. The power expansion coefficients are given in Table \ref{tab.lettable}

In this work eq.(\ref{eq.letionappendixa}) was multiplied by an additional factor which gave better agreement between LET values calculated using this equation and those listed in ICRU Report 73 for ions of $3 \leq Z \leq 18$. An atomic number, $Z$-dependent factor was introduced for ions of atomic numbers $3 \leq Z \leq 18$ and energies $ 1.5 \cdot 10^{-2} MeV/n \leq E/A \leq 10 \;MeV/n$.

\begin{table}[!ht]
\begin{center}
\caption{Power expansion coefficients for LET$_p(E)\;[MeV/cm]$ for proton in water. $E$ is expressed in $MeV$.}
\begin{tabular}{|c|c|c|c|}

\hline
\footnotesize $a_j$ & \footnotesize $1.0\cdot 10^{-3} < E \leq 3.0\cdot 10^{-3}$ & \footnotesize $ 3.0\cdot 10^{-3} < E \leq 1.0\cdot 10^{-2}$ &  \footnotesize  $ 1.0\cdot 10^{-2} < E \leq 3.0\cdot 10^{-2}$ \\
\hline
$a_0$  & $1.29040 \cdot 10^{+2}$   &  $1.32323 \cdot 10^{+2}$  &  $1.97106 \cdot 10^{+2}$ \\
$a_1$  &  $5.11771 \cdot 10^{+4}$  &  $4.92143 \cdot 10^{+4}$  &  $3.09790 \cdot 10^{+4}$ \\
$a_2$  &  $-3.33412 \cdot 10^{+6}$ &  $-3.39801 \cdot 10^{+6}$ &  $-9,03697 \cdot 10^{+5}$\\
$a_3$  &  $-9.43396 \cdot 10^{+6}$ &  $2.02210 \cdot 10^{+8}$  &  $1.80525 \cdot 10^{+7}$ \\
$a_4$  &  $2.57862 \cdot 10^{+10}$ &  $-5.39773 \cdot 10^{+9}$ &  $-1.64848 \cdot 10^{+8}$ \\
\hline
\hline
\footnotesize $a_j$ & \footnotesize $5.0\cdot 10^{-2} < E \leq 7.0\cdot 10^{-2}$ & \footnotesize $  7.0\cdot 10^{-2} < E \leq 1.0\cdot 10^{-1}$ & \footnotesize $1.0\cdot 10^{0} < E \leq 3.0\cdot 10^{0}$ \\
\hline
$a_0$  &  $3.54871 \cdot 10^{+2}$ &  $4.69093 \cdot 10^{+2}$ &  $6.3748 \cdot 10^{+2}$ \\
$a_1$  &  $1.3858 \cdot 10^{+4}$&  $9.00167 \cdot 10^{+3}$ &  $4.64286 \cdot 10^{+3}$ \\
$a_2$  &  $-1.04857 \cdot 10^{+5}$ &  $-5.68095 \cdot 10^{+4}$ &  $-2.85714 \cdot 10^{+4}$ \\
$a_3$  & $0$ &  $0$ &  $0$ \\
$a_4$  & $0$ &  $0$ &  $0$ \\
\hline
\hline
\footnotesize $a_j$ &  \footnotesize  $ 1.0\cdot 10^{-1} < E \leq 3.0\cdot 10^{-1}$ &  \footnotesize $ 3.0\cdot 10^{-1}< E \leq 1.0\cdot 10^{0}$ &  \footnotesize $1.0\cdot 10^{0} < E \leq 3.0\cdot 10^{0}$\\
\hline
$a_0$  &  $7.89045 \cdot 10^{+2}$ &  $9.69567 \cdot 10^{+2}$ & $5.89476 \cdot 10^{+2}$\\
$a_1$  &  $2.92161 \cdot 10^{+3}$ &  $-2.02852 \cdot 10^{+3}$ & $-8.22929 \cdot 10^{+1}$\\
$a_2$  &  $-3.82929 \cdot 10^{+4}$ &  $2.57981 \cdot 10^{+3}$ &  $2.70563 \cdot 10^{+2}$\\
$a_3$  & $1.34528 \cdot 10^{+5}$ & $-1.73575 \cdot 10^{+3}$  &  $-6.81349 \cdot 10^{+1}$\\
$a_4$  &  $-1.60671 \cdot 10^{+5}$ & $4.75799 \cdot 10^{+2}$ &  $6.74312 \cdot 10^{0}$\\
\hline
\hline
\footnotesize $a_j$ & \footnotesize $3.0\cdot 10^{0} < E \leq 1.0\cdot 10^{+1}$ & \footnotesize  $ 1.0\cdot 10^{+1} < E \leq 3.0\cdot 10^{+1}$ & \footnotesize $ 3.0\cdot 10^{+1}< E \leq 1.0\cdot 10^{+2}$\\
\hline
$a_0$  &  $2.70546 \cdot 10^{+2}$  &  $1.15173 \cdot 10^{+2}$  &  $4.53811 \cdot 10^{+1}$\\
$a_1$  &  $-8.22929 \cdot 10^{+1}$ &  $-1.16621 \cdot 10^{+1}$ &  $-1.44762 \cdot 10^{0}$\\
$a_2$  &  $1.32205 \cdot 10^{+1}$  &  $6.13364 \cdot 10^{-1}$  &  $2.38217 \cdot 10^{-2}$\\
$a_3$  &  $-1.04636 \cdot 10^{0}$  & $-1.58255 \cdot 10^{-2}$  & $-1.90613 \cdot 10^{-4}$\\
$a_4$  &  $3.22483 \cdot 10^{-2}$  &  $1.5892 \cdot 10^{-4}$   & $5.90886 \cdot 10^{-7}$\\
\hline
\hline
\footnotesize $a_j$ & \footnotesize $1.0\cdot 10^{+2} < E \leq 3.0\cdot 10^{+2}$ & \footnotesize $3.0\cdot 10^{+2} < E \leq 1.0\cdot 10^{+3}$ &  \footnotesize  $ 1.0\cdot 10^{+3} < E \leq 3.0\cdot 10^{+3}$ \\
\hline
$a_0$  &  $1.76264 \cdot 10^{+1}$  &  $7.04685 \cdot 10^{0}$   &  $3.21900 \cdot 10^{0}$\\
$a_1$  &  $-1.74720 \cdot 10^{-1}$ &  $-1.95139 \cdot 10^{-2}$ &  $-1.80100 \cdot 10^{-3}$ \\
$a_2$  &  $9.29149 \cdot 10^{-4}$  &  $3.3057 \cdot 10^{-5}$   &  $ 1.04100 \cdot 10^{-6}$ \\
$a_3$  &  $-2.40165 \cdot 10^{-6}$ &  $-2.66666 \cdot 10^{-8}$ & $-2.76000 \cdot 10^{-10}$ \\
$a_4$  &  $2.41156 \cdot 10^{-9}$  &  $8.29057 \cdot 10^{-12}$ &  $2.80000 \cdot 10^{-14}$\\
\hline
\hline
\footnotesize $a_j$ & \footnotesize $ 3.0\cdot 10^{+3}< E \leq 1.0\cdot 10^{+4}$\\
\cline{1-2}
$a_0$  &  $2.07404 \cdot 10^{0}$\\
$a_1$  &  $-6.82581 \cdot 10^{-5}$\\
$a_2$  &  $1.96193 \cdot 10^{-8}$\\
$a_3$  & $-1.85227 \cdot 10^{-12}$\\
$a_4$  & $6.25000 \cdot 10^{-17}$\\
\cline{1-2}
\end{tabular}
\label{tab.lettable}
\end{center}
\end{table}

\chapter[Calculation of the $\phi$-function]{Calculation of the $\phi$-function}
\label{ch. appendixb}

To calculate the average dose deposited in a target of radius $a_0$ placed at a distance $t$ from the path of the ion, $\bar{D}(t)$, we divide the volume of the target into sections of the same distance $r$ from the ion's path. The length of the segment of the annular cylindrical ring at which the point dose distribution has a constant value is described by the geometrical function - $\phi(a_0,t,r)$ (\cite{waligorski1988}). From analytical geometry:

\begin{displaymath}
\left\{ \begin{array}{ll}
\phi(a_0,t,r) = 2 \cdot r \cdot \textrm{arctg}(u) \;\quad\quad\quad\quad for\;\; u>0\\
\phi(a_0,t,r) = 0 \quad\quad\quad\quad\quad\quad\quad\quad\qquad for\;\; u=0\\
\phi(a_0,t,r) =  2 \cdot r \cdot (\pi -\textrm{arctg}(-u)) \quad for\;\; u<0
\end{array} \right.
\end{displaymath}
where:
\begin{displaymath}
z=\frac{r^2-a^2_0+t^2}{2t}
\end{displaymath}
and
\begin{displaymath}
tg(u)=\frac{(r^2-z^2)^{1/2}}{z}
\end{displaymath}

\begin{figure}[!ht]
\begin{center}
\includegraphics[width=0.9\textwidth]{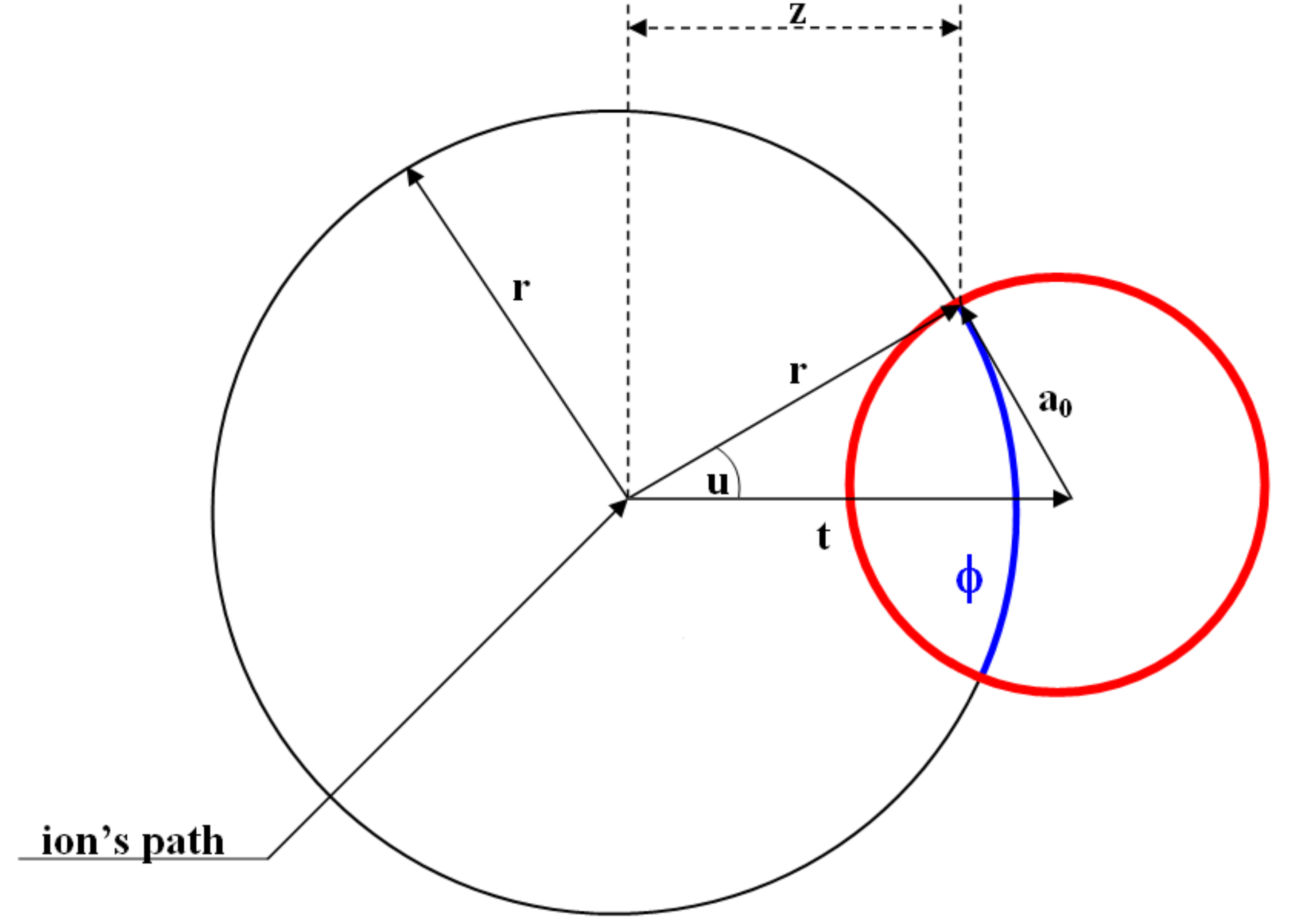}
\end{center}
\caption{Meaning of symbols used in the calculation of the function $\phi(a_0,t,r)$.}
\label{fig.geometry1}
\end{figure}

The meaning of above symbols is explained graphically in Fig. \ref{fig.geometry1}. A similar procedure for calculating $\phi(a_0,t,r)$ was also described by \cite{edmund2007}.

\begin{figure}[!ht]
\begin{center}
\includegraphics[width=0.9\textwidth]{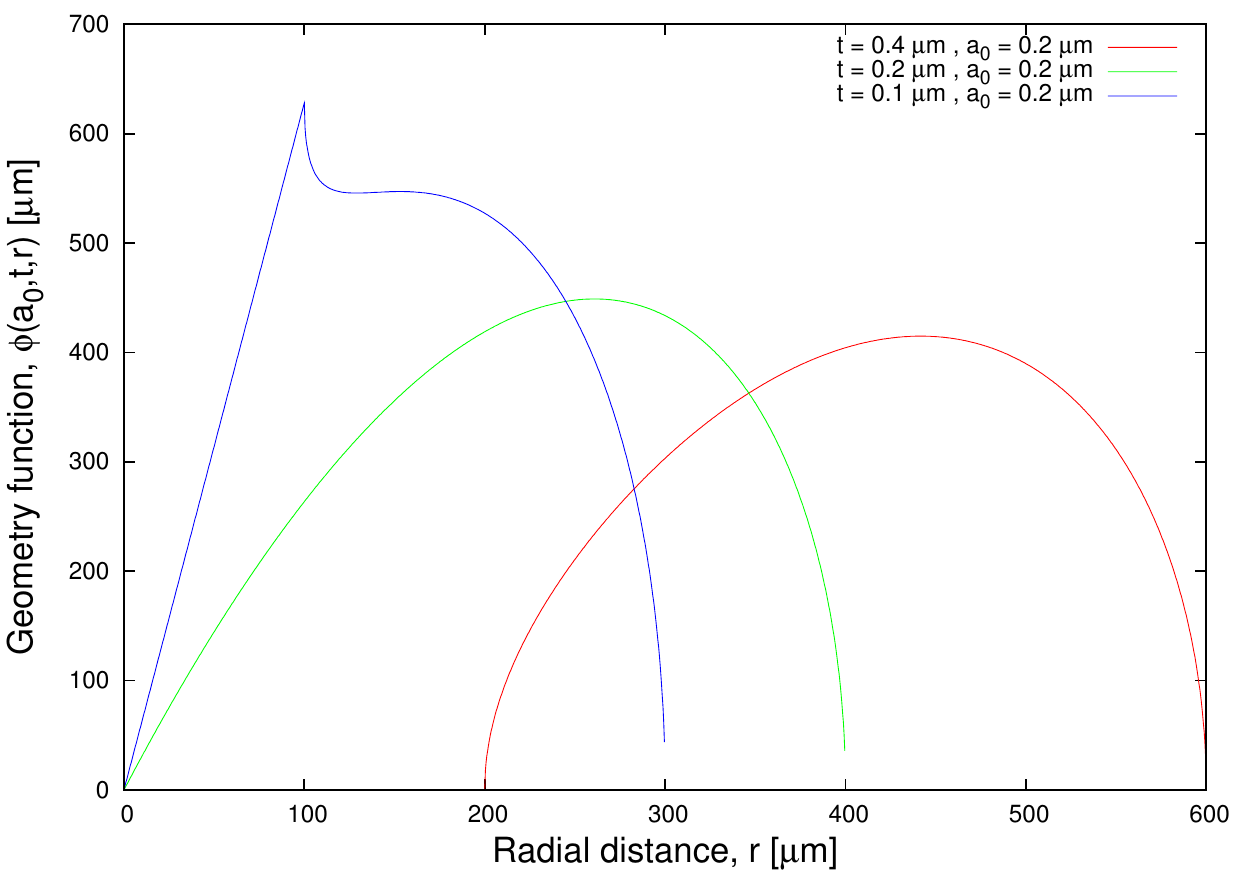}
\end{center}
\caption{Geometry function $\phi(a_0,t,r)$ calculated for three conditions, when $t>a_0$, $t=a_0$, or $t<a_0$.}
\label{fig.geometry2}
\end{figure}

The variation of the geometry function for different cases is presented in Fig. \ref{fig.geometry2}.

\chapter[Numerical approximation of the 'track-width' regime]{Numerical approximation of the 'track-width' regime}
\label{ch. appendixc}

To determine new coefficients present in the formulas used in the original program developed by Robert Katz's group, we initially calculated envelopes of $\sigma/\sigma_0$ as a function of $z^{*2}/\kappa\beta^{2}$ for different values of $m$ ranging between $1.5 \leq m \leq 3.5$. The considered formulae were of the form:

\begin{equation}
\frac{\sigma}{\sigma_{0}} = \left( \frac{Y_A-Y_B}{X_A-X_B}\left(\frac{z^{*2}}{\kappa\beta^{2}}-X_A \right) \right) + Y_A \quad \textrm{for}\quad X_A < z^{*2}/\kappa\beta^{2} \leq X_B ,
\label{eq.sigmaapprox2a}
\end{equation}
and
\begin{equation}
\frac{\sigma}{\sigma_{0}} = \frac{Y_B \cdot c_2}{1-e^{-\left( X_B \cdot c_1 / \frac{z^{*2}}{\kappa\beta^{2}} \right) }} \quad\quad\;\; \textrm{for}\quad z^{*2}/\kappa\beta^{2} \geq X_B ,
\label{eq.sigmaapprox3a}
\end{equation}
where $X_A$, $X_B$, $Y_A$ and $Y_B$ are coordinates of two points, $A(X_A,Y_A$) and $B(X_B,Y_B)$, lying directly on the envelope, and fixed for a given $m$. These points were chosen to ensure a continuous transition between functions approximating the envelope in the 'grain-count' regime, eq.(\ref{eq.sigmaapprox1}), and in the 'track-width' regime, as given by eq.(\ref{eq.sigmaapprox2a}) and eq.(\ref{eq.sigmaapprox3a}). Determination of the position of $A(X_A,Y_A$) was quite simple, because it is a point of transition between 'grain-count' and 'track-width' regimes, where $\sigma/\sigma_0 = 0.98$. Thus, from eq.(\ref{eq.sigmaapprox1}) we find:
\begin{equation}
X_A = -\log(1-0.98^{\frac{1}{m}}),
\label{eq.xa}
\end{equation}
and
\begin{equation}
Y_A = 0.98.
\label{eq.ya}
\end{equation}
The value of the constant $c_1$ in eq.(\ref{eq.sigmaapprox3a}) was determined by a $\chi^2 $ fitting procedure of eq.(\ref{eq.sigmaapprox2a}) and eq.(\ref{eq.sigmaapprox3a}) simultaneously to all envelopes determined for $1.5 \leq m \leq 3.5$ over the region of 'track-width' regime. The value of the constant $c_2$ results form the continuity condition at the point of transition, $B(X_B,Y_B)$, between eq.(\ref{eq.sigmaapprox2a}) and eq.(\ref{eq.sigmaapprox3a}), and can be described by a function of $c_1$. At the point of transition $z^{*2}/\kappa\beta^{2} = X_B$. Comparing the right hand sides of eq.(\ref{eq.sigmaapprox2a}) and eq.(\ref{eq.sigmaapprox3a}) one obtains the condition for $c_2$:
\begin{equation}
c_2=1-e^{-c_1}.
\label{eq.c2}
\end{equation}

\begin{figure}[!ht]
\begin{center}
\includegraphics[width=1.0\textwidth]{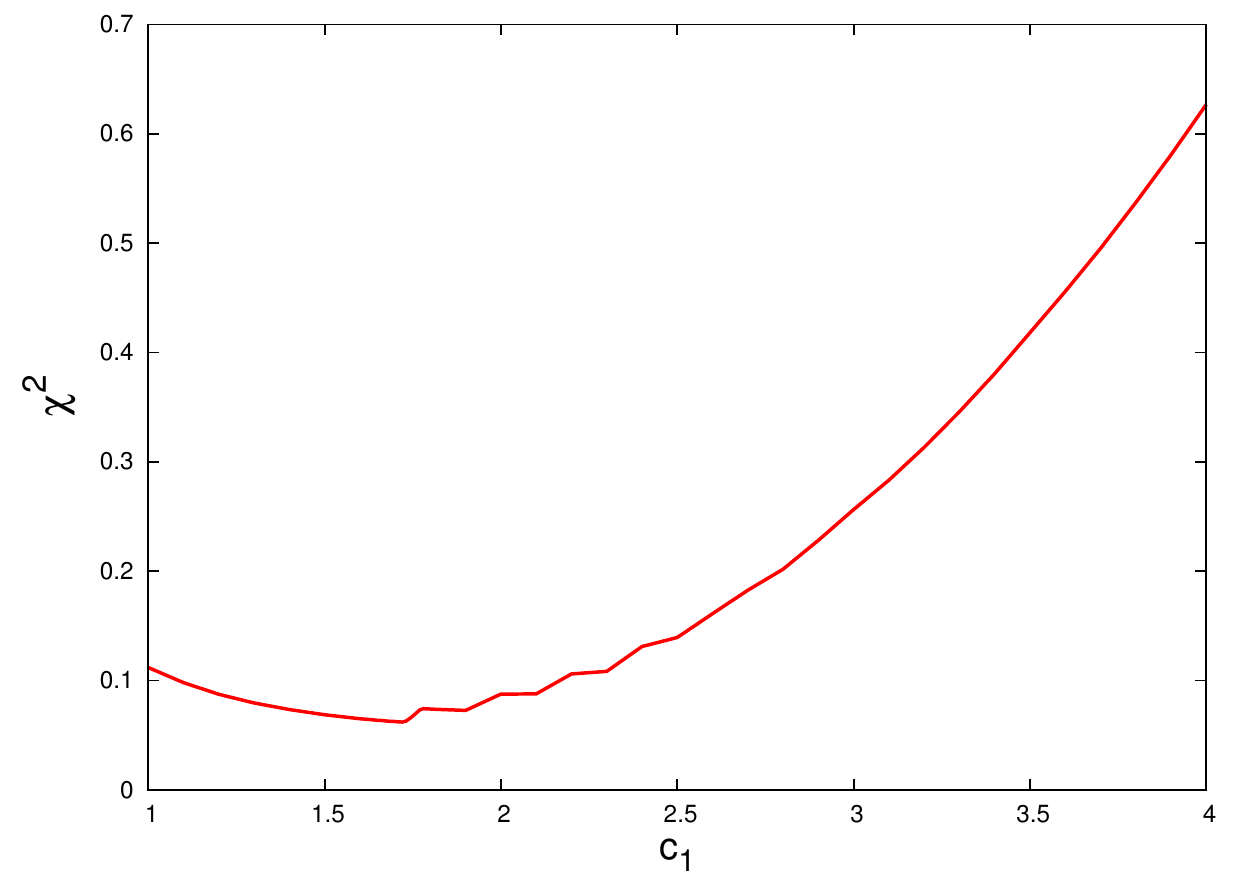}
\end{center}
\caption{ Values of $\chi^2$, eq.(\ref{eq.chi2}), calculated for $1.0 \leq c_1 \leq 4.0 $. }
\label{fig.chisqr2a}
\end{figure}

Neither $c_1$ nor $c_2$ depend on $m$. The coordinates of the point $B(X_B,Y_B)$ together with the value of $c_2$ depend on the choice of $c_1$. Additionally $X_B$, $Y_B$ are characteristic for particular values of $m$. For a given $c_1$ there is only one set of $X_B$, $Y_B$ which gives the best conformity of eq.(\ref{eq.sigmaapprox2a}) and eq.(\ref{eq.sigmaapprox3a}) with the envelope in the 'track-width' regime. In order to find such a value of $c_1$ and the accompanying $X_B$, $Y_B$ and $c_2$ for which eq.(\ref{eq.sigmaapprox2a}) and eq.(\ref{eq.sigmaapprox3a}) will give the best compatibility with the envelope, we defined the coefficient:
\begin{equation}
\chi^{2} = \sum_{m=1.5}^{m=3.5} \; \sum_{\frac{z^{*2}}{\kappa\beta^{2}}=X_A}^{\frac{z^{*2}}{\kappa\beta^{2}}=10^4} \left( \log\left(\frac{\sigma}{\sigma_{0}} \right)_{approx} - \log \left(\frac{\sigma}{\sigma_{0}} \right)_{envelope} \right)^{2},
\label{eq.chi2}
\end{equation}
which was used to estimate the deviation between the approximation, $(\sigma/\sigma_{0})_{approx}$, and the proper values of the envelope, $(\sigma/\sigma_{0})_{envelope}$. Since the 'track-width' regime covers a few orders of magnitude of $\sigma/\sigma_{0}$, to ensure the most accurate fit, $\chi^2$ was based on the difference of logarithmic values of $\sigma/\sigma_{0}$. For a given value of $c_1$, from the points contributed to envelopes calculated for different $m$, we chose a set of $B(X_B,Y_B)$, which gave the best agreement of eq.(\ref{eq.sigmaapprox2a}) and eq.(\ref{eq.sigmaapprox3a}) with these envelopes via the $\chi^2$ minimum. This procedure was repeated for a wide range of $c_1$. In Fig. \ref{fig.chisqr2a} we present results only for $1.0 \leq c_1 \leq 4.0 $, for which we found the lowest values of $\chi^2$.

\begin{figure}[!ht]
\begin{center}
\includegraphics[width=1.0\textwidth]{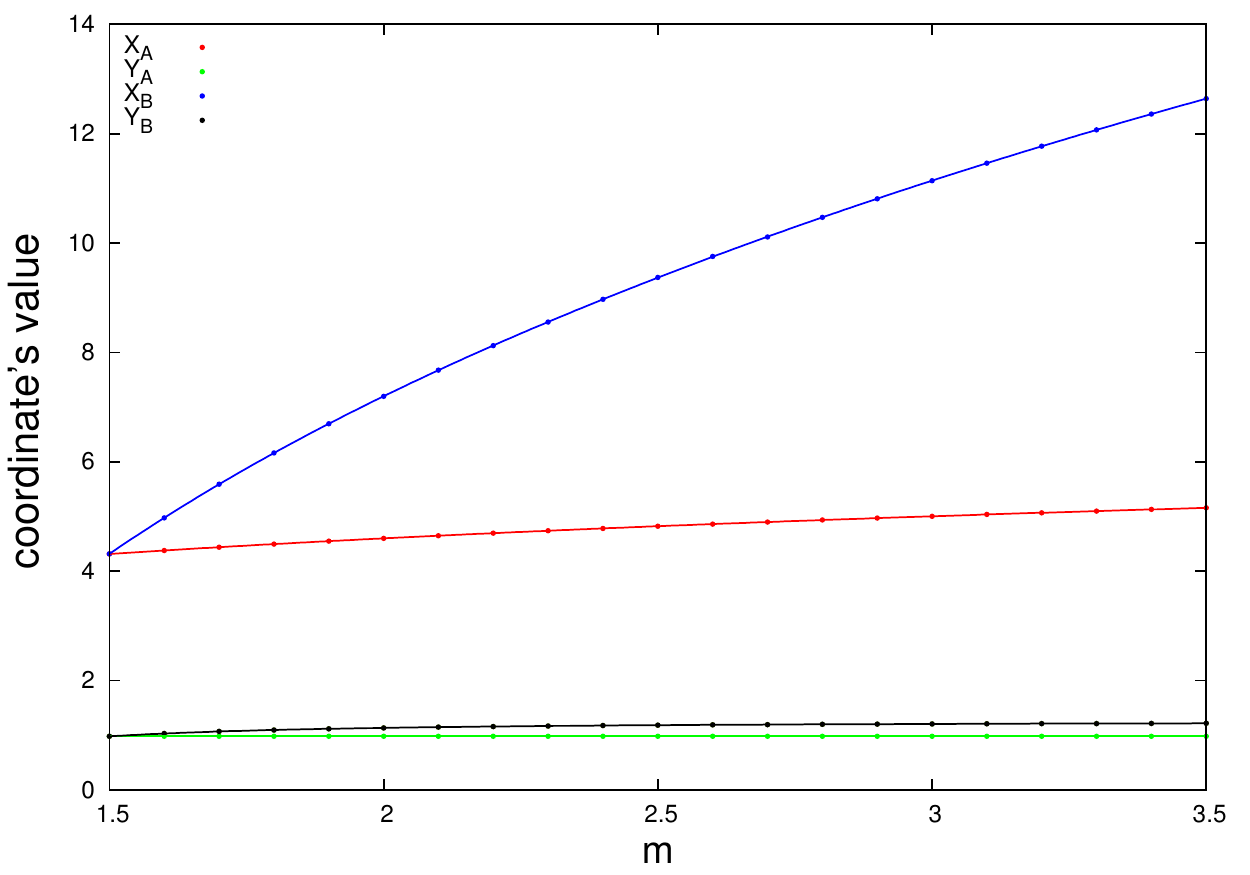}
\end{center}
\caption{Values of the coordinates of $X_A$ (red line), $Y_A$ (green line), $X_B$ (blue line) and $Y_B$ (black line) determined for $1.5 \leq m \leq 3.5$, as described by eq.(\ref{eq.xa}), eq.(\ref{eq.ya}), eq.(\ref{eq.xb}) and eq.(\ref{eq.yb}), respectively. }
\label{fig.xaya}
\end{figure}

Next, we compared the values of $\chi^2$ and we chose $c_1$ with the lowest value of $\chi^2$, namely $c_1 = 1.72$, and we applied it to eq.(\ref{eq.sigmaapprox3a}) in our further calculation. Considering eq.(\ref{eq.c2}) one can calculate that the constant $c_2 = 0.8209$, and eq.(\ref{eq.sigmaapprox3a}) takes the form:
\begin{equation}
\frac{\sigma}{\sigma_{0}} = \frac{Y_B \cdot 0.8209}{1-e^{-\left( X_B \cdot 1.72/ \frac{z^{*2}}{\kappa\beta^{2}} \right) }},
\end{equation}
as introduced in section \ref{ch.approximation}. Respective values of $X_B$ and $Y_B$ characteristic for each $m$, found in our $\chi^2$ optimization, are presented in Fig. \ref{fig.xaya}, and can be described by following functions:
\begin{equation}
X_{B} = \exp \left(g_{0} + \sum_{i=1}^{5}g_{i}\cdot \ln(m)^{i} \right),
\label{eq.xb}
\end{equation}
and
\begin{equation}
Y_{B} = \exp \left(h_{0} + \sum_{i=1}^{5}h_{i}\cdot \ln(m)^{i} \right).
\label{eq.yb}
\end{equation}
The values of power expansion coefficients $g_i$ and $h_i$ are given in Table \ref{tab.gihi}. It should be emphasised that the chosen value of $c_1 = 1.72$ together with accompanying $c_2$, $X_B$ and $Y_B$ values have been determined only for envelopes approximating values of $\sigma/\sigma_{0}$ calculated using the formula of \cite{zhang1985} with $I = 0\;eV$, and are applicable only for $1.5 \leq m \leq 3.5$.
\begin{table}[!ht]
\begin{center}
\caption{Values of coefficients $g_{j}$ and $h_{j}$.}
\begin{tabular}{c c | c c}
\hline
$i$ & $g_i$ & $i$ & $h_i$   \\
\hline
$0$  & -0.223198 & $0$  & -1.03134  \\
$1$  & 6.79973 & $1$  & 5.06363   \\
$2$  & -9.08647 & $2$  & -9.39776  \\
$3$  & 7.62573  &  $3$  & 9.3199 \\
$4$  & -3.42814 & $4$  & -4.75329  \\
$5$  & 0.632902 & $5$  & 0.97801 \\
\hline
\end{tabular}
\label{tab.gihi}
\end{center}
\end{table}

\listoffigures
\listoftables

\addcontentsline{toc}{chapter}{Bibliography}
\bibliographystyle{apa-good}
\bibliography{biblioteka}{}

\end{document}